\documentclass[aip,jmp,reprint,amsmath,amssymb,showpacs,superscriptaddress,footinbib]{revtex4-1}
\usepackage{epsfig,slashed,color,bm,natbib}
\def\vect#1{\mbox{\boldmath $#1$}}
\def\hvect#1{{\hat{\mbox{\boldmath $#1$}}}}
\def\matri#1{{\mbox{\sf  #1}}}

\def\pd#1#2{\frac{\partial #1}{\partial #2}}
\def\bpd#1#2{\frac{\bar{\partial} #1}{\partial #2}}

\def\nd#1#2{{\frac{d #1}{d #2}}}

\def\tfrac#1#2{{\textstyle\frac{#1}{#2}}}

\newcommand{\vpsi}{\varphi}
\def\christoffel#1#2#3{{^0}{\Gamma^{#1}}_{#2 #3}}
\def\zero#1{{^0}\!#1}
\def\czero#1{{^0}#1}

\newcommand{\comma}{, }
\begin{document}
\title{Scale-invariant gauge theories of gravity: theoretical foundations}
\author{A.N.~\surname{Lasenby}}
\email{a.n.lasenby@mrao.cam.ac.uk}
\affiliation{Astrophysics Group, Cavendish Laboratory, JJ Thomson Avenue,
Cambridge CB3 0HE\comma UK}
\affiliation{Kavli Institute for Cosmology, Madingley Road, Cambridge
  CB3 0HA, UK}
\author{M.P.~\surname{Hobson}}
\email{mph@mrao.cam.ac.uk}
\affiliation{Astrophysics Group, Cavendish Laboratory, JJ Thomson Avenue,
Cambridge CB3 0HE\comma UK}
\date{30 August 2016}

\begin{abstract}
We consider the construction of gauge theories of gravity, focussing
in particular on the extension of local Poincar\'e invariance to
include invariance under local changes of scale.  We work exclusively
in terms of finite transformations, which allow for a more transparent
interpretation of such theories in terms of gauge fields in Minkowski
spacetime. Our approach therefore differs from the usual geometrical
description of locally scale-invariant Poincar\'e gauge theory (PGT)
and Weyl gauge theory (WGT) in terms of Riemann--Cartan and
Weyl--Cartan spacetimes, respectively. In particular, we reconsider
the interpretation of the Einstein gauge and also the equations of
motion of matter fields and test particles in these theories. Inspired
by the observation that the PGT and WGT matter actions for the Dirac
field and electromagnetic field have more general invariance
properties than those imposed by construction, we go on to present a
novel alternative to WGT by considering an `extended' form for the
transformation law of the rotational gauge field under local
dilations, which includes its `normal' transformation law in WGT as a
special case. The resulting `extended' Weyl gauge theory (eWGT) has a
number of interesting features that we describe in detail. In
particular, we present a new scale-invariant gauge theory of gravity
that accommodates ordinary matter and is defined by the most general
parity-invariant eWGT Lagrangian that is at most quadratic in the eWGT
field strengths, and we derive its field equations. We also consider the
construction of PGTs that are invariant under local dilations assuming
either the `normal' or `extended' transformation law for the
rotational gauge field, but show that they are special cases of WGT
and eWGT, respectively.
\end{abstract}


\maketitle

\section{Introduction}
\label{sec:intro}

In providing a unified framework for the description of fundamental
interactions, the gauge principle has proved very successful. The
electromagnetic, weak and strong interactions are all described by gauge
theories, and so it is natural to consider the gauge group description
of gravitation.  In the gravitational case, however, in place of
gauging an internal compact semi-simple Lie symmetry group, one must
apply the gauge principle to spacetime symmetries, which act on
spacetime coordinates as well as on the dynamical fields.

All experiments to date show that the non-gravitational fundamental
interactions are consistent with the underlying spacetime symmetry of
the Poincar\'e group, and so in attempting to develop a gauge approach
to gravitation it is most natural to begin by constructing a physical
theory invariant under local Poincar\'e transformations.  As in the
gauging of internal symmetries, this process requires the introduction
of new, compensating gauge fields, which represent
gravitational interactions.

Following an initial attempt by Utiyama\cite{utiyama56}, the idea of
gauging the Poincar\'e group was fully developed by
Kibble\cite{kibble61} (and also considered by Sciama\cite{sciama64}).
The physical model envisaged in Kibble's approach is an underlying
Minkowski spacetime in which a matter field (or fields) $\vpsi$ with
energy-momentum and, in general, spin-angular-momentum is distributed
continuously. The field dynamics are described by a matter action
$S_{\rm M} = \int L_{\rm M}(\vpsi,\partial_\mu \vpsi)\,d^4x$ that is
invariant under global Poincar\'e transformations.  By then demanding
the matter action be invariant with respect to (infinitesimal,
passively interpreted) local Poincar\'e transformations, in which the
ten Poincar\'e group parameters become arbitrary functions of
position, one is led to the introduction of the new field variables
${h_a}^{\mu}$ and ${A^{ab}}_{\mu}$, which are interpreted as
gravitational gauge fields corresponding, respectively, to the
translational and (Lorentz) rotational parts of the Poincar\'e
group. To construct a matter action that is invariant under local
Poincar\'e transformations, the gravitational gauge fields
${h_a}^{\mu}$ and ${A^{ab}}_{\mu}$ are used to assemble a covariant
derivative ${\cal D}_a\vpsi$, and the matter action in the presence of
gravity is typically obtained by the minimal coupling procedure of
replacing partial derivatives in the special-relativistic matter
Lagrangian by covariant ones, to obtain $S_{\rm M} = \int h^{-1}
L_{\rm M}(\vpsi,{\cal D}_a\vpsi)\,d^4x$ (where the factor containing
$h \equiv \mbox{det}({h_a}^\mu)$ is required to make the integrand a
scalar density rather than a scalar). As for any gauge theory, the
form of the covariant derivative defines its structure, and the result
in this case is known as Poincar\'e gauge theory (PGT). Kibble's
approach improved considerably on the earlier work of Utiyama, who
introduced the fields ${A^{ab}}_\mu$ by localising just the Lorentz
symmetry, while the ${h_a}^\mu$ were treated as given functions of
spacetime position (although at a later stage these functions were
regarded as dynamical variables).

In addition to the matter action, the total action must also contain
terms describing the dynamics of the free gravitational gauge fields.
Indeed, the choice of these terms defines the precise form of
Poincar\'e gauge theory under consideration.  Following the normal
procedure used in gauging internal symmetries, Kibble first
constructed covariant field strength tensors for the gauge fields by
commuting covariant derivatives, i.e. by considering $[{\cal
    D}_a,{\cal D}_b]\vpsi$. This procedure yields two field strength
tensors: ${{\cal R}^{ab}}_{cd}(h,A,\partial A)$ corresponding to the
gauge field ${A^{ab}}_\mu$ arising from the (Lorentz) rotational part
of the Poincar\'e group; and ${{\cal T}^a}_{bc}(h,\partial h, A)$
corresponding to the gauge field ${h_a}^\mu$ arising from the
translational part, where (suppressing indices for brevity) we have
indicated the functional dependence of each field strength tensor on
the gauge fields and their derivatives\cite{footnote1}. It is worth noting that
each field strength tensor is linear in the derivatives of the gauge
field to which it corresponds. The free gravitational action then
takes the form $S_{\rm G} = \int h^{-1} L_{\rm G}({{\cal
    R}^{ab}}_{cd},{{\cal T}^a}_{bc})\,d^4x$, where the Lagrangian
$L_{\rm G}$ is a scalar depending on the two field strengths.  The
total action is taken as the sum of the matter and gravitational
actions, and variation of the total action with respect to the gauge
fields ${h_a}^{\mu}$ and ${A^{ab}}_{\mu}$ leads to two coupled
gravitational field equations in which the energy-momentum
${\tau^k}_\mu\equiv \delta{\cal L}_{\rm M}/\delta {h_a}^\mu$ and
spin-angular-momentum ${\sigma_{ab}}^\mu\equiv \delta{\cal L}_{\rm
  M}/\delta {A^{ab}}_\mu$ of the matter field act as sources, where
${\cal L}_{\rm M}\equiv h^{-1}L_{\rm M}(\vpsi,{\cal D}_a\vpsi)$. In
contrast to standard Yang--Mills gauge theory, in PGT one can
construct an invariant for use in $L_{\rm G}$ that is linear in the
derivatives of the gauge fields, namely ${\cal R}\equiv {{\cal
    R}^{ab}}_{ab}$. Indeed, Kibble chose simply to set $L_{\rm G}
\propto {\cal R}$, which defines the so-called Einstein--Cartan (EC)
theory, which, when re-interpreted geometrically, is a direct
generalisation of general relativity to include torsion sourced by the
spin-angular-momentum (if any) of the matter field.

Following Kibble's work, several other approaches to gauging the
Poincar\'e group have been proposed, although they are all closely
related to his original derivation. For example, Kibble's passive
interpretation of the transformations has been criticised by Hehl et
al.\cite{hehl76}, who reproduced Kibble's derivation from the
standpoint of active transformations of the matter fields. In essence,
Kibble considered the `total' variation $\delta\vpsi \equiv
\vpsi'(x')-\vpsi(x)$ of the matter fields under the action of a
passive infinitesimal local Poincar\'e coordinate transformation,
whereas Hehl et al. and others\cite{wiesendanger96} considered instead
the `form' variation $\delta_0\vpsi \equiv \vpsi'(x)-\vpsi(x)$. The
latter approach is closer in spirit to the gauging of internal
symmetries, since the form variation allows one to realise the
representation of the Poincar\'e group as an active transformation in
the space of the matter fields $\vpsi$. Nonetheless, the
transformation properties of the resulting gauge fields, as well as
the form of the covariant derivative, are independent of these
details, so that the final structure of the theory is the
same\cite{footnote1a}.  Hehl et al. further proposed that, in the
matter field transformation $\delta_0\vpsi$, the partial derivative in
the translational generator be replaced by a covariant one to preserve
its geometric meaning after localisation of the Poincar\'e symmetry.
It is easily shown, however, that their new form for the variation
differs from the original $\delta_0\vpsi$ only by a local Lorentz
transformation and, therefore, invariance of a matter action under one
form of the variation implies invariance under the other. In fact, it
is unnecessary to consider only infinitesimal
transformations. Mukunda\cite{mukunda89} considered finite local
passive Poincar\'e transformations and again arrived at a theory
equivalent to Kibble's.  Lasenby, Doran and Gull\cite{lasenby98} also
consider finite transformations, but interpreted actively, and arrive
at a gauge theory of gravity with the same mathematical structure as
Kibble's theory, although formulated in the powerful language of
geometric algebra, which greatly simplifies their subsequent
application of the theory to various astrophysical and cosmological
scenarios. In essence, however, all the above formulations are
equivalent approaches to the localisation of Poincar\'e symmetry.
Finally, it is worth noting that the gauge approach to gravitation is
naturally interpreted as a field theory in Minkowski
spacetime\cite{wiesendanger96,lasenby98}, in the same way as the gauge
field theories describing the other fundamental interactions, and this
is the viewpoint that we shall adopt in this paper. It is more common,
however, to reinterpret the mathematical structure of PGT
geometrically, whereby ${h_a}^\mu$ and ${A^{ab}}_\mu$ are considered
as the components of a vierbein system and a local spin connection,
respectively, in a more general Riemann--Cartan spacetime possessing
non-zero curvature and torsion\cite{hehl76}. For a thorough, modern
account of the gauge approach to gravitation, see
Blagojevic\cite{blagojevic02}, in which many of the issues above are
discussed in some detail.

Although Kibble chose the simple gravitational Lagrangian $L_{\rm G}
\propto {\cal R}$, it is clear that numerous higher-order invariants
can be constructed from the field strength tensors ${{\cal
    R}^{ab}}_{cd}$ and ${{\cal T}^a}_{bc}$, and their inclusion in
$L_{\rm G}$ cannot be ruled out {\em a priori}. Nonetheless, if one
demands that the dynamics of each gauge field must be determined by a
source current which corresponds to the same gauge group generator,
one is naturally led to models with a Lagrangian $L_{\rm G}$ that is
quadratic in the `curvature' and `torsion'. The use of terms quadratic
in the field strengths also mimics conventional gauge theory. In
addition, one can typically ensure coincidence (or small difference)
of predictions of PGT with those of general relativity in macroscopic
domains, where Einstein's theory is satisfactorily verified by
observations and experiments, by adding the linear Hilbert-Einstein
term\cite{hayashi80,obukhov89}. Hence, one is led to consider
\begin{equation}
L_{\rm G} = -\kappa^{-1}(\Lambda + a{\cal R}) + L_{{\cal R}^2} + \kappa^{-1} L_{{\cal T}^2},
\label{pgtlg}
\end{equation}
where $\kappa = 8\pi G/c^4$ is Einstein's gravitational constant,
$\Lambda$ is a cosmological constant, $a$ is a dimensionless free
parameter (which is usually positive with the sign conventions adopted
in this paper\cite{footnote1b}), and
\begin{widetext}
\begin{eqnarray}
L_{{\cal R}^2} & = &
\alpha_1 {\cal R}^2
+ \alpha_2 {\cal R}_{ab}{\cal R}^{ab}
+ \alpha_3 {\cal R}_{ab}{\cal  R}^{ba}
+ \alpha_4 {\cal R}_{abcd}{\cal R}^{abcd}
+ \alpha_5 {\cal R}_{abcd}{\cal R}^{acbd}
+ \alpha_6 {\cal R}_{abcd}{\cal R}^{cdab}, \label{lr2}\\[1mm]
L_{{\cal T}^2} & = & \beta_1 {\cal T}_{abc}{\cal
  T}^{abc} + \beta_2 {\cal T}_{abc}{\cal T}^{bac} +
\beta_3 {\cal T}_a{\cal T}^a,
\label{lt2}
\end{eqnarray}
\end{widetext}
in which the $\alpha_i$ and $\beta_i$ are also dimensionless free
parameters. Note that pseudoscalar terms have
been omitted by requiring parity invariance; parity-odd terms have
been investigated
previously\cite{obukhov89,baekler11,obukhov12,karananas15} and may, in
principle, be included in the free-gravitational Lagrangian, but we do
not consider them here. Also, one may use the generalised
Gauss--Bonnet identity to set one of $\alpha_1$, $\alpha_3$ or
$\alpha_6$ to zero, without loss of generality.
We have
written $L_{\rm G}$ in the form (\ref{pgtlg}) so that, in natural units
($c=\hbar=1$, which we will adopt throughout), it has dimensions of
$(\mbox{length})^{-4}$, and hence the corresponding action $S_{\rm G}$ is
dimensionless; note that in natural units $\kappa$ has dimensions of
$(\mbox{length})^{2}$ and $\Lambda$ has dimensions of $(\mbox{length})^{-2}$.

In general, the equations of motion resulting from (\ref{pgtlg})
contain derivatives of the gauge fields of no higher than second
order. Indeed, restricting the Lagrangian to be at most quadratic in
the field strengths, and hence at most quadratic in the first
derivatives of the gauge fields, ensures one satisfies the so-called
pseudolinearity hypothesis\cite{hehl80}, which suggests that the field
equations should be linear in the highest (i.e. second) order
derivatives of the gauge fields\cite{kuhfuss86}. Most importantly, the
absence from the equations of motion of higher-order time derivatives
of the fields ensures that such theories do not suffer from
Ostrogradsky's instability\cite{woodard15,motohashi15}, wherein the
corresponding Hamiltonian is not bounded from below, so the energy of
the system can take an arbitrarily negative value. While such a system
for a free theory is not pathological, when it is coupled to a normal
(matter) system with positive energy, the total system can `evaporate'
into excitations of positive and negative energy degrees of freedom.
Ostrogradsky's instability is a quite generic feature of
higher-derivative theories and may explain why Nature is described by
theories having second-order equations of motion.  We note that,
although there are 40 dynamical variables ${h_a}^\mu$ (or,
equivalently, the components ${b^a}_\mu$ of the inverse $h$-field)
and ${A^{ab}}_\mu$, the `curvature' and
`torsion' field strengths are defined in terms of their antisymmetric
derivatives and so do not depend on the time derivatives of ${b^a}_0$
and ${A^{ab}}_0$; thus 10 of the field equations represent constraints
on initial data. Among the remaining 30 variables ${b^a}_\alpha$ and
${A^{ab}}_\alpha$ $(\alpha=1,2,3)$, one can impose 10 gauge conditions
by fixing the 10 parameters of the local Poincar\'e symmetry, reducing
the number of independent variables to 20.

The large number of free parameters in $L_{\rm G}$ clearly offers many
possibilities for constructing PGTs. For example, one of the simplest
generalizations of EC theory is the so-called ${\cal R}+{\cal T}^2$
theory, for which $L_{\rm G} = \kappa^{-1}(a{\cal R} + L_{{\cal
    T}^2})$. Like EC theory, however, this generalisation does not
contain a kinetic part for the rotational gauge field ${A^{ab}}_\mu$,
so that the equations of motion couple the translation gauge field
strength ${{\cal T}^a}_{bc}$ (or the torsion in the geometrical
reinterpretation of PGT) {\em algebraically} to the spin density of
the matter field; hence the `torsion' is a non-propagating field,
i.e. it must vanish outside of any matter sources. In a dynamical
sense, therefore, EC and ${\cal R}+{\cal T}^2$ theories are
incomplete: the rotational gauge field ${A^{ab}}_\mu$ is introduced as
an independent field, but the dynamics imposes an algebraic relation
between the $A$-field and the matter spin-density. The $A$-field
becomes fully dynamical only through the quadratic `curvature' terms
in $L_{{\cal R}^2}$, in which case the `torsion' becomes a propagating
field. Consequently, a number of ${\cal R}+{\cal R}^2$-type theories
have been investigated, as have ${\cal R}^2+{\cal T}^2$
theories\cite{vonderheyde76}. It is worth mentioning that the latter
allow the possibility of repulsive torsion interactions at small
distances, which may prevent gravitational collapse to infinite matter
densities\cite{minkevich80}.  A summary of the properties of general
${\cal R}+{\cal R}^2+{\cal T}^2$ theories is given by Obukhov et
al.\cite{obukhov89}. The properties of even more general actions, and
the coupling to matter fields, have been discussed\cite{rubilar98}. In
broad terms, if $L_{\rm G}$ consists of ${\cal R}$ and ${\cal T}^2$
terms, then it contains dimensional constants, which is not an
attractive property for quantisation. On the other hand, the ${\cal
  R}^2$ terms contain dimensionless constants, but the classical limit
of theories without the linear curvature term ${\cal R}$ is
questionable\cite{blagojevic02}.

Despite the advantages of the above framework for constructing $L_{\rm
  G}$, certain general difficulties do arise that go beyond merely the
paucity of guiding physical principles for elucidating the magnitude
or origin of the many free parameters in $L_{\rm G}$, or the algebraic
complexity of the resulting field equations. For example, although the
field equations that follow from the variational principle are of
second differential order, if one substitutes for the rotational gauge
field ${A^{ab}}_\mu$ in terms of ${h_a}^\mu$ one can rewrite them in a
form that contains third-order derivatives, and it becomes difficult
to assess whether they are well-posed\cite{hammond02}. A judicious
choice of the parameters in $L_{\rm G}$ can eliminate such higher
derivatives from the theory\cite{katanaev93}, although the resulting
theories are not without their difficulties, such the lack of a
sensible Newtonian limit or well-defined initial value
problem\cite{hecht91}. Nonetheless, the general structure of the
equations of motion has been analysed to clarify the evolution of
given initial data, i.e. the Cauchy problem. Of the 40 equations of
motion, it is found that 10 represent constraints on the initial data,
as mentioned above, and the remaining 30 define a consistent time
evolution of the dynamical variables provided that the parameters $a$,
$\alpha_i$ and $\beta_i$ in the Lagrangian $L_{\rm G}$ obey certain
conditions\cite{hecht96}.

In assessing various PGTs as candidate theories of gravitation,
however, the issues of greatest interest are renormalisability and
unitarity, both of which can be investigated by examining the particle
content of the theory. This is performed by first linearizing the 40
gauge fields $h_a^\mu$ and ${A^{ab}}_\mu$ about the trivial vacuum
solution to the field equations and retaining only those terms in the
Lagrangian $L_{\rm G}$ that are bilinear in the linearized gauge
fields to obtain $L^{(2)}_{\rm G}$. After transforming to momentum
space, one chooses covariant basis states, so that $L^{(2)}_{\rm G}$,
when considered as a $40\times 40$-matrix in field space, becomes
mostly zeroes. The covariant basis states are constructed using
$k$-space projection operators\cite{rivers64}, which decompose the
linearised gauge fields into tensors that reduce to states of definite
spin-parity $J^P$ in a `rest-frame' where $k^a$ has only a time
component, and are thus irreducible representations of the Poincar\'e
group. One finds that of the 20 independent degrees of freedom in the
gauge fields, two correspond to the massless {\em graviton} (with
spin-parity $J^P=2^+$) and the decomposition of ${A^{ab}}_\alpha$
$(\alpha=1,2,3)$ yields (in general, massive) {\em tordions} (or {\em
  rotons}) with spin-parities of $J^P = 2^\pm, 1^\pm, 0^\pm$, with
$2(5+3+1)=18$ degrees of freedom. The propagator is simply (minus) the
inverse of the terms in the resulting bilinear
Lagrangian\cite{neville78}.

For a theory to be unitary, its particle spectrum must be free from
ghosts (particles with negative definite free-field energy, which
destroys the unitarity of the $S$ matrix) and tachyons (particles
which propagate faster than light and therefore violate causality).
One can check for ghosts or tachyons by examining the residues and
locations of the propagator poles, which generally depend in a
complicated way on all the parameters $a$, $\alpha_i$ and $\beta_i$ in
the original Lagrangian $L_{\rm G}$. The condition for no tachyons is
that the propagator has first-order poles for non-negative $k^2$,
whereas the condition for no ghosts is that the matrix of the residues
at the poles is positive definite. In other words, each element of the
diagonalized propagator must be of the form $-R/(k^2-m^2)$ with $R>0$
and $m^2\geq 0$ (assuming a `mostly minus' metric signature).  These
requirements follow from the spectral representation of the vacuum
expectation value of the time-ordered product\cite{weinberg64}. In
PGT, for the general case in which all the tordion modes are massive,
there exist ranges of values of the parameters in $L_{\rm G}$ (that
divide into 5 broad classes) for which the resulting PGT is indeed
unitary\cite{sezgin80}. If one further requires that the torsion
propagates, one finds that there are 12 six-parameter Lagrangians that
lead to unitary theories\cite{sezgin81}. For the special case in which
there are massless tordion modes, and hence the
possibility of long-range torsion effects, it is important to take
into account the transversality relations between residues (for
example, they are needed to eliminate ghosts from general
relativity). In this case, the results are not so
complete\cite{battiti85,kuhfuss86}. Nonetheless, at least one
3-parameter Lagrangian exists that propagates (only) a massless
$J^P=1^-$ tordion and results in a unitary theory\cite{sezgin81}. It
is worth noting that a generic feature of massless tordion Lagrangians
is the appearance of extra gauge symmetries. The introduction of a
scalar field allows for a still larger choice of unitary
Lagrangians\cite{rovelli83}.

For a theory to be power-counting renormalisable, one requires good
high-energy behaviour of the propagators for its constituent
particles. In particular, in all spin sectors, the propagators for the
graviton should behave as $\sim 1/k^4$, whereas those for the tordions
should be $\sim 1/k^2$, since tordion vertices are better behaved than
those of gravitons\cite{vann73,neville78,sezgin80}. Pure $1/k^4$ poles
can occur in PGT, but are forbidden since they give rise to field
modes with free-field energies not bounded from below
(`dipole-ghosts')\cite{pais50}. Nonetheless, the PGT propagator also
has pole structures of the form
$(k^2-m^2_1)^{-1}(k^2-m^2_2)^{-1}=(m_2^2-m_1^2)^{-1}[(k^2-m_2^2)^{-1}-(k^2-m_1^2)^{-1}]$. These
can be tolerated in lower spin-sectors, since the resulting ghost mode
may be compensated by lower-spin terms in the higher-spin projection
operators. This cannot occur in the spin-2 sector, however, and so any
power-counting renormalizable PGT will have ghosts in the spin-2
sector\cite{neville80}. Conversely, to remove the ghosts, the $\sim
1/k^4$ propagators must cancel, making the theory power-counting
non-renormalizable. Indeed, for unitary massive-tordion Lagrangians, the
graviton propagator goes $\sim 1/k^2$ and the tordion propagator tends
to a constant at high energies. For the massless propagating $J^P=1^-$
tordion Lagrangian, however, the tordion propagator goes as $\sim
1/k^2$, leading to improved ultraviolet behaviour\cite{sezgin81}.
Although power-counting renormalizable PGTs are not unitary, and {\em
  vice-versa}, it still remains possible that unitary PGTs are
renormalizable when higher-order loops are taken into account, but
relatively little is known. It has been shown\cite{stelle77} that, in
the absence of torsion, ${\cal R}+{\cal R}^2$ theories are
renormalizable (but not unitary).  The renormalisability of
EC theory and the quantum
effect of quadratic torsion terms at one-loop level have been
investigated\cite{wiesendanger96,kalmykov95}. More innovative approaches to
quantisation of PGT have been pursued\cite{grignani92,grignani93}, but
are also problematic\cite{strobl93}.

The lack of a clear route to quantising PGT has led to interest in
imposing extra gauge symmetries beyond local Poincar\'e symmetry,
since this could, in principle, lead to surprises that might allow for
a simultaneously unitary and renormalizable
theory\cite{sezgin80,kalmykov97}, without having to resort to
non-local theories\cite{biswas12,biswas14,modesto15}. In particular,
perhaps the most natural extension of PGT is also to demand local
scale invariance, which might resolve some of the problems outlined
above\cite{thooft15}, since such theories contain no absolute energy
scale. It should be noted, however, that scale invariance defines an
unacceptable transformation law for particle masses. If scale
invariance were to hold in Nature it would imply that the mass
spectrum is either continuous (if $m^2 \neq 0$) or all the masses
vanish. Thus scale invariance must be a broken symmetry in a world
with non-vanishing, discrete particle masses.

The most direct approach to constructing gauge theories of gravity
that are invariant under local changes of scale, in addition to local
Poincar\'e transformations, is to gauge the Weyl group W(1,3), which
extends Poincar\'e symmetry to include scale invariance and is a
sub-group of the full conformal group
C(1,3)\cite{bregman73,charap74,kasuya75}. This may be formulated in a
number of ways, e.g. by considering the Weyl transformations as active
or passive, infinitesmial or finite, but they are all essentially
equivalent.  As in PGT, one assumes the physical model of an
underlying Minkowski spacetime in which a continuum matter field (or
fields) $\vpsi$ with energy-momentum and, in general,
spin-angular-momentum is distributed continuously. Now, however, the
field dynamics are described by a matter action $S_{\rm M} = \int
L_{\rm M}(\vpsi,\partial_\mu\vpsi)\,d^4x$ that is invariant under
global Weyl coordinate transformations, which imposes much more severe
constraints on the form of the action. By then demanding the matter
action to be invariant with respect to local Weyl transformations, in
which the eleven Weyl group parameters become arbitrary functions of
position, one is again led to the introduction of gauge fields
${h_a}^{\mu}$ and ${A^{ab}}_{\mu}$ (although the transformation rules
of the former are different to those in PGT) and also a new
gravitational gauge field $B_\mu$ corresponding to the dilation part
of the Weyl group\cite{footnote2,weyl18,weyl31}. Similar to PGT, the
gravitational gauge fields ${h_a}^{\mu}$, ${A^{ab}}_{\mu}$ and $B_\mu$
are used to assemble a covariant derivative ${\cal D}^\ast_a\vpsi$,
and the matter action in the presence of gravity is obtained by the
minimal coupling procedure, so that $S_{\rm M} = \int h^{-1} L_{\rm
  M}(\vpsi,{\cal D}^\ast_a\vpsi)\,d^4x$. It should be noted, however,
that the requirement of local scale invariance imposes tight
constraints on the precise form of $L_{\rm M}$. In particular, since
$h^{-1}$ has a Weyl (or conformal) weight $w(h^{-1})=4$ (see
Section~\ref{sec:lmwgt}), the Lagrangian $L_{\rm M}$ must have a
weight $w(L_{\rm G})=-4$.

As in PGT, the action term describing the dynamics of the free
gravitational gauge fields is assembled by first constructing
covariant field strength tensors for the gauge fields by commuting
covariant derivatives, i.e. by considering $[{\cal D}^\ast_a,{\cal
    D}^\ast_b]\vpsi$. One obtains ${{\cal R}^{ab}}_{cd}(h,A,\partial
A)$ and ${{\cal T}^{\ast a}}_{bc}(h,\partial h, A, B)$ as before (but
where the form of the latter differs from that in PGT) and an
additional field strength ${\cal H}_{ab}(h,\partial B)$ corresponding
to the new gauge field $B_\mu$ arising from the dilation part of the
Weyl group (we have again indicated the functional dependence of each
field strength tensor on the gauge fields and their derivatives,
suppressing indices for brevity).  As in PGT, each field strength
tensor is linear in the derivatives the gauge field to which it
corresponds. The free gravitational action then has the general form
$S_{\rm G} = \int h^{-1} L_{\rm G}({{\cal R}^{ab}}_{cd},{{\cal T}^{\ast
    a}}_{bc},{\cal H}_{ab})\,d^4x$.
As usual, the total action is
taken as the sum of the matter and gravitational actions, and
variation of the total action with respect to the gauge fields
${h_a}^{\mu}$, ${A^{ab}}_{\mu}$ and $B_\mu$ leads to three coupled
gravitational field equations in which the energy-momentum
${\tau^k}_\mu\equiv \delta{\cal L}_{\rm M}/\delta {h_a}^\mu$,
spin-angular-momentum ${\sigma_{ab}}^\mu\equiv \delta{\cal L}_{\rm M}/\delta
{A^{ab}}_\mu$ and dilation current $\zeta^\mu \equiv \delta {\cal
  L}_{\rm M}/\delta B_\mu$ of the matter field act as sources.

The free gravitational Lagrangian $L_{\rm G}$ must also have a Weyl
(conformal) weight $w(L_{\rm G})=-4$, which places tight constraints
on its form.  It is easily shown that $w({{\cal R}^{ab}}_{cd})=w({\cal
  H}_{ab})=-2$ and $w({{\cal T}^{\ast a}}_{bc})=-1$, which means that
$L_{\rm G}$ can be quadratic in ${{\cal R}^{ab}}_{cd}$ and ${\cal
  H}_{ab}$, while terms linear in ${\cal R} \equiv {{\cal
    R}^{ab}}_{ab}$ or quadratic in ${{\cal T}^{\ast a}}_{bc}$ are not
allowed, despite them transforming covariantly under local Weyl
transformations. Thus, in WGT, the general form of $L_{\rm G}$,
possessing terms no higher than quadratic order in the field
strengths, and hence at most quadratic in the first derivatives of
gauge fields, is of ${\cal R}^2+{\cal H}^2$ type:
\begin{widetext}
\begin{equation}
L_{\rm G}   = \alpha_1 {\cal R}^2
+ \alpha_2 {\cal R}_{ab}{\cal R}^{ab}
+ \alpha_3 {\cal R}_{ab}{\cal  R}^{ba}
+ \alpha_4 {\cal R}_{abcd}{\cal R}^{abcd}
+ \alpha_5 {\cal R}_{abcd}{\cal R}^{acbd}
+ \alpha_6 {\cal R}_{abcd}{\cal R}^{cdab}
+ \xi {\cal H}_{ab}{\cal H}^{ab} \equiv  L_{{\cal R}^2} + L_{{\cal H}^2},
\label{lg2}
\end{equation}
\end{widetext}
where the $\alpha_i$ and $\xi$ are dimensionless free parameters. Once
again pseudo-scalar terms have been omitted by requiring parity
invariance, and one may use the generalised Gauss--Bonnet identity to
set one of $\alpha_1$, $\alpha_3$ or $\alpha_6$ to zero, without loss
of generality.
Evidently, local Weyl
invariance removes many of the possibilities that exist in PGT;
essentially the ${\cal R}$ and ${\cal T}^{\ast 2}$ terms, which
possess dimensional constants, are forbidden.

Although the requirement that each term in the total Lagrangian must
have a Weyl weight $w =-4$ clearly places quite restrictive conditions
on its form, one can construct further Weyl-covariant terms with the
appropriate weight for inclusion in the Lagrangian by introducing an
additional massless scalar field (or fields) $\phi$ with Weyl weight
$w({\phi})=-1$, often termed the compensator(s)
\cite{blagojevic02}. This opens up possibilities
for the inclusion of further action terms in which the scalar field is
non-minimally (conformally) coupled to the field strength tensors of
the gravitational gauge fields, combined (usually) with an additional
free kinetic term for $\phi$. For example, terms proportional to
$\phi^2 {\cal R}$ or $\phi^2 L_{{\cal T}^{\ast 2}}$ are Weyl-covariant
with weight $w=-4$ and so may be added to the total Lagrangian. The
inclusion of the term $\phi^2 {\cal R}$ has been widely investigated
\cite{dirac73,omote77}, whereas the phenomenological value of the
  $\phi^2 L_{{\cal T}^{\ast 2}}$ terms has received relatively little
  attention\cite{sijacki82,neeman88}. In any case, the resulting total
  Lagrangian is at most quadratic in the first derivatives of the
  gauge fields, thereby satisfying the pseudolinearity hypothesis that
  field equations be linear in the second-order derivatives of the
  gauge field, and hence ensuring that such theories do not suffer
  from Ostrogradsky's instability.

The inclusion of scalar fields also allows for more flexibility in the
allowed forms of the matter energy-momentum tensor. Let us take as an
example the action for a free Dirac field $\psi$, which has Weyl
weight $w(\psi)=w(\bar{\psi})=-\tfrac{3}{2}$.  This action is not
scale-invariant owing to the mass term $m\bar{\psi}\psi$, so it
appears that one requires the field to be massless, which clearly
cannot describe `ordinary' matter. One may reach the same conclusion
by noting that in WGT the trace of the total matter energy-momentum
tensor must equal the covariant divergence of the dilation
current\cite{blagojevic02} $\zeta^\mu \equiv \delta {\cal L}_{\rm
  M}/\delta B_\mu$. In the case of Dirac matter, the Lagrangian
resulting from gauging the Weyl symmetry, curiously, does not contain
the dilation gauge field $B_\mu$. Hence, for Dirac matter on its own,
the energy-momentum tensor of $\psi$ must be traceless, which again
requires the field to be massless. These difficulties can be
circumvented, however, by making the replacement $m\bar{\psi}\psi \to
\mu\phi\bar{\psi}\psi$ in the Dirac action, where $\mu$ is a
dimensionless parameter but $\mu\phi$ has the dimensions of mass in
natural units. Although the trace of the total energy momentum tensor
of the $\psi$ and $\phi$ fields must still vanish, the energy-momentum
tensor of the Dirac matter field $\psi$ itself need not be traceless,
thereby allowing it to be massive. Indeed, this approach to the
construction of gauge theories of gravity that are scale-invariant
but, at the same time, are able to accommodate `ordinary' matter was
first explored by Dirac\cite{dirac73} in the context of attempting to
establish a deeper physical understanding of his `large numbers
hypothesis' relating microscopic (quantum) and macroscopic
(gravitational) scales. In particular, Dirac hoped that the additional
scalar field $\phi$ might provide a time-dependent coupling of
gravitation to matter, hence allowing the Newtonian gravitational
`constant' $G$ to vary with cosmic epoch. This idea can be naturally
accommodated in the framework of WGT\cite{canuto77} (and extended to
include non-zero torsion, which was assumed to vanish by Dirac).

In general, the introduction of scalar matter fields in WGT, and
scale-invariant theories more generally, is also important since they
provide a natural means for spontaneously breaking the scale symmetry,
which is necessary to achieve an acceptable long-range limit. The
approach most commonly adopted is to use local scale invariance to set
the compensator scalar field $\phi$ to a constant value in the
resulting field equations, which is known as the Einstein
gauge. Setting $\phi=\phi_0$ in the equation of motion for the Dirac
field $\psi$, for example, leads to its interpretation as a massive
field with $m=\mu\phi_0$. It is usually considered that setting
$\phi=\phi_0$ represents the choice of some definite scale in the
theory, thereby breaking scale-invariance. Indeed, it is often given
the physical interpretation of corresponding to some spontaneous
breaking of the scale symmetry (where Nature chooses the gauge). As we
show in Section~\ref{sec:wgteinstein}, however, this interpretation is
questionable. In particular, we demonstrate in WGT that the equations
of motion in the Einstein gauge are identical in form to those
obtained when working in scale-invariant variables, where the latter
involves no breaking of the scale symmetry. This suggests that one
should introduce further scalar fields, in addition to the compensator
field $\phi$, to enable a true physical breaking of the scale
symmetry.

In general, the phenomenology of WGT is far less well understood than
that of PGT.  As for PGT, the presence of ${\cal R}^2$ terms in the
action improves the quantum renormalisability of such theories, but
can again lead to problems with unitarity\cite{blagojevic02}. These
issues, as well as the stability of the ground state, have yet to be
resolved, but the development of a full phenomenological basis for WGT
remains of considerable theoretical interest.

In this paper, however, our primary aim is to present a novel
alternative to WGT by considering a more general `extended'
transformation law for the rotational gauge field under local
dilations, which includes its `normal' transformation law in WGT as a
special case. This is motivated by the observation that the PGT (and
WGT) matter actions both for the Dirac field and the electromagnetic
field are already invariant under local dilations if one assumes this
`extended' transformation law, in the same way as they are for the
`normal' transformation law assumed in WGT. Moreover, under a global
scale transformation, the two transformation laws coincide, and so
both may be considered as an equally valid gauging of global Weyl
scale invariance. The key difference between the two sets of
transformations is that, whereas under the `normal' Weyl
transformations the PGT gauge fields ${h_a}^\mu$ and ${A^{ab}}_\mu$
transform covariantly with weights of $w$, $-1$ and $0$, respectively,
under the `extended' transformations the rotational gauge field
${A^{ab}}_\mu$ transforms inhomogeneously. This also has the
consequence that the transformation properties of the PGT rotational
gauge field strength (or `curvature') ${\cal R}_{abcd}$ and
translational gauge field strength (or `torsion') ${\cal T}_{abc}$ are
treated in a more balanced manner.  The resulting `extended' Weyl
gauge theory (eWGT) has a number of interesting features that we
describe in detail, although we focus here only on its formal
description and postpone a full discussion of its application and
phenomenology to future publications.

Finally, it is worth mentioning that one can also construct gauge
theories of gravity that are invariant under local changes of scale
(assuming either the normal or extended transformation law for the
rotational gauge field), in addition to local Poincar\'e
transformations, {\em without} gauging the Weyl group. Instead, one
works entirely in terms of PGT objects, such as matter fields $\vpsi$
and their PGT covariant derivatives ${\cal D}_a\vpsi$, the PGT gauge
field strengths ${{\cal R}^{ab}}_{cd}$ and ${{\cal T}^a}_{bc}$ (or
even their simpler counterparts in the absence of `torsion'), and a
(compensator) scalar field $\phi$ (as discussed above). These are used
to construct terms in the free gravitational and matter actions that
individually are not necessarily invariant (or even covariant) under
(extended) local scale changes, but by taking the appropriate linear
combinations of such terms, it is possible to arrive at a free
gravitational action and matter action that are locally
scale-invariant (and hence so is the total
action)\cite{adler82,obukhov82,zee83,antoniadis85,mannheim89}.

The remainder of this paper is arranged as follows. In
Section~\ref{sec:wgt}, we establish our general approach to
constructing locally scale-invariant gauge theories of gravity by
giving a brief outline of the mathematical structure of WGT, but in a
way not usually presented in the literature.  In particular, we
maintain throughout the notion of ${h_a}^\mu$, ${A^{ab}}_\mu$ and
$B_\mu$ as gauge fields in Minkowski spacetime, rather than adopting
the more usual geometric interpretation of WGT in terms of the
curvature and torsion of a Weyl--Cartan $Y_4$ spacetime (although we
do give a brief account of the latter); this is in keeping with the
gauge field theories describing the other fundamental interactions and
has some non-trivial consequences. We also work exclusively in terms
of finite local Weyl transformations, rather than their usual
infinitesimal forms; the former allows, in our view, for a more
transparent interpretation of WGT. We also make some observations that
differ from typical accounts of the subject, in particular regarding
the interpretation of the Einstein gauge and also the equations of
motion of matter fields and test particles.  In
Section~\ref{sec:ewgt}, we present our novel alternative to WGT,
obtained by demanding invariance under a more general form for the
transformation law of the rotational gauge field under local
dilations, and describe the general properties of the resulting
`extended' Weyl gauge theory (eWGT); we also discuss its
interpretation in terms of a new spacetime geometry (that we term
$\widehat{Y}_4$), which is an extended form of Weyl--Cartan $Y_4$
spacetime.  In Section~\ref{sec:sigtg}, we present a new
scale-invariant gauge theory of gravity that can accommodate ordinary
matter, and is defined by the most general eWGT Lagrangian that is at
most quadratic in the eWGT field strengths, and we derive its field
equations. In Section~\ref{sec:lsipgt}, we briefly discuss the
construction of PGT actions that are invariant under local dilations.
We conclude in Section~\ref{sec:conc}. In addition, we include six
appendices.  Appendix~\ref{appendix:dirac} presents a re-expression of
the special-relativistic Lagrangian of a Dirac field, which lends
itself to the development of a semi-classical description of a
spinning point particle, which we will use as our model of `ordinary'
matter. Appendix~\ref{app:pgt} provides a brief description of PGT as
a special case, with appropriate notational modifications, of our
discussion of WGT in Section~\ref{sec:wgt}, and
Appendix~\ref{ECtheory} summarises the particular PGT originally
considered by Kibble, namely Einstein--Cartan theory.  In
Appendix~\ref{appendix:dgt}, as an example of a WGT, we describe the
first proposed scale-invariant theory of gravity that can accommodate
`ordinary' matter, which was explored by Dirac\cite{dirac73}, but
extend it to include non-zero torsion. In
Appendix~\ref{appendix:altddagderiv}, we present an alternative
approach to deriving the form of the covariant derivative in eWGT,
which is complementary to that presented in
Section~\ref{sec:ewgt}. Finally, in Appendix~\ref{appendix:notation},
we present a summary of our notation. The phenomenology of eWGT and its application
to astrophysics and cosmology will be described in forthcoming papers.


\section{Weyl gauge theory}
\label{sec:wgt}

As discussed in the Introduction, the minimal and perhaps most natural
extension of PGT is based on gauging the Weyl group W(1,3), which
extends Poincar\'e symmetry to include scale invariance. Thus, in
traditional Weyl gauge theory (WGT), in addition to local Poincar\'e
symmetry, one also demands invariance under local change of scale.  We
give an account of the mathematical structure of WGT here.  In
particular, we will interpret Weyl transformations passively, but
consider them in their more transparent finite form, rather than as
infinitesimal transformations.  The notation used broadly follows
Kibble\cite{kibble61} and Mukunda\cite{mukunda89}.  It is worth noting
that, with appropriate notational modifications, the following
discussion also serves as an account of PGT, which may be considered
as a special case of WGT, as described in Appendix~\ref{app:pgt}.

\subsection{Global Weyl invariance}
\label{sec:globalweyl}

We begin with a Minkowski spacetime ${\cal M}$, labelled using
Cartesian inertial coordinates $x^\mu$, on which the dynamics of some
multiplet of `matter' (i.e. non-gravitational) fields $\vpsi(x)$ is
described by a matter action
\begin{equation}
S_{\rm M} = \int L_{\rm M}(\vpsi,\partial_\mu\vpsi)\,d^4x,
\label{weylsraction}
\end{equation}
that is invariant under global Weyl transformations, which imposes
much more severe constraints on the form of the action than those
required by mere global Poincar\'e invariance.  We denote the
coordinate basis vectors by $\vect{e}_\mu$, which are tangents to the
coordinate curves and for Cartesian inertial coordinates satisfy
$\vect{e}_\mu\cdot\vect{e}_\nu=\eta_{\mu\nu}$, where $\eta_{\mu\nu} =
\mbox{diag}(1,-1,-1,-1)$. At each point of ${\cal M}$, we can also
define a local Lorentz reference frame, unrelated to the coordinates
$x^\mu$, which is represented by an orthonormal tetrad
$\hat{\vect{e}}_a(x)$, such that
$\hvect{e}_a\cdot\hvect{e}_b=\eta_{ab}$. In global Cartesian inertial
coordinates $x^\mu$, we can always choose the tetrad such that it
coincides with a coordinate frame $\vect{e}_\mu(x)$,
i.e. $\hvect{e}_a=\delta_a^\mu\vect{e}_\mu$. We will
adopt the common convention that Latin indices refer to local tetrad
frames, while Greek indices refer to coordinate frames.

We suppose that our multiplet of matter fields $\vpsi(x)$ belongs to
some representation of the Weyl group. More precisely, we replace the
usual Lorentz subgroup $\mbox{SO}(3,1)$ by $\mbox{SL}(2,C)$, so that
spinor fields can be accommodated\cite{footnote3}. Indeed, by
expressing $\vpsi(x)$ in its 2-spinor form, for example, one can
always ensure that its components carry no coordinate (Greek) indices.
Such an approach can be rather cumbersome, however, and so we adopt a
more straightforward notation in which the matter field components
$\{\vpsi^a(x)\}$ simply carry a generic Roman index (or set of
indices) to denote that they are `referred to' a local Lorentz frame.
Thus, matter field components (which may be
expressed in 2-spinor, 4-spinor or 4-tensor form, as is most
convenient) behave as scalars under a change of coordinates, but
transform under a corresponding rotation of the local tetrad frames
(such that they coincide with the new coordinate frame at each point)
or physical dilations according to their $\mbox{SL}(2,C)$ representation and Weyl weight (see below), respectively.  This distinction will
become important shortly, when we consider localising the Weyl
symmetry.

Thus, under a global Weyl transformation
\begin{equation}
{x'}^\mu=e^\rho({\Lambda^\mu}_\nu x^\nu + a^\mu),
\label{weyl}
\end{equation}
the multiplet of matter fields transforms via
\begin{equation}
\vpsi'(x')=e^{w\rho}\matri{S}(\Lambda)\vpsi(x),
\label{weylpsitrans}
\end{equation}
where $\matri{S}(\Lambda)$ is the matrix corresponding to the element
$\Lambda$ of the Lorentz group (or SL(2,C) group) in the
representation to which $\vpsi(x)$ belongs (we have suppressed Lorentz
indices on these objects for notational simplicity), and $w$ is the
Weyl (or conformal) weight of the field $\vpsi(x)$.  Moreover, the
derivatives of the matter fields transform as
\begin{equation}
\partial'_\mu\vpsi'(x')=e^{(w-1)\rho} {\Lambda_\mu}^\nu \matri{S}(\Lambda)\partial_\nu \vpsi(x).
\label{weyldpsitrans}
\end{equation}

It is worth briefly considering in an explicitly geometrical manner
the dilation part of the Weyl transformation, i.e.\ with
${\Lambda^\mu}_\nu = \delta^\mu_\nu$ and $a^\mu=0$ in (\ref{weyl}) and
(\ref{weylpsitrans}).  This transformation may be considered as
consisting of two distinct parts. The first part physically (actively)
dilates the Minkowski spacetime such that the distance between any two
points is increased by a factor $e^\rho$. Adopting global inertial
coordinates, such that $ds^2 = \eta_{\mu\nu}\,dx^\mu\,dx^\nu$, the
metric must therefore transform according to $\eta_{\mu\nu} \to
\gamma_{\mu\nu} = e^{2\rho}\eta_{\mu\nu}$. Hence, under this active
dilation the metric has a Weyl weight of $w=2$. Note, however, that no
coordinate transformation has occurred, since every point in the
spacetime still has its original coordinate labels; one may consider
the coordinates as being `embedded' in the spacetime as it
dilates. The second part of the transformation consists of a
coordinate transformation $x^{\prime\mu} = e^\rho x^\mu$, which
`shrinks' the coordinate system back to its original `size'. Under
this coordinate transformation, $\gamma'_{\mu\nu} =
e^{-2\rho}\gamma_{\mu\nu} = \eta_{\mu\nu}$ and so, overall, the metric
remains unchanged. This distinction between the two parts of the Weyl
dilation will become important when we localise the Weyl symmetry, at
which point they decouple completely.

For invariance of the matter action under a global Weyl
transformation (\ref{weyl}) (for which the Jacobian equals $e^{4\rho}$),
a sufficient condition is that the matter Lagrangian density satisfies
\begin{equation}
L_{\rm M}(\vpsi'(x'),\partial'_\mu \vpsi'(x'))
=e^{-4\rho}L_{\rm M}(\vpsi(x),\partial_\mu \vpsi(x)),
\label{weylinvariance}
\end{equation}
where the arguments on the LHS are given by (\ref{weylpsitrans})
and (\ref{weyldpsitrans}), respectively.

For our later purposes, it is useful to consider the concrete examples
provided by the standard special-relativistic matter Lagrangians for a
massive free scalar field, a massive free Dirac field and the
electromagnetic field, respectively. For a (real) free
scalar field $\phi$ of mass $m$, one has the Klein--Gordon Lagrangian
\begin{equation}
L_{\rm KG} = \tfrac{1}{2}\eta^{\mu\nu}\partial_\mu\phi\,\partial_\nu\phi
-\tfrac{1}{2}m^2\phi^2,
\end{equation}
and under the dilation part of a Weyl transformation
(\ref{weyl}--\ref{weylpsitrans}), one thus obtains
\begin{eqnarray}
L'_{\rm KG} & = &
\tfrac{1}{2}\gamma^{\prime\mu\nu}\partial'_\mu\phi'\,\partial'_\nu\phi'
-\tfrac{1}{2}m^2\phi^{\prime 2} \nonumber\\
& = & \tfrac{1}{2}e^{2(w-1)\rho}\eta^{\mu\nu}\partial_\mu\phi\,\partial_\nu\phi
-\tfrac{1}{2}m^2e^{2w\rho}\phi^2.
\end{eqnarray}
For invariance of the corresponding matter action,  one requires
(\ref{weylinvariance}) to hold. Thus, the free scalar field action is
dilation invariant only for $m=0$ and provided the Weyl weight of the
field is taken as $w(\phi)=-1$.

One can perform a similar analysis for the Lagrangian
for a classical Dirac field $\psi$ of mass $m$, which is
given by\cite{footnote4}
\begin{equation}
L_{\rm D} = \tfrac{1}{2}i
[\bar{\psi}\gamma^\mu\partial_\mu\psi-(\partial_\mu\bar{\psi})\gamma^\mu\psi]
-m\bar{\psi}\psi
\equiv
\tfrac{1}{2}i\bar{\psi}\gamma^\mu{\stackrel{\leftrightarrow}{\partial_\mu}}\psi
- m\bar{\psi}\psi.
\label{diraclm2}
\end{equation}
where the kinetic energy term is written so as to ensure it is always
real (and the sign ensures the terms contribute positive energy on
quantisation). Under a dilation transformation, one obtains
\begin{eqnarray}
L'_{\rm D} & = & \tfrac{1}{2}i
\bar{\psi}'\gamma^\mu{\stackrel{\leftrightarrow}{\partial'_\mu}}\psi'
- m\bar{\psi}'\psi' \nonumber \\
&=& \tfrac{1}{2}i e^{(2w-1)\rho}
\bar{\psi}\gamma^\mu{\stackrel{\leftrightarrow}{\partial_\mu}}\psi
- me^{2w\rho}\bar{\psi}\psi.
\end{eqnarray}
Demanding (\ref{weylinvariance}) to hold, one thus finds that the
Dirac action is dilation invariant only for $m=0$ and provided the
Weyl weight of the Dirac field is taken as
$w(\psi)=w(\bar{\psi})=-\frac{3}{2}$, which is consistent with the
interpretation of $\bar{\psi}\psi$ as a number density. As mentioned
in the Introduction, however, one can circumvent this restriction to a
massless Dirac field by introducing a massless scalar field $\phi$ and
making the replacement $m\bar{\psi}\psi \to \mu\phi \bar{\psi}\psi$ in
the Dirac Lagrangian, where $\mu$ is a dimensionless parameter,
combined usually with an additional kinetic term for $\phi$. One thus obtains
\begin{equation}
L_{\rm M} =
\tfrac{1}{2}i\bar{\psi}\gamma^\mu{\stackrel{\leftrightarrow}{\partial_\mu}}\psi
+ \tfrac{1}{2}\partial_\mu\phi\,\partial^\mu\phi - \mu\phi\bar{\psi}\psi,
\label{diracphilag}
\end{equation}
which is easily shown to be invariant under a dilation transformation
and describes a massless Dirac field interacting with a massless
scalar field. If one then adopts the Einstein gauge $\phi=\phi_0$ in
the resulting field equations, however, one may interpret $\psi$ as a
massive field with $m=\mu\phi_0$.

Finally, we consider the electromagnetic field $A_\mu$
(which should not be confused with the `rotational'
gravitational gauge field ${A^{ab}}_\mu$), for which the
special-relativistic Lagrangian is given by
\begin{equation}
L_{\rm EM} = -\tfrac{1}{4}F_{\mu\nu}F^{\mu\nu} - J^\mu A_\mu
\label{emlag-again}
\end{equation}
where $F_{\mu\nu} \equiv \partial_\mu A_\nu -\partial_\nu A_\mu$ and
$J^\mu$ is some source current density.  Under a dilation
transformation, one thus obtains
\begin{eqnarray}
L'_{\rm EM} & = & -\tfrac{1}{4}F^\prime_{\mu\nu}F^{\prime\mu\nu} - J^{\prime\mu} A^\prime_\mu
\nonumber \\
& = & -\tfrac{1}{4}e^{2(w-1)\rho} F_{\mu\nu}F^{\mu\nu} - e^{(w+w_J)\rho} J^\mu A_\mu,
\label{emlag-again2}
\end{eqnarray}
where $w$ and $w_J$ are the Weyl weights of $A_\mu$ and $J_\mu$
respectively.  Thus we see that the Lagrangian satisfies
(\ref{weylinvariance}), provided that $w(A_\mu)=-1$ and
$w(J_\mu)=-3$. It is worth noting that the Dirac current $J^\mu \equiv \bar{\psi}\gamma^\mu\psi$ satisfies this requirement.

\subsection{Local Weyl invariance}
\label{sec:lmwgt}

In order to gauge the Weyl transformation
(\ref{weyl}--\ref{weylpsitrans}), one makes each of the six Lorentz
rotation parameters in $\Lambda$, the four translation parameters
$a^\mu$ and the dilation parameter $\rho$ into eleven independent
arbitrary functions of $x$.  This leads to a complete decoupling of
the translation parts of the transformation from the rotational and
dilation parts. Thus, it is customary to view $x'$ as arising from a
general coordinate transformation (GCT) (resulting in a consequent
change in the coordinate basis vectors $\vect{e}_\mu(x)$ at each
point), which does not refer at all to the local Lorentz rotation
$\Lambda(x)$ of the orthonormal tetrad $\hvect{e}_a(x)$, or local
(physical) dilation $\rho(x)$, at each point. The local Lorentz
rotation is still an element of $\mbox{SO}(3,1)$ (or more precisely
$\mbox{SL}(2,C)$) for each $x$, and the basic transformation law for
$\vpsi$ involves a local shuffling and scaling of components given by
\begin{equation}
\vpsi'(x')=e^{w\rho(x)}\matri{S}(\Lambda(x))\vpsi(x).
\label{weylpsitrans_gen}
\end{equation}

Since the GCT and the local Lorentz rotation are completely decoupled
it is convenient (and common practice) to use Greek indices on
spacetime coordinates $x^\mu$ and components of the GCT, but use Latin
indices for local Lorentz rotations ${\Lambda^a}_b(x)$.  This means
that any spinor, vector or tensor character that $\vpsi$ has in the
special-relativistic theory now appears in its local Lorentz
transformation law, so that under a GCT each component of $\vpsi$
(carrying only Latin indices) behaves as a scalar field. Consequently,
the partial derivative, $\partial_\mu \vpsi$, is covariant under
GCT. We note that lowering and raising of Latin indices is performed
using the Minkowski metric $\eta_{ab} = \eta^{ab} =
\mbox{diag}(1,-1,-1,-1)$, whereas Greek indices are raised and lowered
using the (flat-space) metric
$\gamma_{\mu\nu}=\vect{e}_\mu\cdot\vect{e}_\nu$ and its inverse
$\gamma^{\mu\nu}$ corresponding to the coordinate system $x^\mu$ used
to label the Minkowski spacetime. In particular, the duals of the
orthonormal tetrad vectors and the coordinate basis vectors are given
by $\hvect{e}^a = \eta^{ab}\hvect{e}_b$ and $\vect{e}^\mu =
\gamma^{\mu\nu}\vect{e}_\nu$, respectively. Moreover, the decoupling of the GCT
from the local Lorentz rotation and local dilation transformations,
means that the metric $\gamma_{\mu\nu}$ of the Minkowski spacetime
obeys the transformation law
\begin{equation}
\gamma'_{\mu\nu}(x') =
\pd{x^\lambda}{{x'}^\mu}\pd{x^\sigma}{{x'}^\nu}\,e^{2\rho(x)}\,
\gamma_{\lambda\sigma}(x).
\end{equation}

From (\ref{weylpsitrans_gen}), we see that the transformation law for the
derivatives of $\vpsi$ under the gauged Weyl transformation is
given by
\begin{widetext}
\begin{equation}
\partial'_\mu \vpsi'(x')=\pd{x^\nu}{{x'}^\mu}e^{w\rho(x)}
         [\matri{S}(\Lambda(x))\partial_\nu\vpsi(x)+\partial_\nu
           \matri{S}(\Lambda(x)) \vpsi(x)+
w\,\partial_\nu\rho\, \matri{S}(\Lambda(x)) \vpsi(x)].
\label{weyldlpsitrans}
\end{equation}
\end{widetext}

In order to take advantage of the property (\ref{weylinvariance}), we
need to construct a covariant derivative that transforms like
(\ref{weyldpsitrans}) under local Weyl transformations. This is
achieved in two steps. Firstly, one defines the
$(\Lambda,\rho)$-covariant derivative of the matter field as\cite{footnote4a}
\begin{eqnarray}
D^\ast_\mu\vpsi(x) &\equiv &[\partial_\mu
+{\textstyle\frac{1}{2}}{A^{ab}}_\mu(x) \Sigma_{ab} +
wB_\mu(x)]\vpsi(x) \nonumber \\
&=& [D_\mu + wB_\mu(x)]\vpsi(x),
\label{weyldmudef}
\end{eqnarray}
where $D_\mu\vpsi(x)$ is the $\Lambda$-covariant derivative
introduced in PGT; the asterisk on the derivative operator on the
left-hand side is intended simply to distinguish it from
$D_\mu\vpsi(x)$, and should not be confused with the operation of
complex conjugation. Most importantly, ${A^{ab}}_\mu(x)$ is the
gauge-field corresponding to the local Lorentz rotations and $B_\mu(x)$
is the gauge-field corresponding to local dilations. The matrices
$\Sigma_{ab} = -\Sigma_{ab}$ are the generators of the
$\mbox{SL}(2,C)$ representation $\matri{S}(\Lambda)$ to which
$\vpsi(x)$ belongs, obeying the usual commutation rules
\begin{equation}
[\Sigma_{ab},\Sigma_{cd}]
=\eta_{ad}\Sigma_{bc}-\eta_{bd}\Sigma_{ac}+\eta_{bc}
\Sigma_{ad}-\eta_{ac}\Sigma_{bd}.
\end{equation}
Given the antisymmetry of the generator matrices $\Sigma_{ab}$,
without loss of generality we may take the $A$ gauge field to share
this antisymmetry property, such that
${A^{ab}}_\mu=-{A^{ba}}_\mu$. The aim of introducing the $A$ and $B$
gauge fields is to compensate for the second and third terms in square
brackets in (\ref{weyldlpsitrans}), so that the
$(\Lambda,\rho)$-covariant derivative transforms under a combined GCT,
local Lorentz rotation and local dilation as
\begin{equation}
D^{\ast\prime}_\mu\vpsi'(x') = \pd{x^\nu}{{x'}^\mu}
e^{w\rho(x)}\matri{S}(\Lambda(x))D^\ast_\nu\vpsi(x).
\label{eqn:dstarprime}
\end{equation}
This is easily achieved by demanding that the rotational gauge field
${A^{ab}}_\mu$ transforms as
\begin{equation}
{{A'}^{ab}}_\mu(x') \!=\! \pd{x^\nu}{{x'}^\mu}
[{\Lambda^a}_c(x){\Lambda^b}_d(x) {A^{cd}}_\nu(x)\!-\!\Lambda^{bc}(x)
\partial_\nu{\Lambda^a}_c(x)], \label{atrans}
\end{equation}
and the dilation gauge field $B_\mu$ transforms as
\begin{equation}
B'_\mu(x')  =  \pd{{x}^\nu}{{x'}^\mu} [B_\nu(x) - \partial_\nu \rho(x)].
\label{btransdef}
\end{equation}
Thus, ${A^{ab}}_\mu$ and $B_\mu$ are covariant vectors under GCT, but
transform inhomogeneously under local Lorentz transformations and
local dilations, respectively, which is typical of gauge fields.  We
also note that ${A^{ab}}_\mu$ is invariant under local dilations
(i.e. has Weyl weight $w=0$).

It is convenient sometimes to write the $(\Lambda,\rho)$-covariant
derivative (\ref{weyldmudef}) as
\begin{equation}
D^\ast_\mu \vpsi(x) = [\partial^\ast_\mu  +
\tfrac{1}{2}{A^{ab}}_\mu(x) \Sigma_{ab}] \vpsi(x),
\label{dstardef}
\end{equation}
where we have defined the derivative operator
\begin{equation}
\partial^\ast_\mu \equiv \partial_\mu + w B_\mu(x).
\label{partialstardef}
\end{equation}
It is straightforward to show that $\partial^\ast_\mu \vpsi(x)$ itself
also transforms covariantly with weight $w$ (i.e. the same weight as
the field $\vpsi$) under local dilations (but does not transform
covariantly under local Poincar\'e transformations).

In the second step, we define a generalised covariant derivative, linearly
related to $D^\ast_\mu\vpsi$, by introducing a new, `translational'
gauge field ${h_a}^\mu(x)$ that transforms via
\begin{equation}
{h'_a}^\mu(x') = \pd{{x'}^\mu}{{x}^\nu}\,
e^{-\rho(x)}{\Lambda_a}^b(x){h_b}^\nu(x).
\label{weylhtrans}
\end{equation}
Thus, the field ${h_a}^\mu$ is a contravariant GCT vector, a local
Lorentz four-vector, and has Weyl weight $w=-1$ under local dilations.
The inverse matrix is usually denoted by ${b^a}_\mu(x)$ (which
has Weyl weight $w=1$) and satisfies
\begin{equation}
{h_a}^\mu {b^a}_\nu =
\delta_\nu^\mu \quad\mbox{and}\quad {h_a}^\mu {b^c}_\mu = \delta_a^c.
\label{hinverse}
\end{equation}
The generalised covariant derivative is then defined as
\begin{eqnarray}
\hspace*{-5mm}{\cal D}^\ast_a\vpsi(x) &\equiv &{h_a}^\mu(x)D^\ast_\mu\vpsi(x)
\nonumber \\
&=& {h_a}^\mu(x)[\partial_\mu \!+\!
{\textstyle\frac{1}{2}}{A^{cd}}_\mu(x) \Sigma_{cd}\! + \! wB_\mu(x)]\vpsi(x),
\label{weylgencovdef}
\end{eqnarray}
which consequently transforms under local Weyl transformations
in the same way as $\partial\vpsi$ does under global Weyl transformations,
namely\cite{footnote5}
\begin{equation}
{\cal D}^{\ast\prime}_a\vpsi'(x') = e^{(w-1)\rho(x)}{\Lambda_a}^b(x) \matri{S}(\Lambda(x)) {\cal D}^\ast_b\vpsi(x).
\label{weylgencovtrans}
\end{equation}

Having achieved our goal of defining a general covariant derivative
with the transformation law (\ref{weylgencovtrans}), we can take advantage
of the property (\ref{weylinvariance}) of the special-relativistic matter
Lagrangian, and we see that under the {\em local} Weyl group one
has
\begin{equation}
L_{\rm M}(\vpsi'(x');{\cal D}^{\ast\prime}_a \vpsi'(x'))
=e^{-4\rho(x)}L_{\rm M}(\vpsi(x);{\cal D}^\ast_a \vpsi(x)).
\end{equation}
Thus the new minimally-coupled matter Lagrangian is a GCT and local
Lorentz scalar, and has a Weyl weight $w=-4$.  However, since $x \to
x'$ is a GCT, we need a scalar density, rather than a scalar, as the
integrand in the matter action.  Moreover, the integrand must have an
overall Weyl weight of zero.  Thus we must multiply $L_{\rm M}(\vpsi;{\cal
  D}^\ast\vpsi)$ by a factor which transforms by the appropriate
Jacobian determinant of the GCT and has a Weyl weight $w=4$.  From the
transformation properties of the $A$-, $B$- and $h$-fields, we see
that the only suitable factor that does not involve derivatives is
$h^{-1}$. Thus, our final minimally-coupled matter action is
\begin{equation}
S_{\rm M} = \int h^{-1}L_{\rm M}(\vpsi,{\cal D}^\ast_a\vpsi)\,d^4x,
\label{weylmatlag_gen}
\end{equation}
which is invariant under local Weyl transformations.  It will be
convenient to denote the integrand of the action by the Lagrangian
density ${\cal L}_{\rm M} \equiv h^{-1}L_{\rm M}(\vpsi,{\cal
  D}^\ast_a\vpsi)$.  We note that the gravitational gauge fields $h$,
$A$ and $B$ contain a total of 44 independent variables.

It should be mentioned, however, that the locally-Weyl-invariant
matter action (\ref{weylmatlag_gen}) is not guaranteed to inherit
invariance properties possessed by the original action
(\ref{weylsraction}) under other types of transformation, so
(\ref{weylmatlag_gen}) may need to be modified to restore any further
required invariances. An important example that illustrates this
phenomenon is provided by the electromagnetic field, and is discussed
in Section~\ref{sec:wgtem}. For the purposes of our general
discussion, however, we will assume the action has the form
(\ref{weylmatlag_gen}).

As mentioned previously, one may also introduce an additional
`compensator' scalar field $\phi$ (or fields) into the matter
action. The introduction of an additional scalar field opens up
possibilities for the inclusion of further terms in the matter action
that non-minimally (conformally) couple $\phi$ to the WGT
gravitational gauge field strengths (see Section~\ref{sec:wgtgfs}). In
particular, terms proportional to $\phi^2 {\cal R}$ or $\phi^2
L_{{\cal T}^{\ast 2}}$ are Weyl-covariant with weight $w=-4$ and so
may be added to $L_{\rm M}$ (where $L_{{\cal T}^{\ast 2}}$ has the
form (\ref{lt2}), but with the PGT translational gauge field strength
replaced by the corresponding quantity in WGT).

\subsection{Minkowski spacetime interpretation of WGT}
\label{sec:minkintwgt}

At this point, it is worth making a few observations that are not
usually presented in the literature. Firstly, we have introduced the
$h$, $A$ and $B$ fields as gravitational gauge fields in Minkowski
spacetime, in keeping with the interpretation of the gauge field
theories describing the other fundamental interactions. We are
therefore still at liberty to perform our calculations in a global
inertial Cartesian coordinate system $x^\mu$. In making this choice,
one may extend the $(\Lambda,\rho)$-covariant derivative $D^\ast_\mu$
to act on any quantity that transforms linearly under local Lorentz
transformations, even if it also carries Greek indices, by simply
ignoring its GCT transformation properties, since the explicit form
for the generators $\Sigma_{ab}$ appearing in $D^\ast_\mu\vpsi$
depends only on the Lorentz transformation properties of
$\vpsi$. Thus, for instance, considering the $h$-field itself, one has
\begin{equation}
D^\ast_\mu {h_a}^\nu  =  \partial^\ast_\mu {h_a}^\nu +
\tfrac{1}{2}{A^{cd}}_\mu(\Sigma^1_{cd}{)_a}^b{h_b}^\nu =
\partial^\ast_\mu {h_a}^\nu - {A^{b}}_{a\mu}{h_b}^\nu,
\label{hderiv}
\end{equation}
where ${(\Sigma^1_{cd})^a}_b =
(\delta_c^a\eta_{db}-\delta_d^a\eta_{cb})$ are the generators of the
vector representation of the Lorentz group. Similarly, for the inverse
$h$-field,
\begin{equation}
D^\ast_\mu {b^a}_\nu  =  \partial^\ast_\mu {b^a}_\nu + {A^a}_{b\mu}{b^b}_\nu.
\label{bderiv}
\end{equation}

It may prove useful occasionally, however, to work in a different
coordinate system in Minkowski spacetime, such as spherical polar
coordinates. In this case, we are obliged to generalise the
$(\Lambda,\rho)$-covariant derivative so that it is capable of being applied
to fields carrying Greek indices, which have definite tensor behaviour
under GCT, in addition to $\mbox{SL}(2,C)$ behaviour. In this case,
the $(\Lambda,\rho)$-covariant derivative in (\ref{dstardef}) becomes
\begin{equation}
D^\ast_\mu \equiv \partial^\ast_\mu + {^0}{\Gamma^\sigma}_{\rho\mu}
{\matri{X}^\rho}_\sigma + \tfrac{1}{2}{A^{ab}}_\mu\Sigma_{ab} =
{^0}\nabla^\ast_\mu + \tfrac{1}{2}{A^{ab}}_\mu\Sigma_{ab},
\label{extcovdweyl}
\end{equation}
where ${^0}\nabla^\ast_\mu \equiv {^0}\nabla_\mu + wB_\mu$,
${^0}{\Gamma^\lambda}_{\mu\nu} \equiv
\tfrac{1}{2}\gamma^{\lambda\rho}(\partial_\mu
\gamma_{\nu\rho}+\partial_\nu \gamma_{\mu\rho}-\partial_\rho
\gamma_{\mu\nu})$ is the metric connection corresponding to the
(flat-space) metric $\gamma_{\mu\nu}$ defined by the coordinate
system, and ${\matri{X}^\rho}_\sigma$ are the $\mbox{GL}(4,R)$
generator matrices appropriate to the GCT tensor character of the
field to which $D^\ast_\mu$ is applied, and $w$ is its Weyl weight.
Once again taking the $h$-field as an example, one thus has
\begin{equation}
D^\ast_\mu {h_a}^\nu  =  \partial^\ast_\mu {h_a}^\nu
+ {^0}{\Gamma^\nu}_{\rho\mu} {h_a}^\rho - {A^{b}}_{a\mu}{h_b}^\nu.
\label{dladef}
\end{equation}
It should be remembered, however, that in Minkowski spacetime, the
connection coefficients ${^0}{\Gamma^\sigma}_{\rho\mu}$ are not dynamical
entities, but are fixed by our choice of coordinate system.  Moreover,
the curvature and torsion tensors derived from the connection both
vanish in Minkowski spacetime, meaning that the derivatives
${^0}\nabla_\mu$ and ${^0}\nabla_\nu$ always commute, i.e.
\begin{equation}
[^{0}\nabla_\mu,^{0}\nabla_\nu]\vpsi=0,
\end{equation}
irrespective of the GCT tensor character of $\vpsi$. Unless otherwise
stated, however, from now on we will adopt a global Cartesian inertial
coordinate system $x^\mu$ in our Minkowski spacetime, which greatly
simplifies calculations.

Finally, some further comments about the interpretation of the
translational gauge field are worthwhile. It is a rank-2 tensor field,
which we will denote by $\vect{h}$, defined on Minkowski
spacetime. Thus, formally, it provides a linear mapping from any two
vectors in Minkowski spacetime to the real numbers. In the usual
manner, the components of $\vect{h}$ that we have considered hitherto
are given by ${h_a}^\mu
\equiv\vect{h}(\hvect{e}_a,\vect{e}^\mu)$. Similarly, the inverse of
$\vect{h}$, which we denote by $\vect{b}$, is also a rank-2 tensor
field defined on Minkowski spacetime, and the components that we
have considered so far are given by ${b^a}_\mu
\equiv\vect{b}(\hvect{e}^a,\vect{e}_\mu)$. Other components of
$\vect{h}$ and $\vect{b}$ may be obtained in the usual manner by
acting with them on different combinations of the tetrad and
coordinate basis vectors and their duals. The two tensors $\vect{h}$
and $\vect{b}$ are inverses in the sense that the components
${h_a}^\mu$ and ${b^a}_\mu$ satisfy the conditions (\ref{hinverse}).

Given their transformation properties, the components of $\vect{h}$
and $\vect{b}$ are able to convert between GCT tensor behaviour and
the corresponding $\mbox{SL}(2,C)$ behaviour, i.e. they can be used to
replace Greek indices with Latin ones and vice-versa (although
$\mbox{SL}(2,C)$ spinors cannot be converted into any GCT tensorial
objects). As an example, let us consider the simplest case of a vector
field $\vect{J}$ defined on the Minkowski spacetime, which may be
written in terms of its components in the coordinate basis or its dual
as $\vect{J}=J^\mu\vect{e}_\mu=J_\mu\vect{e}^\mu$.  Replacing the
Greek index with a Latin one using the components of $\vect{h}$, we
may obtain the quantities ${\cal J}_a \equiv {h_a}^\mu J_\mu =
h_{a\mu}J^\mu$ or ${\cal J}^a \equiv {h^a}_\mu J^\mu
=h^{a\mu}J_\mu$. Adopting the viewpoint that $\vect{h}$ is a tensor
(gauge) field in Minkowski spacetime, and thus attaching to it no
geometric significance {\em a priori}, one must regard these as the
components in the local tetrad basis of a {\em new} vector field
$\bm{\mathcal{J}} = \vect{h}(\cdot,\vect{J})$\cite{footnote6}. It is also worth
noting that the components of $\vect{h}$ may also be used to convert
Latin indices into Greek ones. For example, consider the vector
$\vect{h}(\bm{\mathcal{J}},\cdot)$, which has contravariant components
in the coordinate basis given by ${h_a}^\mu\mathcal{J}^a$; these are
the components of another {\em new} vector, and are, in general, {\em
  not} equal to the original $J_\mu$.

Similar conclusions follow for the inverse tensor $\vect{b}$. For
example, if we place $\bm{\mathcal{J}}$ into the first `slot' of
$\vect{b}$, we recover the {\em original} vector $\vect{J}=
\vect{b}(\bm{\mathcal{J}},\cdot)$, since its covariant components in
the coordinate basis are ${b^a}_\mu \mathcal{J}_a = J_\mu$, where in
the last equality we have used the reciprocity relations
(\ref{hinverse}). These components can also be written $J_\mu =
b_{a\mu}\mathcal{J}^a$, and their contravariant counterparts as $J^\mu
= {b_a}^\mu\mathcal{J}^a = b^{a\mu}\mathcal{J}_a$. In all these cases,
the components of $\vect{b}$ have been used to convert a Latin index
into a Greek one. One may also perform the reverse operation:
consider, for example, the vector $\vect{b}(\cdot,\vect{J})$; this has
contravariant components in the tetrad basis given by ${b^a}_\mu
J^\mu$, but these are {\em not}, in general, equal to $\mathcal{J}^a$.

The index conversion properties of the components of $\vect{h}$ and
$\vect{b}$ trivially extend to objects possessing multiple indices:
for example, if $\vect{T}$ is a rank-2 tensor field, then one obtains
the {\em new} rank-2 tensor field $\bm{\mathcal{T}}$ with covariant
components (say) in the tetrad basis given by $\mathcal{T}_{ab} =
{h_a}^\mu {h_b}^\nu T_{\mu\nu}$. Using the reciprocity relations
(\ref{hinverse}), one may therefore also write, for example,
$T_{\mu\nu} = {b^a}_\mu {b^b}_\nu \mathcal{T}_{ab}$. One may also, of
course, convert just some subset of the indices on a higher-rank
object, and each case corresponds to the components of a {\em
  different} object, referred to the corresponding combination of
tetrad and coordinate basis vectors\cite{footnote7}.

The interpretation presented above contrasts sharply with the (more
commonly adopted) geometric interpretation of WGT (which we describe
in detail in Section~\ref{sec:geowgt}), in which ${h_a}^\mu$ is
identified as a vierbein that directly relates the coordinate basis
vectors $\vect{e}_\mu$ and local tetrad frame vectors $\hvect{e}_a$;
in this case, the components ${h_a}^\mu J_\mu$, for example, are
interpreted as the components $J_a$ of the {\em original} vector field
$\vect{J}$ in the local Lorentz basis.
Moreover, in the Minkowski
spacetime gauge field interpretation, the RHS of (\ref{dladef}) need
{\em not} vanish, which should be contrasted with the usual `tetrad
postulate' that applies in the geometric interpretation of WGT.

\subsection{Gauge field strengths in WGT}
\label{sec:wgtgfs}

The gauge field strengths in WGT are defined in the usual way in terms
of the commutator of the covariant derivatives. Considering first the
$(\Lambda,\rho)$-covariant derivative, one finds
\begin{equation}
[D^\ast_\mu,D^\ast_\nu]\vpsi =
\tfrac{1}{2}{R^{ab}}_{\mu\nu}\Sigma_{ab}\vpsi + wH_{\mu\nu}\vpsi,
\label{dstarmucomm}
\end{equation}
where we have defined the `rotational' and `dilation' field strength
tensors of the gauge fields ${A^{ab}}_\mu$ and $B_\mu$, respectively,
as
\begin{eqnarray}
\hspace*{-7mm}{R^{ab}}_{\mu\nu}  & \equiv & \partial_\mu {A^{ab}}_\nu \!-\! \partial_\nu {A^{ab}}_\mu
\!+\!{A^a}_{c\mu}{A^{cb}}_\nu \!-\! {A^a}_{c\nu}{A^{cb}}_\mu,
\label{rfsdef} \\
H_{\mu\nu} & \equiv & \partial_\mu B_\nu - \partial_\nu B_\mu.
\label{dfsdef}
\end{eqnarray}
Both ${R^{ab}}_{\mu\nu}$ and $H_{\mu\nu}$ transform covariantly under
GCT and local Lorentz rotations in accordance with their respective
index structures, and are invariant under local dilations (i.e. each
has Weyl weight $w=0$).

Of greater relevance, however, is the commutator of two generalised
covariant derivatives.  Since ${\cal D}^\ast_a\vpsi ={h_a}^\mu
D^\ast_\mu\vpsi$, this commutator differs from (\ref{dstarmucomm}) by
an additional term containing the derivatives of ${h_a}^\mu$, and
reads
\begin{equation}
[{\cal D}^\ast_c,{\cal D}^\ast_d]\vpsi =
\tfrac{1}{2}{{\cal R}^{ab}}_{cd}\Sigma_{ab}\vpsi + w{\cal H}_{cd}\vpsi
- {{\cal T^\ast}^a}_{cd}{\cal D}^\ast_a\vpsi,
\label{weylfsdefs}
\end{equation}
where ${{\cal R}^{ab}}_{cd} \equiv
{h_a}^{\mu}{h_b}^{\nu}{R^{ab}}_{\mu\nu}$ and ${\cal H}_{cd} = {h_c}^\mu
{h_d}^\nu H_{\mu\nu}$, and the `translational' field strength of the
gauge field ${h_a}^\mu$ is given by
\begin{equation}
{{\cal T^\ast}^a}_{bc} \equiv {h_b}^{\mu}{h_c}^{\nu} {T^{\ast
    a}}_{\mu\nu} \equiv {h_b}^{\mu}{h_c}^{\nu}(D^\ast_\mu {b^a}_\nu
-D^\ast_\nu {b^a}_\mu),
\label{cstardef}
\end{equation}
which transforms covariantly under local Lorentz transformations in
accordance with its index structure.  It is worth noting
(\ref{cstardef}) can also be written
\begin{equation}
{{\cal T}^{\ast a}}_{bc} = -{b^a}_\mu({\cal D}^\ast_b{h_c}^\mu
- {\cal D}^\ast_c{h_b}^\mu).
\label{cstardef2}
\end{equation}
One may easily show that the translational field strength in WGT is
related to the corresponding field strength ${{\cal T}^{a}}_{bc}$
in PGT (see Appendix~\ref{app:pgt}), by
\begin{equation}
{{\cal T}^{\ast a}}_{bc} = {{\cal T}^{a}}_{bc} + \delta^a_c{\cal
  B}_b-\delta^a_b{\cal B}_c,
\label{tstarfromt}
\end{equation}
where ${\cal B}_a = {h_a}^\mu B_\mu$. It is also convenient to define ${\cal T}^\ast_b \equiv {{\cal
    T}^{\ast a}}_{ba} = {\cal T}_b + 3 {\cal B}_b$, where $ {\cal
  T}_b$ is the corresponding PGT quantity.

It is easy to show that ${{\cal R}^{ ab}}_{cd}$, ${\cal H}_{cd}$ and
${{\cal T}^{\ast a}}_{cd}$ transform covariantly under local
dilations, with weights $w({{\cal R}^{ab}}_{cd})=w({\cal H}_{cd})=-2$
and $w({{\cal T}^{\ast a}}_{cd})=-1$ respectively. We further note
that the functional dependencies of the three field strengths tensors
on the gauge fields and their first derivatives are ${{\cal
    R}^{ab}}_{cd}(h, A,\partial A)$, ${{\cal T}^{\ast
    a}}_{bc}(h,\partial h, A,B)$ and ${\cal H}_{ab}(h,\partial B)$.
When WGT is re-interpreted geometrically (see
Section~\ref{sec:geowgt}), the field strengths ${{\cal R}^{ab}}_{cd}$,
${\cal H}_{ab}$ and ${{\cal T}^{\ast a}}_{bc}$ are related to the {\em
  curvature} and {\em torsion} tensors of a Weyl--Cartan $Y_4$
spacetime, but here are to be regarded simply as local tensors defined
in Minkowski spacetime.

\subsection{Alternative form of covariant derivative in WGT}
\label{sec:wgtcovdalt}

An immediate use of the expression (\ref{cstardef2}) is that it can be
rearranged to give an important expression for the $A$ gauge field.
Defining the quantities
\begin{equation}
{c^{\ast a}}_{bc} \equiv {h_b}^\mu
{h_c}^\nu (\partial^\ast_\mu {b^a}_\nu-\partial^\ast_\nu {b^a}_\mu),
\label{riccicoeffsdef}
\end{equation}
the relevant relationship is easily found to be
$A_{ab\mu} = {b^c}_\mu {\cal A}_{abc}$, where
\begin{equation}
{\cal A}_{abc} = \tfrac{1}{2}(c^\ast_{abc}+c^\ast_{bca}-c^\ast_{cab})
-\tfrac{1}{2}({\cal T}^\ast_{abc}+{\cal T}^\ast_{bca}-{\cal T}^\ast_{cab}).
\label{afromht}
\end{equation}
It is important to note that this is an {\em identity} that is valid
{\em independently} of any equations of motion satisfied by the gauge
fields $h$, $A$ and $B$.  In the special case where the equations of
motion result in ${\cal T}^\ast_{abc}$ being independent of ${\cal
  A}_{abc}$, (\ref{afromht}) gives an explicit expression for the
$A$-field in terms of the $h$-field, its first derivatives and the
$B$-field. In a such a case, the $A$-field is no longer an independent
gauge field and the resulting theory can, if desired, be written
entirely in terms of the other gauge fields, i.e. one moves from a
`first-order' to a `second-order' formalism.

It will prove convenient for our later discussion to rewrite the
expression (\ref{afromht}) as
\begin{equation}
{\cal A}_{abc} = \zero{{\cal A}^\ast_{abc}}(h,\partial h,B)
+{\cal K}^\ast_{abc} (h,\partial h,A,B),
\label{astarzerodef}
\end{equation}
in which we have defined the quantities $\zero{{\cal A}^\ast_{abc}}
\equiv \tfrac{1}{2}(c^\ast_{abc}+c^\ast_{bca}-c^\ast_{cab})$ that, as
indicated, are fully determined by the $h$-field, its first
derivatives, and the $B$-field. We have also defined ${\cal
  K}^\ast_{abc} \equiv -\tfrac{1}{2} ({\cal T}^\ast_{abc}+{\cal
  T}^\ast_{bca}-{\cal T}^\ast_{cab})$, which when re-interpreted
geometrically is analogous to the {\em contortion} tensor of a
Weyl--Cartan spacetime $Y_4$.  It is worth noting that both
$\zero{{\cal A}^\ast_{abc}}$ and ${\cal K}^\ast_{abc}$ are
antisymmetric in their first two indices\cite{footnote8}.

Under a local Weyl transformation, the quantities $\zero{{\cal
    A}^\ast_{abc}}$ transform in the same way as ${\cal
  A}^\ast_{abc}$, whereas ${\cal K}^\ast_{abc}$ transform as the
components of a local tensor with weight $w=-1$. Thus, the quantities
$\zero{{\cal A}^\ast_{abc}}$ can serve equally well as ${\cal
  A}^\ast_{abc}$ in the construction of a covariant derivative.  We
may therefore construct the `reduced' $(\Lambda,\rho)$-covariant
derivative
\begin{equation}
\zero{D}^\ast_\mu\vpsi \equiv (\partial^\ast_\mu
+{\textstyle\frac{1}{2}} \zero{{A^{\ast ab}}_\mu}\Sigma_{ab})\vpsi,
\label{dstarmuzerodef}
\end{equation}
which transforms according to (\ref{eqn:dstarprime}) under a local
Weyl transformation, in the same way as $D^\ast_\mu\vpsi$, but depends
only on the $h$ field, its first derivatives, and the
$B$-field. Similarly, one can define the corresponding generalised
covariant derivative ${^0}{\cal D}^\ast_a \equiv {h_a}^\mu
\,\zero{D}^\ast_\mu$, which consequently transforms as in
(\ref{weylgencovtrans}), as required. The `full' generalised covariant
derivative is given in terms of the `reduced' one by the
alternative form
\begin{equation}
\mathcal{D}^\ast_a\vpsi = (\czero{\mathcal{D}}^\ast_a +
\tfrac{1}{2}{\mathcal{K}^{\ast bc}}_a\Sigma_{bc})\vpsi,
\label{wgtcovderivalt},
\end{equation}
which makes clear that, if one (covariantly) sets the WGT torsion (and
hence contortion) to zero, then $\mathcal{D}^\ast_a$ reduces to
$\czero{\mathcal{D}}^\ast_a$.

\subsection{Bianchi identities in WGT}
\label{sec:wgtbianchi}

It will also be convenient for our later development to obtain the
Bianchi identities satisfied by the gravitational gauge field
strengths ${{\cal R}^{ab}}_{cd}$, ${{\cal T}^{\ast a}}_{bc}$ and
${\cal H}_{ab}$ in WGT. These may be straightforwardly derived from
the Jacobi identity applied to the generalised covariant derivative,
namely
\begin{equation}
[{\cal D}^\ast_a,[{\cal D}^\ast_b,{\cal D}^\ast_c]]\vpsi +
[{\cal D}^\ast_c,[{\cal D}^\ast_a,{\cal D}^\ast_b]]\vpsi +
[{\cal D}^\ast_b,[{\cal D}^\ast_c,{\cal D}^\ast_a]]\vpsi =0.
\end{equation}
Inserting the form (\ref{weylgencovdef}) for the generalised covariant
derivative into the above Jacobi identity, one quickly finds the three
Bianchi identities
\begin{eqnarray}
{\cal D}^\ast_{[a}{{\cal R}^{de}}_{bc]}-{{\cal T}^{\ast
    f}}_{[ab} {{\cal R}^{de}}_{c]f} & = & 0, \label{wgtbi1} \\
{\cal D}^\ast_{[a}{{\cal T}^{\ast d}}_{bc]}-{{\cal T}^{\ast
    e}}_{[ab} {{\cal T}^{\ast d}}_{c]e}-{{\cal R}^{d}}_{[abc]}+ {\cal H}_{[ab}\delta^d_{c]}
 & = & 0, \label{wgtbi2} \\
{\cal D}^\ast_{[a}{{\cal H}}_{bc]} -{{\cal T}^{\ast
    e}}_{[ab} {{\cal H}}_{c]e} & = & 0. \label{wgtbi3}
\end{eqnarray}
It is worth noting that for the special case in which the `torsion' is
totally antisymmetric, such that ${\cal T}^\ast_{abc}= \epsilon_{abcd}
t^d$ for some vector $t^d$, then the second term on
the LHS of (\ref{wgtbi2}) vanishes.

By contracting over the indices $a$ and $d$ in the `${\cal R}$-identity'
(\ref{wgtbi1}), once obtains the once-contracted Bianchi
identity
\begin{equation}
{\cal D}^\ast_{a}{{\cal R}^{ae}}_{bc}-2
{\cal D}^\ast_{[b}{{\cal R}^{e}}_{c]}
-2{{\cal T}^{\ast
    f}}_{a[b} {{\cal R}^{ae}}_{c]f}
-{{\cal T}^{\ast
    f}}_{bc} {{\cal R}^{e}}_{f}
 =  0. \label{wgtcbi1}
\end{equation}
Contracting again over $b$ and $e$, one then finds the
twice-contracted Bianchi identity
\begin{equation}
{\cal D}^\ast_{a}({{\cal R}^{a}}_{c}
-\tfrac{1}{2}\delta^a_c{\cal R})
+{{\cal T}^{\ast
    f}}_{bc} {{\cal R}^{b}}_{f}
+\tfrac{1}{2}{{\cal T}^{\ast
    f}}_{ab} {{\cal R}^{ab}}_{cf}
 =  0. \label{wgtcbi2}
\end{equation}

Turning to the `${\cal T}$-identity' (\ref{wgtbi2}) and contracting
over $a$ and $d$, one obtains the further once-contracted Bianchi
identity
\begin{equation}
{\cal D}^\ast_{a}{{\cal T}^{\ast a}}_{bc}
+2{\cal D}^\ast_{[b}{\cal T}^\ast_{c]}
+{{\cal T}^{\ast e}}_{bc}{\cal T}^\ast_e
+2{{\cal R}}_{[bc]}
+ 2{\cal H}_{bc}
=  0. \label{wgtcbi3}
\end{equation}
One should note that (\ref{wgtcbi3}) simplifies considerably if the
torsion trace ${\cal T}^\ast_a$ vanishes, which clearly includes the
special case in which the torsion is totally antisymmetric.
It is clear that the `${\cal H}$-identity' (\ref{wgtbi3}) has no
non-trivial contractions.

\subsection{Free gravitational action in WGT}
\label{sec:wgtfga}

In addition to the locally-Weyl-invariant matter action
(\ref{weylmatlag_gen}), one must also include in the total action
some further terms, which describe the dynamics of the free
gravitational gauge fields. From ${\cal R}_{abcd}$, ${\cal
  T}^\ast_{abc}$ and ${\cal H}_{ab}$, one can construct a free
gravitational action of the general form
\begin{equation}
S_{\rm G} = \int h^{-1} L_{\rm G}({\cal R}_{abcd},{\cal T}^\ast_{abc},{\cal
  H}_{ab})\,d^4x,
\label{wgtgaction}
\end{equation}
where the requirement of local scale invariance imposes the constraint
that $L_{\rm G}$ must be a relative scalar with Weyl weight $w(L_{\rm
  G})=-4$, so that the Lagrangian density ${\cal L}_{\rm G} \equiv
h^{-1} L_{\rm G}$ has weight zero (i.e scale invariant).  It is
trivial to show that ${\cal R}_{abcd}$, ${\cal H}_{ab}$ and ${\cal
  T}^\ast_{abc}$ transform covariantly under local dilations with Weyl
weights $w({\cal R}_{abcd})=w({\cal H}_{ab})=-2$ and $w({\cal
  T}^\ast_{abc})=-1$ respectively\cite{footnote9}. Hence $L_{\rm G}$
can be quadratic in ${\cal R}_{abcd}$ and ${\cal H}_{ab}$, but terms
linear in ${\cal R} \equiv {{\cal R}^{ab}}_{ab}$ or quadratic in
${\cal T}^\ast_{abc}$ are not allowed. Similarly, higher-order terms
in ${\cal R}_{abcd}$ and ${\cal H}_{ab}$ are forbidden; in principle
one could include quartic terms in ${\cal T}^\ast_{abc}$, but we will
not consider them here. Thus, in WGT, the general form of $L_{\rm G}$,
possessing terms no higher than quadratic order in the field
strengths, is of the form
\begin{equation}
L_{\rm G} = L_{{\cal R}^2} + L_{{\cal H}^2},
\label{wgtgravlagdef}
\end{equation}
where the expressions for $L_{{\cal R}^2}$ and $L_{{\cal H}^2}$ are
given in (\ref{lg2}).  It may be shown, however, that
the field strength ${\cal R}_{abcd}$ satisfies a form of Gauss--Bonnet
identity\cite{nieh80}, such that the combination
\begin{equation}
{\cal R}^2-4{\cal R}_{ab}{\cal R}^{ba}+{\cal R}_{abcd}{\cal R}^{cdab}
\label{gbid}
\end{equation}
contributes a total derivative to the action (in $D \le 4$
dimensions), and so has no effect on the resulting field
equations. Hence one may set any one of $\alpha_1$, $\alpha_3$ or
$\alpha_6$ in (\ref{lg2}) to zero, without loss of generality (at
least classically), but we will retain all these terms for the
moment. As mentioned in Section~\ref{sec:lmwgt}, we note that one may
also include in the total Lagrangian terms of the generic form
$\phi^2{\cal R}$ and $\phi^2 L_{{\cal T}^{\ast 2}}$, where $L_{{\cal
    T}^{\ast 2}}$ is given in (\ref{lt2}), but with ${\cal T}_{abc}
\to {\cal T}^\ast_{abc}$. However, since these terms depend upon the
scalar (compensator) field $\phi$, in addition to the gravitational
gauge fields, we do not consider them to be part of the {\em free}
gravitational action, but instead regard them as belonging to the
matter Lagrangian $L_{\rm M}$ (see Section~\ref{sec:wgtdirac}).

The precise form of WGT under consideration depends on the form of the
free gravitational Lagrangian density ${\cal L}_{\rm G}$.  As an
illustration, in Appendix~\ref{sec:diracgrav} we give a brief account
of a scale-invariant theory of gravity suggested by Dirac (but
extended to include torsion), which can accommodate `ordinary' matter
via the inclusion of an additional `compensator' scalar field. In this
theory, $L_{\rm G} = L_{{\cal H}^2}$ and $L_{\rm M}$ contains a term
proportional to $\phi^2{\cal R}$.

\subsection{Field equations in WGT}
\label{sec:wgtfieldeqns}

The total action $S_{\rm T}$ is simply the sum of the matter and free
gravitational actions. In the free gravitational sector, the form of
the Lagrangian (\ref{wgtgravlagdef}) induces a
dependence on $h$, $A$, $\partial A$ and $\partial B$
(suppressing indices for brevity).  Note, in particular, that the
absence of ${\cal T}^{\ast}_{abc}$ from (\ref{wgtgravlagdef}) means
that it does not contain derivatives of the $h$-field.  In the matter
sector, covariant derivatives of the matter field $\vpsi$ induce a
dependence on $\vpsi$, $\partial\vpsi$, $h$, $A$ and $B$.  We will
also consider here the inclusion of an additional `compensator' scalar
field $\phi$ in the matter action, and further admit the possibility
that it may include a kinetic term for the scalar field that contains
derivatives of $\phi$.  Moreover, if the matter action includes a term
proportional to $\phi^2{\cal R}$, then this brings an additional
dependence on $\partial A$.  Finally, if one includes terms of the
generic form $\phi^2L_{{\cal T}^{\ast 2}}$, then it produces an
additional dependence on $\partial h$.  Consequently, we will take the
total Lagrangian density to be
\begin{equation}
{\cal L}_{\rm T} \!=\! {\cal L}_{\rm G}(h,A,\partial A,\partial B) +
{\cal L}_{\rm M}(\vpsi,\partial\vpsi,\phi,\partial\phi,h,\partial h,
A,\partial A, B),
\label{wgtlagtot}
\end{equation}
where we have indicated the functional dependencies in the
most general case.

Since ${\cal L}_{\rm T}$ is at most quadratic in the field strength
tensors ${\cal R}_{abcd}(h,A,\partial A)$, ${\cal
  T}^\ast_{abc}(h,\partial h,A,B)$ and ${\cal H}_{ab}(h,\partial B)$,
and each of these is linear in the first derivative of the
corresponding gauge field, the resulting equations of motion are, in
general, linear in the second-order derivatives of the $h$, $A$ and
$B$ gauge fields, and contain no higher-derivative terms; such
theories therefore do not suffer from Ostrogradsky's instability. It
is worth noting, however, that if one substitutes for ${\cal A}_{abc}$
using (\ref{astarzerodef}), one obtains equations of motion that
contain up to third-order derivatives of the $h$-field, which can lead
to difficulties in the interpretation of initial-value problems.

Variation of $S_{\rm T}$ with respect to the gauge fields ${h_a}^{\mu}$,
${A^{ab}}_{\mu}$ and $B_\mu$ leads to three coupled gravitational
field equations for the gauge fields. In general, these read
\begin{subequations}
\label{eqn:wgenfe}
\begin{eqnarray}
{t^a}_\mu + {\tau^a}_\mu  & = & 0, \label{eqn:wgenfe1}\\
{s_{ab}}^\mu + {\sigma_{ab}}^\mu  & = & 0,\label{eqn:wgenfe2}\\
j^\mu+ \zeta^\mu & = & 0.\label{eqn:wgenfe3}
\end{eqnarray}
\end{subequations}
In the free gravitational sector, we have defined ${t^a}_\mu \equiv
\delta {\cal L}_{\rm G}/\delta {h_a}^\mu = \partial{\cal L}_{\rm
  G}/\partial {h_a}^\mu$, ${s_{ab}}^\mu \equiv \delta {\cal L}_{\rm
  G}/\delta {A^{ab}}_\mu$ and $j^\mu \equiv \delta {\cal L}_{\rm
  G}/\delta B_\mu$, respectively. In the matter sector, the
energy-momentum ${\tau^a}_\mu\equiv \delta{\cal L}_{\rm M}/\delta
{h_a}^\mu$, spin-angular-momentum ${\sigma_{ab}}^\mu\equiv \delta{\cal
  L}_{\rm M}/\delta {A^{ab}}_\mu$ and dilation current
$\zeta^\mu\equiv\delta{\cal L}_{\rm M}/\delta B_\mu=\partial{\cal
  L}_{\rm M}/\partial B_\mu$ act as sources\cite{footnote10}. It is
easy to show that all the quantities in
(\ref{eqn:wgenfe1})--(\ref{eqn:wgenfe3}) transform covariantly under
local Lorentz rotations and GCTs in accordance with their respective
index structures, and are also covariant under local dilations with
Weyl weights $w=1$, $w=0$ and $w=0$, respectively.  It is, in fact,
more convenient to work with quantities carrying only Latin indices
and write the general field equations as
\begin{subequations}
\label{eqn:weylgenfe}
\begin{eqnarray}
{t^a}_b + {\tau^a}_b  & = & 0, \label{eqn:weylgenfe1}\\
{s_{ab}}^c + {\sigma_{ab}}^c  & = & 0,\label{eqn:weylgenfe2}\\
j^a+ \zeta^a & = & 0,\label{eqn:weylgenfe3}
\end{eqnarray}
\end{subequations}
where ${t ^a}_b \equiv {t^a}_\mu {h_b}^\mu$, ${s_{ab}}^c \equiv
{s_{ab}}^\mu {b^c}_\mu$ and $j^a\equiv j^\mu {b^a}_\mu$, and similarly
for the matter sector\cite{footnote11}. All the quantities in
(\ref{eqn:weylgenfe1})--(\ref{eqn:weylgenfe3}) are clearly invariant
under GCT, transform covariantly under local Lorentz rotations in
accordance with their respective index structures, and are also
covariant under local dilations with Weyl weights $w=0$, $w=1$ and
$w=1$, respectively.

Varying $S_{\rm T}$ with respect to $\vpsi$ and $\phi$ leads to the
matter field equations. Considering just the $\vpsi$-equation for the
moment, since ${\cal L}_{\rm G}$ does not depend on $\vpsi$ one has simply
\begin{equation}
\frac{\delta {\cal L}_{\rm M}}{\delta\vpsi}
\equiv \pd{{\cal L}_{\rm M}}{\vpsi}
- \partial_\mu\left(\pd{{\cal L}_{\rm M}}{(\partial_\mu\vpsi)}\right)
=0.
\end{equation}
One may straightforwardly show that this equation can be recast in
terms of the $(\Lambda,\rho)$-covariant derivative (\ref{weyldmudef}) as
\begin{equation}
\frac{\delta {\cal L}_{\rm M}}{\delta\vpsi}
\equiv \frac{\bar{\partial}{{\cal L}_{\rm M}}}{\partial\vpsi}
- D^\ast_\mu\left(\pd{{\cal L}_{\rm M}}{(D^\ast_\mu\vpsi)}\right)
=0,
\label{eomdmuform}
\end{equation}
where $\bar{\partial}{\cal L}_{\rm M}/\partial \vpsi \equiv [\partial
  {\cal L}_{\rm M}(\vpsi,{\cal D}^\ast_\mu u,\ldots)/\partial
  \vpsi]_{u=\vpsi}$, so that $\vpsi$ and ${\cal D}^\ast_\mu \vpsi$ are
treated as independent variables, rather than $\vpsi$ and
$\partial_\mu\vpsi$. Even more convenient for our purposes is to
rewrite (\ref{eomdmuform}) in terms of the generalised covariant derivative
(\ref{weylgencovdef}), which yields
\begin{equation}
\bpd{L_{\rm M}}{\vpsi}-{\cal D}^\ast_a\left(\pd{L_{\rm M}}{({\cal D}^\ast_a\vpsi)}\right)= hD^\ast_\mu(h^{-1}{h_a}^\mu)\pd{L_{\rm M}}{({\cal D}^\ast_a\vpsi)},
\label{peqn}
\end{equation}
where we have defined ${\cal L}_{\rm M} = h^{-1}L_{\rm M}$ and $\bar{\partial}L_{\rm
  M}/\partial \vpsi \equiv [\partial L_{\rm M}(\vpsi,{\cal
    D}^\ast_au,\ldots)/\partial \vpsi]_{u=\vpsi}$, so that $\vpsi$ and
${\cal D}^\ast_a\vpsi$ are treated as independent variables.
One can develop the RHS of (\ref{peqn}) further, and put it into
manifestly covariant form, by noting that
\begin{equation}
hD^\ast_\mu(h^{-1}{h_a}^\mu)={b^b}_\mu({\cal D}^\ast_b{h_a}^\mu-{\cal
  D}^\ast_a{h_b}^\mu) = {{\cal T}^{\ast b}}_{ab} \equiv {\cal T}^\ast_a,
\label{htrelation}
\end{equation}
where we have used the expression (\ref{cstardef2}) for the field
strength of the translational gauge field. An analogous procedure can be
applied to the $\phi$ field, and so the matter equations of motion can
be written in the convenient forms
\begin{subequations}
\label{eqn:wgtmattereoms}
\begin{eqnarray}
\frac{\delta L_{\rm M}}{\delta\vpsi} =
\bpd{L_{\rm M}}{\vpsi}-({\cal D}^\ast_a + {\cal T}^\ast_a)
\left(\pd{L_{\rm M}}{({\cal D}^\ast_a\vpsi)}\right) & = & 0, \label{weylpeqn} \\
\frac{\delta L_{\rm M}}{\delta\phi} =
\bpd{L_{\rm M}}{\phi}-({\cal D}^\ast_a + {\cal T}^\ast_a)
\left(\pd{L_{\rm M}}{({\cal D}^\ast_a\phi)}\right) & = & 0.
\label{weylpeqnphi}
\end{eqnarray}
\end{subequations}
It is worth noting that these field equations, derived from the
minimal-coupling procedure applied to the action $S_{\rm M}$, are {\em
  not}, in general, equivalent to those obtained simply by applying
the minimal-coupling procedure to the field equations
directly\cite{saa96,lasenby98}, which would be given by
(\ref{weylpeqn}) and (\ref{weylpeqnphi}) {\em without} the terms
containing ${\cal T}^\ast_a$.  We also note that the general forms
(\ref{weylpeqn}) and (\ref{weylpeqnphi}) are valid only if the action
(\ref{weylmatlag_gen}) has not subsequently been modified to satisfy
any further required invariances that were lost in the localisation of
the Weyl symmetry; this is illustrated in Section~\ref{sec:wgtem}.

\subsection{Conservation laws in WGT}
\label{sec:wgtconslaws}

The conservation laws for WGT are also easily obtained. Contrary to
the approach usually adopted in the literature, it is most
useful to present them in terms of quantities carrying only Latin
indices. Invariance of $S_{\rm G}$ under (infinitesimal) local Lorentz
(or $\mbox{SL}(2,C)$) rotations, GCTs and local dilations,
respectively, lead to the following manifestly covariant conservation
laws among the contributions of the free gravitational sector to the
gravitational field equations:
%
\begin{subequations}
\label{eqn:wgtconslaws}
\begin{eqnarray}
({\cal D}^\ast_c + {\cal T}^\ast_c) (h{s_{ab}}^c) + ht_{[ab]} \!& = & \!0,
\label{eqn:wgtcons1}\\
\hspace*{-10mm}({\cal D}^\ast_c \!+\! {\cal T}^\ast_c) (h{t^c}_d)
\!-\! h({s_{ab}}^c{{\cal R}^{ab}}_{cd}
\!-\! {t^c}_b {{\cal T}^{\ast b}}_{cd} \!+\! j^c {\cal H}_{cd}) \!& = & \!0,
\label{eqn:wgtcons2} \\
({\cal D}^\ast_c + {\cal T}^\ast_c) (hj^c)- h{t^c}_c \!& =& \!0.
\label{eqn:wgtcons3}
\end{eqnarray}
\end{subequations}
Similarly, the local Lorentz, GCT and local dilation invariance
properties of $S_{\rm M}$ lead, respectively, to the following manifestly
covariant identities among the contributions of the matter sector to
the gravitational field equations:
\begin{widetext}
\begin{subequations}
\label{eqn:wgtlmcons}
\begin{eqnarray}
({\cal D}^\ast_c + {\cal T}^\ast_c)(h{\sigma_{ab}}^c)
+ h\tau_{[ab]} +\tfrac{1}{2}\frac{\delta L_{\rm M}}{\delta\vpsi}
\Sigma_{ab}\vpsi & = & 0, \label{wgtlmcons1}\\
({\cal D}^\ast_c + {\cal T}^\ast_c)(h{\tau^c}_d)
- h({\sigma_{ab}}^c {{\cal R}^{ab}}_{cd}
- {\tau^c}_b {{\cal T}^{\ast b}}_{cd}
+ \zeta^c {\cal H}_{cd}) +
\frac{\delta L_{\rm M}}{\delta\phi}
{\cal D}^\ast_d \phi
+ \frac{\delta L_{\rm M}}{\delta\vpsi}
{\cal D}^\ast_d\vpsi & = & 0,\label{wgtlmcons2}\\
({\cal D}^\ast_c + {\cal T}^\ast_c)(h\zeta^c) -
h{\tau^c}_c - \frac{\delta L_{\rm M}}{\delta\phi} \phi + \frac{\delta
  L_{\rm M}}{\delta\vpsi} w\vpsi & = & 0. \label{wgtlmcons3}
\end{eqnarray}
\end{subequations}
\end{widetext}
Thus, one is assured with the help of the matter field equations that
the gravitational field equations become consistent. Moreover, it is
worth noting that the above sets of conservation laws hold for {\em
  any subset} of terms in $L_{\rm G}$ and $L_{\rm M}$, respectively,
that is covariant with weight $w=-4$ under local Weyl transformations.

\subsection{Dirac matter field in WGT}
\label{sec:wgtdirac}

So far, we have not introduced a specific form for the matter
Lagrangian. To illustrate the coupling of matter to the gravitational
gauge fields, let us consider the case of a Dirac field, which will be
useful for our later developments.  Following our discussion in
Section~\ref{sec:globalweyl}, we begin with the special-relativistic
Lagrangian (\ref{diracphilag}) for a classical Dirac field $\psi$
coupled to a (compensator) scalar field $\phi$. For the moment,
however, we will consider only the $\psi$-dependent terms and hence
omit the kinetic term for $\phi$ to obtain
\begin{equation}
L_{\rm D} =
\tfrac{1}{2}i\bar{\psi}\gamma^\mu{\stackrel{\leftrightarrow}{\partial_\mu}}\psi
- \mu\phi\bar{\psi}\psi,
\label{diraclm3}
\end{equation}
where $\mu$ is a dimensionless parameter, but $\mu\phi$ has the
dimensions of mass in natural units. The action constructed from
(\ref{diraclm3}) is invariant under global Weyl transformations,
provided $\psi$ and $\phi$ have the Weyl weights $w=-\frac{3}{2}$ and
$w=-1$, respectively; we note that the
action is also invariant under the global phase transformation $\psi
\to \psi e^{i\alpha}$.

Following our general procedure for constructing a
locally-Weyl-invariant matter action (\ref{weylmatlag_gen}) and
applying it to (\ref{diraclm3}), the appropriate form for the
corresponding Lagrangian density is
\begin{equation}
{\cal L}_{\rm D} = h^{-1} L_{\rm D} =
h^{-1}(\tfrac{1}{2}i
\bar{\psi}\gamma^a{\stackrel{\leftrightarrow}{{\cal D}^\ast_a}}\psi -
\mu\phi\bar{\psi}\psi),
\label{weyldiracaction}
\end{equation}
where, on each $\gamma$-matrix, one has simply replaced the Greek
index with a Roman one, without altering the matrix (since the
$\gamma$-matrices naturally refer to the local Lorentz basis). It is
clear that (\ref{weyldiracaction}) has also inherited the invariance
of the original Dirac action (\ref{diraclm3}) under a global phase
transformation.  The first point to note is that, since both $\psi$
and $\bar{\psi}$ have Weyl weight $w=-\tfrac{3}{2}$, the dilation
gauge field $B_\mu$ interacts in the same manner with each of them,
thereby ruling out the interpretation of $B_\mu$ as the
electromagnetic potential. Moreover, rather curiously, $B_\mu$
vanishes completely from (\ref{weyldiracaction}). Consequently, the
WGT covariant derivative operator ${\cal D}^\ast_a$ in the kinetic
term in (\ref{weyldiracaction}) can be replaced by its PGT counterpart
${\cal D}_a$, so that
\begin{equation}
{\cal L}_{\rm D} = h^{-1}(\tfrac{1}{2}i
\bar{\psi}\gamma^a{\stackrel{\leftrightarrow}{{\cal D}_a}}\psi
- \mu\phi\bar{\psi}\psi).
\label{diracactionagain}
\end{equation}
Put another way, the kinetic term in the Dirac action in PGT is {\em
  already} also invariant to local dilations.

To illustrate our general approach, however, for the moment we will
consider the manifestly WGT-covariant form (\ref{weyldiracaction}) of
the Lagrangian density.  Varying the corresponding action with respect
to $\bar{\psi}$, one obtains a field equation of the form
(\ref{weylpeqn}) (with $\vpsi$ replaced by $\bar{\psi}$), which is
immediately found to read
\begin{equation}
i\gamma^a({\cal D}^\ast_a+\tfrac{1}{2}{\cal T}^\ast_a)\psi-\mu\phi\psi = 0.
\label{weylcovardirac2}
\end{equation}
An equivalent adjoint field equation in $\bar{\psi}$ is obtained by
varying the action with respect to $\psi$.  Since ${\cal
  D}^\ast_a\psi$ and ${{\cal T}}^{\ast}_{a}$ transform covariantly
under local dilations with weights $w=-\tfrac{5}{2}$ and $w=-1$
respectively, we see immediately that the LHS of
(\ref{weylcovardirac2}) does indeed transform covariantly under local
dilations with weight $w=-\tfrac{5}{2}$. We note that, as expected,
the field equation (\ref{weylcovardirac2}) differs from that which
would be obtained by applying the minimal coupling procedure directly
at the level of the field equation; the latter would not contain the
term proportional to ${\cal T}^\ast_a$. Since the Dirac Lagrangian
density (\ref{weyldiracaction}) may also be expressed as
(\ref{diracactionagain}), however, one may use the
PGT version of (\ref{weylpeqn}) (obtained by removing all asterisks;
see Appendix~\ref{app:pgt}) to show that the field equation
(\ref{weylcovardirac2}) can be rewritten more simply in terms of
corresponding PGT quantities as
\begin{equation}
i\gamma^a({\cal D}_a+\tfrac{1}{2}{\cal T}_a)\psi-\mu\phi\psi = 0.
\label{weylcovardirac}
\end{equation}
Although this equation is still WGT covariant, this is no longer
manifest.  The covariance of (\ref{weylcovardirac}) is easily checked
directly by recalling that ${\cal D}_a\psi={\cal
  D}^\ast_a\psi+\tfrac{3}{2}{\cal B}_a\psi$ and ${\cal T}_a ={{\cal
    T}}^{\ast}_a-3{\cal B}_a$ to recover (\ref{weylcovardirac2}).  We
note that (\ref{weylcovardirac2}) and (\ref{weylcovardirac}) are
valid for any choice of the free gravitational Lagrangian density
${\cal L}_{\rm G}$, but the expression for ${\cal T}_a$, or ${\cal
  T}^\ast_a$, will depend on the form of ${\cal L}_{\rm G}$.

It is of interest to rewrite (\ref{weylcovardirac2}) using the
reduced covariant derivative discussed in
Section~\ref{sec:wgtcovdalt}.  Substituting the expression
(\ref{wgtcovderivalt}) into (\ref{weylcovardirac}) and using the fact
that the Lorentz group generator matrices for Dirac spinors are given
by $\Sigma_{ab} = \tfrac{1}{4}[\gamma_a,\gamma_b]$, after a short
calculation one finds that
\begin{equation}
i\gamma^a({^0}{\cal D}^\ast_a-\tfrac{1}{4}{\cal T}^\ast_{[abc]}\Sigma^{bc})\psi-\mu\phi\psi = 0.
\label{weylcovardiracalt2}
\end{equation}
This form of the Dirac equation is still manifestly WGT-covariant, but
reveals that only the total antisymmetric part of the WGT `torsion'
explicitly affects the dynamics of the Dirac field $\psi$. In other
words, if ${\cal T}^\ast_{[abc]}=0$ then (\ref{weylcovardiracalt2})
reduces to a form expressible entirely in terms of the `reduced'
covariant derivative (which depends only on the $h$-field, its
derivatives and the $B$-field), and is thus equivalent to the Dirac
equation obtained in the absence of torsion completely.
Alternatively, one can perform the analogous procedure on
(\ref{weylcovardirac}) to obtain the form
\begin{equation}
i\gamma^a({^0}{\cal D}_a-\tfrac{1}{4}{\cal
  T}_{[abc]}\Sigma^{bc})\psi-\mu\phi\psi = 0,
\label{weylcovardiracalt}
\end{equation}
in terms of the corresponding PGT quantities. Although
(\ref{weylcovardiracalt}) is not manifestly WGT-covariant, it provides
the simplest expression of the dynamics of the $\psi$ field for
calculational purposes, and shows analogously that only the total
antisymmetric part of the PGT `torsion' has an explicit effect.

From the Lagrangian density (\ref{diracactionagain}), we may also
calculate the energy-momentum and spin-angular-momentum tensors of the
Dirac field $\psi$. In terms of the corresponding quantities carrying only
Latin indices, one finds
\begin{eqnarray}
h{\tau^a}_b & = & \tfrac{1}{2}i\bar{\psi}\gamma^a{\stackrel{\leftrightarrow}{{\cal D}_b}}\psi
- \delta_b^a L_{\rm D},
\label{diracemt}\\
h{\sigma_{abc}} & = & \tfrac{1}{4}i\bar{\psi}\gamma_{[a}\gamma_b\gamma_{c]}\psi,
\label{diracsam}
\end{eqnarray}
where we have used the fact that $\Sigma_{ab} =
\tfrac{1}{4}[\gamma_a,\gamma_b]$ for Dirac spinors to establish the
second result.  It is easily shown that the energy-momentum tensor is
asymmetric and has the trace $\tau = h^{-1}\mu\phi\bar{\psi}\psi$,
whereas the spin-angular-momentum tensor is totally antisymmetric,
such that $\sigma_{abc} = \sigma_{[abc]}$. Consequently, all
contractions of the spin-angular-momentum tensor vanish for Dirac
matter.  As one might expect from the above discussion,
(\ref{diracemt}) may be rewritten in manifestly WGT-covariant form by
simply making the replacement ${\cal D}_a \to {\cal D}^\ast_a$ throughout; the
expression (\ref{diracsam}) is already manifestly WGT covariant.

Finally, we note that one might also consider adding the following to
the matter Lagrangian density (\ref{diracactionagain}): kinetic and
self-interaction terms for $\phi$ with Weyl weight $w=-4$ and, as
discussed in Sections~\ref{sec:lmwgt} and \ref{sec:wgtfga}, terms
proportional to $\phi^2{\cal R}$ and $\phi^2L_{{\cal T}^{\ast 2}}$,
which directly couple $\phi$ non-minimally (conformally) to the
gravitational gauge field strengths, where $L_{{\cal T}^{\ast 2}}$ is
given by (\ref{lt2}), but with ${{\cal T}^a}_{bc}$ replaced by ${{\cal
    T}^{\ast a}}_{bc}$. Thus, in the most general case, one might
consider
%
\begin{eqnarray}
{\cal L}_{\rm M} = h^{-1}[\tfrac{1}{2}i\bar{\psi}\gamma^a
{\stackrel{\leftrightarrow}{{\cal D}_a}}\psi - \mu\phi\bar{\psi}\psi
&&+ \tfrac{1}{2}\nu ({\cal D}^\ast_a\phi) ({\cal D}^{\ast a} \phi) -
\lambda\phi^4 \nonumber \\&&-
  a\phi^2{\cal R} + \phi^2L_{{\cal T}^{\ast 2}}],
\label{weylcovardirac3}
\end{eqnarray}
%
where $\mu$, $\nu$, $\lambda$ and $a$ are dimensionless constants
(usually positive), and
there are three further dimensionless constants $\beta_1$, $\beta_2$
and $\beta_3$ in $L_{{\cal T}^{\ast 2}}$. Since none of the additional
terms in (\ref{weylcovardirac3}) depend on $\psi$, the results derived
above are not altered, but clearly the equation of motion for $\phi$
will become significantly more complicated.

\subsection{Electromagnetic field in WGT}
\label{sec:wgtem}

It is worthwhile also considering the coupling of the electromagnetic
field to the gravitational gauge fields, since this illustrates some
important differences from the Dirac field discussed above.
The standard special-relativistic Lagrangian for the electromagnetic
field $A_\mu$ (which should not be confused with the `rotational'
gravitational gauge field ${A^{ab}}_\mu$) is given by
\begin{equation}
L_{\rm M} = -\tfrac{1}{4}F_{\mu\nu}F^{\mu\nu} - J^\mu A_\mu,
\label{emlagwgt}
\end{equation}
where the electromagnetic field strength tensor
$F_{\mu\nu}\equiv\partial_\mu A_\nu -\partial_\nu A_\mu$ and $J^\mu$
is some source current density. We showed earlier that the action
constructed from the Lagrangian (\ref{emlagwgt}) is invariant under
global Weyl transformations, provided $w(A_\mu)=-1$ and
$w(J_\mu)=-3$. Moreover, it is also invariant under the
electromagnetic gauge transformation $A_\mu \to A_\mu + \partial_\mu
\chi$, where $\chi$ is any scalar field, provided the source
current density satisfies the continuity equation $\partial_\mu J^\mu
= 0$, which embodies the conservation of charge.

Unlike the Dirac field, in this case the dynamical field $A_\mu$
carries a coordinate index. Our first task, which was unnecessary for
the Dirac field, is therefore to convert this to a Roman index by
constructing the covariant field ${\cal A}_a = {h_a}^\mu
A_\mu$. Similarly, we introduce the covariant current density ${\cal
  J}_a = {h_a}^\mu J_\mu$. This leads us, however, to a subtle
point. In the special relativistic case (\ref{emlagwgt}), the field
components can equally well refer to the Cartesian coordinate basis
$\vect{e}_\mu$ or the local Lorentz basis $\hvect{e}_a$, which can be
chosen to coincide. In principle, however, the field components should
properly refer to the latter. Thus, once the Weyl transformation is
localised, for consistency one requires the covariant fields to have
the Weyl weights $w({\cal A}_a)=-1$ and $w({\cal J}_a)=-3$;
consequently, the fields carrying a Greek index have weights
$w(A_\mu)=0$ and $w(J_\mu)=-2$.

Following the general procedure for constructing a
locally-Weyl-invariant matter action (\ref{weylmatlag_gen}) leads to
the Lagrangian density
\begin{equation}
{\cal L}_{\rm M} = h^{-1}L_{\rm M}
=-h^{-1}(\tfrac{1}{4} \widehat{{\cal F}}^\ast_{ab}\widehat{{\cal F}}^{\ast ab}
+ {\cal J}^a {\cal A}_a),
\label{emlagcovwgt}
\end{equation}
where we have defined $\widehat{{\cal F}}^\ast_{ab} \equiv {\cal
  D}^\ast_a{\cal A}_b - {\cal D}^\ast_b{\cal A}_a$. Since
$w(\widehat{{\cal F}}^\ast_{ab})=-2$, ${\cal
  L}_{\rm M}$ has an overall Weyl weight of zero, as required.
Unfortunately, the Lagrangian density (\ref{emlagcovwgt}) is
inappropriate for describing the coupling of the electromagnetic and
gravitational gauge fields, since the corresponding action does not
inherit the necessary invariance under electromagnetic gauge
transformations. This is easily seen by noting that
\begin{equation}
\widehat{{\cal F}}^\ast_{ab} = {\cal F}_{ab}-{{\cal T}^{\ast c}}_{ab}{\cal A}_c,
\label{fijhatstar}
\end{equation}
where we have defined ${\cal F}_{ab} \equiv {h_a}^\mu {h_b}^\nu
F_{\mu\nu}$, in which $F_{\mu\nu} \equiv D^\ast_\mu A_\nu -D^\ast_\nu A_\mu =
\partial_\mu A_\nu -\partial_\nu A_\mu$, and ${{\cal T}^{\ast
    c}}_{ab}$ is the WGT field strength tensor of the $h$
gravitational gauge field, as defined in (\ref{cstardef}). It is
apparent that ${\cal F}_{ab}$ does inherit the original
electromagnetic gauge invariance of $F_{\mu\nu}$, whereas the second
term on the RHS of (\ref{fijhatstar}) is clearly not invariant under
the electromagnetic gauge transformation, and so neither is
$\widehat{{\cal F}}^\ast_{ab}$. Moreover, ${\cal F}_{ab}$ has a Weyl
weight $w( {\cal F}_{ab})=-2$. Thus, substituting (\ref{fijhatstar})
into (\ref{emlagcovwgt}) and discarding the offending terms, we arrive
instead at the Lagrangian density
\begin{equation}
{\cal L}_{\rm M} = h^{-1}L_{\rm M}
=-h^{-1}(\tfrac{1}{4} {\cal F}_{ab}{\cal F}^{ab}
+ {\cal J}^a {\cal A}_a),
\label{emlagcov2wgt}
\end{equation}
which has an overall Weyl weight of zero, as required, provided
$w({\cal J}_a)=-3$. It is worth noting that this form for ${\cal
  L}_{\rm M}$ is identical to that obtained in PGT, as should be clear
from the above discussion. Hence, as we found for the Dirac field, the
action for the electromagnetic field in PGT is {\em already} invariant
under local changes of scale. Note that ${\cal L}_{\rm M}$ does not
depend on the $A$ or $B$ gauge fields.

The equation of motion for the electromagnetic field is given by
$\delta {\cal L}_{\rm EM}/\delta A_\mu = 0$, which is easily shown to
be equivalent to $\delta {\cal L}_{\rm EM}/\delta {\cal A}_a = 0$, and
in manifestly WGT-covariant form this reads
\begin{equation}
({\cal D}^\ast_a+{\cal T}^\ast_a){\cal F}^{ac}-\tfrac{1}{2}{{\cal
      T}^{\ast c}}_{ab}{\cal F}^{ab} = {\cal J}^c.
\label{emeomwgtstar}
\end{equation}
It is worth noting that this equation does not have the generic form
(\ref{weylpeqn}), owing to the term containing ${{\cal T}^{\ast
    c}}_{ab}$ on the LHS.  This additional term arises because the
demands of electromagnetic gauge invariance of the action dictated the
use of the Lagrangian density (\ref{emlagcov2wgt}), rather than
(\ref{emlagcovwgt}). Had we (erroneously) used the latter, the
resulting equation of motion would not have included this extra term,
and would indeed have taken the general form (\ref{weylpeqn}).  We
note that the electromagnetic field strength tensor also satisfies a
differential (Bianchi) identity, which is easily expressible in
manifestly WGT-covariant form as
\begin{equation}
{\cal D}^\ast_{[a}{\cal F}_{bc]} -{{\cal T}^{\ast d}}_{[ab}{\cal F}_{c]d}= 0.
\label{emdiffidwgt}
\end{equation}
Note that all contractions of this identity are trivial.

Since the Lagrangian (\ref{emlagcov2wgt}) is identical to that in PGT,
one immediately finds that the equation of motion (\ref{emeomwgtstar})
and differential identity (\ref{emdiffidwgt}) satisfied by the
electromagnetic field can be rewritten in terms of corresponding PGT
quantities by simply removing asterisks as
\begin{eqnarray}
({\cal D}_a+{\cal T}_a){\cal F}^{ac}-\tfrac{1}{2}{{\cal
      T}^c}_{ab}{\cal F}^{ab} & = & {\cal J}^c, \label{emeomwgt} \\
{\cal D}_{[a}{\cal F}_{bc]} -{{\cal T}^{d}}_{[ab}{\cal F}_{c]d} & = & 0.
\label{emdiffid}
\end{eqnarray}
The covariance of these equations under local dilations is easily
checked directly by noting that ${\cal D}_a{\cal F}^{ac}=({\cal
  D}^\ast_a+2{\cal B}_a){\cal F}^{ac}$ and ${\cal T}_a ={{\cal
    T}^{\ast}}_a-3{\cal B}_a$, which on substitution recovers
(\ref{emeomwgtstar}) and (\ref{emdiffidwgt}).

As we did for the Dirac equation in the previous section, it is
interesting to rewrite the electromagnetic field equation
(\ref{emeomwgtstar}) and Bianchi identity (\ref{emdiffidwgt}) in terms of the
reduced covariant derivative discussed in
Section~\ref{sec:wgtcovdalt}. Thus, substituting
(\ref{wgtcovderivalt}) into (\ref{emeomwgtstar}) and (\ref{emdiffidwgt}), one
may show straightforwardly that the electromagnetic field equation and
Bianchi identity become, respectively,
\begin{equation}
{^0}{\cal D}^\ast_a{\cal F}^{ac} = {\cal J}^c
\qquad\mbox{and}\qquad
{^0}{\cal D}^\ast_{[a}{\cal F}_{bc]}= 0.
\label{emwgtaltforms}
\end{equation}
These equations are still manifestly WGT-covariant, but have precisely
the same forms as would be obtained in the absence of torsion. Thus,
irrespective of the nature of the WGT torsion, it plays no explicit
role in the dynamics of the electromagnetic field. Alternatively, one
can perform an analogous procedure to rewrite (\ref{emeomwgt}) and
(\ref{emdiffid}) in terms of the reduced PGT covariant derivative as
\begin{equation}
{^0}{\cal D}_a{\cal F}^{ac} = {\cal J}^c
\qquad\mbox{and}\qquad
{^0}{\cal D}_{[a}{\cal F}_{bc]}= 0,
\label{empgtaltforms}
\end{equation}
which are identical to (\ref{emwgtaltforms}) with all asterisks
removed.  Although not manifestly WGT-covariant, the equations
(\ref{empgtaltforms}) provide the simplest expression of the dynamics
of the electromagnetic field for calculational purposes, and shows
that the PGT torsion has no direct effect.

To complete the description of the electromagnetic field in WGT, we
derive its energy-momentum and spin-angular-momentum tensors.  Setting
${\cal J}^a=0$ in (\ref{emlagcov2wgt}), one finds
\begin{equation}
h{\tau^a}_b  = \tfrac{1}{4}\delta^a_b{\cal F}^{cd}{\cal F}_{cd}
-{\cal F}^{ac}{\cal F}_{bc},
\label{pgtememt}
\end{equation}
whereas ${\sigma_{ab}}^c=0$ immediately. We see that ${\tau^a}_b$ is
the natural generalisation of the standard energy-momentum tensor for
the electromagnetic field in the absence of gravity. Moreover,
${\tau}_{ab}$ is symmetric and traceless, as one would expect for a field with
vanishing spin-angular-momentum and zero mass. We note that the expression
(\ref{pgtememt}) is already manifestly WGT (and PGT) covariant.

Finally, one may straightforwardly include the interaction of the
electromagnetic and Dirac fields, both of which are coupled to
gravity, by noting that the covariant current density of the Dirac
field is simply ${\cal J}^a = q\bar{\psi}\gamma^a\psi$, which has the
required Weyl weight $w({\cal J}^a)=-3$. Thus, one
simply makes this substitution into (\ref{emlagcov2wgt}) and adds the
result to (\ref{weylcovardirac3}).

\subsection{Einstein gauge and scale-invariant variables}
\label{sec:wgteinstein}

As mentioned in the Introduction, one may use the scale gauge freedom
to set the scalar field $\phi$ to a constant (provided $\phi\neq 0$),
which is known as the Einstein gauge.  Although it is most common
simply to set $\phi(x)=1$, this can be misleading and so we instead
set $\phi(x)=\phi_0$, where $\phi_0$ is a constant with the same units
as $\phi$.  Adopting the Einstein gauge can be viewed simply as a
mathematical convenience, but it is often also given the physical
interpretation of corresponding to some spontaneous breaking of the
scale symmetry. Indeed, it is usually considered that setting
$\phi=\phi_0$ represents the choice of some definite scale in the
theory, thereby breaking scale-invariance.

The forms of the field equations in WGT, which we discussed in general
in Section~\ref{sec:wgtfieldeqns}, can be simplified considerably by
adopting the Einstein gauge. There is, however, an alternative
approach, which is not usually followed in the literature, where one
recasts the theory in terms of scale-invariant field
variables\cite{kasuya75}. We show here that, for {\em any} WGT
Lagrangian density of the form considered in (\ref{wgtlagtot}), this
procedure yields equations of motion in the new field variables that
are {\em identical} in form to those obtained after setting
$\phi(x)=\phi_0$ in the equations of motion for the original
fields. This therefore provides a rather different interpretation of
the Einstein gauge than that traditionally assumed, since the approach
using scale-invariant variables involves no breaking of the scale
symmetry. Indeed, one may regard adopting the Einstein gauge merely as
a shortcut to obtaining the forms of the equations of motion for the
scale-invariant field variables.

To demonsrate this equivalence, we begin by introducing the new matter
and gravitational gauge fields
\begin{subequations}
\label{wgtsivardef}
\begin{eqnarray}
\widehat{\vpsi}&\equiv& \left(\frac{\phi}{\phi_0}\right)^{-w}\vpsi,\\
{\widehat{h}_a}^{\phantom{a}\mu} &\equiv& \left(\frac{\phi}{\phi_0}\right)^{-1}{h_a}^\mu,\\
{\widehat{A}^{ab}}_{\phantom{ab}\mu} &\equiv& {A^{ab}}_\mu,\\
\widehat{B}_\mu &\equiv& B_\mu -\partial_\mu \ln\left(\frac{\phi}{\phi_0}\right),
\end{eqnarray}
\end{subequations}
where $w$ is the Weyl weight of the original matter field $\vpsi$. It
is straightforward to show that each of these new variables transforms
covariantly with Weyl weight $w=0$ under local dilations, and is hence
scale-invariant. It is also convenient to define the further
scale-invariant variables
${\widehat{\mathcal{A}}^{ab}}_{\phantom{ab}c} \equiv
{\widehat{h}_a}^{\phantom{a}\mu}{\widehat{A}^{ab}}_{\phantom{ab}\mu}$
and $\widehat{\mathcal{B}}_a \equiv {\widehat{h}_a}^{\phantom{a}\mu}
\widehat{B}_\mu$.

One immediately finds that the covariant derivative of the scalar
field $\phi$ may be written as ${\cal D}^\ast_a\phi =
-(\phi^2/\phi_0)\widehat{\mathcal{B}}_a$, whereas the covariant derivative of
some general field $\chi$ with Weyl weight $w$ may be written
\begin{eqnarray}
\hspace*{-5mm}{\cal D}^\ast_a\chi &\equiv& ({\cal D}_a + w {\cal B}_a)\chi
\nonumber\\
&=&
\left(\frac{\phi}{\phi_0}\right)^{1-w} (\widehat{\mathcal{D}}_a + w
\widehat{\mathcal{B}}_a)\widehat{\chi} \equiv
\left(\frac{\phi}{\phi_0}\right)^{1-w} \widehat{\mathcal{D}}^\ast_a
\widehat{\chi},
\label{scaleinvdiff}
\end{eqnarray}
where $\widehat{\chi}=(\phi/\phi_0)^{-w}\chi$ and we have defined the
derivative operators $\widehat{\mathcal{D}}_a \equiv
\widehat{h}_a^{\phantom{a}\mu} D_\mu$ and
$\widehat{\mathcal{D}}^\ast_a = \widehat{\mathcal{D}}_a + w
\widehat{\mathcal{B}}_a$, the latter of which preserves the Weyl weight of
the quantity on which it acts. From (\ref{scaleinvdiff}), one sees
that, aside from the overall multiplicative factor
$(\phi/\phi_0)^{1-w}$, the scale-invariant quantity
$\widehat{\mathcal{D}}^\ast_a \widehat{\chi}$ has the {\em same}
functional dependency on $\widehat{\chi}$,
$\widehat{h}_a^{\phantom{a}\mu}$,
${\widehat{A}^{ab}}_{\phantom{ab}\mu}$ and $\widehat{B}_\mu$,
respectively, as ${\cal D}^\ast_a\chi$ does on $\chi$, ${h_a}^\mu$,
${A^{ab}}_\mu$ and $B_\mu$. Moreover, one also quickly finds that the
WGT gauge field strengths defined in (\ref{weylfsdefs}) may be written
as ${{\cal R}^{ab}}_{cd} = (\phi/\phi_0)^2
\widehat{\mathcal{R}}^{ab}_{\phantom{ij}cd}$, ${{\cal T}^{\ast
    a}}_{bc} = (\phi/\phi_0)\widehat{\mathcal{T}}^{\ast
  a}_{\phantom{abc}bc}$ and ${\cal H}_{ab} = (\phi/\phi_0)^2
\widehat{\mathcal{H}}_{ab}$, where each quantity with a caret is
scale-invariant and has the {\em same} functional dependence on
$\widehat{h}_a^{\phantom{a}\mu}$,
${\widehat{A}^{ab}}_{\phantom{ab}\mu}$ and $\widehat{B}_\mu$ as the
corresponding original quantity does on ${h_a}^\mu$, ${A^{ab}}_\mu$
and $B_\mu$, respectively.

Noting that $h^{-1} = (\phi/\phi_0)^{-4} \widehat{h}^{-1}$, one is
thus led to the important conclusion that the Lagrangian density
${\cal L}_{\rm T}$ in (\ref{wgtlagtot}) may be written as (suppressing
indices for brevity)
\begin{equation}
{\cal L}_{\rm T} \!=\! {\cal L}_{\rm G}(\widehat{h},\widehat{A},\partial
\widehat{A},\partial \widehat{B}) + {\cal L}_{\rm
  M}(\widehat{\vpsi},\partial\widehat{\vpsi},\phi_0,0,\widehat{h},\partial
\widehat{h}, \widehat{A}, \partial \widehat{A}, \widehat{B}).
\label{wgtlagtotsi}
\end{equation}
In other words, when written in terms of the scale-invariant field
variables, the Lagrangian density ${\cal L}_{\rm T}$ (indeed each term
separately) has the {\em same} functional form as it does in terms of
the original variables with $\phi=\phi_0$. Thus, if $\chi$ represents
$\vpsi$, ${h_a}^\mu$, ${A^{ab}}_\mu$ or $B_\mu$, one may immediately
conclude that (each term in) the equation of motion $\delta{\cal
  L}_{\rm T}/\delta\widehat{\chi}=0$ has the {\em same} functional
form as (the corresponding term in) $\left.\delta{\cal L}_{\rm
  T}/\delta\chi\right|_{\phi=\phi_0}=0$, but with $\vpsi$,
${h_a}^\mu$, ${A^{ab}}_\mu$ and $B_\mu$ replaced by their
scale-invariant counterparts.

For the $\phi$-field equation, however, the situation is more subtle,
since (\ref{wgtlagtotsi}) does not depend explicitly on
$\phi$. Nonetheless, one can still show that the same equivalence
holds as follows. Working firstly in terms of the original variables,
one begins by noting that $\delta{\cal L}_{\rm T}/\delta\phi =
\delta{\cal L}_{\rm M}/\delta\phi$, since ${\cal L}_{\rm G}$ does not
depend on $\phi$ or its derivative. As discussed in
Section~\ref{sec:wgtconslaws}, the invariance of the matter action
$S_{\rm M}$ under local dilations leads to the WGT conservation law
(\ref{wgtlmcons3}), which may be straightforwardly rewritten as
\begin{equation}
\phi_0\left.\frac{\delta{\cal L}_{\rm
    M}}{\delta\phi}\right|_{\phi=\phi_0} = -{h_a}^\mu
\left.\frac{\delta{\cal L}_{\rm M}}{\delta
  {h_a}^\mu}\right|_{\phi=\phi_0} +
\partial_\mu\left(\frac{\delta{\cal L}_{\rm
    M}}{\delta B_\mu}\right)_{\phi=\phi_0},
\label{phieqnorigvar}
\end{equation}
where we have made use of the $\vpsi$-field equation $\delta{\cal
  L}_{\rm T}/\delta\vpsi = \delta{\cal L}_{\rm M}/\delta\vpsi=0$ and
set $\phi=\phi_0$ in the resulting expression. Turning now to the
expression (\ref{wgtlagtotsi}) for ${\cal L}_{\rm T}$ in terms of
scale-invariant variables and applying the chain rule for partial
derivatives, one may show after a lengthy but straightforward
calculation that
\begin{equation}
\phi \frac{\delta{\cal L}_{\rm
    M}}{\delta\phi} = -{\widehat{h}_a}^{\phantom{a}\mu}
\frac{\delta{\cal L}_{\rm M}}{\delta
  {\widehat{h}_a}^{\phantom{a}\mu}} +
\partial_\mu\left(\frac{\partial{\cal L}_{\rm
    M}}{\partial \widehat{B}_\mu}\right),
\label{phieqnsivar}
\end{equation}
where we have again made use of the $\vpsi$-field equation. Since
${\cal L}_{\rm M}$ does not depend on the derivatives of
$\widehat{B}_\mu$, one may replace $\partial{\cal L}_{\rm M}/\partial
\widehat{B}_\mu$ by $\delta {\cal L}_{\rm M}/\delta \widehat{B}_\mu$
in the above equation. Then, comparing (\ref{phieqnorigvar}) and
(\ref{phieqnsivar}) and making use of our earlier observations
regarding the other field equations, one sees that the $\phi$-field
equation written in terms of the scale-invariant variables also has
the {\em same} functional form as the $\phi$-field equation in terms of
the original variables with $\phi=\phi_0$.

For the remainder of our discussion, we will continue to use the
original field variables for the most part, and occasionally simplify
field equations by adopting the Einstein gauge $\phi(x)=\phi_0$, since
this approach is more familiar and straightforward. Nonetheless, the
equivalence of the results so obtained with those expressed in terms
of scale-invariant variables should be borne in mind throughout.

\subsection{Motion of test particles in WGT}
\label{sec:WGTmassivemotion}

We may determine the equation of motion of a massive matter test
particle in WGT from the Dirac Lagrangian (\ref{diraclm2}), but
modified to accommodate local scale-invariance by simply introducing
the massless scalar field $\phi$ and making the replacement $m \to
\mu\phi$, as in (\ref{diraclm3}). One begins by constructing the
appropriate action for a spin-$\frac{1}{2}$ point particle, and then
makes the full classical approximation in which the particle spin is
neglected. This procedure is outlined for a free particle in
Appendix~\ref{appendix:dirac}.  To include gravitational effects in
the point particle action (\ref{pointparticleaction}), one needs only
to make the replacements $p_\mu \to p_a \equiv{h_a}^\mu p_\mu$,
$\dot{x}^\mu \to {v}^a \equiv {b^a}_\mu \dot{x}^\mu$ and
$m\to\mu\phi$, which gives\cite{footnote12}
\begin{equation}
S = - \int d\lambda\, [p_a v^{a}-\tfrac{1}{2}e(p_a p^a - \mu^2\phi^2)],
\label{WGTppaction}
\end{equation}
where the three dynamical variables are the particle 4-momentum
$p^a(\lambda)$, 4-velocity $v^a(\lambda)$ and the einbein $e(\lambda)$
along the worldline, which is parameterised by $\lambda$. We note that
the Weyl weights of the quantities appearing in (\ref{WGTppaction})
  are $w(p^a)=-1$, $w(v^a)=0$, $w(e)=1$, $w(\lambda)=1$ and
  $w(\phi)=-1$, so that the action is indeed scale-invariant.

Varying the action (\ref{WGTppaction}) with respect to the three
dynamical variables, respectively, one obtains the equations of motion
\begin{eqnarray}
v^a & = & ep^a, \label{WGTppeom1}\\
\dot{p}_a & = & {c^c}_{ab} v^bp_c + e\mu^2 \phi\,\partial_a\phi, \label{WGTpppdot}\\
p^2 &=& \mu^2\phi^2.\label{WGTppeom3}
\end{eqnarray}
where $p^2 \equiv p_a p^a$ and the quantities ${c^c}_{ab}$ are the PGT
counterparts of those defined in (\ref{riccicoeffsdef}) (obtained by
removing all asterisks). It is straightforward
to show that (\ref{WGTpppdot}) can be written in a manifestly
WGT-covariant manner as
\begin{equation}
v^c({\cal D}^\ast_cp_a - {\cal
  T}^\ast_{cab}p^b) =e\mu^2 \phi\,{\cal D}^\ast_a\phi.
\label{WGTppeom1a}
\end{equation}
At first sight, (\ref{WGTppeom1a}) appears to show that `torsion'
enters the particle equation of motion directly. As one might suspect
from our discussion in Sections~\ref{sec:wgtdirac} and
\ref{sec:wgtem}, however, appearances can be deceptive. Indeed, one
may show directly that (\ref{WGTppeom1a}) can be rewritten as
\begin{equation}
v^c\,{^0}{\cal D}^\ast_cp_a=e\mu^2 \phi\,{^0}{\cal D}^\ast_a\phi,
\label{WGTppeom1.5}
\end{equation}
where ${^0}{\cal D}^\ast_c$ is the reduced WGT covariant
derivative, defined in (\ref{dstarmuzerodef}). The equation of motion
(\ref{WGTppeom1.5}) is  again manifestly WGT-covariant, but
takes its simplest form
when one chooses $e=1/(\mu\phi)$, in which case $v^2=1$ and $\lambda$
corresponds to the proper time $\tau$ along the worldline; in this
case the equation of motion becomes
\begin{equation}
\phi\,v^b\,{^0}{\cal D}^\ast_bv_a= (\delta_a^b - v_av^b){^0}{\cal D}^\ast_b\phi.
\label{WGTppeom}
\end{equation}
For ease of calculation, one can rewrite this equation in terms of the
reduced PGT-covariant derivative
as\cite{footnote12aa,brans61,puetzfeld15}
\begin{equation}
\phi\,v^b\,{^0}{\cal D}_bv_a= (\delta_a^b - v_av^b)\partial_b\phi,
\label{WGTppeom2}
\end{equation}
which is again WGT covariant, but not manifestly so. Moreover, if one uses
local scale invariance to impose the  Einstein gauge $\phi=\phi_0$ (a
constant), then (\ref{WGTppeom2}) reduces to
\begin{equation}
v^b\, {^0}{\cal D}_bv_a =0.
\label{WGTppeomfinal}
\end{equation}

When reinterpreted geometrically (see Section~\ref{sec:geowgt} and
Appendix~\ref{app:pgt}), the quantities ${^0}{A^{ab}}_\mu$ appearing
in ${^0}{\cal D}_b$ contain the same geometrical information as the
metric connection. Thus, (\ref{WGTppeomfinal}) describes the gauge
theory equivalent of {\em geodesic} motion (as opposed, for example,
to motion along autoparallels, which would correspond to the equation
of motion $v^b\, {\cal D}_bv_a =0$).
We note that this conclusion differs from some previous
studies\cite{rosen82,mirabotalebi08}, in which the WGT conservation
laws applied to the energy momentum tensor of dust yield an equation
of motion for a matter test particle that contains an additional
electromagnetic-like Lorentz force term proportional to $(q_{\rm
  W}/m){\cal H}_{ab}v^b$, where $q_{\rm W}$ is the particle's Weyl
charge and $m$ is its mass. Such a term is absent from the equation of
motion (\ref{WGTppeomfinal}), since the latter is derived by
considering the Dirac action, which does not depend on the dilation
gauge field $B_\mu$. It is feasible that by constructing a point
particle action based instead upon (say) the action for a scalar
field, which does depend on the dilation gauge field, one would arrive
at an equation of motion that does contain the Lorentz-force term
above. We do not consider this further here, however, since in our
opinion `normal' matter is more appropriately described by Dirac
particles.

By setting $\mu=0$ in the action (\ref{WGTppaction}) and choosing the
einbein $e=1$, such that $v^2=0$, one finds that the equation of
motion in the massless case (for example, a massless neutrino) is
again given by (\ref{WGTppeomfinal}), even without imposing the
Einstein gauge. One may also arrive at a similar conclusion for the
motion of photons by directly considering the dynamics of the
electromagnetic field. As discussed in Section~\ref{sec:wgtem}, the EM
field tensor in WGT again satisfies the field equation and Bianchi
identity given in (\ref{empgtaltforms}), which have precisely the same
form as those obtained in the absence of torsion.  Consequently, one
may immediately infer that the equation of motion for photons is also
given by (\ref{WGTppeomfinal}), so that they too follow the gauge
theory equivalent of {\em geodesic} motion. It is worth noting that
the lack of a Lorentz-force term is again related to the absence of
the dilation gauge field $B_\mu$ from the WGT action for the
electromagnetic field.

To deduce the geodesic equation of motion (\ref{WGTppeomfinal}), it
was necessary to impose the Einstein gauge condition $\phi=\phi_0$. As
one might suspect from our discussion in the previous section,
however, this is unnecessary and an equivalent result may be obtained
directly in terms of scale-invariant variables, as we now
demonstrate. Returning to the action (\ref{WGTppaction}), let us
introduce the new scale-invariant variables
\begin{subequations}
\label{wgtppsivardef}
\begin{eqnarray}
\widehat{\lambda} &\equiv& \left(\frac{\phi}{\phi_0}\right)\lambda,\\
\widehat{p}_a&\equiv& {\widehat{h}_a}^{\phantom{a}\mu} p_\mu =
\left(\frac{\phi}{\phi_0}\right)^{-1}p_a,\\
\widehat{v}^a&\equiv& {\widehat{b}^a}_{\phantom{a}\mu} \nd{x^\mu}{\widehat{\lambda}}
=v^a,\\
\widehat{e} &\equiv& \left(\frac{\phi}{\phi_0}\right)e,
\end{eqnarray}
\end{subequations}
where $\widehat{p}_a$, $\widehat{v}^a$ and $\widehat{e}$ are to be
considered as functions of $\widehat{\lambda}$, such that for example
$\widehat{p}_a(\widehat{\lambda}) \equiv
{\widehat{h}_a}^{\phantom{a}\mu}
p_\mu(\lambda(\widehat{\lambda}))$. It is then straightforward to show
that (\ref{WGTppaction}) can be rewritten as
\begin{equation}
S = - \int d\widehat{\lambda}\, [\widehat{p}_a \widehat{v}^{a}-
\tfrac{1}{2}\widehat{\epsilon}(\widehat{p}_a \widehat{p}^a - \mu^2\phi_0^2)],
\label{WGTppactionsi}
\end{equation}
where we have introduced the new scale-invariant einbein
$\widehat{\epsilon} =
[1-\widehat{\lambda}(dx^\mu/d\widehat{\lambda})\partial_\mu\ln(\phi/\phi_0)]\widehat{e}$. Thus,
in terms of the scale-invariant variables $\widehat{\lambda}$,
$\widehat{p}_a$, $\widehat{v}^a$ and $\widehat{\epsilon}$,
respectively, the action (\ref{WGTppactionsi}) has the same functional
form as it does in terms of the original variables $\lambda$, $p_a$,
$v^a$ and $e$ with $\phi=\phi_0$. Moreover, in the definitions of the
scale-invariant variables (\ref{wgtppsivardef}), each occurrence of the
gravitational gauge field ${h_a}^\mu$ or its inverse ${b^a}_\mu$ in
the original variables has merely been replaced by its scale-invariant
counterpart ${\widehat{h}_a}^{\phantom{a}\mu}$ or
${\widehat{b}^a}_{\phantom{a}\mu}$.  Thus, the calculation presented
above follows through in a similar manner, but in terms of the
corresponding scale-invariant variables and with $\phi=\phi_0$. One
therefore arrives at a result analogous to (\ref{WGTppeomfinal}) but
written entirely in terms of scale-invariant variables, namely
\begin{equation}
\widehat{v}^b\, {^0}\widehat{\cal D}_b\widehat{v}_a =0,
\label{WGTppeomfinalsi}
\end{equation}
where we have defined the derivative operator ${^0}\widehat{\cal D}_a
\equiv {\widehat{h}_a}^{\phantom{a}\mu}\,{^0}{\cal D}_\mu$.

\subsection{Reduced WGT}
\label{sec:reducedwgt}

In Section~\ref{sec:wgtcovdalt}, we introduced the `reduced' WGT
covariant derivative operator ${^0}D^\ast_\mu$ in
(\ref{dstarmuzerodef}), to which the `full' WGT covariant derivative
(\ref{weyldmudef}) reduces in the case that the WGT torsion ${\cal
  T}^\ast_{abc}$ vanishes (which is a properly WGT-covariant
condition). In the context of WGT, we will use the term `reduced' to
refer to versions of quantities in which, by construction, the
rotational gauge field is not an independent field, but is determined
by the other gauge fields $h$ and $B$ through the expression
(\ref{afromht}) with ${\cal T}^\ast_{abc} \equiv 0$. Hence such
quantities may be written entirely in terms of these other gauge
fields, and are denoted by a zero superscript preceding the kernel
letter.

One can use the covariant derivative ${^0}D^\ast_\mu$ to build an
alternative class of scale-invariant gravitational gauge theories,
which we term `reduced WGT', that depend only the $h$ and $B$
gravitational gauge fields and correspond mathematically to imposing
the condition of vanishing WGT torsion directly at the level of the
action. Such theories have been studied previously, most notably by
Dirac~\cite{dirac73}, and are of some phenomenological interest.

One begins by defining the `reduced' gauge field strengths in the
usual manner by considering the commutator of reduced covariant
derivatives. One finds that
\begin{equation}
[\zero{D^\ast_\mu},\zero{D^\ast_\nu}]\vpsi=\tfrac{1}{2} \zero{{R^{\ast ab}}_{\mu\nu}}\Sigma_{ab}\vpsi + wH_{\mu\nu}\vpsi,
\end{equation}
where we have defined the `reduced' field strength tensor
$\zero{{R^{\ast ab}}_{\mu\nu}}(h,\partial h,\partial^2 h,B,\partial
B)$, which is again given by the formula (\ref{rfsdef}), but with
${A^{ab}}_\mu$ replaced by $\zero{{A^{\ast ab}}_\mu}$, and hence
depends only on the $h$ and $B$ fields and their derivatives, as
indicated. Considering instead the commutator of `reduced' generalised
covariant derivatives, one obtains
\begin{equation}
[{^0}{\cal D}^\ast_c,{^0}{\cal D}^\ast_d]\vpsi=\tfrac{1}{2}
\czero{{{\cal R}^{\ast ab}}_{cd}}\Sigma_{ab}\vpsi + w{\cal H}_{cd}\vpsi,
\label{weylzerotcomm}
\end{equation}
where $\czero{{{\cal R}^{\ast ab}}_{cd}} = {h_c}^\mu {h_d}^\mu
\,\zero{{R^{\ast ab}}_{\mu\nu}}$. Unlike (\ref{weylfsdefs}), the
commutator (\ref{weylzerotcomm}) has no term containing a `reduced'
translational field strength of the $h$ gauge field, since
$\czero{{{\cal T}^{\ast a}}_{bc}} \equiv
{h_b}^{\mu}{h_c}^{\nu}(\zero{D}^\ast_\mu {b^a}_\nu
-\zero{D}^\ast_\nu{b^a}_\mu) = 0$. As one might expect, the Bianchi
identities satisfied by the reduced field strength tensors are
identical to those given in Section~\ref{sec:wgtbianchi}, but with the
replacements ${\cal D}^\ast_a \to {^0}{\cal D}^\ast_a$, ${\cal
  R}_{abcd} \to {^0}{\cal R}^\ast_{abcd}$ and ${\cal T}^\ast_{abc} \to
{^0}{\cal T}^\ast_{abc} \equiv 0$.

Since the `full' generalised covariant derivative is given in terms of the
`reduced' one by (\ref{wgtcovderivalt}), it is
straightforward to show that the `full' field strength tensor ${{\cal
    R}^{ab}}_{cd}$ appearing in (\ref{weylfsdefs}) is related to its
`reduced' counterpart by
\begin{widetext}
\begin{subequations}
\label{riemann0andricci0}
\begin{eqnarray}
{{\cal R}^{ab}}_{cd} &=& \czero{{{\cal R}^{\ast ab}}_{cd}}
+ \czero{{\cal D}^\ast_c}{{\cal K}^{\ast ab}}_d
- \czero{{\cal D}^\ast_d}{{\cal K}^{\ast ab}}_c
+{{\cal K}^{\ast a}}_{ec}{{\cal K}^{\ast eb}}_d - {{\cal K}^{\ast
    a}}_{ed}{{\cal K}^{\ast eb}}_c \label{riemann0}\\
{{\cal R}^{a}}_{c} & = & \czero{{{\cal R}^{\ast a}}_{c}}
+ \czero{{\cal D}^\ast_c}{{\cal K}^{\ast ab}}_b
- \czero{{\cal D}^\ast_b}{{\cal K}^{\ast ab}}_c
+{{\cal K}^{\ast a}}_{ec}{{\cal K}^{\ast eb}}_b - {{\cal K}^{\ast a}}_{eb}{{\cal K}^{\ast eb}}_c,
\label{ricci0}\\
{\cal R} & = & \czero{{\cal R}^\ast}+\tfrac{1}{4}{\cal T}^{\ast abc}{\cal T}^\ast_{abc}
+ \tfrac{1}{2}{\cal T}^{\ast abc}{\cal T}^\ast_{bac}-{\cal T}^{\ast a}{\cal T}^\ast_a
- 2 \,\czero{{\cal D}^\ast_a}{\cal T}^{\ast a}.
\label{randr0}
\end{eqnarray}
\end{subequations}
\end{widetext}
where, for completeness, we have also given explicitly the
relationships between the contractions of ${{\cal R}^{ab}}_{cd}$ and
$\czero{{{\cal R}^{\ast ab}}_{cd}}$.

For reduced WGT, the free gravitational Lagrangian density ${\cal
  L}_{\rm G} = h^{-1}L_{\rm G}$, where by analogy with
(\ref{wgtgravlagdef}),
\begin{equation}
L_{\rm G} = L_{\czero{{\cal R}}^{\ast 2}} + L_{{\cal H}^2},
\label{wgtnotlg}
\end{equation}
which is based on (\ref{lg2}) with ${{\cal R}^{ab}}_{cd}$ replaced by
$\czero{{{\cal R}^{\ast ab}}_{cd}}$.  As previously, one can simplify
(\ref{wgtnotlg}) still further (in $D \le 4$ dimensions), since
$\czero{{\cal R}}^\ast_{abcd}$ and its contractions also satisfy a
Gauss--Bonnet identity of the form (\ref{gbid}), but with ${\cal
  R}_{abcd} \to \czero{{\cal R}}^\ast_{abcd}$. Thus, one can set to
zero any one of the parameters $\alpha_i$ in (\ref{wgtnotlg}) with no
loss of generality (at least classically). A typical form for the
matter Lagrangian density ${\cal L}_{\rm M}$ is that given in
(\ref{weylcovardirac3}), but with appropriate
modifications\cite{footnote12a}, namely
\begin{eqnarray}
{\cal L}_{\rm M} = h^{-1}[\tfrac{1}{2}i\bar{\psi}\gamma^a
\,{\stackrel{\leftrightarrow}{{^{0}{\cal D}}_a}}\psi \!\!-\!\!
\mu\phi\bar{\psi}\psi
&&+ \tfrac{1}{2}\nu ({^{0}{\cal D}}^\ast_a\phi) ({^{0}{\cal D}}^{\ast a} \phi) -
\lambda\phi^4 \nonumber \\&&
\hspace{1.5cm} - a\phi^2\,{^{0}{\cal R}}^\ast ],
\label{reducedweylcovardirac3}
\end{eqnarray}

The total Lagrangian density thus has the following functional
dependencies in the most general case,
\begin{eqnarray}
{\cal L}_{\rm T} & = & {\cal L}_{\rm G}(h,\partial h, \partial^2 h, B,
\partial B) \nonumber \\
&&
\hspace{0.5cm}
+ {\cal L}_{\rm M}(\vpsi,\partial\vpsi,\phi,\partial\phi,h,\partial h,
\partial^2 h, B, \partial B).\phantom{AAA}
\label{reducedwgtlagtot}
\end{eqnarray}
The resulting field equations for the gravitational gauge fields will
clearly have the same generic structure as those given in
Section~\ref{sec:wgtfieldeqns}, although without the $A$-field
equation (\ref{eqn:weylgenfe2}), but the specific forms for each term
are {\em not}, in general, obtained from the corresponding `full' WGT
expressions simply by replacing the `full' covariant derivative and
field strength tensors with their `reduced' counterparts. Nonetheless,
the matter field equations will still take the form
(\ref{weylpeqn})--(\ref{weylpeqnphi}), but with the replacements
${\cal D}^\ast_a \to {^0}{\cal D}^\ast_a$ and ${\cal T}^\ast_{a} \to
{^0}{\cal T}^\ast_{a} \equiv 0$. One should also note, however, that
the gravitational and matter Lagrangians are both, in general,
quadratic in second derivatives of the $h$-field, and so the resulting
field equations are typically linear in fourth-order derivatives of
$h$; such theories typically suffer from Ostrogradsky's instability,
although this needs to be investigated on a case-by-case basis.

The conservation laws in reduced WGT do have the same form as those
given in Section~\ref{sec:wgtconslaws} for WGT, but with the
replacements ${\cal D}^\ast_a \to {^0}{\cal D}^\ast_a$, ${\cal
  T}^\ast_{abc} \to {^0}{\cal T}^\ast_{abc} \equiv 0$ and $s_{abc}
\equiv 0 \equiv\sigma_{abc}$. Furthermore, by performing analogous
calculations to those presented in Sections~\ref{sec:wgteinstein} and
\ref{sec:WGTmassivemotion}, respectively, one may show that our
conclusions in WGT regarding the interpretation of the Einstein gauge
and the motion of test particles also apply in reduced WGT.

Finally, as alluded to above, it is important to distinguish reduced
WGT, in which the condition of vanishing WGT torsion is imposed
directly at the level of the action, from instead setting the
(properly WGT-covariant) condition ${\cal T}^\ast_{abc}= 0$ in the
field equations of WGT. In the latter case, the notion of the reduced
covariant derivative operator ${^0}D^\ast_\mu$ in
(\ref{dstarmuzerodef}) is still useful, as we have seen in
Sections~\ref{sec:wgtdirac}, \ref{sec:wgtem} and
\ref{sec:WGTmassivemotion}, but the basic field equations remain
linear in second-order derivatives of the gauge fields; although one
can substitute for the rotational gauge field to obtain equations that
contain higher-order derviatives, the theory does not suffer from
Ostrogradsky's instability. In essence, the two approaches correspond
to `second-order' and `first-order' variational formalisms,
respectively. Indeed, they are analogous to the metric and Palatini
variations used in the study of geometric theories of modified
gravity, where in the latter method the metric (Levi--Civita)
connection is subsequently imposed in the resulting field equations;
the two methods yield the same (second-order) field equations only for
the Lovelock action\cite{exirifard08,borunda08,dadich11}, which
coincides with the Einstein--Hilbert action (plus a cosmological
constant term) in $D=4$ dimensions.

\subsection{Geometric interpretation of WGT}
\label{sec:geowgt}

So far, we have been firm in our resolve to regard ${h_a}^\mu$,
${A^{ab}}_\mu$ and $B_\mu$ purely as gauge fields in Minkowski
spacetime, and have avoided attaching any geometric interpretation to
them. Although we will maintain this viewpoint in this paper, it is
nonetheless common practice to reinterpret WGT in geometric terms, and
we therefore give a brief account of this reinterpretation here.

At the heart of the geometric interpretation of WGT (and PGT) is the
identification of ${h_a}^\mu$ as the components of a vierbein system
in a more general spacetime. Thus, at any point $x$ in the spacetime,
one demands that the orthonormal tetrad frame vectors $\hvect{e}_a(x)$
and the coordinate frame vectors $\vect{e}_\mu(x)$ are related
by\cite{footnote13}
\begin{equation}
\hvect{e}_a = {h_a}^\mu\vect{e}_\mu,\qquad \vect{e}_\mu =
{b^a}_\mu\hvect{e}_a,
\label{geotetrad}
\end{equation}
with similar relationships holding between the dual basis vectors
$\hvect{e}^a(x)$ and $\vect{e}^\mu(x)$ in each set. For any other
vector $\vect{J}$, written in the coordinate basis as (say)
$J_\mu\vect{e}^\mu$, one then identifies the quantities $J_a =
{h_a}^\mu J_\mu$, for example, as the components of the {\em same}
vector, but in the tetrad basis. This is a fundamental difference from
the Minkowski spacetime viewpoint presented earlier, in which ${\cal
  J}_a={h_a}^\mu J_\mu$ would be regarded as the components in the
tetrad basis of a {\em new} vector field ${\cal J}$.

The identification of ${h_a}^\mu$ as the components of a vierbein
system has a number of far-reaching consequences. Firstly, the
index-conversion properties of ${h_a}^\mu$ and ${b^a}_\mu$ are
extended. It is straightforward to show, for example, that ${h_a}^\mu
J^a=J^\mu$ and ${b^a}_\mu J^\mu=J^a$. Moreover, any contraction over
Latin (Greek) indices can be replaced by one over Greek (Latin)
indices. None of these operations is admissible when the $h$ and $A$
fields are viewed purely as gauge fields in Minkowski
spacetime.

Perhaps the most important consequence of identifying ${h_a}^\mu$ as
the components of a vierbein system is that the inner product of the
coordinate basis vectors becomes
\begin{equation}
\vect{e}_\mu \cdot \vect{e}_\nu = \eta_{ab}{b^a}_\mu {b^b}_\nu \equiv g_{\mu\nu}.
\label{metricdef}
\end{equation}
Thus, in this geometric interpretation, one must work in a more
general spacetime with metric $g_{\mu\nu}$.
Conversely, since the tetrad basis vectors still form an orthonormal set,
one has
\begin{equation}
\hvect{e}_a \cdot \hvect{e}_b = \eta_{ab} = g_{\mu\nu}{h_a}^\mu {h_b}^\nu.
\label{etagdef}
\end{equation}
From (\ref{metricdef}), one also finds that $h^{-1}=\sqrt{-g}$ (where
we are working with a metric signature of $-2$).  Under a (local,
physical) dilation, the spacetime metric and $h$-field have Weyl
weights $w(g_{\mu\nu})=2$ and $w({h_a}^\mu)=-1$ respectively, and so
(\ref{metricdef}) and (\ref{etagdef}) imply that $w(\eta_{ab})=0$, as
expected. From (\ref{geotetrad})--(\ref{etagdef}), one
immediately finds that the $h$-field and its inverse are directly
related by index raising/lowering, so there no need to distinguish
between them by using different kernel letters. Consequently, the
standard practice, which we will follow here, is to notate ${h_a}^\mu$
and ${b^a}_\mu$ as ${e_a}^\mu$ and ${e^a}_\mu$, respectively.

One is also led naturally to the interpretation of ${A^{ab}}_\mu$ as
the components of the `spin-connection' that encodes the rotation
of the local tetrad frame between points $x$ and $x+\delta x$, which
is accompanied by a local change in the standard of length between the
two points, which is encoded by $B_\mu$. Thus, the operation of
parallel transport for some vector $J^a$ of weight $w$ is defined as
\begin{equation}
\delta J^a = - ({A^a}_{b\mu}+wB_\mu\delta^a_b) J^b \,\delta x^\mu,
\label{weylparallel}
\end{equation}
which is required to compare vectors
$J^a(x)$ and $J^a(x+\delta x)$ at points $x$ and $x+\delta x$, determined with
respect to the tetrad frames $\hvect{e}_a(x)$ and
$\hvect{e}_a(x+\delta x)$ respectively.
Hence, in general, a vector not only changes its direction on parallel
transport around a closed loop, but also its length. The expression
(\ref{weylparallel}) establishes the correct form for the related
$(\Lambda,\rho)$-covariant derivative, e.g.
\begin{equation}
D^\ast_\mu J^a = \partial_\mu J^a + wB_\mu J^a +  {A^a}_{b\mu}J^b =
\partial^\ast_\mu J^a +  {A^a}_{b\mu}J^b,
\end{equation}
where we have used the partial derivative operator $\partial_\mu^\ast$
defined in (\ref{dstardef}). Moreover, the existence of tetrad frames
at each point of the spacetime implies the existence of the Lorentz
metric $\eta_{ab}$ at each point. Then demanding that $\eta_{ab}$ is
invariant under parallel transport, and recalling that
$w(\eta_{ab})=0$, requires the spin-connection to be antisymmetric,
i.e.  ${A^{ab}}_\mu=-{A^{ba}}_\mu$, as previously.

Substantial differences between the Minkowski spacetime gauge field
viewpoint and the geometric interpretation do occur, however, when
generalising the $(\Lambda,\rho)$-covariant derivative to apply to fields
with definite GCT tensor behaviour. First, in the geometric
interpretation, one can in general no longer construct a global
inertial Cartesian coordinate system in the more general
spacetime. Thus, one must rely on arbitrary coordinates and so define
the `total' covariant derivative
\begin{equation}
\Delta^\ast_\mu \equiv \partial^\ast_\mu + {\Gamma^\sigma}_{\rho\mu} {\matri{X}^\rho}_\sigma
+ \tfrac{1}{2}{A^{ab}}_\mu\Sigma_{ab} = \nabla^\ast_\mu + D^\ast_\mu - \partial^\ast_\mu,
\label{weylextcovd2}
\end{equation}
where $\nabla^\ast_\mu = \partial^\ast_\mu + {\Gamma^\sigma}_{\rho\mu}
{\matri{X}^\rho}_\sigma$ and ${\matri{X}^\rho}_\sigma$ are the
$\mbox{GL}(4,R)$ generator matrices appropriate to the GCT tensor
character of the field to which $\Delta^\ast_\mu$ is applied.  If a
field $\psi$ carries only Latin indices, then $\nabla^\ast_\mu\psi
=\partial^\ast_\mu\psi$ and so $\Delta^\ast_\mu\psi=D^\ast_\mu\psi$;
conversely, if a field $\psi$ carries only Greek indices, then
$D^\ast_\mu\psi =\partial^\ast_\mu\psi$ and so
$\Delta^\ast_\mu\psi=\nabla^\ast_\mu\psi$. When acting on an object of
weight $w$, for all these derivative operators the resulting object
also transforms covariantly with the same weight $w$.

Most importantly, in a dynamical spacetime, the affine connection
coefficients ${\Gamma^\sigma}_{\rho\mu}$ are themselves dynamical
variables, no longer fixed by our choice of cooordinate system.  They
are, however, necessarily related to the spin-connection and dilation
vector since, provided $J^a$ has weight $w=1$, the tetrad components
of a vector with coordinate components $J^\mu$ should, when parallel
transported from $x$ to $x+\delta x$, be equal to $J^a+\delta J^a$,
i.e.
\begin{equation}
J^a+\delta J^a = (J^\mu+\delta J^\mu)\,{e^a}_\mu(x+\delta x).
\label{vreln}
\end{equation}
In other words, the quantities ${A^{ab}}_\mu$ and
${\Gamma^\sigma}_{\rho\mu}$ represent the same geometrical object in
two different frames, and hence there remain 44 gravitational field
variables in all. From (\ref{vreln}), we obtain the relation
\begin{equation}
\Delta^\ast_\mu {e^a}_\nu  \equiv
\partial^\ast_\mu {e^a}_\nu - {\Gamma^\sigma}_{\nu\mu}{e^a}_\sigma
+ {A^a}_{b\mu}{e^a}_\nu =0,
\label{weyltetradp}
\end{equation}
which relates $A$ and $\Gamma$ (and $B$); in particular, we note that
$w({\Gamma^\sigma}_{\nu\mu})=0$. The relation (\ref{weyltetradp}) is
sometimes known as the `tetrad postulate', but note that it always
holds. It is straightforward to show that $A$ or $\Gamma$ may be
written explicitly in terms of the other as
\begin{eqnarray}
{\Gamma^\lambda}_{\nu\mu} & = &
{e_a}^\lambda (\partial^\ast_\mu {e^a}_\nu +
{A^a}_{b\mu} {e^b}_\nu),\label{gammaofa}\\
{A^a}_{b\mu} & = &
{e^a}_\lambda (\partial^\ast_\mu {e_b}^\lambda
+ {\Gamma^\lambda}_{\nu\mu}{e_b}^\nu).
\label{aofgamma}
\end{eqnarray}

Using (\ref{metricdef}) and (\ref{weyltetradp}), one
finds that $\nabla^\ast_\sigma g_{\mu\nu}=0$, and so this derivative
operator commutes with raising and lowering of coordinate
indices. Equivalently, one may write this semi-metricity condition as
\begin{equation}
\nabla_\sigma g_{\mu\nu} = -2B_\sigma g_{\mu\nu},
\label{wgtsemimet}
\end{equation}
which shows that the spacetime has, in general, a Weyl--Cartan
$Y_4$ geometry. Hence, in general, the connection is neither metric
compatible nor torsion-free. Moreover, substituting (\ref{aofgamma}) into the
expressions (\ref{rfsdef}) and (\ref{cstardef}) for the gauge field
strengths ${R^{ab}}_{\mu\nu}$ and ${T^{\ast a}}_{\mu\nu}$, one finds that
%
%
%
\begin{eqnarray}
{R^\rho}_{\sigma\mu\nu} & = &
2(\partial_{[\mu}{\Gamma^\rho}_{|\sigma|\nu]}
+{\Gamma^\rho}_{\lambda[\mu}{\Gamma^\lambda}_{|\sigma|\nu]})- H_{\mu\nu}\delta_\sigma^\rho,\phantom{AAA}
\label{wgtcurvature}\\
{T^{\ast\lambda}}_{\mu\nu}
& = & 2{\Gamma^\lambda}_{[\nu\mu]},
\label{wgttorsion}\\
H_{\mu\nu} & = & 2\partial_{[\mu} B_{\nu]},
\label{wgthtensor}
\end{eqnarray}
where ${R^\rho}_{\sigma\mu\nu} = {e_a}^\rho {e^b}_\sigma
{R^a}_{b\mu\nu}$ and ${T^{\ast\lambda}}_{\mu\nu} =
{e_a}^\lambda{T^{\ast a}}_{\mu\nu}$. Thus, although we recognise
${T^{\ast\lambda}}_{\mu\nu}$ as (minus) the torsion tensor of the
$Y_4$ spacetime, we see that ${R^\rho}_{\sigma\mu\nu}$ is not simply
its Riemann tensor. Rather, the Riemann tensor of the $Y_4$ spacetime
is given by
\begin{equation}
{\widetilde{R}^{\rho}}_{\phantom{\rho}\sigma\mu\nu} \equiv {R^\rho}_{\sigma\mu\nu} +
H_{\mu\nu}\delta_\sigma^\rho.
\label{wgtrtildedef}
\end{equation}
One should note that, although $\widetilde{R}_{\rho\sigma\mu\nu}$ is
antisymmetric in $(\mu,\nu)$, it is no longer antisymmetric in
$(\rho,\sigma)$ (indeed $\widetilde{R}_{(\rho\sigma)\mu\nu} =
g_{\rho\sigma}H_{\mu\nu}$) and does not satisfy the familiar cyclic and
Bianchi identities of the Riemann tensor in a Riemannian $V_4$
spacetime.  One may also show that, with the given arrangements of
indices, both ${\widetilde{R}^{\rho}}_{\phantom{\rho}\sigma\mu\nu}$
(or ${R^{\rho}}_{\sigma\mu\nu}$) and ${T^{\ast\lambda}}_{\mu\nu}$
transform covariantly with weight $w=0$ under a local dilation.  It is
also worth noting that $\widetilde{R}_{\mu\nu} \equiv
{\widetilde{R}_{\mu\lambda\nu}}^{\phantom{\mu\lambda\nu}\lambda}
= R_{\mu\nu}- H_{\mu\nu}$ and $\widetilde{R} \equiv
{\widetilde{R}^{\mu}}_{\phantom{\mu}\mu} = R$. As one might expect,
the quantities (\ref{wgtcurvature})--(\ref{wgtrtildedef}) arise
naturally in the expression for the commutator of two derivative
operators acting on a vector $J^\rho$ (say) of Weyl weight $w$, which
is given by
\begin{equation}
[\nabla^\ast_\mu,\nabla^\ast_\nu] J^\rho =
{\widetilde{R}^{\rho}}_{\phantom{\rho}\sigma\mu\nu}J^\sigma +
wH_{\mu\nu} J^\rho - {T^{\ast\sigma}}_{\mu\nu}\nabla^\ast_\sigma V^\rho.
\label{wgtnablacomm}
\end{equation}

Finally, from (\ref{wgtsemimet}), one
also finds that the affine connection must satisfy
\begin{equation}
{\Gamma^\lambda}_{\mu\nu} = \christoffel{\ast\lambda}{\mu}{\nu}
+ {K^{\ast\lambda}}_{\mu\nu},
\label{weylaffinec}
\end{equation}
where the first term on the RHS reads
\begin{eqnarray}
\christoffel{\ast\lambda}{\mu}{\nu} &=&
\tfrac{1}{2}g^{\lambda\rho}(\partial^\ast_\mu g_{\nu\rho}+\partial^\ast_\nu
g_{\mu\rho}-\partial^\ast_\rho g_{\mu\nu}) \nonumber \\
&=&
\christoffel{\lambda}{\mu}{\nu}
+ \delta^\lambda_\nu B_\mu +
\delta^\lambda_\mu B_\nu - g_{\mu\nu}B^\lambda,
\label{diracconnection}
\end{eqnarray}
in which $\christoffel{\lambda}{\mu}{\nu}\equiv
\tfrac{1}{2}g^{\lambda\rho}(\partial_\mu g_{\nu\rho}+\partial_\nu
g_{\mu\rho}-\partial_\rho g_{\mu\nu})$ is the standard metric
(Christoffel) connection and ${K^{\ast\lambda}}_{\mu\nu}$ is the
$Y_4$ contortion tensor
\begin{equation}
{K^{\ast\lambda}}_{\mu\nu}=-\tfrac{1}{2}({T^{\ast\lambda}}_{\mu\nu}-
{{{T^\ast}_\nu}^{\lambda}}_\mu + {{T^\ast}_{\mu\nu}}^{\lambda}),
\label{wgtcontortiondef}
\end{equation}
for which it is worth noting that $K^\ast_{\lambda\mu\nu} = -
K^\ast_{\mu\lambda\nu}$.

The result (\ref{weylaffinec}) is the analogue of the
expression (\ref{astarzerodef}) in the gauge theory viewpoint.
Indeed, it is also of interest to consider briefly the geometric interpretation of
the quantities ${^0}{A^{\ast ab}}_\mu$, introduced in
(\ref{astarzerodef}), which depend only on the $h$-field, its first
derivatives and the $B$-field.  Following an analogous argument to
that given above, but considering instead the reduced covariant
derivative ${^0}D^\ast_\mu$, as defined in (\ref{dstarmuzerodef}), one finds
that ${^0}{A^{\ast ab}}_\mu$ and the connection
${^0}{\Gamma^{\ast \sigma}}_{\rho\mu}$ represent the same geometrical object
in two different frames, and one obtains a `reduced' form of the
tetrad postulate (\ref{weyltetradp}) given by
\begin{equation}
\zero{\Delta^\ast_\mu} {e^a}_\nu  \equiv
\partial^\ast_\mu {e^a}_\nu - \czero{{\Gamma^{\ast \sigma}}_{\nu\mu}}{e^a}_\sigma
+ \zero{{A^{\ast a}}_{b\mu}{e^b}_\nu} =0.
\label{wgttetradp0}
\end{equation}
It thus follows that the relationships (\ref{gammaofa}) and
(\ref{aofgamma}) again hold with the replacements
${\Gamma^\sigma}_{\nu\mu} \to \czero{{\Gamma^{\ast\sigma}}_{\nu\mu}}$
and ${A^a}_{b\mu} \to \zero{{A^{\ast a}}_{b\mu}}$, from which one can
directly derive (\ref{diracconnection}). One also obtains the
metricity condition ${^0}\nabla^\ast_\sigma g_{\mu\nu}=0$. Finally,
the expression (\ref{wgtcurvature}) for the curvature also holds, but
with ${R^{\rho}}_{\sigma\mu\nu} \to \zero{{R^{\ast
      \rho}}_{\sigma\mu\nu}}$ and ${\Gamma^\sigma}_{\nu\mu} \to
\czero{{\Gamma^{\ast \sigma}}_{\nu\mu}}$, whereas (\ref{wgttorsion})
becomes simply $\czero{{T^{\ast \lambda}}_{\mu\nu}} = 0$, indicating
the absence of torsion, as expected. The expression
(\ref{wgtnablacomm}) is also valid, but with $\nabla^\ast_\mu \to
{^0}\nabla^\ast_\mu$, ${R^\rho}_{\sigma\mu\nu} \to \zero{{R^{\ast
      \rho}}_{\sigma\mu\nu}}$ and ${T^{\ast \sigma}}_{\mu\nu} \to
{^0}{T^{\ast\sigma}}_{\mu\nu} = 0$.

Finally, it is worth noting that, in any given gravitational theory,
one can choose to interpret some quantities as
geometric properties of the underlying spacetime and others as fields
residing in that spacetime.  Indeed, one can place the dividing line
anywhere, with the gauge theory approach advocated in this paper and
the geometric interpretation just described representing extreme ends
of this range of possibilities: the former interprets all quantities
as fields residing in a background Minkowski spacetime, whereas
the latter interprets all gravitational quantities in terms of
geometric properties of the underlying spacetime. This is discussed
further in Appendix~\ref{app:pgt}, in the context of PGT.


\section{Extended Weyl gauge theory}
\label{sec:ewgt}

We now consider a novel alternative to standard Weyl gauge theory,
which has a number of interesting features.  Recall that, in
Sections~\ref{sec:wgtdirac} and \ref{sec:wgtem} respectively, we
showed that the PGT matter actions for the (massless) Dirac field and
the electromagnetic field are already invariant under local dilations
(and are, indeed, identical to their WGT counterparts), provided the
gravitational gauge fields ${h_a}^\mu$ and ${A^{ab}}_\mu$ transform
under local Weyl transformations as (\ref{weylhtrans}) and
(\ref{atrans}),
respectively. Considering just the local
dilation part of the transformation\cite{footnote14}, for the fields
${h_a}^\mu$ and ${A^{ab}}_\mu$ one has
\begin{eqnarray}
{h'_a}^\mu(x)& = & e^{-\rho(x)}{h_a}^\mu(x),
\label{weylhtrans2} \\
{A'^{ab}}_\mu (x) & = & {A^{ab}}_\mu(x)
\label{weylatrans2},
\end{eqnarray}
showing that they transform covariantly with weights $-1$ and $0$,
respectively (where we have temporarily reinstated the explicit
notational dependence of the fields on spacetime position $x$).

It is straightforward to show, however, that the Dirac and
electromagnetic field actions in PGT (and WGT) remain invariant under
local dilations, even if one assumes a more general `extended'
transformation law for the $A$-field, whilst still retaining the
original $h$-field transformation law, such that (\ref{weylhtrans2})
and (\ref{weylatrans2}) are replaced by
\begin{eqnarray}
{h'_a}^\mu & = & e^{-\rho}{h_a}^\mu
\label{eweylhtrans2} \\
{A'^{ab}}_\mu & = & {A^{ab}}_\mu + \theta
({b^a}_\mu{\cal P}^b-{b^b}_\mu{\cal P}^a),
\label{eweylatrans}
\end{eqnarray}
where $P_\nu \equiv \partial_\nu \rho$, ${\cal P}_a \equiv {h_a}^\nu
P_\nu$ and $\theta$ is an arbitrary parameter that can take any
value\cite{footnote14a}. Moreover, under a global scale
transformation, the `extended' transformation laws
(\ref{eweylhtrans2}) and (\ref{eweylatrans}) for the $h$ and $A$
fields reduce to the same form as the `normal' ones
(\ref{weylhtrans2}) and (\ref{weylatrans2}), and may be considered as
an equally valid gauging of global scale invariance.  These
observations provide compelling motivation for exploring the
properties of the gauge theory of local Weyl transformations in which
the $h$ and $A$ fields obey the transformations (\ref{eweylhtrans2})
and (\ref{eweylatrans}) under local dilations; we call this extended
Weyl gauge theory (eWGT).

A complementary motivation for exploring eWGT is that the `extended'
transformation laws (\ref{eweylhtrans2}) and (\ref{eweylatrans}) lead
to transformation properties for the PGT curvature ${{\cal
    R}^{ab}}_{cd}$ and torsion ${{\cal T}^a}_{bc}$ that are on a more
equal footing with one another than those that result from
the standard WGT transformations laws (\ref{weylhtrans2}) and
(\ref{weylatrans2}). Under the latter set of transformations, ${{\cal
    R}^{ab}}_{cd}$ and ${{\cal T}^a}_{bc}$ behave in very different
ways.  The PGT curvature ${{\cal R}^{ab}}_{cd}$ transforms covariantly
with weight $w=-2$ and also acts as the field strength tensor of the
$A$ gauge field in WGT.  By contrast, the PGT torsion ${{\cal
    T}^a}_{bc}$ transforms inhomogeneously as
\begin{equation}
{{\cal T}^{\prime a}}_{bc} = e^{-\rho}({{\cal T}^a}_{bc}+{\cal
  P}_b\delta_c^a - {\cal P}_c\delta_b^a),
\label{eqn:tortrans}
\end{equation}
and one must instead introduce ${{\cal T}^{\ast
    a}}_{bc}$ as the field strength tensor for the $h$ gauge field in
WGT. Indeed, the transformations (\ref{weylhtrans2}) and
(\ref{weylatrans2}) have the consequence of inducing PGT torsion.
Considering the case where the $h$-field originally has vanishing PGT
torsion ${{\cal T}^a}_{bc}=0$, the inhomogeneous nature of the
transformation law (\ref{eqn:tortrans}) means that the local dilation
leads to ${{\cal T}^{\prime a}}_{bc}\neq 0$.

Assuming the `extended' transformation laws (\ref{eweylhtrans2}) and
(\ref{eweylatrans}), however, the PGT curvature and torsion transform
under local dilations as
\begin{widetext}
\begin{eqnarray}
{{\cal R}^{\prime ab}}_{cd} & = & e^{-2\rho}\{{{\cal R}^{ab}}_{cd} +
2\theta\delta^{[a}_d({\cal D}_c-\theta{\cal P}_c){\cal P}^{b]} -
2\theta\delta^{[a}_c({\cal D}_d-\theta{\cal P}_d){\cal P}^{b]} -
2\theta{\cal P}^{[a}{{\cal T}^{b]}}_{cd} -
2\theta^2\delta^{[a}_c\delta^{b]}_d {\cal P}^e {\cal P}_e \},
\label{rspecialt} \\
{{\cal T}^{\prime a}}_{bc} & = & e^{-\rho}\{{{\cal T}^a}_{bc}+2(1-\theta){\cal
  P}_{[b}\delta_{c]}^a\}.\label{eqn:tortransgeneral}
\end{eqnarray}
\end{widetext}
Thus, for general values of $\theta$, neither ${{\cal R}^{ab}}_{cd}$
nor ${{\cal T}^{\dagger a}}_{bc}$ transforms covariantly, but for
$\theta=0$ one recovers the `normal' transformation law
(\ref{weylatrans2}) for the $A$-field, which results in a covariant
transformation law for the PGT curvature under local dilations,
whereas for $\theta=1$ one obtains the `special' transformation law for
the $A$-field,
\begin{equation}
{A'^{ab}}_\mu = {A^{ab}}_\mu + ({b^a}_\mu{\cal P}^b-
{b^b}_\mu{\cal P}^a),
\label{specialt}
\end{equation}
which leads to a covariant transformation law with weight $w=-1$ for
the PGT torsion. By considering the `extended' $A$-field
transformation law (\ref{eweylatrans}), one can therefore accommodate
both cases in a balanced manner. The corresponding transformation laws
for the contractions of the PGT curvature and torsion are given
explicitly in Section~\ref{sec:elsipgt}, and those for the `reduced'
PGT curvature ${^0}{{\cal R}^{ab}}_{cd}$ and its contractions (see
Appendix~\ref{app:pgt}) are given in Section~\ref{sec:elsipgtnot}; the
`reduced' PGT torsion ${^0}{{\cal T}^a}_{bc}$ vanishes identically.

\subsection{Extended local Weyl invariance}
\label{sec:elwi}

As usual, our first task is to define a method for converting a matter
action invariant under global Weyl transformations into one that is
invariant under local Weyl transformations, but where the $h$ and $A$
fields are now assumed to transform under local dilations as
(\ref{eweylhtrans2}) and (\ref{eweylatrans}), respectively. From now
on we will call this combination `extended local Weyl
transformations', although it should be noted that we are not
extending the form of GCT, local Lorentz or local scale
transformations under consideration, but merely assuming an
alternative (or `extended') form for the transformation of the
$A$-field under local dilations. In particular, we require the action
to be invariant for arbitrary values of the parameter $\theta$ in
(\ref{eweylatrans}). Following the standard gauge-theory approach,
we construct a new covariant derivative that
transforms under these transformations in the same way as the standard
partial derivative transforms under global Weyl transformations,
namely according to (\ref{weyldpsitrans}). As in WGT, we construct the
covariant derivative in two stages.

In the first step, we construct a
$(\Lambda,\rho)$-covariant derivative $D_\mu^\dagger\vpsi$ by
introducing a dilation vector gauge field (called $V_\mu$ to
distinguish it from the dilation gauge field $B_\mu$ in WGT). In
constructing this derivative, the $A$-field transformation law
(\ref{eweylatrans}) suggests the introduction of a modified $A$-field
of the form
\begin{equation}
{A^{\dagger ab}}_\mu  \equiv  {A^{ab}}_\mu + ({\cal V}^a{b^b}_\mu
- {\cal V}^b{b^a}_\mu),
\label{adaggerdef}
\end{equation}
where ${\cal V}_a = {h_a}^\mu V_\mu$. Clearly, by construction,
${A^{\dagger ab}}_\mu$ is also antisymmetric in $a$ and $b$. It is
worth noting that we do not consider ${A^{\dagger ab}}_\mu$ to be a
fundamental field, but merely a shorthand for the combination of the
gauge fields ${h_a}^\mu$ (or its inverse), ${A^{ab}}_\mu$ and $V_\mu$
on the RHS of (\ref{adaggerdef}). As in WGT, one must also include
terms in the covariant derivative to cancel terms arising from
$\partial_\mu \vpsi'$ and, in this case, also from the transformation
of ${A^{\dagger ab}}_\mu$. By demanding that $D^\dagger_\mu \vpsi$
transforms in the same way as (\ref{eqn:dstarprime}) under
extended local Weyl transformations, one finds that the
appropriate form for the covariant derivative is\cite{footnote15}
\begin{equation}
D^\dagger_\mu \vpsi \equiv (\partial_\mu + \tfrac{1}{2} {A^{\dagger
    ab}}_\mu\Sigma_{ab} - wV_\mu -\tfrac{1}{3}w T_\mu)\vpsi,
\label{ewgtcovdiv}
\end{equation}
in which $w$ is the Weyl weight of the field $\vpsi$ and
$T_\mu={b^a}_\mu {\cal T}_a$, where ${\cal T}_a$ is the trace of the
PGT torsion (see Appendix~\ref{app:pgt}), obtained by removing the
asterisks from equation (\ref{htrelation}). Here we consider $T_\mu$
merely as a shorthand for this corresponding function of the gauge
fields ${h_a}^\mu$ and ${A^{ab}}_\mu$. Under the extended
local scale transformations (\ref{eweylhtrans2})--(\ref{eweylatrans}), it is
straightforward to show that $T^\prime_\mu = T_\mu + 3(1-\theta)P_\mu$
and so (\ref{ewgtcovdiv}) does indeed transform covariantly with Weyl
weight $w$ provided that the dilation gauge field transforms
(inhomogeneously, as expected) according to
\begin{equation}
V'_\mu = V_\mu +\theta P_\mu,
\label{eqn:vtransform}
\end{equation}
which is the `extended' counterpart to (\ref{btransdef}) in WGT.  This
transformation law also ensures that the modified $A$-field
(\ref{adaggerdef}) is invariant under the extended scale
transformations (\ref{eweylhtrans2})--(\ref{eweylatrans}),
i.e. ${A^{\dagger\prime ab}}_\mu = {A^{\dagger ab}}_\mu$ and so it has
a weight $w=0$. Note how this emulates the invariance property
(\ref{weylatrans2}) of the original $A$-field under normal scale
transformations.

An alternative approach to obtaining the form of the covariant
derivative (\ref{ewgtcovdiv}) and the $V$-field transformation law
(\ref{eqn:vtransform}) is given in
Appendix~\ref{appendix:altddagderiv}, where we demonstrate that the
Weyl weight $w$ of the field $\vpsi$ on which the covariant derivative
acts does not need to be inserted into (\ref{ewgtcovdiv}) `by hand',
but instead arises naturally, in contrast to standard WGT, and
(\ref{ewgtcovdiv}) is picked out as the unique form for the eWGT
covariant derivative.

It is important to note that the covariant derivative
(\ref{ewgtcovdiv}) does not explicitly contain the parameter
$\theta$. Consequently, it does {\em not} reduce to the standard WGT
covariant derivative $D_\mu^\ast\vpsi$ (with $B_\mu$ replaced by
$V_\mu$) in the case $\theta=0$, in which one recovers the `normal'
local Weyl transformations.  Indeed, in this case, one sees from
(\ref{eqn:vtransform}) that one can covariantly set $V_\mu=0$, so that
${A^{\dagger ab}}_\mu = {A^{ab}}_\mu$ and the covariant derivative
(\ref{ewgtcovdiv}) may be written simply as
$\left.D^\dagger_\mu\vpsi\right|_{\theta=0} =
(D_\mu-\frac{1}{3}wT_\mu)\vpsi$, in which the role played by the usual
dilation gauge field $B_\mu$ in WGT has been taken up instead by the
trace of the PGT torsion; such a replacement has been discussed
previously\cite{obukhov82}. Conversely, in the case
$\theta=1$, for which one recovers the `special' transformation
(\ref{specialt}) for the $A$-field, one can covariantly set the PGT
torsion to zero, and hence $T_\mu = 0$, so that the covariant
derivative (\ref{ewgtcovdiv}) may be written as
$\left.D^\dagger_\mu\vpsi\right|_{\theta=1} \equiv (\partial_\mu +
\tfrac{1}{2} {A^{\dagger ab}}_\mu\Sigma_{ab} - wV_\mu)\vpsi$.

For the remainder of our discussion, however, we will consider the
general case in which the parameter $\theta$ may take any value.  As
for any gauge theory, the corresponding covariant derivative
(\ref{ewgtcovdiv}) defines the structure of eWGT. Indeed, in
principle, one could forgo the preceding discussion and simply start
with the definition (\ref{ewgtcovdiv}), together with
(\ref{adaggerdef}) and the transformation laws of ${h_a}^\mu$,
${A^{ab}}_\mu$, and $V_\mu$, to construct eWGT. By analogy with WGT,
we note that it is sometimes convenient to write the
$(\Lambda,\rho)$-covariant derivative (\ref{ewgtcovdiv}) as
\begin{equation}
D^\dagger_\mu \vpsi = (\partial^\dagger_\mu +
\tfrac{1}{2}{A^{\dagger ab}}_\mu \Sigma_{ab}) \vpsi,
\label{ddagdef}
\end{equation}
where we have defined the derivative operator
\begin{equation}
\partial^\dagger_\mu
\equiv \partial_\mu - w (V_\mu+\tfrac{1}{3}T_\mu).
\label{partialdaggerdef}
\end{equation}
It is straightforward to show that $\partial^\dagger_\mu \vpsi(x)$
itself also transforms covariantly with weight $w$ (i.e. the same
weight as the field $\vpsi$) under local dilations assuming the
extended transformation laws (\ref{eweylhtrans2})--(\ref{eweylatrans})
(but does not transform covariantly under local Poincar\'e
transformations).

In the second step of the gauging process, precisely as in WGT, we
define a generalised covariant derivative
\begin{equation}
{\cal D}^\dagger_a \vpsi \equiv {h_a}^\mu D^\dagger_\mu \vpsi,
\end{equation}
which transforms covariantly with weight $w-1$, as desired. Having
achieved our aim of constructing an appropriate covariant derivative,
we can now straightforwardly convert a matter action invariant under
global Weyl transformations into one that is invariant under extended
local Weyl transformations, which is given by
\begin{equation}
S_{\rm M} = \int h^{-1} L_{\rm M} (\vpsi,{\cal D}^\dagger_a \vpsi)\,d^4x.
\label{eweylmatlag_gen}
\end{equation}
Once again, it is convenient to denote the integrand of the action by
the Lagrangian density ${\cal L}_{\rm M} \equiv h^{-1} L_{\rm
  M}(\vpsi,{\cal D}^\dagger \vpsi)$. We note that the gravitational
gauge fields $h$, $A$ and $V$ in eWGT contain a total of 44
independent variables, as in WGT. As was the case in WGT,
(\ref{eweylmatlag_gen}) is not guaranteed to inherit invariance
properties possessed by original action under other types of
transformation, so (\ref{eweylmatlag_gen}) may need to be modified to
satisfy any further required invariances.

As in WGT, one may also introduce an additional `compensator' scalar
field $\phi$ (or fields) into the matter action, which opens up
possibilities for the inclusion of further terms in the matter action
that non-minimally (conformally) couple $\phi$ to the eWGT
gravitational gauge field strengths (see
Section~\ref{sec:ewgtgfs}). In particular, terms proportional to
$\phi^2 {\cal R}^\dagger$ or $\phi^2 L_{{\cal T}^{\dagger 2}}$ are
extended Weyl-covariant with weight $w=-4$ and so may be added to
$L_{\rm M}$.

\subsection{Relationship to other approaches}

Before continuing with the formal development of eWGT, it is worth
commenting briefly on the relationship of our approach to others in
the literature. In particular, by introducing the modified $A$-field
in (\ref{adaggerdef}), one might ask whether the modified
transformation law in (\ref{eweylatrans}) corresponds to some
`deformation' of the action of the Weyl group, which underlies the
physical symmetries we are assuming. Indeed, it might seem that such a
deformation does occur if one takes the view that specifying the group
action leads uniquely to a given set of gauge fields and
transformation laws; in some approaches these do follow directly from
the structure constants of the associated Lie group.

There have, for example, been several attempts in the literature to
carry out local gauging not just for the Weyl group, but for the full
conformal group, i.e.\ with special conformal translations included as
well as rotations, translations and dilations. An early approach to
this by Kaku et al.\cite{kaku77} considers the representation of the
full conformal group $\mbox{O}(4,2)$ via the $4\times4$ Dirac matrix
algebra of $\mbox{SU}(2,2)$, to which $\mbox{O}(4,2)$ is locally
isomorphic. In this approach, the structure constants of the group to
be locally gauged feed through via the Maurer--Cartan equations to
give the `curvatures' corresponding to the different group
elements. These equations are therefore fixed by the initial group
structure. 

Of course, such a method is very different to that which we carry out
here. We emphasise in particular that after the decoupling of the
translation and dilational parts of the transformation described at
the start of Section~\ref{sec:lmwgt}, the form of scaling we describe
here corresponds to `Weyl scaling', as discussed, for example, in
Section~2.2 of Blagojevic\cite{blagojevic02}. In this case, all fields
are rescaled according to their Weyl weight, but without any
accompanying coordinate changes. This is still exactly the symmetry
for which we are providing a local gauging via our modified $A$-field
(\ref{adaggerdef}), and modified transformation laws
(\ref{eweylatrans}), and thus no `deformation' has taken place of the
underlying symmetry. In particular the associated group is still `Weyl
scaling', despite the changes in the implementation of it.

It would, however, still be of interest to attempt to relate the
proposed new transformations and extended $A$-field found here with
attempts at full conformal gauging, such as carried out by Kaku et
al., since some intriguing relationships between the two approaches do
exist. In particular, both Kaku et al., and later
Wheeler\cite{wheeler91}, find that the gauge field corresponding to
special conformal transformations, which in their approaches is an
additional `vierbein-like' field similar to the field corresponding to
translations, does not propagate, and can be eliminated via its
equations of motion. This leaves a version of `Weyl scaling' as the
only further symmetry remaining in addition to translational and
rotational symmetries. 

The interest for our approach lies in the fact that, during the
gauging process, additional components of the spin connection are
found that are of similar form to the new piece of the $A$-field we
are proposing. An explicit example of this can be seen in equation (6)
of Kaku et al., in which a piece $e_{b\mu}b_a-e_{a\mu}b_b$ is added to
the spin connection $\omega_{\mu a b}$ (their $b$-field is the
`dilatation field', which in our case corresponds to ${\cal V}$, while
their $e_{a\mu}$ is a version of either our $h$-field or its inverse;
it is difficult to make an exact correspondence due to the very
different way the translation gauge field is treated between their
approach and ours.)

Furthermore, one of the problems of gauging the conformal group, which
the Kaku et al.\ approach is trying to overcome, lies in the fact that
the global group generators in a coordinate representation involve a
naked 4D position vector, $x^\mu$, and a local equivalent of this does
not exist. However, if we nevertheless examine global conformal
transformations, then Section~4.1 of Blagojevic\cite{blagojevic02}
shows how variation under these leads to a modified Lorentz spin
connection containing an extra term of the form
$c_{b\mu}x_a-c_{a\mu}x_b$, where here $c_{a\mu}$ is the extra
`vierbein-type' field corresponding to the special conformal
transformations. Here $x_a$ is a global surrogate for ${\cal V}_a$
(noting the fact that the generator of global dilations is $x \cdot
\nabla$), and so this is again suggestive that our proposed additional
term in the Lorentz spin connection given in (\ref{adaggerdef}), and
the modified transformation laws (\ref{eweylatrans}) (involving an
arbitrary new parameter $\theta$, which in the end does not appear in
the covariant derivative itself), may be linked with the symmetries of
the full conformal group, and may even provide a new route through to
its successful gauging. We leave this topic to future research,
however, and now continue with our development of eWGT.

\subsection{Minkowski spacetime interpretation of eWGT}

The comments made in Section~\ref{sec:minkintwgt} also hold in eWGT,
in particular the freedom to choose global inertial Cartesian
coordinates in the Minkowski spacetime and, if required, the way in
which the covariant derivative can be extended to act on quantities
with a definite tensor behaviour under GCT. In particular, the
$(\Lambda,\rho)$-covariant derivative in (\ref{ewgtcovdiv}) becomes
\begin{equation}
D^\dagger_\mu = \partial^\dagger_\mu + {^0}{\Gamma^\sigma}_{\rho\mu}
{\matri{X}^\rho}_\sigma + \tfrac{1}{2}{A^{\dagger ab}}_\mu\Sigma_{ab} =
{^0}\nabla^\dagger_\mu + \tfrac{1}{2}{A^{\dagger ab}}_\mu\Sigma_{ab},
\label{extcovdweyle}
\end{equation}
where ${^0}\nabla^\dagger_\mu \equiv {^0}\nabla_\mu - w(V_\mu+\tfrac{1}{3}T_\mu)$,
${^0}{\Gamma^\lambda}_{\mu\nu} \equiv
\tfrac{1}{2}\gamma^{\lambda\rho}(\partial_\mu
\gamma_{\nu\rho}+\partial_\nu \gamma_{\mu\rho}-\partial_\rho
\gamma_{\mu\nu})$ is the metric connection corresponding to the
(flat-space) metric $\gamma_{\mu\nu}$ defined by the coordinate
system, and ${\matri{X}^\rho}_\sigma$ are the $\mbox{GL}(4,R)$
generator matrices appropriate to the GCT tensor character of the
field to which $D^\dagger_\mu$ is applied, and $w$ is its Weyl weight.

\subsection{Gauge field strengths in eWGT}
\label{sec:ewgtgfs}

As in WGT, we define the gauge field strengths in terms of the
commutator of covariant derivatives. Considering first the
eWGT $(\Lambda,\rho)$-covariant derivative, one finds that
\begin{equation}
[D^\dagger_\mu,D^\dagger_\nu]\vpsi = \tfrac{1}{2}{R^{\dagger
ab}}_{\mu\nu}\Sigma_{ab}\vpsi - w H^\dagger_{\mu\nu}\vpsi,
\label{rdaggerdefine}
\end{equation}
which is clearly of a similar form to the corresponding result
(\ref{dstarmucomm}) in WGT, but the eWGT field strengths have very
different forms to their WGT counterparts, in particular
their dependencies on the gauge fields.

The eWGT rotational field strength tensor is found to have the form
\begin{widetext}
\begin{subequations}
\label{eqn:rdaggerdefmaster}
\begin{eqnarray}
{R^{\dagger ab}}_{\mu\nu}  & \equiv & \partial_\mu {A^{\dagger ab}}_\nu
- \partial_\nu {A^{\dagger ab}}_\mu
+{A^{\dagger a}}_{c\mu}{A^{\dagger cb}}_\nu - {A^{\dagger
    a}}_{c\nu}{A^{\dagger cb}}_\mu,
\label{eqn:rdaggerdef}\\
& = & {R^{ab}}_{\mu\nu}
+ 4{b^{[b}}_{[\nu} D_{\mu]} {\cal V}^{a]}
+4 {\cal V}^{[a} V_{[\mu} {b^{b]}}_{\nu]}
-2 {\cal V}^e {\cal V}_e {b^{[a}}_\mu {b^{b]}}_\nu
+2 {\cal V}^{[a}{T^{b]}}_{\mu\nu} ,\label{eqn:rdaggerdef2}
\end{eqnarray}
\end{subequations}
\end{widetext}
where ${R^{ab}}_{\mu\nu}$ is the rotational gauge field strength (curvature) in
PGT (and WGT), ${T^a}_{\mu\nu}$ is the PGT translational gauge field
(torsion), and $D_\mu$ is the PGT covariant derivative. The eWGT dilation gauge field strength reads
\begin{equation}
H^\dagger_{\mu\nu}
= \partial_\mu(V_\nu+\tfrac{1}{3}T_\nu)
- \partial_\nu(V_\mu+\tfrac{1}{3}T_\mu).
\label{hdaggerdef}
\end{equation}
As in WGT, both ${R^{\dagger ab}}_{\mu\nu}$ and $H^\dagger_{\mu\nu}$
transform covariantly under GCT and local Lorentz rotations in
accordance with their respective index structures, and are invariant
under extended local dilations.

Considering instead the commutator of two generalised covariant
derivatives, one finds
\begin{equation}
[{\cal D}^\dagger_c,{\cal D}^\dagger_d]\vpsi = \tfrac{1}{2}{{\cal R}^{\dagger
ab}}_{cd}\Sigma_{ab}\vpsi - w {\cal H}^\dagger_{cd}\vpsi
- {{\cal T}^{\dagger a}}_{cd}{\cal D}^\dagger_a \vpsi,
\label{ewgtdacomm}
\end{equation}
where ${{\cal R}^{\dagger ab}}_{cd}
= {h_c}^\mu {h_d}^\nu {R^{\dagger ab}}_{\mu\nu}$ and
${\cal H}^\dagger_{cd} = {h_c}^\mu {h_d}^\nu H^\dagger_{\mu\nu}$, and
the translational field strength is now given by
\begin{equation}
{{\cal T}^{\dagger a}}_{bc} \equiv {h_b}^\mu {h_c}^\nu
(D^\dagger_\mu {b^a}_\nu - D^\dagger_\nu {b^a}_\mu)
\equiv {h_b}^\mu {h_c}^\nu {T^{\dagger a}}_{\mu\nu}.
\label{ewgttorsiondef}
\end{equation}
It is easy to show that, similarly to WGT, ${{\cal R}^{\dagger
    ab}}_{cd}$, ${\cal H}^\dagger_{cd}$ and ${{\cal T}^{\dagger
    a}}_{cd}$ transform covariantly under extended local dilations
with weights $w({{\cal R}^{\dagger ab}}_{cd})=w({\cal
  H}^\dagger_{cd})=-2$ and $w({{\cal T}^{\dagger a}}_{cd})=-1$
respectively.

For later convenience, it is worth noting that the
explicit forms of ${{\cal R}^{\dagger ab}}_{cd}$ and its contractions
are given in terms of their PGT (and WGT) counterparts by
\begin{widetext}
\begin{subequations}
\label{rdaggerrelations}
\begin{eqnarray}
{{\cal R}^{\dagger ab}}_{cd} & = & {{\cal R}^{ab}}_{cd}
+ 2\delta^{[b}_d({\cal D}_c + {\cal V}_c){\cal V}^{a]}
- 2\delta^{[b}_c({\cal D}_d + {\cal V}_d){\cal V}^{a]}
-2 {\cal V}^e {\cal V}_e \delta^{[a}_c \delta^{b]}_d
+2 {\cal V}^{[a}{{\cal T}^{b]}}_{cd} , \label{rdaggerijkl}\\
{{\cal R}^{\dagger a}}_{c} & = &  {{\cal R}^{a}}_{c} + 2({\cal
  D}_c+\tfrac{1}{2}{\cal T}_c + {\cal V}_c){\cal V}^a +
\delta^a_c({\cal D}_b -2{\cal V}_b){\cal V}^b -{{\cal T}^a}_{cb}{\cal
  V}^b,\label{riccitdagger}\\
 {\cal R}^\dagger & = & {\cal R} + 6({\cal D}_a+\tfrac{1}{3}{\cal
   T}_a-{\cal V}_a){\cal V}^a,\label{riccisdagger}
\end{eqnarray}
\end{subequations}
\end{widetext}
where ${\cal D}_a$ is the general PGT covariant derivative.
Similarly, ${{\cal T}^{\dagger a}}_{bc}$ is given in terms of its
PGT counterpart by
\begin{equation}
{{\cal T}^{\dagger a}}_{bc}  =  {{\cal T}^a}_{bc}
+\tfrac{1}{3}(\delta^a_b {\cal T}_c-\delta^a_c{\cal T}_b).
\label{tdaggerikl}
\end{equation}
It is particularly important to note from
(\ref{tdaggerikl}) that the trace of the eWGT torsion vanishes
identically, namely
\begin{equation}
{\cal T}^\dagger_b \equiv {{\cal T}^{\dagger a}}_{ba} = 0,
\label{zerotorsiontrace}
\end{equation}
so that ${{\cal T}^{\dagger a}}_{bc}$ is completely trace-free
(contraction on any pair of indices yields zero).  As we will see
below, the property (\ref{zerotorsiontrace}) has some desirable and
interesting consequences for eWGT. Clearly, the dilational gauge field
strength ${\cal H}^\dagger_{ab}$ is antisymmetric in $a$ and $b$ and
hence also has no non-trivial contractions.

The functional dependencies of the three field strengths tensors on
the gauge fields and their derivatives are ${{\cal R}^{\dagger
    ab}}_{cd}(h,A,\partial A, V, \partial V)$, ${{\cal T}^{\dagger
    a}}_{bc}(h,\partial h, A)$ and ${\cal H}^\dagger_{ab}(h,\partial
h, \partial^2 h, A, \partial A, \partial V)$. It is important to note
that these functional dependencies differ markedly from those in
standard WGT. In particular, ${{\cal R}^{\dagger ab}}_{cd}$ depends on
the dilation gauge field $V$ and it first derivatives, whereas its WGT
counterpart has no dependence on the WGT dilation gauge field $B$. By
contrast, the translational field strength ${{\cal T}^{\dagger
    a}}_{bc}$ does not depend on the dilation gauge field, whereas its
WGT counterpart does.
Finally, the most profound difference occurs for
the dilation field strength ${\cal H}^\dagger_{ab}$, which depends on
$\partial h$, $\partial^2 h$, $A$ and $\partial A$, in addition to the
dependency, in common with its WGT counterpart, on $h$ and the first
derivative of the dilation gauge field.
The dependence of ${\cal
  H}^\dagger_{ab}$ on the second derivative $\partial^2 h$ is a
particularly unusual feature as compared with typical gauge theories,
and is discussed in detail in Section~\ref{sec:ewgtsecondorder}.

\subsection{Alternative form of covariant derivative in eWGT}
\label{sec:ewgtaltcovd}

In a similar manner to that discussed in Section~\ref{sec:wgtcovdalt}
for WGT, one may obtain an alternative form for the covariant
derivative in eWGT. By analogy with (\ref{astarzerodef}) in WGT, it
  is straightforward to show that
\begin{equation}
{\cal A}^\dagger_{abc} = \zero{{\cal A}^\dagger_{abc}}(h,\partial h,A,V)
+{\cal K}^\dagger_{abc} (h,\partial h,A),
\label{adaggerzerodef}
\end{equation}
in which we have defined the quantities $\zero{{\cal A}^\dagger_{abc}}
\equiv \tfrac{1}{2}(c^\dagger_{abc}+c^\dagger_{bca}-c^\dagger_{cab})$,
where ${c^{\dagger\,c}}_{ab} \equiv {h_a}^\mu {h_b}^\nu
(\partial^\dagger_\mu {b^c}_\nu-\partial^\dagger_\nu {b^c}_\mu)$, and
${\cal K}^\dagger_{abc} \equiv -\tfrac{1}{2} ({\cal
  T}^\dagger_{abc}+{\cal T}^\dagger_{bca}-{\cal T}^\dagger_{cab})$. As
before, both $\zero{{\cal A}^\dagger_{abc}}$ and ${\cal
  K}^\dagger_{abc}$ are antisymmetric in their first two indices.
There is a fundamental difference with WGT, however, since
$\zero{{\cal A}^\dagger_{abc}}$ itself depends on the $A$ gauge field,
whereas this was not the case previously. Less importantly, ${\cal
  K}^\dagger_{abc}$ is independent of the dilation gauge field, unlike
its WGT counterpart\cite{footnote16}.

Under an extended local Weyl transformation, the quantities $\zero{{\cal
    A}^\dagger_{abc}}$ transform in the same way as ${\cal
  A}^\dagger_{abc}$, whereas ${\cal K}^\dagger_{abc}$ transform as the
components of a local tensor with weight $w=-1$. Thus, one can
construct the `reduced' $(\Lambda,\rho)$-covariant derivative
\begin{equation}
\zero{D}^\dagger_\mu\vpsi \equiv (\partial^\dagger_\mu
+{\textstyle\frac{1}{2}} \zero{{A^{\dagger ab}}_\mu}\Sigma_{ab})\vpsi,
\label{ddaggermuzerodef}
\end{equation}
which transforms in the same way as $D^\dagger_\mu\vpsi$ under an
extended local Weyl transformation. The key difference in eWGT,
however, is that $\zero{D}^\dagger_\mu\vpsi$ still depends on the $A$
gauge field, and so is less deserving of the description `reduced' than
its counterpart in WGT.  One can define the corresponding generalised
covariant derivative ${^0}{\cal D}^\dagger_a \equiv {h_a}^\mu
\,\zero{D}^\dagger_\mu$, which consequently transforms in the same way
as ${\cal D}^\dagger_a$, as required. The `full' generalised covariant
derivative is given in terms of the `reduced' one by the alternative
form
\begin{equation}
\mathcal{D}^\dagger_a\vpsi = (\czero{\mathcal{D}}^\dagger_a +
\tfrac{1}{2}{\mathcal{K}^{\dagger bc}}_a\Sigma_{bc})\vpsi,
\label{ewgtcovderivalt}
\end{equation}
which illustrates that, if one (covariantly) sets the eWGT torsion (and
hence contortion) to zero, then $\mathcal{D}^\dagger_a$ reduces to
$\czero{\mathcal{D}}^\dagger_a$.

\subsection{Bianchi identities in eWGT}
\label{sec:ewgtbianchi}

One may calculate the Bianchi identities satisfied by the
gravitational gauge field strengths ${{\cal R}^{\dagger ab}}_{cd}$,
${{\cal T}^{\dagger a}}_{bc}$ and ${\cal H}^\dagger_{ab}$ in eWGT by
applying the Jacobi identity to the generalised covariant derivative
${\cal D}^\dagger_a$ , in a similar manner to that used in
Section~\ref{sec:wgtbianchi} for WGT.

One quickly finds the three Bianchi identities
\begin{subequations}
\label{ewgtbianchiids}
\begin{eqnarray}
{\cal D}^\dagger_{[a}{{\cal R}^{\dagger de}}_{bc]}-{{\cal T}^{\dagger
    f}}_{[ab} {{\cal R}^{\dagger de}}_{c]f} & = & 0, \label{ewgtbi1} \\
\hspace*{-9mm}{\cal D}^\dagger_{[a}{{\cal T}^{\dagger d}}_{bc]}-{{\cal T}^{\dagger
    e}}_{[ab} {{\cal T}^{\dagger d}}_{c]e}-{{\cal R}^{\dagger
  d}}_{[abc]}+ {\cal H}^\dagger_{[ab}\delta^d_{c]}
 & = & 0, \label{ewgtbi2} \\
{\cal D}^\dagger_{[a}{{\cal H}^\dagger}_{bc]} -{{\cal T}^{\dagger
    e}}_{[ab} {{\cal H}^{\dagger}}_{c]e} & = & 0. \label{ewgtbi3}
\end{eqnarray}
\end{subequations}
Similar to WGT, for the special case in which the `torsion' is
totally antisymmetric, such that ${\cal T}^\dagger_{abc}= \epsilon_{abcd} t^d$
for some vector $t^d$, then one may show that the second term on the
LHS of (\ref{ewgtbi2}) vanishes.

By contracting over the indices $a$ and $d$ in the `${\cal R}$-identity'
(\ref{ewgtbi1}), once obtains the once-contracted Bianchi
identity
\begin{equation}
{\cal D}^\dagger_{a}{{\cal R}^{\dagger ae}}_{bc}-2
{\cal D}^\dagger_{[b}{{\cal R}^{\dagger e}}_{c]}
-2{{\cal T}^{\dagger
    f}}_{a[b} {{\cal R}^{\dagger ae}}_{c]f}
-{{\cal T}^{\dagger
    f}}_{bc} {{\cal R}^{\dagger e}}_{f}
 =  0. \label{ewgtcbi1}
\end{equation}
Contracting again over $b$ and $e$, one then finds the
twice-contracted Bianchi identity
\begin{equation}
{\cal D}^\dagger_{a}({{\cal R}^{\dagger a}}_{c}
-\tfrac{1}{2}\delta^a_c{\cal R}^\dagger)
+{{\cal T}^{\dagger
    f}}_{bc} {{\cal R}^{\dagger b}}_{f}
+\tfrac{1}{2}{{\cal T}^{\dagger
    f}}_{ab} {{\cal R}^{\dagger ab}}_{cf}
 =  0. \label{ewgtcbi2}
\end{equation}

Turning to the `${\cal T}$-identity' (\ref{ewgtbi2}) and contracting
over $a$ and $d$, one obtains the further once-contracted Bianchi
identity
\begin{equation}
{\cal D}^\dagger_{a}{{\cal T}^{\dagger a}}_{bc}+2{{\cal
    R}^{\dagger}}_{[bc]}
+ 2{\cal H}^\dagger_{bc}
=  0. \label{ewgtcbi3}
\end{equation}
It is worth noting that this is somewhat simpler than the
corresponding Bianchi identity (\ref{wgtcbi3}) in WGT, because of the
condition (\ref{zerotorsiontrace}) for the automatic vanishing of the
eWGT torsion trace.  Thus, even in the most general case, the
covariant divergence of the translation gauge field strength depends only the
rotational and dilational gauge field strengths.
It is clear that the `${\cal H}$-identity'
(\ref{ewgtbi3}) has no non-trivial contractions.

\subsection{Free gravitational action in eWGT}
\label{sec:ewgtfga}

The construction of the free gravitational action in eWGT is analogous
to the approach adopted in WGT and discussed in
Section~\ref{sec:wgtfga}. In principle, from ${\cal
  R}^\dagger_{abcd}$, ${\cal T}^\dagger_{abc}$ and ${\cal
  H}^\dagger_{ab}$, one can construct a free gravitational action of
the general form
\begin{equation}
S_{\rm G} = \int h^{-1} L_{\rm G}({\cal R}^\dagger_{abcd},
{\cal T}^\dagger_{abc},{\cal  H}^\dagger_{ab})\,d^4x,
\label{ewgtgaction}
\end{equation}
where the requirement of local scale invariance imposes the constraint
that $L_{\rm G}$ must be a relative scalar with Weyl weight $w(L_{\rm
  G})=-4$. Hence, once again, $L_{\rm G}$ can be quadratic in ${\cal
  R}^\dagger_{abcd}$ and ${\cal H}^\dagger_{ab}$, but terms linear in
${\cal R}^\dagger \equiv {{\cal R}^{\dagger ab}}_{ab}$ or quadratic in
${\cal T}^\dagger_{abc}$ are not allowed. Similarly, higher-order terms in
${\cal R}^\dagger_{abcd}$ and ${\cal H}^\dagger_{ab}$ are forbidden; in principle one
could include quartic terms in ${\cal T}^\dagger_{abc}$, but we will not
consider them here.

Thus, by analogy with WGT, the general form of $L_{\rm G}$, possessing
terms no higher than quadratic order in the field strengths, is of the
form
\begin{equation}
L_{\rm G} = L_{{\cal R}^{\dagger 2}} + L_{{\cal H}^{\dagger 2}},
\label{ewgtgravlagdef}
\end{equation}
where the expressions for $L_{{\cal R}^{\dagger 2}}$ and $L_{{\cal
    H}^{\dagger 2}}$ are given in (\ref{lg2}), but written in terms of
the eWGT field strengths, i.e.  making the replacements ${\cal
  R}_{abcd} \to {\cal R}^\dagger_{abcd}$ and ${\cal H}_{ab} \to {\cal
  H}^\dagger_{ab}$. Performing a similar derivation to the WGT
case\cite{nieh80}, one can show that the eWGT field strength ${\cal
  R}^\dagger_{abcd}$ satisfies a form of the Gauss--Bonnet identity
such that the combination (\ref{gbid}), with the replacement ${\cal
  R}_{abcd} \to {\cal R}^\dagger_{abcd}$, contributes a total
derivative to the action (in $D \le 4$ dimensions), and so has no
effect on the resulting field equations. Hence one may set any one of
$\alpha_1$, $\alpha_3$ or $\alpha_6$ in (\ref{lg2}) to zero, without
loss of generality (at least classically), but we will retain all
these terms for the moment.

We note that, in the total Lagrangian
(see Section~\ref{sec:ewgtfecons}), one may also include terms of the
generic form $\phi^2{\cal R}^\dagger$ and $\phi^2 L_{{\cal T}^{\dagger
    2}}$, where $L_{{\cal T}^{\dagger 2}}$ is given in (\ref{lt2})
with ${\cal T}_{abc} \to {\cal T}^\dagger_{abc}$ (but with no term
proportional to $\beta_3$, because the eWGT torsion trace vanishes identically).
Since these terms depend upon the scalar (compensator) field
$\phi$, in addition to the gravitational gauge fields, we do not
consider them to be part of the {\em free} gravitational action, but
instead regard them as belonging to the matter Lagrangian $L_{\rm M}$.

\subsection{Field equations in eWGT}
\label{sec:ewgtfecons}

As in WGT, the total action $S_{\rm T}$ is simply the sum of the
matter and free gravitational actions.  In the free gravitational
sector, the form of the gravitational Lagrangian
(\ref{ewgtgravlagdef}) induces a dependence on all three gauge fields
and their first derivatives, plus a dependence on $\partial^2 h$
(suppressing indices for brevity).  The dependence of the
gravitational Lagrangian on $\partial h$ makes eWGT more similar to
PGT than WGT, but the dependence on $\partial^2 h$ is unique to eWGT
and arises from the term proportional to ${\cal H}^\dagger_{ab}{\cal
  H}^{\dagger ab}$ in (\ref{ewgtgravlagdef}).  In the matter sector,
covariant derivatives of the matter field $\vpsi$ induce a dependence
on $\vpsi$, $\partial\vpsi$, $h$, $\partial h$, $A$ and $V$; the
dependence on $\partial h$ is new compared to WGT and arises from the
term containing ${\cal T}_a$ in ${\cal D}^\dagger_a\vpsi$. We will
also consider here the inclusion of an additional `compensator' scalar
field $\phi$ in the matter action, and further admit the possibility
that it may include a kinetic term for the scalar field that contains
derivatives of $\phi$.  Moreover, if the matter action includes a term
proportional to $\phi^2{\cal R}^\dagger$, then this brings an
additional dependence on $\partial A$ and $\partial V$; the latter is
new compared to WGT.  Finally, unlike in WGT, the inclusion of a term
of the generic form $\phi^2L_{{\cal T}^{\dagger 2}}$ produces no
additional functional dependencies.  Consequently, we take the
total Lagrangian density to be
\begin{eqnarray}
{\cal L}_{\rm T} &=&
{\cal L}_{\rm G}(h, \partial h, \partial^2 h, A,\partial A,V,\partial
V) \nonumber \\&&+
{\cal L}_{\rm M}(\vpsi,\partial\vpsi,\phi,\partial\phi,h,\partial
h,A,\partial A,V,\partial V),
\label{ltotewgt}
\end{eqnarray}
where we have indicated the functional dependencies in the most
general case. It is worth reiterating that these dependencies differ
considerably from those in WGT, as indicated in (\ref{wgtlagtot}).

Although ${\cal L}_{\rm T}$ is at most quadratic in the field strength
tensors ${\cal R}^\dagger_{abcd}(h,A,\partial A,V,\partial V)$, ${\cal
  T}^\dagger_{abc}(h,\partial h,A)$ and ${\cal
  H}^\dagger_{ab}(h,\partial h, \partial^2h, A,\partial A, \partial
V)$, the last of these depends (linearly) on the second derivative
$\partial^2 h$ of the translational gauge field, as indicated. Thus,
unlike in WGT, if the term proportional to ${\cal H}^{\dagger ab}{\cal
  H}^\dagger_{ab}$ is included in ${\cal L}_{\rm G}$, then the
corresponding terms it generates in the resulting field equations will
typically be linear in fourth-order derivatives of $h$. We discuss
this issue in detail in Section~\ref{sec:ewgtsecondorder}.

Variation of $S_{\rm T}$ with respect to ${h_a}^{\mu}$, ${A^{ab}}_{\mu}$,
and $V_\mu$ leads to three coupled gravitational field
equations. As previously, it is convenient to work with quantities
carrying only Latin indices and write the general field equations as
\begin{subequations}
\label{eqn:eweylgenfe}
\begin{eqnarray}
{t^a}_b + {\tau^a}_b & = & 0, \label{eqn:eweylgenfe1}\\
{s_{ab}}^c + {\sigma_{ab}}^c & = & 0,\label{eqn:eweylgenfe2}\\
j^a + \zeta^a & = &0,\label{eqn:eweylgenfe3}
\end{eqnarray}
\end{subequations}
where ${t^a}_b \equiv {t^a}_\mu {h_b}^\mu$, ${s_{ab}}^c \equiv
{s_{ab}}^\mu {b^c}_\mu$ and $j^a\equiv j^\mu {b^a}_\mu$, and similarly
for the matter sector. In the gravitational sector, ${t^a}_\mu \equiv
\delta {\cal L}_{\rm G}/\delta {h_a}^\mu$, ${s_{ab}}^\mu \equiv \delta
       {\cal L}_{\rm G}/\delta {A^{ab}}_\mu$ and $j^\mu \equiv \delta
       {\cal L}_{\rm G}/\delta V_\mu$ respectively. In the matter
       sector, the energy-momentum ${\tau^a}_\mu\equiv \delta{\cal
         L}_{\rm M}/\delta {h_a}^\mu$, spin-angular-momentum
       ${\sigma_{ab}}^\mu\equiv \delta{\cal L}_{\rm M}/\delta
       {A^{ab}}_\mu$ and dilation current $\zeta^\mu\equiv\delta{\cal
         L}_{\rm M}/\delta V_\mu$ of the matter fields act as sources.

All the quantities in (\ref{eqn:eweylgenfe1})--(\ref{eqn:eweylgenfe3})
are clearly invariant under GCT, and it is straightforward to show
that they all transform covariantly under local Lorentz rotations in
accordance with their respective index structures.  Under extended
local dilations, the quantities in (\ref{eqn:eweylgenfe2}) and
(\ref{eqn:eweylgenfe3}) are all covariant with weight $w=1$. Contrary
to WGT, however, the individual quantities ${t^a}_b$ and ${\tau^a}_b$
in (\ref{eqn:eweylgenfe1}) are {\em not}, in general, covariant under
extended local dilations. This rather unusual feature is a result of
the transformation law (\ref{eweylatrans}) for ${A^{ab}}_\mu$
containing one of the other gauge fields, namely the inverse $h$-field
${b^a}_\mu$. Indeed, under an extended local dilation,
\begin{subequations}
\label{eqn:emtnoncov}
\begin{eqnarray}
t'_{ab} & = & t_{ab} + 2\theta({s_{ac}}^c{\cal P}_b
-s_{cba}{\cal P}^c), \label{eqn:emtnoncov1}\\
\tau'_{ab} & = & \tau_{ab} + 2\theta({\sigma_{ac}}^c{\cal P}_b
-\sigma_{cba}{\cal P}^c), \label{eqn:emtnoncov2}
\end{eqnarray}
\end{subequations}
although it is worth noting that the traces ${t^a}_a$ and ${\tau^a}_a$
do transform covariantly, with weight $w=0$ (i.e. invariant).  Hence
the non-covariant part of the transformation results only from the
antisymmetric part of the energy-momentum tensor. Nonetheless,
although neither $t_{ab}$ nor $\tau_{ab}$ is individually covariant,
one sees that the full $h$-field equation (\ref{eqn:eweylgenfe1}) {\em
  is} covariant, since the non-covariant terms in
(\ref{eqn:emtnoncov1}) and (\ref{eqn:emtnoncov2}) cancel by virtue of
the $A$-field equation (\ref{eqn:eweylgenfe2}). Moreover, the
transformations (\ref{eqn:emtnoncov1}) and (\ref{eqn:emtnoncov2})
suggest\cite{footnote17} how to construct covariant forms of
$t_{ab}$ and $\tau_{ab}$. From the transformation property
(\ref{eqn:vtransform}) of the vector gauge field $V_\mu$, we see
immediately that, under an extended local dilation, the quantity
\begin{equation}
\tau^\dagger_{ab} \equiv \tau_{ab} + 2\sigma_{cba}{\cal V}^c -
     2{\sigma_{ac}}^c{\cal V}_b,
\label{covemtdef}
\end{equation}
does transform covariantly with weight $w=0$ (i.e. it is
invariant). It is also worth noting that ${\tau^{\dagger a}}_a =
{\tau^a}_a$. One may construct a covariant $t^\dagger_{ab}$ in a
similar way. Indeed, the gravitational field equation
(\ref{eqn:eweylgenfe1}) can then be replaced by
\begin{equation}
{t^{\dagger a}}_b  + {\tau^{\dagger a}}_b =0.
\label{eqn:eweylgenfe1dagger}
\end{equation}

Varying $S_{\rm T}$ with respect to $\vpsi$ and $\phi$ leads to the
matter field equations. These are easily found from those obtained for
WGT in Section~\ref{sec:wgtfieldeqns} and amount to replacing ${\cal
  D}^\ast_a$ with ${\cal D}^\dagger_a$.  An important difference does
arise, however, since ${\cal T}^\dagger_a \equiv 0$. Consequently, in
eWGT, one finds that the matter equations of motion can be written in
the form
\begin{subequations}
\label{eqn:eweylgenfeall}
\begin{eqnarray}
\frac{\delta L_{\rm M}}{\delta\vpsi} =
\bpd{L_{\rm M}}{\vpsi}-{\cal D}^\dagger_a
\left(\pd{L_{\rm M}}{({\cal D}^\dagger_a\vpsi)}\right) & = & 0, \label{eweylpeqn} \\
\frac{\delta L_{\rm M}}{\delta\phi} =
\bpd{L_{\rm M}}{\phi}-{\cal D}^\dagger_a
\left(\pd{L_{\rm M}}{({\cal D}^\ast_a\phi)}\right) & = & 0,
\label{eweylpeqnphi}
\label{eqn:eweylgenfe4}
\end{eqnarray}
\end{subequations}
where $\bar{\partial}L_{\rm M}/\partial \vpsi \equiv [\partial L_{\rm
    M}(\vpsi,{\cal D}^\dagger_au,\phi,{\cal
    D}^\dagger_a\phi,\ldots)/\partial \vpsi]_{u=\vpsi}$, so that
$\vpsi$ and ${\cal D}^\dagger_a\vpsi$ are treated as independent
variables, and similarly for the $\phi$ field equation. This result
has the important consequence that, contrary to the general case in
WGT (and PGT), the matter field equations derived from the
minimal-coupling procedure applied to the action $S_{\rm M}$ are {\em
  equivalent} to those obtained simply by applying the
minimal-coupling procedure to the field equations directly,
independent of the forms of ${\cal L}_{\rm G}$ and ${\cal L}_{\rm
  M}$. We should note, however, that the general form
(\ref{eweylpeqn}) is only valid if the action (\ref{eweylmatlag_gen})
has not subsequently been modified to restore any further required
invariances that were lost in the localisation of the extended Weyl
symmetry.

\subsection{Conservation laws in eWGT}
\label{sec:ewgtconslaws}

The conservation laws for eWGT may also be straightforwardly obtained.
Invariance of $S_{\rm G}$ under (infinitesimal) local Lorentz
rotations, GCTs and extended local dilations, respectively, lead
(after a lengthy calculation) to the following manifestly covariant
conservation laws in the free gravitational sector:
\begin{eqnarray}
{\cal D}^\dagger_c(h{s_{ab}}^c) + ht^\dagger_{[ab]} & = & 0,\label{eqn:ewgtcons1}\\
\hspace*{-7mm}{\cal D}^\dagger_c(h{t^{\dagger c}}_d)\! -\! h({s_{ab}}^c {{\cal R}^{\dagger ab}}_{cd}
\!-\! {t^{\dagger c}}_b {{\cal T}^{\dagger b}}_{cd} \!+\! j^{\dagger c}
{\cal H}_{cd}) & = & 0,
\label{eqn:ewgtcons2}
\end{eqnarray}
where we have defined the manifestly covariant quantity
\begin{equation}
j^{\dagger a} \equiv j^a -2{s^{ab}}_b.
\label{jdaggerdef}
\end{equation}
Comparing these conservation laws with the equivalent results in
standard WGT, we see that (\ref{eqn:wgtcons1}) and
(\ref{eqn:ewgtcons1}) have analogous forms, whereas the forms
(\ref{eqn:wgtcons2}) and (\ref{eqn:ewgtcons2}) differ slightly in that
the latter does not contain the field strength tensor
$H^\dagger_{ab}$ of the vector gauge field (\ref{hdaggerdef}), but
instead the simpler object $H_{ab} = {h_a}^\mu {h_b}^\nu(\partial_\mu
V_\nu-\partial_\nu V_\mu)$, which is easily verified also to be covariant
under extended local dilations; further, this object is multiplied by
$j^{\dagger a}$ rather than simply by $j^a$. The largest
difference with WGT occurs, however, when we consider invariance of
$S_{\rm G}$ under (infinitesimal) local dilations, which in eWGT leads to
the {\em two} further manifestly covariant identities
\begin{eqnarray}
D^\dagger_c (hj^{\dagger c}) & = & 0, \label{eqn:ewgtva}\\
h{t^{\dagger c}}_c & = & 0. \label{eqn:ewgtphih}
\end{eqnarray}
Thus, whereas in WGT one obtains the single differential conservation
equation (\ref{eqn:wgtcons3}), in eWGT one also obtains an additional
algebraic relation.
The relation (\ref{eqn:ewgtphih}) shows that the trace of the
gravitational sector's contribution to the $h$-field equation
(\ref{eqn:eweylgenfe1dagger}) vanishes. In practice, the relation
(\ref{eqn:ewgtphih}) provides a very useful check (albeit only
partial) on the derivation of the gravitational sector's contribution to
the $h$-field equation.

Turning to the matter sector, the local Lorentz, GTC and extended local dilation invariance properties of $S_{\rm M}$ lead, respectively,
to the corresponding manifestly covariant identities:
\begin{widetext}
\begin{subequations}
\label{eqn:ewgtmconsall}
\begin{eqnarray}
{\cal D}^\dagger_c(h{\sigma_{ab}}^c) + h\tau^\dagger_{[ab]}
+\tfrac{1}{2}\frac{\delta L_{\rm M}}{\delta\vpsi}
\Sigma_{ab}\vpsi & = & 0,
\\
{\cal D}^\dagger_c(h{\tau^{\dagger c}}_d) - h({\sigma_{ab}}^c {{\cal R}^{\dagger ab}}_{cd}
- {\tau^{\dagger c}}_b {{\cal T}^{\dagger b}}_{cd}
+ \zeta^{\dagger c} {\cal H}_{cd}) +
\frac{\delta L_{\rm M}}{\delta\phi}
{\cal D}^\dagger_d \phi
+ \frac{\delta L_{\rm M}}{\delta\vpsi}
{\cal D}^\dagger_d\vpsi & = & 0,\\
{\cal D}^\dagger_c(h\zeta^{\dagger c}) & = & 0,\label{eqn:ewgtmcons3}\\
h{\tau^{\dagger c}}_c +\frac{\delta L_{\rm M}}{\delta\phi} \phi - \frac{\delta
  L_{\rm M}}{\delta\vpsi} w\vpsi & = & 0,
\label{eqn:ewgtmcons4}
\end{eqnarray}
\end{subequations}
\end{widetext}
where we have defined the quantity
\begin{equation}
\zeta^{\dagger a} \equiv \zeta^a -2{\sigma^{ab}}_b,
\label{zetadaggerdef}
\end{equation}
which we will consider in more detail below.  Thus, as in PGT and WGT,
one is assured with the help of the matter field equations that the
gravitational field equations become consistent.  Moreover, the
relation (\ref{eqn:ewgtmcons4}) shows that the trace of the matter
sector's contribution to the $h$-equation is simply related to the
$\phi$ and $\vpsi$ matter field equations; this
provides a useful (partial) check on the
derivation of the matter sector's contribution to the $h$-field equation.
It is worth
noting that the above sets of conservation laws hold
for {\em any subset} of terms in $L_{\rm G}$ and $L_{\rm M}$, respectively, that
is covariant with weight $w=-4$ under extended local Weyl transformations.

\subsection{Alternative variational principle in eWGT}
\label{sec:ewgtaltvarp}

The identification of the covariant energy-momentum tensor
(\ref{covemtdef}) and its gravitational sector counterpart
$t^\dagger_{ab}$, and the covariant currents (\ref{jdaggerdef}) and
(\ref{zetadaggerdef}), raises the question of whether these quantities
may be arrived at more directly from an alternative variational
principle. It transpires that this is indeed the case, and they arise
naturally if one simply makes a change of field variables from the set
$\vpsi$, $\phi$, $h_a^\mu$, ${A^{ab}}_\mu$, $V_\mu$ and their
derivatives, with which we have worked so far, to the new set $\vpsi$,
$\phi$, $h_a^\mu$, ${A^{\dagger ab}}_\mu$, $V_\mu$ and their
derivatives (i.e. we replace ${A^{ab}}_\mu$ by ${A^{\dagger ab}}_\mu$
in the set). One should note that we are simply making a change of
field variables here, rather than considering ${A^{\dagger ab}}_\mu$
to be an independent field variable; in other words, we still consider
${A^{\dagger ab}}_\mu$ to be given in terms of $h_a^\mu$,
${A^{ab}}_\mu$, $V_\mu$ by its defining relationship
(\ref{adaggerdef}), rather than an independent quantity whose
relationship to the other variables would be determined from the
variational principle. It is worth recalling that the quantities
${A^{\dagger ab}}_\mu$ (unlike ${A^{ab}}_\mu$) transform covariantly (with
weight $w=0$, i.e. invariant) under extended local dilations, and so
we might expect a more straightforward expression of the covariance of
${\cal L}_{\rm M}$ and ${\cal L}_{\rm G}$ when they are written in
terms of ${A^{\dagger ab}}_\mu$ rather than ${A^{ab}}_\mu$.

On making this change of field variables, a significant simplification
does indeed occur. This is most easily demonstrated by introducing the
derivative operators (see Appendix~\ref{appendix:altddagderiv})
\begin{equation}
{\cal D}^\natural_a\vpsi \equiv {h_a}^\mu D^\natural_\mu \vpsi \equiv
 {h_a}^\mu (\partial_\mu +
\tfrac{1}{2}{A^{\dagger bc}}_\mu\Sigma_{bc})\vpsi.
\label{dnatdefagain}
\end{equation}
One should note that these derivatives do {\em not}
transform covariantly under extended local dilations. Nonetheless, by
analogy with (\ref{ewgttorsiondef}), one can still define the
quantities
\begin{equation}
{{\cal T}^{\natural a}}_{bc} \equiv {h_b}^\mu {h_c}^\nu
(D^\natural_\mu {b^a}_\nu - D^\natural_\nu {b^a}_\mu),
\label{tnatdef}
\end{equation}
and the
corresponding trace ${\cal T}^\natural_b \equiv {{\cal T}^{\natural
    a}}_{ba}$. It is straightforward to show that ${\cal T}^\natural_b
= {\cal T}_b + 3{\cal V}_b$, where ${\cal T}_b$ is the trace of the
PGT torsion. Hence the full eWGT covariant derivative can be written as
\begin{equation}
{\cal D}^\dagger_a\vpsi = ({\cal D}^\natural_a-\tfrac{1}{3}w{\cal
  T}^\natural_a)\vpsi,
\label{dnatplust}
\end{equation}
which is expressible wholly in terms of the fields
$h$, $\partial h$ and $A^\dagger$
(suppressing indices). Since (\ref{dnatplust}) does not depend
explicitly on the dilation gauge field $V_\mu$ or its derivatives,
then neither will any Lagrangian constructed from eWGT covariant
derivatives or field strengths, when working in terms of the new set
of field variables.
Thus, the functional dependencies in
(\ref{ltotewgt}) become
\begin{eqnarray}
{\cal L}_{\rm T} &=&
{\cal L}_{\rm G}(h, \partial h, \partial^2 h, A^\dagger,\partial A^\dagger) \nonumber\\&&+
{\cal L}_{\rm M}(\vpsi,\partial\vpsi,\phi,\partial\phi,h,\partial
h,A^\dagger,\partial A^\dagger),
\label{ewgtlagtdag}
\end{eqnarray}
in which there is {\em no} explicit dependence on the dilation gauge
field $V_\mu$ or its derivatives. Thus, eWGT can be expressed in terms
of just the two combinations $h$ and $A^\dagger$ of the gauge fields. In this
respect, eWGT is structurally more similar to PGT, whilst possessing
(a more general version of) the local scale invariance of WGT.

Working in terms of the new field variables, ${\cal L}_{\rm T}$ is at
most quadratic in the field strength tensors ${\cal
  R}^\dagger_{abcd}(h,A^\dagger,\partial A^\dagger)$, ${\cal
  T}^\dagger_{abc}(h,\partial h,A^\dagger)$ and ${\cal
  H}^\dagger_{ab}(h,\partial h, \partial^2h, A^\dagger,\partial
A^\dagger)$, where once again the last of these depends (linearly) on
the second derivative $\partial^2 h$ of the translational gauge field,
as indicated. Hence, the term proportional to ${\cal
  H}^\dagger_{ab}{\cal H}^{\dagger ab}$ in ${\cal L}_{\rm G}$ (if
included) will
again generate terms in the equations of motion that are typically
linear in fourth-order derivatives of $h$. As mentioned previously, we
discuss this issue in detail in Section~\ref{sec:ewgtsecondorder}.

As a notational convenience, when working with the new set of
field variables, we denote the variational derivative of a
Lagrangian ${\cal L}$ with respect to any one of the fields $\chi$ by
$(\delta{\cal L}/\delta\chi)_\dagger$ to distinguish it from the
variational derivative $\delta{\cal L}/\delta\chi$ obtained previously
using the original set of field variables. Straightforward (but
lengthy) application of the chain rule for partial derivatives shows
that for (say) the matter Lagrangian one obtains
\begin{subequations}
\label{ewgtshortcuts}
\begin{eqnarray}
\left(\frac{\delta{\cal L}_{\rm M}}{\delta {h_c}^\mu}\right)_\dagger & = &
{\tau^c}_\mu + 2{\sigma_{ab}}^c {\cal V}^a b^b_\mu - 2{\sigma^{cb}}_b
V_\mu
 =
{\tau^{\dagger c}}_\mu, \phantom{AB}\label{emtshortcut} \\
\hspace*{-15mm}\left(\frac{\delta{\cal L}_{\rm M}}{\delta {A^{\dagger
      ab}}_\mu}\right)_\dagger
& = & {\sigma_{ab}}^\mu, \\
\left(\frac{\delta{\cal L}_{\rm M}}{\delta V_\mu}\right)_\dagger & = &
\zeta^\mu - 2{h_a}^\mu {\sigma^{ab}}_b
= \zeta^{\dagger \mu} = 0, \label{zetashortcut}\\
\left(\frac{\delta{\cal L}_{\rm M}}{\delta\vpsi}\right)_\dagger & = &
\frac{\delta{\cal L}_{\rm M}}{\delta\vpsi}, \\
\left(\frac{\delta{\cal L}_{\rm M}}{\delta \phi}\right)_\dagger & = &
\frac{\delta{\cal L}_{\rm M}}{\delta \phi}, \label{phishortcut}
\end{eqnarray}
\end{subequations}
and directly analogous results hold for the free gravitational Lagrangian.
Thus we see that straightforward variation with respect to the new set
of field variables leads directly to the covariant energy-momentum
tensor (\ref{covemtdef}) and the covariant current (\ref{jdaggerdef})
(and their counterparts in the free gravitational sector)
that we identified earlier; the latter of
which always vanishes.  We note further that the spin-angular momentum
tensor ${\sigma_{ab}}^\mu$ and the variational derivatives with
respect to $\vpsi$ and $\phi$ (and their counterparts from the
gravitational sector) are not altered by working in terms of the new
set of field variables. Indeed, the results
(\ref{emtshortcut})--(\ref{phishortcut}) (and their counterparts for
the free gravitational sector) lead to valuable short-cuts in
calculating the equations of motion in terms of either the old or new
set of variables, since holding ${A^{\dagger ab}}_\mu$ constant in the
derivatives (rather than just ${A^{ab}}_\mu$, as done previously)
vastly reduces the number of terms to be calculated.

The relation (\ref{zetashortcut}) that defines the covariant dilation
current is worthy of further comment. Since the matter Lagrangian does not
depend explicitly on the dilation gauge field $V_\mu$ or its
derivative, then $\zeta^{\dagger \mu}=0$ and so
\begin{equation}
\zeta^a = 2{\sigma^{ab}}_b,
\label{viscona}
\end{equation}
and similarly for the free gravitational Lagrangian, so
\begin{equation}
j^a = 2{s^{ab}}_b.
\label{viscona2}
\end{equation}
Thus, when working in terms of the original set of fields,
the contribution of the matter sector to the $V$-field equation
(\ref{eqn:eweylgenfe3}) is merely twice the relevant contraction of
its contribution to the $A$-field equation
(\ref{eqn:eweylgenfe2}). The result (\ref{viscona2}) shows that the
same is true for the gravitational sector.  Finally, we note that,
combining the relations (\ref{eqn:ewgtphih}) and
(\ref{eqn:ewgtmcons4}) and the relations (\ref{viscona}) and
(\ref{viscona2}), shows that the resulting theory has only {\em three}
independent field equations, namely the $h$-equation
(\ref{eqn:eweylgenfe1dagger}), the $A$-equation (or, equivalently, the
$A^\dagger$-equation) (\ref{eqn:eweylgenfe2}) and the $\vpsi$-equation
(\ref{eweylpeqn}).  In this respect, eWGT is again more
similar to PGT than standard WGT.

For the remainder of this paper, unless otherwise stated, we will work
in terms of the original set of fields $\vpsi$, $\phi$, ${h_a}^\mu$,
${A^{ab}}_\mu$, $V_\mu$ and their derivatives, but we will use
the relations (\ref{emtshortcut}) and (\ref{zetashortcut}) to simplify
our calculations. Moreover, we will use the relations
(\ref{eqn:ewgtphih}), (\ref{eqn:ewgtmcons4}), (\ref{viscona}) and
(\ref{viscona2}) as a check on our derivations of the two independent
gravitational field equations (\ref{eqn:eweylgenfe1dagger}) and
(\ref{eqn:eweylgenfe2}).


\subsection{Dirac matter field in eWGT}
\label{sec:ewgtdirac}

To illustrate the coupling of matter to the gravitational gauge fields
in eWGT, we again consider a Dirac field, and begin with the
special-relativistic Lagrangian (\ref{diraclm3}), which was the
starting point for our discussion in
Section~\ref{sec:wgtdirac}. Following our general procedure for
constructing a matter action (\ref{eweylmatlag_gen}) that is invariant
under extended Weyl transformations and
applying it to (\ref{diraclm3}), the appropriate form for the
corresponding Lagrangian density is
\begin{equation}
{\cal L}_{\rm D} = h^{-1} L_{\rm D} =
h^{-1}(\tfrac{1}{2}i
\bar{\psi}\gamma^a{\stackrel{\leftrightarrow}{{\cal D}^\dagger_a}}\psi -
\mu\phi\bar{\psi}\psi),
\label{eweyldiracaction}
\end{equation}
which is simply the WGT Lagrangian density (\ref{weyldiracaction}),
but with ${\cal D}^\ast_a \to {\cal D}^\dagger_a$. Similar to WGT,
since both $\psi$ and $\bar{\psi}$ have Weyl weight $w=-\tfrac{3}{2}$,
the dilation gauge field $V_\mu$ interacts in the same manner with
each of them, thereby ruling out the interpretation of $V_\mu$ as the
electromagnetic potential. Moreover, $V_\mu$ similarly vanishes
completely from the action (\ref{eweyldiracaction}). Thus, as occurred
in WGT, (\ref{eweyldiracaction}) is identical to the covariant Dirac
Lagrangian in PGT, given in (\ref{diracactionagain}). In other words,
the kinetic term in the Dirac action in PGT is {\em already} also
invariant to extended local dilations. Indeed, this was one of our
original motivations for exploring eWGT.

Varying the action corresponding to (\ref{eweyldiracaction}) with respect
to $\bar{\psi}$, one obtains a field equation of the form
(\ref{eweylpeqn}) (with $\vpsi$ replaced by $\bar{\psi}$), which is
immediately found to read
\begin{equation}
i\gamma^a{\cal D}^\dagger_a\psi-\mu\phi\psi= 0.
\label{eweylcovardirac2}
\end{equation}
An equivalent adjoint field equation in $\bar{\psi}$ is obtained by
varying the action with respect to $\psi$.  In particular, we verify
that the field equation (\ref{eweylcovardirac2}) derived from minimal
coupling at the level of the action is equivalent to that which would
be obtained simply by applying the minimal-coupling procedure to the
field equation directly, independent of the form of ${\cal L}_{\rm
  G}$. Since the Dirac Lagrangian density (\ref{eweyldiracaction}) may
also be expressed as (\ref{diracactionagain}), however, one can
rewrite (\ref{eweylcovardirac2}) more simply in terms of PGT
quantities as in (\ref{weylcovardirac}).  The covariance of
(\ref{weylcovardirac}) under extended local dilations is easily
checked directly by noting that $\gamma^a ({\cal
  D}_a+\tfrac{1}{2}{\cal T}_a)\psi = \gamma^a {\cal D}^\dagger_a\psi$,
where we have used the fact that for a Dirac spinor $w=-\tfrac{3}{2}$
and $\gamma^a \Sigma_{ba} = \tfrac{1}{4}\gamma^a[\gamma_b,\gamma_a] =
-\tfrac{3}{2}\gamma_b$. We also note that (\ref{eweylcovardirac2}) and
(\ref{weylcovardirac}) are valid for any choice of the gravitational
Lagrangian density ${\cal L}_{\rm G}$, but the expression for ${\cal
  T}_a$ in (\ref{weylcovardirac}) will depend on the form of ${\cal
  L}_{\rm G}$.

Following the discussion in Section~\ref{sec:wgtdirac}, one may
rewrite (\ref{weylcovardirac}), and hence (\ref{eweylcovardirac2}), in
terms of the reduced PGT covariant derivative as
(\ref{weylcovardiracalt}), which again reveals that only the total
antisymmetric part of the PGT `torsion' explicitly affects the
dynamics of the Dirac field $\psi$. Nonetheless, unlike in WGT, one
must note that the presence of torsion (even in the case ${\cal
  T}_{[abc]}=0$) will still induce physical effects on $\psi$,
since the PGT torsion enters into the definition of ${\cal R}^\dagger_{abcd}$
on which depend the gravitational field equations that determine the
$h$ and $B$ fields. Alternatively, one can rewrite
(\ref{eweylcovardirac2}) in manifestly eWGT-covariant form in terms of
the reduced eWGT covariant derivative, and one quickly finds that
(\ref{weylcovardiracalt}) is again satisfied, but with the
replacements ${^0}{\cal D}_a \to {^0}{\cal D}^\dagger_a$ and ${\cal
  T}_{[abc]} \to {\cal T}^\dagger_{[abc]}$. Thus, only the total
antisymmetric part of the eWGT `torsion' explicitly affects the
dynamics of the $\psi$ field.

From (\ref{weylcovardiracalt}), the energy-momentum and
spin-angular-momentum tensors of the massive Dirac field $\psi$ are
again given by the expressions (\ref{diracemt}) and (\ref{diracsam}),
respectively. The expression (\ref{diracsam}) for the
spin-angular-momentum tensor ${\sigma_{ab}}^c$ is already manifestly
eWGT covariant. For the energy-momentum tensor ${\tau^a}_b$, one may
show that (\ref{diracemt}) can be rewritten as
\begin{equation}
{\tau^a}_b  =
\tfrac{1}{2}ih^{-1}\bar{\psi}\gamma^a{\stackrel{\leftrightarrow}{{\cal D}^\dagger_b}}\psi
-\delta_b^a {\cal L}_{\rm D}
-2{\sigma_{cb}}^a{\cal V}^c,
\end{equation}
but, as anticipated from our discussion in
Section~\ref{sec:ewgtfecons}, this quantity is {\em not} covariant
with respect to extended local dilations. Indeed, one may show that
${\tau^a}_b$ transforms as
\begin{equation}
{\tau^{\prime a}}_b = {\tau^a}_b - 2\theta {\sigma_{cd}}^a
{\cal P}^c\delta^d_b.
\label{diracemttrans}
\end{equation}
Recalling that ${\sigma_{ab}}^b=0$ for Dirac matter, the
transformation (\ref{diracemttrans}) agrees with the general result
(\ref{eqn:emtnoncov2}). Following the discussion in
Section~\ref{sec:ewgtfecons}, however, we can construct the
eWGT-covariant energy-momentum tensor
\begin{equation}
{\tau^{\dagger a}}_{b} = {\tau^a}_{b} + 2{\sigma_{cb}}^a{\cal V}^c.
\end{equation}

Finally, we note that one might also consider adding further terms to
the matter Lagrangian density (\ref{diracactionagain}) in an analogous
manner to that discussed in Section~\ref{sec:wgtdirac} for WGT. Hence,
we base our massive Dirac field matter Lagrangian on
(\ref{weylcovardirac3}), but appropriately generalised to be eWGT
covariant.  Thus, in the most general case, one might consider
%
\begin{eqnarray}
{\cal L}_{\rm M} = h^{-1}[\tfrac{1}{2}i\bar{\psi}\gamma^a
{\stackrel{\leftrightarrow}{{\cal D}_a}}\psi - \mu\phi\bar{\psi}\psi
&&+ \tfrac{1}{2}\nu ({\cal D}^\dagger_a\phi) ({\cal D}^{\dagger a} \phi)
- \lambda\phi^4 \nonumber \\&&-
  a\phi^2{\cal R}^\dagger + \phi^2L_{{\cal T}^{\dagger 2}}],
\label{eweylcovardirac3}
\end{eqnarray}
%
where $\mu$, $\nu$, $\lambda$ and $a$ are dimensionless constants
(usually positive), and
there are two further dimensionless constants $\beta_1$ and $\beta_2$
in $L_{{\cal T}^{\dagger 2}}$.  It is worth noting that
$\mu\phi$ has the dimensions of mass in natural units.


\subsection{Electromagnetic field in eWGT}
\label{sec:ewgtem}

The dynamics of the electromagnetic field in eWGT follow by direct
analogy from our discussion in Section~\ref{sec:wgtem} in the context
of WGT, with the replacements $\widehat{{\cal F}}^\ast_{ab} \to
\widehat{{\cal F}}^\dagger_{ab}$, ${{\cal T}^{\ast c}}_{ab} \to {{\cal
    T}^{\dagger c}}_{ab}$ and ${\cal D}^\ast_a \to {\cal
  D}^\dagger_a$, and recalling that ${\cal T}^{\dagger}_{a} \equiv 0$.
Thus, as in WGT (and PGT), one arrives at the conclusion that the
appropriate Lagrangian density for the EM field is given by
(\ref{emlagcov2wgt}), namely
\begin{equation}
{\cal L}_{\rm M} = h^{-1}L_{\rm M}
=-h^{-1}(\tfrac{1}{4} {\cal F}_{ab}{\cal F}^{ab}
+ {\cal J}^a {\cal A}_a),
\label{emlagcov2ewgt}
\end{equation}
which has an overall Weyl weight of zero, as required, provided
$w({\cal J}_a)=-3$. Hence, the electromagnetic Lagrangian in PGT is
{\em already} covariant under extended local dilations, which again
was one of our original motivations for exploring eWGT.

Since the Lagrangian (\ref{emlagcov2ewgt}) is identical to that in PGT,
then the equation of motion for the electromagnetic field is again
given by (\ref{emeomwgt}), which may be straightforwardly recast as
\begin{equation}
{\cal D}^\dagger_a{\cal F}^{ac}-\tfrac{1}{2}{{\cal
      T}^{\dagger c}}_{ab}{\cal F}^{ab} = {\cal J}^c.
\label{emeomewgtstar}
\end{equation}
which is manifestly eWGT covariant. One should note that this differs
slightly from the corresponding form (\ref{emeomwgtstar}) in WGT,
since ${\cal T}^{\dagger}_{a} \equiv 0$. Also, the differential
identity (\ref{emdiffid}) satisfied by the electromagnetic field
strength tensor may also directly be shown to be covariant under local
dilations, since it can be easily rewritten as
\begin{equation}
{\cal D}^\dagger_{[a}{\cal F}_{bc]} -{{\cal T}^{\dagger d}}_{[ab}{\cal F}_{c]d}= 0.
\label{ewgtembi}
\end{equation}

Since (\ref{emeomwgt}) and (\ref{emdiffid}) also hold in eWGT, then so
too do the alternative forms given in (\ref{empgtaltforms}) in terms
of the reduced covariant derivative in PGT, which show that the PGT
torsion plays no explicit role in the dynamics of the electromagnetic
field. One may also rewrite (\ref{empgtaltforms}) in terms of the
reduced eWGT covariant derivative, and one finds that
(\ref{empgtaltforms}) is again satisfied, but with the replacement
${^0}{\cal D}_a \to {^0}{\cal D}^\dagger_a$. Thus the eWGT torsion
similarly has no direct effect of the EM field dynamics.  As mentioned
above for the Dirac field, however, the presence of torsion will still
induce physical effects (unlike in WGT), since the reduced covariant
derivative ${^0}{\cal D}^\dagger_a$ depends on the gauge fields, which
are determined by gravitational field equations; these, in turn,
depend on the `curvative' tensor ${\cal R}^\dagger_{abcd}$ or its
contractions, which contain the torsion.

The energy-momentum tensor ${\tau^a}_b$ of the EM field has the same
form as in PGT and WGT, namely that given in (\ref{pgtememt}), which
is already manifestly eWGT covariant. Indeed, from (\ref{covemtdef}),
one sees immediately that ${\tau^{\dagger a}}_b = {\tau^a}_b$, since
the EM field spin-angular-momentum tensor ${\sigma_{ab}}^c$ vanishes.

Finally, one may straightforwardly include the interaction of the
electromagnetic and Dirac fields, both of which are coupled to
gravity, in the same way as in WGT, as discussed in
Section~\ref{sec:wgtem}.


\subsection{Einstein gauge and scale-invariant variables}
\label{sec:ewgteinstein}

As was the case for WGT, the forms of the field equations in eWGT can
be simplified considerably by using the scale gauge freedom to impose
the Einstein gauge, in which the scalar field is set to a constant
everywhere, $\phi=\phi_0$ (provided $\phi$ does not vanish). Moreover,
as occurred for WGT (see Section~\ref{sec:wgteinstein}), we show here
that the resulting field equations are {\em identical} in form to
those obtained when working in terms of a new set of scale-invariant
field variables. Thus, this again provides an alternative
interpretation of the Einstein gauge, since the approach using
scale-invariant variables involves no breaking of the scale symmetry.

It is convenient to work in terms of the alternative set of field
variables $\vpsi$, $\phi$, $h_a^\mu$ and ${A^{\dagger ab}}_\mu$ discussed
in Section~\ref{sec:ewgtaltvarp}, in which the total Lagrangian
density ${\cal L}_{\rm T}$ does not depend explicitly on the
dilational gauge field $V_\mu$ or its derivatives. We begin by
introducing the new scale-invariant matter field and gravitational
gauge fields
\begin{equation}
\widehat{\vpsi}\equiv \left(\frac{\phi}{\phi_0}\right)^{-w}\vpsi,
\quad
{\widehat{h}_a}^{\phantom{a}\mu} \equiv \left(\frac{\phi}{\phi_0}\right)^{-1}{h_a}^\mu,
\quad
{\widehat{A}^{\dagger ab}}_{\phantom{ab}\mu} \equiv {A^{\dagger ab}}_\mu,
\label{ewgtsivarsdef}
\end{equation}
where $w$ is the Weyl weight of the original matter field $\vpsi$.  It
is also convenient to define the further scale-invariant variables
${\widehat{\mathcal{A}}^{\dagger ab}}_{\phantom{\dagger ab}c} \equiv
{\widehat{h}_a}^{\phantom{a}\mu}{\widehat{A}^{\dagger ab}}_{\phantom{\dagger ab}\mu}$.

As shown in Section~\ref{sec:ewgtaltvarp}, the eWGT covariant
derivative of some general field $\chi$ with Weyl weight $w$ can be
written in the form ${\cal D}^\dagger_a\chi = ({\cal
  D}^\natural_a-\tfrac{1}{3}w{\cal T}^\natural_a)\chi$, where ${\cal
  D}_a^\natural$ and ${\cal T}^\natural_a$ are defined in
(\ref{dnatdefagain}) and (\ref{tnatdef}), respectively. In an analogous
manner, we may define the quantities $\widehat{{\cal D}}_a^\natural$
and $\widehat{{\cal T}}^\natural_a$, for which each occurrence of
${h_a}^\mu$ and ${A^{\dagger ab}}_\mu$ in (\ref{dnatdef}) and
(\ref{tnatdef}) is replaced by its scale-invariant counterpart
${\widehat{h}_a}^{\phantom{a}\mu}$ and ${\widehat{A}^{\dagger
    ab}}_{\phantom{ab}\mu}$, respectively. Thus, $\widehat{{\cal
    D}}_a^\natural = {\widehat{h}_a}^{\phantom{a}\mu} D_\mu^\natural$
and a short calculation shows that $\widehat{{\cal T}}^\natural_a$
transforms covariantly with Weyl weight $w=0$ under local dilations,
i.e.\ it is scale-invariant. One then immediately finds that the eWGT
covariant derivative of the scalar field $\phi$ may be written as
${\cal D}^\dagger_a\phi = \frac{1}{3}(\phi^2/\phi_0)\widehat{{\cal T}}^\natural_a$,
whereas the covariant derivative of some general field $\chi$ with
Weyl weight $w$ may be written
\begin{equation}
{\cal D}^\dagger_a\chi =
\left(\frac{\phi}{\phi_0}\right)^{1-w} \widehat{\mathcal{D}}^\dagger_a
\widehat{\chi},
\label{ewgtscaleinvdiff}
\end{equation}
where $\widehat{\chi}=(\phi/\phi_0)^{-w}\chi$ and we have defined the
derivative operator $\widehat{\mathcal{D}}^\dagger_a =
\widehat{\mathcal{D}}^\natural_a - \tfrac{1}{3}w
\widehat{\mathcal{T}}^\natural_a$, which preserves the Weyl weight of
the quantity on which it acts. From (\ref{ewgtscaleinvdiff}), one sees
that, aside from the overall multiplicative factor
$(\phi/\phi_0)^{1-w}$, the scale-invariant quantity
$\widehat{\mathcal{D}}^\dagger_a \widehat{\chi}$ has the {\em same}
functional dependency on $\widehat{\chi}$,
$\widehat{h}_a^{\phantom{a}\mu}$, and ${\widehat{A}^{\dagger
    ab}}_{\phantom{ab}\mu}$, respectively, as ${\cal D}^\dagger_a\chi$
does on $\chi$, ${h_a}^\mu$ and ${A^{\dagger ab}}_\mu$. Similarly, one
also quickly finds that the eWGT gauge field strengths defined in
(\ref{weylfsdefs}) may be written as ${{\cal R}^{\dagger ab}}_{cd} =
(\phi/\phi_0)^2 \widehat{\mathcal{R}}^{\dagger ab}_{\phantom{ij}cd}$,
${{\cal T}^{\dagger a}}_{bc} =
(\phi/\phi_0)\widehat{\mathcal{T}}^{\dagger a}_{\phantom{abc}bc}$ and
${\cal H}^\dagger_{ab} = (\phi/\phi_0)^2
\widehat{\mathcal{H}}^\dagger_{ab}$, where each quantity with a caret
is scale-invariant and has the {\em same} functional dependence on
$\widehat{h}_a^{\phantom{a}\mu}$ and ${\widehat{A}^{\dagger
    ab}}_{\phantom{ab}\mu}$ as the corresponding original quantity
does on ${h_a}^\mu$ and ${A^{\dagger ab}}_\mu$, respectively.

Thus, noting that $h^{-1} = (\phi/\phi_0)^{-4} \widehat{h}^{-1}$, one
is led to the important conclusion, analogous to that discussed in
Section~\ref{sec:wgteinstein} for WGT, that the Lagrangian density
${\cal L}_{\rm T}$ in (\ref{ewgtlagtdag}) may be written as (suppressing
indices for brevity)
\begin{eqnarray}
{\cal L}_{\rm T} &=& {\cal L}_{\rm G}(\widehat{h},
\partial\widehat{h},\partial^2\widehat{h},\widehat{A}^\dagger,\partial
\widehat{A}^\dagger) \nonumber \\&&+ {\cal L}_{\rm
  M}(\widehat{\vpsi},\partial\widehat{\vpsi},\phi_0,0,,\widehat{h},\partial
\widehat{h}, \widehat{A}^\dagger, \partial \widehat{A}^\dagger).
\label{ewgtlagtotsi}
\end{eqnarray}
Specifically, when written in terms of the scale-invariant field
variables, the Lagrangian density ${\cal L}_{\rm T}$ (indeed each term
separately) has the {\em same} functional form as it does in terms of
the original variables with $\phi=\phi_0$. Thus, if $\chi$ represents
$\vpsi$, ${h_a}^\mu$, or ${A^{\dagger ab}}_\mu$, one may immediately
conclude that (each term in) the equation of motion $\delta{\cal
  L}_{\rm T}/\delta\widehat{\chi}=0$ has the {\em same} functional
form as (the corresponding term in) $\left.\delta{\cal L}_{\rm
  T}/\delta\chi\right|_{\phi=\phi_0}=0$, but with $\vpsi$, ${h_a}^\mu$
and ${A^{\dagger ab}}_\mu$ replaced by their scale-invariant
counterparts (\ref{ewgtsivarsdef}).

As mentioned in Section~\ref{sec:ewgtaltvarp}, the equations of motion
for the $\vpsi$, $h$ and $A$ (or $A^\dagger$) fields are the only
three {\em independent} field equations of eWGT. One can also obtain
equation of motion for the $V$ and $\phi$ fields, but the relations
(\ref{eqn:ewgtphih}), (\ref{eqn:ewgtmcons4}), (\ref{viscona}) and
(\ref{viscona2}) show that these correspond merely to contractions
of the $A$-equation and $h$-equation, respectively. Thus, somewhat
more straightforwardly than in WGT, one may immediately conclude that
the equivalence discussed above also holds for the $V$ and $\phi$
equations of motion.

It is worth mentioning that one may arrive at similar conclusions
to those reached above, if one instead begins by working in terms of the
`standard' set of field variables $\vpsi$, $\phi$, $h_a^\mu$,
${A^{ab}}_\mu$ and $V_\mu$, rather than the alternative set introduced
in Section~\ref{sec:ewgtaltvarp}. In this case, in place of
(\ref{ewgtsivarsdef}), one introduces the scale-invariant matter field
and gravitational gauge fields
\begin{subequations}
\label{ewgtsivarsdef2all}
\begin{eqnarray}
\widehat{\vpsi}&\equiv& \left(\frac{\phi}{\phi_0}\right)^{-w}\vpsi,\\
{\widehat{h}_a}^{\phantom{a}\mu} &\equiv &\left(\frac{\phi}{\phi_0}\right)^{-1}{h_a}^\mu,\\
{\widehat{A}^{\dagger ab}}_{\phantom{ab}\mu} &\equiv& {A^{ab}}_\mu + ({\cal V}^a{b^b}_\mu
- {\cal V}^b{b^a}_\mu),\\
\widehat{V}_\mu &\equiv& V_\mu +\tfrac{1}{3}T_\mu + \partial_\mu\ln\left(\frac{\phi}{\phi_0}\right),
\label{ewgtsivarsdef2}
\end{eqnarray}
\end{subequations}
where it is clear from the last definition that $\widehat{V}_\mu =
\frac{1}{3}\widehat{T}^\natural_\mu$.  It is also convenient to define
the further scale-invariant variables ${\widehat{\mathcal{A}}^{\dagger
    ab}}_{\phantom{\dagger ab}c} \equiv
{\widehat{h}_a}^{\phantom{a}\mu}{\widehat{A}^{\dagger
    ab}}_{\phantom{\dagger ab}\mu}$ and $\widehat{\mathcal{V}}_a\equiv
{\widehat{h}_a}^{\phantom{a}\mu}\widehat{V}_\mu$. The situation is,
however, slightly different to that outlined above. Specifically, the
equations of motion in terms of $\vpsi$, $h_a^\mu$, ${A^{ab}}_\mu$ and
$V_\mu$ arrived at by adopting the Einstein gauge, must first be
rewritten in terms of $\vpsi$, $h_a^\mu$ and the combinations
${A^{\dagger ab}}_\mu$ and $V_\mu+\tfrac{1}{3}T_\mu$ (which will
always be possible). Only then will they be identical in form to the
full field equations written in terms of the scale-invariant variables
$\widehat{\vpsi}$, $\widehat{h}_a^{\phantom{a}\mu}$,
${\widehat{A}^{\dagger ab}}_{\phantom{\dagger ab}\mu}$ and
$\widehat{V}_\mu$, respectively.


\subsection{Motion of test particles in eWGT}
\label{sec:ewgttestp}
\label{sec:eWGTmassivemotion}

To determine the equation of motion of a massive test particle in
eWGT, we follow the procedure outlined in
Section~\ref{sec:WGTmassivemotion} for WGT, but modify it to
accommodate extended local scale-invariance. Thus, we again adopt the
fully classical point particle action (\ref{WGTppaction}), in which
the Weyl weights of the dynamical variables, namely the particle
4-momentum $p^a(\lambda)$, 4-velocity $v^a(\lambda)$ and the einbein
$e(\lambda)$ along the worldline parameterised by $\lambda$, are the
same as in WGT, so that the action is scale-invariant.

Varying the action (\ref{WGTppaction}) with respect to the three
dynamical variables, respectively, one again obtains the equations of
motion (\ref{WGTppeom1})--(\ref{WGTppeom3}). In this case, however, it
is straightforward to show that (\ref{WGTpppdot}) can be written in a
manifestly eWGT-covariant manner as $v^c({\cal D}^\dagger_cp_a - {\cal
  T}^\dagger_{cab}p^b) =e\mu^2 \phi\,{\cal D}^\dagger_a\phi$. By analogy with our treatment in WGT, one
may rewrite this equation in terms of the reduced eWGT covariant
derivative, defined in (\ref{ddaggermuzerodef}), as $v^c\,{^0}{\cal
  D}^\dagger_cp_a=e\mu^2 \phi\,{^0}{\cal D}^\dagger_a\phi$, which is again
manifestly eWGT-covariant. As in WGT, this takes its simplest form when one
chooses $e=1/(\mu\phi)$, in which case $v^2=1$ and $\lambda$
corresponds to the proper time $\tau$ along the worldline; in this case
the equation of motion becomes
\begin{equation}
\phi\,v^b\,{^0}{\cal D}^\dagger_bv_a= (\delta_a^b - v_av^b){^0}{\cal D}^\dagger_b\phi.
\label{eWGTppeom}
\end{equation}
Once again, for ease of calculation, this may be rewritten in terms of
the reduced PGT-covariant derivative to yield (\ref{WGTppeom2}), which
is also eWGT covariant, but not manifestly so. Precisely as in WGT, if
one uses local scale invariance to impose the Einstein gauge
$\phi=\phi_0$ (a constant), then (\ref{WGTppeom2}) reduces to same
result as that given in (\ref{WGTppeomfinal}) for WGT, which again
describes the gauge theory equivalent of {\em geodesic} motion.
Unlike in WGT, however, the presence of PGT torsion may still induce
physical effects, since the $h$ and $B$ fields are determined by
gravitational field equations that depend on the ‘curvature’ tensor
${\cal R}^\dagger_{abcd}$, into which the PGT torsion enters.

As in WGT, by setting $\mu=0$ in the action (\ref{WGTppaction}) and
choosing the einbein $e=1$, such that $v^2=0$, one finds that the
equation of motion in the massless case (for example, a massless
neutrino) is also given by (\ref{WGTppeomfinal}), even without
imposing the Einstein gauge. One may also arrive at a similar
conclusion for the motion of photons by directly considering the
dynamics of the electromagnetic field. As discussed in
Section~\ref{sec:ewgtem}, the EM field tensor in eWGT again satisfies
the field equation and Bianchi identity given in
(\ref{empgtaltforms}), which have precisely the same form as those
obtained in the absence of torsion.  Consequently, one may immediately
infer that the equation of motion for photons is also given by
(\ref{WGTppeomfinal}), so that they too follow the gauge theory
equivalent of {\em geodesic} motion.

Finally, as in WGT, the imposition of the Einstein gauge is not
necessary to arrive at the geodesic equation of motion
(\ref{WGTppeomfinal}).  One may again introduce the set of
scale-invariant variables (\ref{wgtppsivardef}) and rewrite the action
(\ref{WGTppaction}) in the form (\ref{WGTppactionsi}). Thus, one
arrives once more at the geodesic equation of motion
(\ref{WGTppeomfinalsi}), which is written entirely in terms of
scale-invariant variables.


\subsection{Reduced eWGT}
\label{sec:reducedewgt}

In Section~\ref{sec:ewgtaltcovd}, we introduced the `reduced' eWGT
covariant derivative operator ${^0}D^\dagger_\mu$ in
(\ref{ddaggermuzerodef}), to which the `full' eWGT covariant
derivative (\ref{ewgtcovdiv}) reduces in the case that the eWGT
torsion ${\cal T}^\dagger_{abc}$ vanishes (which is a properly
eWGT-covariant condition). In the context of eWGT, the term `reduced'
refers to versions of quantities in which, by construction, the
expression (\ref{adaggerzerodef}) holds with ${\cal T}^\dagger_{abc}
\equiv 0$ (and hence ${\cal K}^\dagger_{abc} \equiv 0$). In contrast
to WGT, however, such quantities still depend on the rotational gauge
field $A$, and hence cannot be written entirely in terms of the other
gauge fields. Such quantities are again denoted by a zero superscript
preceding the kernel letter, but are perhaps less deserving of the
description `reduced' than their counterparts in WGT (or PGT; see
Appendix~\ref{app:pgt}).

Nonetheless, as was the case in WGT, one can use the covariant
derivative ${^0}D^\dagger_\mu$ to build an alternative class of
scale-invariant gravitational gauge theories, which we term `reduced
eWGT'. They correspond mathematically to imposing the condition of
vanishing eWGT torsion directly at the level of the action, but differ
from reduced WGT in that they still depend on all the gravitational
gauge fields, in particular the $A$ gauge field.

In the usual manner, one begins by defining the `reduced' gauge field
strengths by considering their commutator, which gives
\begin{equation}
[\zero{D^\dagger_\mu},\zero{D^\dagger_\nu}]\vpsi=\tfrac{1}{2} \zero{{R^{\dagger ab}}_{\mu\nu}}\Sigma_{ab}\vpsi + wH^\dagger_{\mu\nu}\vpsi,
\end{equation}
where we have defined the `reduced' field strength tensor $\zero{{R^{\dagger
      ab}}_{\mu\nu}}(h,\partial h,\partial^2 h,A,\partial A,V,\partial
V)$, which is again given by the formula (\ref{rfsdef}), but with
${A^{ab}}_\mu$ replaced by $\zero{{A^{\dagger ab}}_\mu}$. Unlike in
WGT, however, $\zero{{R^{\dagger ab}}_{\mu\nu}}$ depends on all the
gauge fields and their derivatives, most particularly the $A$ gauge
field; indeed its functional dependencies are more complicated than
the `full' field strength tensor ${R^{\dagger
    ab}}_{\mu\nu}(h,A,\partial A,V,\partial V)$. Considering instead
the commutator of `reduced' generalised covariant derivatives, one
obtains
\begin{equation}
[{^0}{\cal D}^\dagger_c,{^0}{\cal D}^\dagger_d]\vpsi=\tfrac{1}{2}
\czero{{{\cal R}^{\dagger ab}}_{cd}}\Sigma_{ab}\vpsi + w{\cal H}^\dagger_{cd}\vpsi,
\label{eweylzerotcomm}
\end{equation}
where $\czero{{{\cal R}^{\dagger ab}}_{cd}} = {h_c}^\mu {h_d}^\mu
\,\zero{{R^{\dagger ab}}_{\mu\nu}}$, and the commutator
(\ref{eweylzerotcomm}) has no term containing a `translational' field
strength of the $h$ gauge field (or `torsion'), since $\czero{{{\cal
      T}^{\dagger a}}_{bc}} \equiv
{h_b}^{\mu}{h_c}^{\nu}(\zero{D}^\dagger_\mu {b^a}_\nu
-\zero{D}^\dagger_\nu{b^a}_\mu) = 0$.  As one might expect,
 the Bianchi
identities satisfied by the reduced field strength tensors are
identical to those given in Section~\ref{sec:ewgtbianchi}, but with the
replacements ${\cal D}^\dagger_a \to {^0}{\cal D}^\dagger_a$,
${\cal R}_{abcd} \to {^0}{\cal R}^\dagger_{abcd}$ and ${\cal
  T}^\dagger_{abc} \to {^0}{\cal T}^\dagger_{abc} = 0$.

Since the `full' generalised covariant derivative is given in terms of
the `reduced' one by (\ref{ewgtcovderivalt}), it follows that the field strength
tensor ${{\cal R}^{\dagger ab}}_{cd}$ appearing in (\ref{ewgtdacomm}),
and its contractions, may be written in the forms
(\ref{riemann0})--(\ref{randr0}), but with the replacements on the RHS
$\czero{{{\cal R}^{ab}}_{cd}} \to \czero{{{\cal R}^{\dagger
      ab}}_{cd}}$, $\czero{\mathcal{D}}_a \to
\czero{\mathcal{D}}^\dagger_a$, ${\mathcal{K}_{abc}} \to
      {\mathcal{K}^{\dagger}_{abc}}$ and ${\mathcal{T}_{abc}} \to
      {\mathcal{T}^{\dagger}_{abc}}$, and recalling that
      ${\mathcal{T}^{\dagger}_{a}} \equiv 0$.

For reduced eWGT, the free gravitational Lagrangian density ${\cal
  L}_{\rm G} = h^{-1}L_{\rm G}$, where by analogy with
(\ref{ewgtgravlagdef}),
\begin{equation}
L_{\rm G} = L_{\czero{{\cal R}}^{\dagger 2}} + L_{{\cal H}^{\dagger 2}},
\label{ewgtnotlg}
\end{equation}
which is based on (\ref{lg2}) with ${{\cal R}^{ab}}_{cd} \to
\czero{{{\cal R}^{\dagger ab}}_{cd}}$ and ${{\cal H}}_{ab} \to
\czero{{{\cal R}^{\dagger}}_{ab}}$. As was the case for WGT, one can
simplify (\ref{ewgtnotlg}) still further (in $D \le 4$ dimensions),
since $\czero{{\cal R}}^\dagger_{abcd}$ and its contractions also
satisfy a Gauss--Bonnet identity of the form (\ref{gbid}), but with
${\cal R}_{abcd} \to \czero{{\cal R}}^\dagger_{abcd}$. Thus, one can
set to zero any one of the parameters $\alpha_i$ in (\ref{ewgtnotlg})
with no loss of generality (at least classically).  Unlike in WGT,
however, the Lagrangian (\ref{ewgtnotlg}) still depends on all the
gauge fields and their derivatives, in particular the $A$-field.
A typical form for the
matter Lagrangian density ${\cal L}_{\rm M}$ is that given in
(\ref{eweylcovardirac3}), but with appropriate
modifications\cite{footnote17a}, namely
\begin{eqnarray}
{\cal L}_{\rm M} = h^{-1}[\tfrac{1}{2}i\bar{\psi}\gamma^a
\,{\stackrel{\leftrightarrow}{{^{0}{\cal D}}_a}}\psi \!\!-\!\!
\mu\phi\bar{\psi}\psi
&&+ \tfrac{1}{2}\nu ({^{0}{\cal D}}^\dagger_a\phi) ({^{0}{\cal D}}^{\dagger a} \phi) -
\lambda\phi^4 \nonumber \\&&
\hspace{1.5cm} - a\phi^2\,{^{0}{\cal R}}^\dagger ].
\label{reducedweylcovardirac3}
\end{eqnarray}

The total Lagrangian density thus has the following functional
dependencies in the most general case,
\begin{eqnarray}
{\cal L}_{\rm T} & = & {\cal L}_{\rm G}(h,\partial h, \partial^2 h, A,
\partial A, V, \partial V) \nonumber \\
&&
\hspace{0.2cm}
+ {\cal L}_{\rm M}(\vpsi,\partial\vpsi,\phi,\partial\phi,h,\partial h,
\partial^2 h, A, \partial A, V, \partial V).\phantom{AAA}
\label{reducedewgtlagtot}
\end{eqnarray}
In terms of the alternative set of variables discussed in
Section~\ref{sec:ewgtaltvarp}, one may equally write
\begin{eqnarray}
{\cal L}_{\rm T} & = & {\cal L}_{\rm G}(h,\partial h, \partial^2 h, A^\dagger,
\partial A^\dagger) \nonumber \\
&&
\hspace{0.2cm}
+ {\cal L}_{\rm M}(\vpsi,\partial\vpsi,\phi,\partial\phi,h,\partial h,
\partial^2 h, A^\dagger, \partial A^\dagger),\phantom{AAA}
\label{reducedewgtlagtot2}
\end{eqnarray}
which does not depend explicitly on the $V$ gauge field or its
derivatives. In each case, the resulting field equations for the
gravitational gauge fields (including the $A$-field equation, which
was absent in reduced WGT) will clearly have the same generic
structure as those given in Sections~\ref{sec:ewgtfecons} and
\ref{sec:ewgtaltvarp}, respectively, although once again the specific
forms for each term are {\em not}, in general, obtained from the
corresponding `full' eWGT expressions simply by replacing the `full'
covariant derivative and field strength tensors with their `reduced'
counterparts. For the matter field equations, however, the forms
(\ref{weylpeqn})--(\ref{weylpeqnphi}) are still valid, but with the
replacement ${\cal D}^\dagger_a \to {^0}{\cal D}^\dagger_a$.  As in
WGT, however, the gravitational and matter Lagrangians are both, in
general, quadratic in second derivatives of the $h$-field (indeed,
{\em every} term in ${\cal L}_{\rm G}$ and the term proportional to
$\phi^2\,\czero{\cal R}^\dagger$ in ${\cal L}_{\rm M}$ depends on
$\partial^2 h$, whereas in `full' eWGT, only $L_{{\cal H}^{\dagger
    2}}$ has this dependency). Thus, the resulting field equations
will typically be linear in fourth-order derivatives of $h$, and such
theories typically suffer from Ostrogradsky's instability, although
this needs to be investigated on a case-by-case basis.

The conservation laws in reduced eWGT will have the same form as those
given in Section~\ref{sec:ewgtconslaws}, but with the replacements
${\cal D}^\dagger_a \to {^0}{\cal D}^\dagger_a$, ${\cal
  R}^\dagger_{abcd} \to {^0}{\cal R}^\dagger_{abcd}$ and ${\cal
  T}^\dagger_{abc} \to {^0}{\cal T}^\dagger_{abc} \equiv
0$. Furthermore, by performing analogous calculations to those
presented in Sections~\ref{sec:ewgteinstein} and
\ref{sec:eWGTmassivemotion}, respectively, one may show that our
conclusions in eWGT regarding the interpretation of the Einstein gauge
and the motion of test particles also apply in reduced eWGT.

As was discussed for WGT in Section~\ref{sec:reducedewgt}, it is
important to distinguish reduced eWGT, in which the condition of
vanishing eWGT torsion is imposed directly at the level of the action,
from instead setting the (properly eWGT-covariant) condition ${\cal
  T}^\dagger_{abc}= 0$ in the field equations of eWGT. In the latter
case, aside from terms generated by the term proportional to ${\cal
  H}^\dagger_{ab}{\cal H}^{\dagger ab}$ in $L_{\rm G}$ (if included),
the basic field equations remain linear in second-order derivatives of
the gauge fields. Although one can substitute for the rotational gauge
field to obtain terms that contain higher-order derviatives, such
theories do not suffer from Ostrogradsky's instability if the term
proportional to ${\cal H}^\dagger_{ab}{\cal H}^{\dagger ab}$ is
excluded from $L_{\rm G}$.


\subsection{Geometric interpretation of eWGT}
\label{sec:ewgtgeo}

We have developed eWGT as a gauge theory of gravity in Minkowski
spacetime, but one can reinterpret eWGT in geometric terms, along
similar lines to WGT, as we presented in Section~\ref{sec:geowgt}.
The central tenet of the geometrical interpretation remains the
identification of ${h_a}^\mu$ as the components of a vierbein system
in a more general spacetime (see equation~\ref{geotetrad}). As in WGT,
this identification leads to the results (\ref{metricdef}) and
(\ref{etagdef}) relating the spacetime metric $g_{\mu\nu}$, Lorentz
metric $\eta_{ab}$ and the inverse $h$-field ${b^a}_\mu$,
and consequently the result $h^{-1}=\sqrt{-g}$ also remains
valid. Under a (local, physical) extended dilation, the spacetime
metric and $h$-field have Weyl weights $w(g_{\mu\nu})=2$ and
$w({h_a}^\mu)=-1$ respectively, as in WGT, and so (\ref{metricdef})
and (\ref{etagdef}) again imply that $w(\eta_{ab})=0$, as expected.
As in WGT, one immediately finds that the $h$-field and its inverse
are directly related by index raising/lowering, so there no need to
distinguish betweem them by using different kernel letters. We
therefore again notate ${h_a}^\mu$ and ${b^b}_\nu$ by ${e_a}^\mu$ and
${e^b}_\nu$, respectively.

In eWGT, unlike WGT, it is the {\em combination} ${A^{\dagger
    ab}}_\mu$ in (\ref{adaggerdef}) of the three gauge fields
${e_a}^\mu$, ${A^{ab}}_\mu$, and $V_\mu$ that is naturally interpreted
as the components of the spin-connection that encodes the rotation of
the local tetrad frame between points $x$ and $x+\delta x$. As in WGT,
this is accompanied by a local change in the standard of length
between the two points, but in contrast to WGT this is encoded not
only by the dilation gauge field $V_\mu$, but also by the trace of the
PGT torsion $T_\mu$, which depends on ${h_a}^\mu$ and ${A^{ab}}_\mu$.
Thus, both the rotation of the local tetrad frame and the local change
in the standard of length between two points are encoded by all three
gauge fields. The operation of parallel transport for some vector
$J^a$ of weight $w$ is therefore defined as
\begin{equation}
\delta J^a = - [{A^{\dagger a}}_{b\mu}-w(V_\mu+\tfrac{1}{3}T_\mu)\delta^a_b] J^b
\,\delta x^\mu.
\label{eweylparallel}
\end{equation}
As in WGT, in general, a vector not only changes its direction on parallel
transport around a closed loop, but also its length. The expression
(\ref{eweylparallel}) establishes the correct form for the related
$(\Lambda,\rho)$-covariant derivative, e.g.
\begin{eqnarray}
D^\dagger_\mu J^a &=& \partial_\mu J^a - w(V_\mu+\tfrac{1}{3}T_\mu) J^a
+  {A^{\dagger a}}_{b\mu}J^b \nonumber \\
&=& \partial^\dagger_\mu J^a +  {A^{\dagger
    a}}_{b\mu}J^b,
\end{eqnarray}
where we have used the partial derivative operator $\partial_\mu^\dagger$
defined in (\ref{partialdaggerdef}). As in WGT, one can easily deduce that
${A^{\dagger ab}}_\mu=-{A^{\dagger ba}}_\mu$, as previously.

Owing to the necessity of using arbitrary coordinates in the more
general spacetime, one must generalise the $(\Lambda,\rho)$-covariant
derivative to apply to fields with definite GCT tensor
behaviour. Following the approach used in WGT, one defines the `total'
covariant derivative
\begin{equation}
\Delta^\dagger_\mu \equiv \partial^\dagger_\mu + {\Gamma^\sigma}_{\rho\mu} {\matri{X}^\rho}_\sigma
+ \tfrac{1}{2}{A^{ab}}_\mu\Sigma_{ab} = \nabla^\dagger_\mu + D^\dagger_\mu - \partial^\dagger_\mu,
\label{eweylextcovd2}
\end{equation}
where $\nabla^\dagger_\mu = \partial^\dagger_\mu + {\Gamma^\sigma}_{\rho\mu}
{\matri{X}^\rho}_\sigma$ and ${\matri{X}^\rho}_\sigma$ are the
$\mbox{GL}(4,R)$ generator matrices appropriate to the GCT tensor
character of the field to which $\Delta^\dagger_\mu$ is applied.  If a
field $\psi$ carries only Latin indices, then $\nabla^\dagger_\mu\psi
=\partial^\dagger_\mu\psi$ and so $\Delta^\dagger_\mu\psi=D^\dagger_\mu\psi$;
conversely, if a field $\psi$ carries only Greek indices, then
$D^\dagger_\mu\psi =\partial^\dagger_\mu\psi$ and so
$\Delta^\dagger_\mu\psi=\nabla^\dagger_\mu\psi$. When acting on an object of
weight $w$, for all these derivative operators the resulting object
also transforms covariantly with the same weight $w$.

The affine connection coefficients ${\Gamma^\sigma}_{\rho\mu}$ again
become dynamical variables, but are necessarily related to the
spin-connection and dilation current, since (\ref{vreln}) should still
hold provided $J^a$ has weight $w=1$. This result then yields the
relation
\begin{equation}
\Delta^\dagger_\mu {e^a}_\nu  \equiv
\partial^\dagger_\mu {e^a}_\nu - {\Gamma^\sigma}_{\nu\mu}{e^a}_\sigma
+ {A^{\dagger a}}_{b\mu}{e^b}_\nu =0,
\label{eweyltetradp}
\end{equation}
which relates $A$ and $\Gamma$ (and $V$ and $T$). Comparing this
result with the equivalent relation (\ref{weyltetradp}) in WGT, we see
that (\ref{eweyltetradp}) is obtained simply by making the
replacements $\partial^\ast_\mu \to \partial^\dagger_\mu$ and $A^{ab}_\mu
\to {A^{\dagger ab}}_\mu$. Consequently, the results (\ref{gammaofa})
and (\ref{aofgamma}) explicitly relating $A$ and $\Gamma$ are simply
replaced by expressions of the same form but with the above
substitutions. As in WGT, this replacement strategy enables one
straightforwardly to derive the further relationships in the geometric
interpretation of eWGT that correspond to the results
(\ref{wgtsemimet})--(\ref{wgtcontortiondef}) in WGT.

In particular, using (\ref{metricdef}) and (\ref{eweyltetradp}), one finds that
$\nabla^\dagger_\sigma g_{\mu\nu}=0$, and so this derivative operator
commutes with raising and lowering of coordinate
indices. Equivalently, one may write this as the semi-metricity
condition
\begin{equation}
\nabla_\sigma g_{\mu\nu} = 2(V_\sigma + \tfrac{1}{3}T_\sigma)g_{\mu\nu}.
\label{ewgtsemimet}
\end{equation}
Comparing this result with the corresponding semi-metricity condition
(\ref{wgtsemimet}) in WGT, we see that it has a similar
form\cite{footnote18}. In addition to depending on the dilation gauge
field $V_\mu$, however, the eWGT version also depends on the trace of
the PGT torsion, which is itself a function of ${e_a}^\mu$ and
${A^{ab}}_\mu$. Therefore, the spacetime may be considered as having
some extended form of Weyl--Cartan geometry. To our knowledge,
spacetimes with the particular semi-metricity condition
(\ref{ewgtsemimet}) have not been studied previously; in what follows
we refer to them as $\widehat{Y}_4$ spacetimes. Moreover, one finds
that
%

%
\begin{eqnarray}
{R^{\dagger\rho}}_{\sigma\mu\nu} & = &
2(\partial_{[\mu}{\Gamma^\rho}_{|\sigma|\nu]}
+{\Gamma^\rho}_{\lambda[\mu}{\Gamma^\lambda}_{|\sigma|\nu]})+ H^\dagger_{\mu\nu}\delta_\sigma^\rho,\phantom{AAA}
\label{ewgtgeocurvature}\\
{T^{\dagger\lambda}}_{\mu\nu}
& = & 2{\Gamma^\lambda}_{[\nu\mu]},
\label{ewgtgeotorsion}\\
H^\dagger_{\mu\nu} & = & \partial_\mu (V_\nu +\tfrac{1}{3}T_\nu) -
\partial_\nu (V_\mu +\tfrac{1}{3}T_\mu).
\end{eqnarray}
where ${R^{\dagger \rho}}_{\sigma\mu\nu} = {e_a}^\rho {e^b}_\sigma
{R^{\dagger a}}_{b\mu\nu}$ and ${T^{\dagger\lambda}}_{\mu\nu} =
{e_a}^\lambda{T^{\dagger a}}_{\mu\nu}$. We thus recognise
(\ref{ewgtgeotorsion}) as (minus) the torsion tensor of the
$\widehat{Y}_4$ spacetime. One further finds that the trace of the
torsion vanishes, ${T^{\dagger\lambda}}_{\mu\lambda}=0$, so that the
affine connection has the additional symmetry property
\begin{equation}
{\Gamma^\lambda}_{\mu\lambda} = {\Gamma^\lambda}_{\lambda\mu}.
\end{equation}
From (\ref{ewgtgeocurvature}), we see that ${R^{\dagger
    \rho}}_{\sigma\mu\nu}$ is not simply its Riemann tensor. Rather,
the Riemann tensor of the $\widehat{Y}_4$ spacetime is given by
\begin{equation}
{\widehat{R}^\rho}_{\phantom{\rho}\sigma\mu\nu}
\equiv {R^{\dagger \rho}}_{\sigma\mu\nu} -
H^\dagger_{\mu\nu}\delta_\sigma^\rho.
\label{ewgtrhatdef}
\end{equation}
One should note that, although $\widehat{R}_{\rho\sigma\mu\nu}$ is
antisymmetric in $(\mu,\nu)$, it is not antisymmetric in
$(\rho,\sigma)$, since
$\widehat{R}_{(\rho\sigma)\mu\nu}=-g_{\rho\sigma}H^\dagger_{\mu\nu}$. Indeed,
it does not satisfy the usual cyclic and Bianchi identities of the
Riemann tensor in a Riemannian $V_4$ spacetime.  One may also show
that, with the given arrangements of indices, both
${\widehat{R}^{\rho}}_{\phantom{\rho}\sigma\mu\nu}$ (or
${R^{\dagger\rho}}_{\sigma\mu\nu}$) and
${T^{\dagger\lambda}}_{\mu\nu}$ transform covariantly with weight
$w=0$ under a local dilation.  It is also worth noting that
$\widehat{R}_{\mu\nu} \equiv
{\widehat{R}_{\mu\lambda\nu}}^{\phantom{\mu\lambda\nu}\lambda} =
R^\dagger_{\mu\nu}+ H^\dagger_{\mu\nu}$ and $\widehat{R} \equiv
{\widehat{R}^{\mu}}_{\phantom{\mu}\mu} = R^\dagger$. In a similar
manner to WGT, the quantities
(\ref{ewgtgeocurvature})--(\ref{ewgtrhatdef}) arise naturally in the
expression for the commutator of two derivative operators acting on a
vector $J^\rho$ (say) of Weyl weight $w$, which is given by
\begin{equation}
[\nabla^\dagger_\mu,\nabla^\dagger_\nu] J^\rho =
{\widehat{R}^{\rho}}_{\phantom{\rho}\sigma\mu\nu}J^\sigma -
wH^\dagger_{\mu\nu} J^\rho - {T^{\dagger\sigma}}_{\mu\nu}\nabla^\dagger_\sigma V^\rho.
\label{ewgtnablacomm}
\end{equation}

One also finds that the affine connection must satisfy
\begin{equation}
{\Gamma^\lambda}_{\mu\nu} = \christoffel{\dagger\lambda}{\mu}{\nu}
+ {K^{\dagger\lambda}}_{\mu\nu},
\label{eweylaffinec}
\end{equation}
where the first term on the RHS reads
\begin{eqnarray}
\christoffel{\dagger\lambda}{\mu}{\nu} &=&
\tfrac{1}{2}g^{\lambda\rho}(\partial^\dagger_\mu g_{\nu\rho}+\partial^\dagger_\nu
g_{\mu\rho}-\partial^\dagger_\rho g_{\mu\nu}) \nonumber \\
&=&
\christoffel{\lambda}{\mu}{\nu}
- \delta^\lambda_\nu (V_\mu+\tfrac{1}{3}T_\mu) -
\delta^\lambda_\mu (V_\nu+\tfrac{1}{3}T_\nu) \nonumber \\
&&\hspace{35mm}+ g_{\mu\nu}(V^\lambda
+\tfrac{1}{3}T^\lambda),\phantom{ABC}
\label{ediracconnection}
\end{eqnarray}
in which $\christoffel{\lambda}{\mu}{\nu}$ is the standard metric
(Christoffel) connection and ${K^{\dagger\lambda}}_{\mu\nu}$ is the
$\widehat{Y}_4$ contortion tensor
\begin{equation}
{K^{\dagger\lambda}}_{\mu\nu}=-\tfrac{1}{2}({T^{\dagger\lambda}}_{\mu\nu}-
{{{T^\dagger}_\nu}^{\lambda}}_\mu + {{T^\dagger}_{\mu\nu}}^{\lambda}).
\end{equation}

In direct comparison with our treatment of WGT, the result
(\ref{eweylaffinec}) is the analogue of the expression
(\ref{adaggerzerodef}) in the gauge theory viewpoint.
We may consider the geometric interpretation of the quantities
${^0}{A^{\dagger ab}}_\mu$, introduced in (\ref{adaggerzerodef}), but
note that they depend on all the gauge fields; this contrasts with PGT
and WGT, for which the corresponding quantities do not depend on the
$A$-field. Following an analogous argument to
that given above, but considering instead the reduced covariant
derivative ${^0}D^\dagger_\mu$, as defined in (\ref{ddaggermuzerodef}), one finds
that ${^0}{A^{\dagger ab}}_\mu$ and the connection
${^0}{\Gamma^{\dagger \sigma}}_{\rho\mu}$ represent the same geometrical object
in two different frames, and one obtains a `reduced' form of the
tetrad postulate (\ref{eweyltetradp}) given by
\begin{equation}
\zero{\Delta^\dagger_\mu} {e^a}_\nu  \equiv
\partial^\dagger_\mu {e^a}_\nu - \czero{{\Gamma^{\dagger \sigma}}_{\nu\mu}}{e^a}_\sigma
+ \zero{{A^{\dagger a}}_{b\mu}{e^b}_\nu} =0.
\label{ewgttetradp0}
\end{equation}
It thus follows that the relationships (\ref{gammaofa}) and
(\ref{aofgamma}) again hold with the replacements $\partial^\ast_\mu \to
\partial^\dagger_\mu$, ${\Gamma^\sigma}_{\nu\mu} \to
\czero{{\Gamma^{\dagger \sigma}}_{\nu\mu}}$ and ${A^a}_{b\mu} \to
\zero{{A^{\dagger a}}_{b\mu}}$, from which one can directly derive
(\ref{ediracconnection}). One also obtains the metricity condition
${^0}\nabla^\dagger_\sigma g_{\mu\nu}=0$. Finally, the expression
(\ref{ewgtgeocurvature}) for the curvature also holds, but with
${R^{\dagger \rho}}_{\sigma\mu\nu} \to \zero{{R^{\dagger \rho}}_{\sigma\mu\nu}}$
and ${\Gamma^\sigma}_{\nu\mu} \to \czero{{\Gamma^{\dagger
      \sigma}}_{\nu\mu}}$, whereas (\ref{ewgtgeotorsion}) becomes simply
$\czero{{T^{\dagger \lambda}}_{\mu\nu}} = 0$, indicating the absence of
torsion, as expected. The expression (\ref{ewgtnablacomm}) is also
valid, but with $\nabla^\dagger_\mu \to {^0}\nabla^\dagger_\mu$,
${R^\rho}_{\sigma\mu\nu} \to \zero{{R^{\dagger \rho}}_{\sigma\mu\nu}}$ and
${T^{\dagger \sigma}}_{\mu\nu} \to {^0}{T^{\dagger\sigma}}_{\mu\nu} = 0$.


\section{Quadratic extended Weyl gauge theory of gravity}
\label{sec:sigtg}

The new scale-invariant gauge theory of gravity that we now consider
in detail has a total Lagrangian $L_{\rm T}=L_{\rm G}+L_{\rm M}$
corresponding to the general eWGT form (\ref{ltotewgt}).

In the free gravitational sector, it is defined by the most general
parity-invariant\cite{footnote19} eWGT Lagrangian
(\ref{ewgtgravlagdef}) that is at most quadratic in the eWGT field
strengths, namely $L_G = L_{{\cal R}^{\dagger 2}} + L_{{\cal
    H}^{\dagger 2}}$, where
\begin{widetext}
\begin{eqnarray}
L_{{\cal R}^{\dagger 2}} & = & \alpha_1 {\cal R}^{\dagger 2}
+ \alpha_2 {\cal R}^\dagger_{ab}{\cal R}^{\dagger ab}
+ \alpha_3 {\cal R}^\dagger_{ab}{\cal  R}^{\dagger ba}
+ \alpha_4 {\cal R}^\dagger_{abcd}{\cal R}^{\dagger abcd}
+ \alpha_5 {\cal R}^\dagger_{abcd}{\cal R}^{\dagger acbd}
+ \alpha_6 {\cal R}^\dagger_{abcd}{\cal R}^{\dagger cdab}, \label{ewgtlr2}\\
L_{{\cal H}^{\dagger 2}} & = & \tfrac{1}{2}\xi{\cal H}^\dagger_{ab} {\cal H}^{\dagger ab},
\label{ewgtlh2}
\end{eqnarray}
\end{widetext}
in which $\alpha_i$ and $\xi$ are dimensionless parameters and the
factor of $\tfrac{1}{2}$ has been introduced for later convenience. As
discussed in Section~\ref{sec:ewgtfga}, the eWGT field strength ${\cal
  R}^\dagger_{abcd}$ satisfies a form of the Gauss--Bonnet identity
such that, with no loss of generality (up to a classically unimportant
boundary term in $D \le 4$ dimensions), one is free to set {\em one}
of $\alpha_1$, $\alpha_3$ or $\alpha_6$ equal to zero. In what
follows, we will retain all these terms, but this freedom should be
borne in mind during our analysis\cite{footnote20}.

In the matter sector, we base our Lagrangian on
(\ref{eweylcovardirac3}), which builds on Dirac's original suggestion
for accommodating `ordinary' matter in scale-invariant theories of
gravity.  In particular, we adopt the form $L_{\rm M} = L_\vpsi +
L_\phi + \phi^2L_{{\cal R}^\dagger} + \phi^2L_{{\cal T}^{\dagger 2}}$,
where
\begin{eqnarray}
L_\vpsi & = & L_\vpsi(\vpsi,\partial\vpsi,h,\partial h,A,V,\phi), \label{ewgtlpsi}\\
L_\phi & = & \tfrac{1}{2}\nu{\cal D}^\dagger_a\phi\,{\cal D}^{\dagger
  a}\phi -\lambda\phi^4, \\
L_{{\cal R}^\dagger} & = & -\tfrac{1}{2}a{\cal R}^\dagger, \\
L_{{\cal T}^{\dagger 2}} & = &  \beta_1{\cal T}^\dagger_{abc}{\cal
  T}^{\dagger abc} + \beta_2 {\cal T}^\dagger_{abc}{\cal
  T}^{\dagger bac},\label{ewgtlt2}
\end{eqnarray}
in which $\nu$, $\lambda$, $a$ and $\beta_i$ are again dimensionless
parameters
(with $\nu$, $\lambda$ and $a$ usually positive)
and some
factors of $\tfrac{1}{2}$ have been introduced for later
convenience\cite{footnote21}. For the moment we allow the form of
$L_\vpsi$ for the matter field $\vpsi$ to remain general and possibly
include a dependence on the scalar field $\phi$.

In this section, we limit ourselves to deriving the field equations for
the above theory and commenting briefly on their general
structure. The phenomenological content of these field equations,
including their application and solution in various astrophysical and
cosmological situations, will be discussed in forthcoming papers.

\subsection{Field equations}
\label{sec:sigtgfeqns}

Recalling the discussion of Section~\ref{sec:ewgtfecons}, the theory
has only {\em three} independent field equations, namely the
$h$-equation (\ref{eqn:eweylgenfe1dagger}), the $A$-equation
(\ref{eqn:eweylgenfe2}) and the $\vpsi$-equation (\ref{eweylpeqn}). We
calculate the explicit forms of the first two of these equations
(recall we are leaving $L_\vpsi$ general for the moment), without
setting to zero any of the dimensionless parameters in
(\ref{ewgtlr2})--(\ref{ewgtlt2}). This entails a very long, but
nevertheless straightforward, calculation, the results of which are
given below.  In addition, although they are not independent equations
of motion, we will also calculate the $V$-field equation
(\ref{eqn:eweylgenfe3}) and the $\phi$-field equation
(\ref{eqn:eweylgenfe4}) to check that they are indeed related simply
to the contracted $A$-equation and the contracted $h$-equation,
respectively, through the results (\ref{eqn:ewgtphih}),
(\ref{eqn:ewgtmcons4}), (\ref{viscona}) and (\ref{viscona2}). This
provides a useful, albeit partial, check on our calculations.

\bigskip
\subsubsection{The $h$-field equation}

The $h$-field equation (\ref{eqn:eweylgenfe1dagger}) takes
the form
\begin{equation}
 {(t^\dagger_{{\cal R}^2})^a}_b + {(t^\dagger_{{\cal H}^2})^a}_b
+ {(\tau^\dagger_\vpsi)^a}_b + {(\tau^\dagger_\phi)^a}_b
+ {(\tau^\dagger_{\cal R})^a}_b + {(\tau^\dagger_{{\cal T}^2})^a}_b=0,
\label{sigtg:heqn}
\end{equation}
where, in the free gravitational sector, one finds
\begin{widetext}
\begin{eqnarray}
h{(t^\dagger_{{\cal R}^2})^a}_b
& = & \alpha_1 {\cal R}^\dagger(4{{\cal R}^{\dagger a}}_b-\delta^a_b
{\cal R}^\dagger) \nonumber \\
& & + \alpha_2[2({\cal R}^{\dagger ca}{\cal R}^\dagger_{cb}
\!\!-\!\! {\cal R}^{\dagger cd}{{\cal R}^{\dagger a}}_{cdb})
- \delta_b^a {\cal R}^{\dagger}_{cd} {\cal R}^{\dagger cd}] \nonumber \\
& & + \alpha_3[2({\cal R}^{\dagger ac}{\cal R}^\dagger_{cb}
\!\!-\!\! {\cal R}^{\dagger dc}{{\cal R}^{\dagger a}}_{cdb})
- \delta_b^a {\cal R}^{\dagger}_{cd} {\cal R}^{\dagger dc}] \nonumber \\
& & + \alpha_4[4{\cal R}^{\dagger cdea} {\cal R}^\dagger_{cdeb}
\!-\delta^a_b{\cal R}^{\dagger cdef}{\cal R}^\dagger_{cdef}] \nonumber \\
& & + \alpha_5[2({\cal R}^{\dagger acde}\!\!-\!\!{\cal R}^{\dagger ecda})
{\cal R}^\dagger_{cdeb} \!- \delta_b^a
{\cal R}^{\dagger cedf}{\cal R}^\dagger_{cdef}] \nonumber \\
& & + \alpha_6[4{\cal R}^{\dagger eacd} {\cal R}^\dagger_{cdeb}
-\delta^a_b{\cal R}^{\dagger efcd}{\cal R}^\dagger_{cdef}],\label{sigtg:heqn1}\\ [1mm]
h {(t^\dagger_{{\cal H}^2})^a}_b & = & \tfrac{1}{2}\xi
[4{\cal H}^\dagger_{bc}{\cal H}^{\dagger ac}-\delta_b^a
{\cal H}^\dagger_{cd}{\cal H}^{\dagger cd}+ \tfrac{4}{3}{\cal
  D}^\dagger_b({\cal D}^\dagger_c{\cal H}^{\dagger
  ca}-\tfrac{1}{2}{{\cal T}^{\dagger a}}_{cd}{\cal H}^{\dagger cd})],
\label{sigtg:heqn2}
\end{eqnarray}
and in the matter sector, in addition
to ${(\tau^\dagger_\vpsi)^a}_b = {h_b}^\mu(\delta L_\vpsi/\delta
{h_a}^\mu)_\dagger$, one has
%
\begin{eqnarray}
h{(\tau^\dagger_\phi)^a}_b & = &  \tfrac{1}{2}\nu
[\tfrac{4}{3} {\cal D}^{\dagger a}\phi\,{\cal D}^\dagger_b\phi-\tfrac{1}{3}\delta_b^a{\cal D}^{\dagger c}\phi\,{\cal D}^\dagger_c\phi
+\tfrac{2}{3}\phi
(\delta_b^a{\cal D}^\dagger_c {\cal D}^{\dagger c}\phi -
{\cal D}^\dagger_b {\cal D}^{\dagger a}\phi)] + \lambda \delta_b^a \phi^4
\label{sigtg:heqn3}\\[1mm]
h{(\tau^\dagger_{\cal R})^a}_b & = & -a\phi^2({{\cal R}^{\dagger
    a}}_b-\tfrac{1}{2}\delta^a_b {\cal R}^\dagger), \label{sigtg:heqn4}\\ [1mm]
h{(\tau^\dagger_{{\cal T}^2})^a}_b & = &
\beta_1[\phi^2(4 {\cal T}^{\dagger cda}{\cal T}^\dagger_{cdb}
-2{\cal T}^{\dagger ade}{\cal T}^\dagger_{bde}
- \delta_b^a{\cal T}^{\dagger cde}{\cal T}^\dagger_{cde})
+4{\cal D}^\dagger_c(\phi^2 {{\cal T}^\dagger_b}^{ca})]
\nonumber \\
& & +\beta_2
[\phi^2(2 {\cal T}^{\dagger dca}{\cal T}^\dagger_{cdb}
+2 {\cal T}^{\dagger adc}{\cal T}^\dagger_{cdb}
-2{\cal T}^{\dagger acd}{\cal T}^\dagger_{cbd}
-\delta_b^a
{\cal T}^{\dagger dce}{\cal T}^\dagger_{cde})+4
{\cal D}^\dagger_c(\phi^2
{\cal T}^{\dagger [c\phantom{b}\,a]}_{\phantom{\dagger [cc}b})].
\label{sigtg:heqn5}
\end{eqnarray}
\end{widetext}
In particular, we note that ${(t^\dagger_{{\cal H}^2})^a}_b$ in
(\ref{sigtg:heqn2}) is linear in fourth-order derivatives of the
$h$-field. As discussed in Section~\ref{sec:ewgtfecons}, this occurs
because the dilation field strength ${\cal H}^\dagger_{ab}$ is itself
linear in the second-order derivatives of the $h$-field. Moreover,
since ${\cal H}^\dagger_{ab}$ contains first-order derivatives of the
$h$, $A$ and $V$ gauge fields, ${(t^\dagger_{{\cal H}^2})^a}_b$ also
depends on third-order derivatives of these fields. We will shortly discuss this
issue in detail in Section~\ref{sec:ewgtsecondorder}.

\subsubsection{The $A$-field equation}

The $A$-field equation (\ref{eqn:eweylgenfe2}) has the form
\begin{eqnarray}
\hspace*{-8mm}{(s_{{\cal R}^2})_{ab}}^c\! &+&
  \!{(s_{{\cal H}^2})_{ab}}^c \nonumber \\
&+& \!{(\sigma_\vpsi)_{ab}}^c
\!+\! {(\sigma_\phi)_{ab}}^c \!+\! {(\sigma_{\cal R})_{ab}}^c \!+\!
{(\sigma_{{\cal T}^2})_{ab}}^c =0,
\label{sigtg:aeqn}
\end{eqnarray}
where, in the free gravitational sector,
\begin{eqnarray}
h{(s_{{\cal R}^2})_{ab}}^c & = & 4\alpha_1
(\delta_{[a}^c {\cal D}^\dagger_{b]}
+ \tfrac{1}{2}{{\cal T}^{\dagger c}}_{ab}){\cal R}^\dagger \nonumber \\
& & + 4\alpha_2(\delta_{[d}^c {\cal D}^\dagger_{e]}
+ \tfrac{1}{2}{{\cal T}^{\dagger c}}_{de}){{\cal R}^{\dagger\;\, d}_{[a}}\delta_{b]}^e
\nonumber \\
& & + 4\alpha_3(\delta_{[d}^c {\cal D}^\dagger_{e]}
+ \tfrac{1}{2}{{\cal T}^{\dagger c}}_{de}){{\cal R}^{\dagger d}}_{[a}\delta_{b]}^e
\nonumber \\
& & +4\alpha_4 (\delta^c_{[d}{\cal D}^\dagger_{e]} +
  \tfrac{1}{2} {{\cal T}^{\dagger c}}_{de}) {{\cal
      R}^\dagger_{ab}}^{de} \nonumber \\
& & +4\alpha_5(\delta^c_{[d}{\cal D}^\dagger_{e]} +
  \tfrac{1}{2} {{\cal T}^{\dagger c}}_{de}) {\cal R}^{\dagger
    [d\phantom{[ab]}e]}_{\phantom{\dagger r}\,\,[ab]} \nonumber \\
& & +4\alpha_6(\delta^c_{[d}{\cal D}^\dagger_{e]} +
  \tfrac{1}{2} {{\cal T}^{\dagger c}}_{de})
{{\cal R}^{\dagger de}}_{ab}, \label{sigtg:aeqn1}\\[1mm]
h{(s_{{\cal H}^2})_{ab}}^c & = &
\tfrac{2}{3}\xi\eta_{f[b}\delta_{a]}^c({\cal
    D}^\dagger_e {\cal H}^{\dagger ef}-\tfrac{1}{2}{{\cal T}^{\dagger
      f}}_{de}{\cal H}^{\dagger de}), \label{sigtg:aeqn2}
\end{eqnarray}
and in the matter sector, in addition to ${(\sigma^\dagger_\vpsi)_{ab}}^c =
{b^c}_\mu\, \delta L_\vpsi/\delta {A^{ab}}_\mu$, one has
\begin{eqnarray}
h{(\sigma_{\phi})_{ab}}^c & = &
-\tfrac{1}{3}\nu\phi\,\delta_{[a}^c{\cal D}^\dagger_{b]}\phi,
\label{sigtg:aeqn3} \\[1mm]
h{(\sigma_{\cal R})_{ab}}^c & = & -a
(\delta_{[a}^c {\cal D}^\dagger_{b]}
+\tfrac{1}{2}{{\cal T}^{\dagger c}}_{ab})\phi^2, \label{sigtg:aeqn4}\\[1mm]
\hspace*{-10mm}h{(\sigma_{{\cal T}^2})_{ab}}^c & = &
\!-4\beta_1\phi^2{{{\cal T}^{\dagger}}_{[ab]}}^c
\!\!+\!2\beta_2\phi^2({{\cal T}^{\dagger
    c}}_{ab} \!+{{{\cal T}^{\dagger}}_{[ab]}}^c).\phantom{AB}
\label{sigtg:aeqn5}
\end{eqnarray}
We note that ${(s_{{\cal H}^2})_{ab}}^c$ in (\ref{sigtg:aeqn2}) is
linear in third-order derivatives of the $h$-field but, as mentioned
above we, we will shortly discuss this issue in detail in
Section~\ref{sec:ewgtsecondorder}. It is also worth noting in
(\ref{sigtg:aeqn5}) that the translational field strength
satisfies the identity
${\cal T}^\dagger_{cab}+2{\cal T}^\dagger_{[ab]c}-3{\cal T}^\dagger_{[cab]}=0$.

As one might expect, the contributions to the $A$-field equation
(\ref{sigtg:aeqn}) arising from the Riemann-squared terms in the
action (namely those in (\ref{sigtg:aeqn1}) proportional to
$\alpha_4$, $\alpha_5$ and $\alpha_6$, respectively) have a form
similar to the LHS of the equation of motion (\ref{emeomewgtstar}) for
the electromagnetic field. Indeed, recalling that ${\cal R}^\dagger_{abcd}$ is
antisymmetric in $a$ and $b$, the term proportional to $\alpha_4$ in
(\ref{sigtg:aeqn2}) is completely analogous to the LHS of
(\ref{emeomewgtstar}). Moreover, the term proportional to $\alpha_6$
in (\ref{sigtg:aeqn1}) can be brought into the same form by working in terms of
the adjoint Riemann tensor $\overline{\cal R}^\dagger_{abcd} \equiv
{\cal R}^\dagger_{cdab}$. Since ${\overline{\cal R}^{\dagger\,a}}_{b}
= {{\cal R}^{\dagger}_b}^a$, one also sees that terms in (\ref{sigtg:aeqn1})
proportional to $\alpha_2$ and $\alpha_3$, respectively, may be
written in a symmetrical fashion in terms of the Ricci tensor and its
adjoint; this is also true for the term proportional to $\alpha_1$,
since $\overline{\cal R}^\dagger = {\cal R}^\dagger$. The term
proportional to $\alpha_5$, however, cannot be written in a simpler or
more symmetric form using adjoints. Nonetheless, this is to be
expected, since only the term proportional to $\alpha_5$ in the
Lagrangian (\ref{ewgtlr2}) cannot be written as the product of the Riemann,
or one of its contractions, and the corresponding adjoint quantity.
Finally, we note also that the Bianchi identity (\ref{ewgtbi1}) satisfied by
the Riemman tensor (and its adjoint) is analogous to the identity
(\ref{ewgtembi}) satisfied by the EM field strength tensor.

Similar comments to the above apply to the contribution
(\ref{sigtg:aeqn2}) to the $A$-field equation, which contains a factor
identical in form to the LHS of EM field equation
(\ref{emeomewgtstar}), but in terms of the dilation field strength
tensor ${\cal H}^\dagger_{ab}$. Moreover, ${\cal H}^\dagger_{ab}$
satisfies the Bianchi identity (\ref{ewgtbi3}), which is again
directly analogous to the identity (\ref{ewgtembi}) satisfied by the
EM field strength tensor. The above observations taken together
suggest that one might search for solutions of the $A$-field equation
(\ref{sigtg:aeqn}) using techniques derived from
electromagnetism. This is indeed the case and will be discussed in a
forthcoming paper.

\subsubsection{The $V$-field equation}

The $V$-field equation (\ref{eqn:eweylgenfe3}) has the form
\begin{equation}
(j_{{\cal R}^2})_a + (j_{{\cal H}^2})_a
+ (\zeta_\vpsi)_a + (\zeta_\phi)_a + (\zeta_{\cal R})_a + (\zeta_{{\cal T}^2})_a = 0,
\label{sigtg:veqn}
\end{equation}
where, in the free gravitational sector,
\begin{eqnarray}
h(j_{{\cal R}^2})_a & = & -12\alpha_1{\cal D}^{\dagger}_a{\cal R}^\dagger
\nonumber \\
& & - 4\alpha_2[{\cal D}^\dagger_b({{\cal R}^{\dagger \,b}_a}
    +\tfrac{1}{2}\delta_a^b{\cal R}^\dagger)-\tfrac{1}{2}{{\cal
    T}^{\dagger b}}_{ac} {{\cal R}^{\dagger c}_b}] \nonumber \\
& & - 4\alpha_3[{\cal D}^\dagger_b({{\cal R}^{\dagger
    b}}_a+\tfrac{1}{2}\delta_a^b{\cal R}^\dagger)-\tfrac{1}{2}{{\cal
    T}^{\dagger b}}_{ac} {{\cal R}^{\dagger c}}_b] \nonumber \\
& & - 8\alpha_4[{\cal D}^\dagger_b{\cal R}^{\dagger\,
    b}_a-\tfrac{1}{2}{\cal T}^{\dagger bcd}{\cal R}^{\dagger}_{abcd}]
\nonumber \\
& & -2\alpha_5[{\cal D}^\dagger_b({\cal R}^{\dagger \,b}_a \!+\! {{\cal
    R}^{\dagger b}}_a)\!-\!{\cal T}^{\dagger bcd}({\cal R}^{\dagger}_{cbad}
\!+\!{\cal R}^{\dagger}_{cabd})]\nonumber \\
& & - 8\alpha_6[{\cal D}^\dagger_b{{\cal R}^{\dagger
    b}}_a-\tfrac{1}{2}{\cal T}^{\dagger bcd}{\cal R}^{\dagger}
    _{cdab}],\label{sigtg:veqn1}\\
h(j_{{\cal H}^2})_a & = & -2\xi[{\cal D}^\dagger_b{{\cal H}^{\dagger b}}_a-\tfrac{1}{2}
{{\cal T}^{\dagger}}_{acd}{\cal H}^{\dagger cd}],\label{sigtg:veqn2}
\end{eqnarray}
and in the matter sector, in addition to $(\zeta_\vpsi)_a =
b_{a\mu}\,\delta L_\vpsi/\delta V_\mu$,
\begin{eqnarray}
h(\zeta_\phi)_a & = & \nu \phi {\cal D}^{\dagger}_a\phi,
\label{sigtg:veqn3} \\
h(\zeta_{\cal R})_a & = & 3a{\cal D}^{\dagger}_a\phi^2,\label{sigtg:veqn4}\\
h(\zeta_{{\cal T}^2})_a & = & 0.\label{sigtg:veqn5}
\end{eqnarray}

By comparing the expressions (\ref{sigtg:veqn1})--(\ref{sigtg:veqn5})
with those given in (\ref{sigtg:aeqn1})--(\ref{sigtg:aeqn5}), we see
that, as anticipated, the $V$-equation (\ref{sigtg:veqn}) is indeed
identical to (twice) the $A$-equation (\ref{sigtg:aeqn}) contracted on
the indices $b$ and $c$, in accordance with the general results
(\ref{viscona}) and (\ref{viscona2}), and so $j^\dagger_a$ and
$\zeta^\dagger_a$ vanish identically for each part of the Lagrangian.

\subsubsection{The $\phi$-field equation}

Finally, the $\phi$-field equation (\ref{eqn:eweylgenfe4}) has the form
\begin{equation}
\partial_\phi L_{{\cal R}^2} + \partial_\phi L_{{\cal H}^2} +
\partial_\phi L_\vpsi + \delta_\phi L_\phi + \partial_\phi L_{\cal R}
+ \partial_\phi L_{{\cal T}^2} = 0,
\label{sigtg:phieqn}
\end{equation}
where $\partial_\phi \equiv \partial/\partial\phi$ and $\delta_\phi
\equiv \delta/\delta\phi$, since only $L_\phi$ depends on the
derivatives of $\phi$. In the free gravitational sector,
\begin{eqnarray}
\partial_\phi L_{{\cal R}^2} & = & 0,\label{sigtg:phieqn1}\\
\partial_\phi L_{{\cal H}^2} & = & 0,\label{sigtg:phieqn2}
\end{eqnarray}
and, in the matter sector,
\begin{eqnarray}
 \delta_\phi L_\phi & = & -\nu {\cal D}^\dagger_a({\cal D}^{\dagger
   a}\phi) - 4\lambda\phi^3.\label{sigtg:phieqn3} \\
\partial_\phi L_{\cal R} & = & -a\phi^2{\cal R}^\dagger, \label{sigtg:phieqn4}\\
\partial_\phi L_{{\cal T}^2} & = & 2\beta_1\phi {\cal T}^\dagger_{abc}{\cal
  T}^{\dagger abc} + 2\beta_2 \phi {\cal T}^\dagger_{abc}{\cal
  T}^{\dagger bac}.\label{sigtg:phieqn5}
\end{eqnarray}

By comparing the expressions
(\ref{sigtg:phieqn1})--(\ref{sigtg:phieqn5}) with those given in
(\ref{sigtg:heqn1})--(\ref{sigtg:heqn5}), we see that, as anticipated,
the $\phi$-equation (\ref{sigtg:phieqn}) is indeed identical to the
$h$-equation (\ref{sigtg:heqn}) contracted on the indices $a$ and $b$
and divided through by $-\phi$, in accordance with the general results
(\ref{eqn:ewgtphih}) and (\ref{eqn:ewgtmcons4}), assuming
the $\vpsi$-equation $\delta_\vpsi L_\vpsi=0$ is satisfied.

Since the number of terms in the $\phi$-equation is quite small, it is
worth writing it out as a single equation, which reads
\begin{eqnarray}
\hspace*{-10mm}\nu {\cal D}^\dagger_a({\cal D}^{\dagger a}\phi) &+& 4\lambda\phi^3
- \partial_\phi L_\vpsi \nonumber \\
&&\hspace*{-17mm}+ \phi(a{\cal R}^\dagger- 2\beta_1{\cal T}^\dagger_{abc}{\cal
  T}^{\dagger abc} - 2\beta_2 {\cal T}^\dagger_{abc}{\cal
  T}^{\dagger bac}) = 0.
\end{eqnarray}

\subsection{Higher derivative terms in the field equations}
\label{sec:ewgtsecondorder}

As we have mentioned several times, and demonstrated above, if one
includes in the free gravitational Lagrangian the term proportional to
${\cal H}^\dagger_{ab}{\cal H}^{\dagger ab}$, namely (\ref{ewgtlh2}), the
corresponding terms it generates in the $h$-field equation are linear
in fourth-order derivatives of the $h$-field, and also contain
third-order derivatives of all three gauge fields $h$, $A$ and
$V$. Moreover, the terms generated in the $A$-field equation are
linear in third-order derivatives of the $h$-field (we need not
consider the field equations for $\phi$ and $V$, since they are
related simply to contractions of the $h$ and $A$ field equations,
respectively).

At first sight, this would seem to indicate that theories containing
the ${\cal H}^\dagger_{ab}{\cal H}^{\dagger ab}$ term in the
Lagrangian suffer from Ostrogradsky's instability. This conclusion is
not clear cut, however, since in applying such theories to particular
physical systems, which we will discuss in forthcoming papers, we have
found in every case that the field equations organise themselves into
combinations of coupled second-order equations in the gauge
fields. Specifically, one finds the terms containing third- or
fourth-order derivatives correspond to the derivative of already known
expressions, and so contain no new information. Moreover, on
linearising the set of field equations displayed above, which we will
also discuss in a forthcoming publication, we find that the same
behaviour occurs generally, independent of any application of the
theory to a particular physical system. Since the leading order of the
field equations is preserved by linearisation, this suggests that the
full non-linear set of field equations derived above should also enjoy
this general property and hence not suffer from Ostrogradsky's
instability, although we do not yet have a direct proof in this
case. Consequently, we have chosen to retain the terms resulting from
the inclusion of the ${\cal H}^\dagger_{ab}{\cal H}^{\dagger ab}$ in
the field equations given above, so that they are readily available if
required.

The obvious alternative approach to ensuring that Ostrogradsky's
instability does not occur is simply to omit the term proportional to
${\cal H}^\dagger_{ab}{\cal H}^{\dagger ab}$ from the free
gravitational Lagrangian (by setting $\xi=0$). Indeed, one can argue
that the inclusion of the ${\cal H}^\dagger_{ab}{\cal H}^{\dagger ab}$
term as the kinetic term for the $V$ gauge field is not well
motivated. As discussed in Section~\ref{sec:ewgtaltvarp}, eWGTs of the
general form considered here can be written entirely in terms of the
two combinations $h$ and $A^\dagger$ of the gauge fields, such that
there is {\em no} explicit dependence on the dilation gauge field $V$
or its derivatives. Moreover, the natural field strength tensors for
the $h$ and $A^\dagger$ fields are ${\cal T}^\dagger_{abc}$ and ${\cal
  R}^\dagger_{abcd}$, respectively, and the total Lagrangian already
contains all possible (parity-invariant) terms quadratic in these
field strengths {\em without} including the ${\cal
  H}^\dagger_{ab}{\cal H}^{\dagger ab}$ term. The field equations in
the special case $\xi=0$ still display a rich phenomenology, and there
is already a good deal to be explored in them. In particular, we note
that in this case the linearisation process mentioned above reveals the
further desirable property that one can obtain a second-order equation
in $V$ alone, which can then be used to solve for $A$ alone and thence
for the $h$ function alone, with sensible boundary conditions being
available at each stage. Thus, in this case, the theory is inherently
fully solvable. These issues will be discussed fully in a forthcoming
paper.

\subsection{Einstein gauge}

As discussed in Section~\ref{sec:ewgteinstein}, the field equations
can be simplified considerably by adopting the Einstein gauge,
provided $\phi \neq 0$. To obtain the corresponding forms of the field
equations, one merely sets $\phi=\phi_0$ and makes the replacement
$\left.{\cal D}^\dagger_a\phi\right|_{\phi=\phi_0} = \phi_0({\cal V}_a
+ \tfrac{1}{3}{\cal T}_a)$ throughout, but we will not carry out this
procedure explicitly here. As we showed earlier, however, if the
resulting field equations are written in terms of $\vpsi$, $h_a^\mu$
and the combinations ${A^{\dagger ab}}_\mu$ and
$V_\mu+\tfrac{1}{3}T_\mu$ (which will always be possible), they will
be identical in form to the full field equations above expressed in
terms of the scale-invariant variables $\widehat{\vpsi}$,
$\widehat{h}_a^{\phantom{a}\mu}$, ${\widehat{A}^{\dagger
    ab}}_{\phantom{\dagger ab}\mu}$ and $\widehat{V}_\mu$,
respectively, so that there is no need to consider the condition
$\phi=\phi_0$ as representing a spontaneously broken symmetry.

\section{Locally scale-invariant Poincar\'e gauge theory}
\label{sec:lsipgt}

In Sections~\ref{sec:wgt}--\ref{sec:sigtg}, we have focussed our
attention on locally scale-invariant gauge theories for which local
Poincar\'e invariance is extended to include invariance under local
changes of scale by gauging the Weyl group.  In Section~\ref{sec:wgt},
we assumed the `normal' form for the transformation of the rotational
gauge field under local dilations to obtain the well-known Weyl gauge
theory (WGT). This approach provides a systematic means of
constructing both free gravitational and matter actions in which each
term (and hence the full action in each case) is invariant under local
dilations, in addition to GCT and local Lorentz rotations, by the
introduction of a new (vector) gravitational gauge field related to
local dilations and the construction of an associated and more general
covariant derivative. Moreover, in Section~\ref{sec:ewgt}, we
presented a novel alternative to WGT by considering a more general
`extended' form of the transformation law of the rotational gauge
field under local dilation, which includes the `normal' transformation
law of WGT as a special case. The resulting `extended' Weyl gauge
theory (eWGT) again relies on the introduction of a new (vector)
gravitational gauge field related to these `extended' local dilations
and the construction of an even more general associated covariant
derivative.

As mentioned in the Introduction, however, one may construct PGTs (see
Appendix~\ref{app:pgt} for a summary of PGT) that are {\em already}
invariant under local dilations {\em without} the need to introduce an
additional gravitational gauge field. This is possible assuming either
the `normal' or `extended' transformation law for the rotational gauge
field; we will focus on the latter here, which clearly includes the
former as a special case. The construction of such
locally-scale-invariant PGTs is achieved by requiring the free
parameters in the free gravitational action $S_{\rm G}$ and matter
action $S_{\rm M} $ to obey certain relationships, so that, although
each term in the actions may not be individually invariant under
(extended) local dilations, $S_{\rm G}$ and $S_{\rm M}$ do each enjoy
this property and hence so does the total action\cite{footnote22}. In
Section~\ref{appendix:lsipgt}, we discuss the construction of such
PGTs, and in Section~\ref{appendix:lsipgtzt} we consider locally
scale-invariant reduced PGT.

\subsection{Local scale invariance in PGT}
\label{sec:elsipgt}
\label{appendix:lsipgt}

As discussed in Section~\ref{sec:ewgt}, under a local dilation the
`extended' transformations laws of the $h$ and $A$ gauge fields are
given by ${{h'}_a}^\mu = e^{-\rho} {h_a}^\mu$ and ${{A'}^{ab}}_\mu =
{A^{ab}}_\mu + \theta ({b^a}_\mu{\cal P}^b-{b^b}_\mu{\cal P}^a)$,
respectively, where ${\cal P}_a={h_a}^\mu\partial_\mu\rho$. Here
$\theta=0$ recovers the `normal' transformations laws
(\ref{weylhtrans2})--(\ref{weylatrans2}) assumed in WGT and $\theta=1$
corresponds to the `special' transformation law (\ref{specialt}) for
the $A$ field.

Consequently, under a local dilation, the PGT covariant derivative
transforms inhomogeneously as
\begin{equation}
{\cal D}_c^\prime \vpsi^\prime
= e^{(w-1)\rho}[{\cal D}_c \vpsi + w {\cal P}_c\vpsi
+\theta{\cal P}^{[b}\delta_c^{a]}\Sigma_{ab}\vpsi],
\label{pgtdprime}
\end{equation}
where $w$ is the Weyl weight of the field $\vpsi$ and $\Sigma_{ab}$
are the generator matrices of the Lorentz group representation to
which $\vpsi$ belongs. Thus, one finds that the PGT rotational
gauge field strength and its contractions transform as
\begin{widetext}
\begin{subequations}
\label{rspecialall}
\begin{eqnarray}
{{\cal R}^{\prime ab}}_{cd} & = & e^{-2\rho}\{{{\cal R}^{ab}}_{cd} +
2\theta\delta^{[a}_d({\cal D}_c-\theta{\cal P}_c){\cal P}^{b]} -
2\theta\delta^{[a}_c({\cal D}_d-\theta{\cal P}_d){\cal P}^{b]} -
2\theta{\cal P}^{[a}{{\cal T}^{b]}}_{cd} -
2\theta^2\delta^{[a}_c\delta^{b]}_d {\cal P}^e {\cal P}_e \},
\label{rspecialtagain}\\
{{\cal R}^{\prime a}}_{c} & = & e^{-2\rho}\{{{\cal R}^{a}}_{c} -2\theta{\cal D}_c{\cal P}^a - \theta\delta_c^a{\cal D}_b{\cal P}^b +
2\theta^2 {\cal P}^a{\cal P}_c -2\theta^2\delta_c^a{\cal P}^2 +
\theta{\cal P}^b{{\cal T}^a}_{cb} - \theta {\cal P}^a{\cal T}_c\},\\
{\cal R}^\prime & = & e^{-2\rho}\{{\cal R} -6\theta({\cal D}_a +\tfrac{1}{3}{\cal T}_a){\cal P}^a-6\theta^2{\cal P}^2\},
\end{eqnarray}
\end{subequations}
\end{widetext}
where ${\cal P}^2 \equiv {\cal P}_a{\cal P}^a$, and the translational
gauge field strength and its contractions transform as
\begin{subequations}
\label{eqn:tortransall}
\begin{eqnarray}
{{\cal T}^{\prime a}}_{bc} & = & e^{-\rho}\{{{\cal T}^a}_{bc}+2(1-\theta){\cal
  P}_{[b}\delta_{c]}^a\},\label{eqn:tortransagain}\\
{{\cal T}^{\prime}}_{b} & = & e^{-\rho}\{{{\cal T}}_{b}+3(1-\theta){\cal P}_b\}.
\end{eqnarray}
\end{subequations}
Thus neither ${{\cal R}^{ab}}_{cd}$ nor ${{\cal T}^{a}}_{bc}$ is
covariant under a general extended Weyl transformation with arbitrary
$\theta$. As mentioned in Section~\ref{sec:ewgt}, however, one sees
that ${{\cal R}^{ab}}_{cd}$ is covariant with weight $w=-2$ under
`normal' $(\theta=0)$ transformations, whereas ${{\cal T}^{a}}_{bc}$
is covariant with weight $w=-1$ under `special' transformations
$(\theta=1)$.

\subsubsection{Free gravitational action}

To construct a free gravitational action that is invariant under local
dilations, one requires the gravitational Lagrangian $L_{\rm G}$ to
transform covariantly with weight $w=-4$, at least up to a surface
term. Assuming the extended transformation law for the $A$-field, with
$\theta\neq 0$, a long but straightforward calculation shows that
there is {\em no} PGT Lagrangian of the form $L_{{\cal R}^2}$ in
(\ref{lr2}) that transforms covariantly (aside from the trivial case
$\alpha_i=0$ for all $i$) and so $L_{\rm G}$ must vanish. Nonetheless,
by contrast, in the special case of the `normal' transformation law
$(\theta=0)$ for the $A$-field, $L_{\rm G}$ may be {\em any} PGT
Lagrangian of the form $L_{{\cal R}^2}$ in (\ref{lr2}).

\subsubsection{Matter action}

To construct an appropriate matter action, we follow the rationale
presented in Section~\ref{sec:wgtdirac} and base our Lagrangian
density on (\ref{weylcovardirac3}), but appropriately modified to use
only PGT quantities. Thus, in the most general case, one might
consider\cite{footnote23}
%
\begin{eqnarray}
{\cal L}_{\rm M}  =  h^{-1}[\tfrac{1}{2}i\bar{\psi}\gamma^a
{\stackrel{\leftrightarrow}{{\cal D}_a}}\psi - \mu\phi\bar{\psi}\psi
&&+ \tfrac{1}{2}\nu ({\cal D}_a\phi) ({\cal D}^a \phi) - \lambda\phi^4
\nonumber \\&&-
  a\phi^2{\cal R} + \phi^2L_{{\cal T}^2}],\label{sipgtlm}
\label{sipgtlmldefs}
\end{eqnarray}
%
where $\mu$, $\nu$, $\lambda$ and $a$ are dimensionless parameters
(usually positive), and
there are three further dimensionless parameters $\beta_1$, $\beta_2$
and $\beta_3$ in $L_{{\cal T}^2}$.

Let us denote (\ref{sipgtlmldefs}) by ${\cal L}_{\rm M} \equiv {\cal
  L}_\psi + {\cal L}_\phi + \phi^2{\cal L}_{\cal R} + \phi^2{\cal
  L}_{{\cal T}^2}$. As discussed in Section~\ref{sec:ewgtdirac}, ${\cal
  L}_\psi$ is invariant under extended local dilations, but one may
show that the other constituent parts of ${\cal L}_{\rm M}$ transform
inhomogeneously as
\begin{widetext}
\begin{subequations}
\label{sipgtlmtrans}
\begin{eqnarray}
{\cal L}'_\phi
& = & {\cal L}_\phi
+\tfrac{1}{2}\nu h^{-1}(\phi^2{\cal P}^2-2\phi{\cal P}^a{\cal D}_a\phi), \\
(\phi^2{\cal L}_{\cal R})'
& = & \phi^2{\cal L}_{\cal R}
+a\phi^2h^{-1}\theta[6({\cal D}_a+\tfrac{1}{3}{\cal T}_a){\cal P}^a + 6\theta{\cal P}^2], \\
(\phi^2{\cal L}_{{\cal T}^2})' & = & \phi^2{\cal L}_{{\cal T}^2}
+(2\beta_1+\beta_2+3\beta_3)\phi^2 h^{-1}(1-\theta)
[2{\cal P}_a{\cal T}^a+3(1-\theta){\cal P}^2].
\end{eqnarray}
\end{subequations}
\end{widetext}
In order to make
${\cal L}_{\rm M}$ invariant, one must therefore impose some
conditions on the (dimensionless) parameters it contains. Indeed, a
short calculation reveals a substantial simplification is obtained
by setting
\begin{equation}
\tfrac{1}{2}\nu = -6a\theta = 3(\theta-1)(2\beta_1+\beta_2+3\beta_3),
\label{pgtsicondition}
\end{equation}
in which case ${\cal L}_{\rm M}$ transforms as
\begin{equation}
{\cal L}'_{\rm M} = {\cal L}_{\rm M} + 6a\theta h^{-1} ({\cal
  D}_a+{\cal T}_a)(\phi^2 {\cal P}^a).
\label{eqn:lmtranslaw}
\end{equation}
It is a simple matter to show that the second term on the RHS is a
total derivative, and so the action based on (\ref{sipgtlm}) is
invariant under extended local dilations with a {\em particular value}
of $\theta$, provided the dimensionless parameters $\nu$, $a$ and
$\beta_i$ $(i=1,2,3)$ satisfy the conditions
(\ref{pgtsicondition}). It is worth noting that for a positive value
of $\theta$, the parameters $\nu$ and $a$ must be of opposite sign,
which is different to the usual case in WGT and eWGT, where they are
both positive.

It is also worth noting the two special cases of a `normal'
transformation $(\theta=0)$, for which one requires
$\nu=0=2\beta_1+\beta_2+3\beta_3$, and a `special' transformation
$(\theta=1)$, for which $\nu=0=a$. Thus, in both special cases, the
kinetic term for the scalar field $\phi$ is inadmissible. Moreover, in
the former case ($\theta=0$), one obtains a generalisation of the
gravitational theory proposed by Obukhov\cite{obukhov82}, who set
$\nu=\beta_1=\beta_2=\beta_3=0$. Adopting our more general conditions
$\nu=0=2\beta_1+\beta_2+3\beta_3$, one may further show that the
matter Lagrangian (\ref{sipgtlmldefs}) can be written as
\begin{equation}
{\cal L}_{\rm M}  =  h^{-1}[\tfrac{1}{2}i\bar{\psi}\gamma^a
{\stackrel{\leftrightarrow}{{\cal D}_a}}\psi - \mu\phi\bar{\psi}\psi
- \lambda\phi^4 - a\phi^2{\cal R} + \phi^2L_{{\cal T}^{\ast 2}}],
\label{wgtspecialcase}
\end{equation}
with $2\beta_1+\beta_2+3\beta_3=0$ in $L_{{\cal T}^{\ast 2}}$,
which is merely a special case of the general WGT matter Lagrangian
(\ref{weylcovardirac3}).

Since the condition (\ref{pgtsicondition}) depends on the value of the
parameter $\theta$, it is clear that no single matter Lagrangian of
the form (\ref{sipgtlmldefs}) results in an action that is invariant
under an extended local dilation for {\em arbitrary values} of
$\theta$. Nonetheless, it is possible to construct such a Lagrangian
using only PGT quantities by making only a minor modification to
(\ref{sipgtlmldefs}). Specifically, making the replacement ${\cal
  D}_a\phi \to ({\cal D}_a + \tfrac{1}{3}{\cal T}_a)\phi$ (and
similarly for ${\cal D}^a\phi$) in the kinetic term for $\phi$, one
may show that one recovers the transformation law
(\ref{eqn:lmtranslaw}) under an extended local dilation, provided only
that $\tfrac{1}{2}\nu + 6a=0$ and $2\beta_1+\beta_2+3\beta_3=0$ {\em
  separately}. Thus, in this case, the parameters $\nu$ and $a$ must
always be of opposite sign, rather than the usual case in WGT and
eWGT, where both are positive. One may further show that these
conditions allow one to write the resulting Lagrangian as
\begin{widetext}
\begin{equation}
{\cal L}_{\rm M} = h^{-1}[\tfrac{1}{2}i\bar{\psi}\gamma^a
{\stackrel{\leftrightarrow}{{\cal D}_a}}\psi -
\mu\phi\bar{\psi}\psi + \tfrac{1}{2}\nu
({\cal D}^\dagger_a\phi) ({\cal D}^{\dagger a} \phi)
- \lambda\phi^4 + \tfrac{1}{12}\nu\phi^2{\cal R}^\dagger + \phi^2L_{{\cal
    T}^{\dagger 2}}] - \tfrac{1}{2}\nu h^{-1} ({\cal
  D}_a+{\cal T}_a)(\phi^2 {\cal V}^a),
\label{ewgtspecialcase}
\end{equation}
\end{widetext}
with $2\beta_1+\beta_2=0$ in $L_{{\cal T}^{\dagger 2}}$ (in which
there is no term proportional to $\beta_3$).  Since the final term on
the RHS is a total derivative, this Lagrangian is equivalent merely to
a particular special case of the eWGT matter Lagrangian
(\ref{eweylcovardirac3}), but one where (unusually) the kinetic term
for the $\phi$-field and the $\phi^2 {\cal R}^\dagger$ term have the
same sign.

\subsubsection{Einstein gauge, scale-invariant variables and motion of
  test particles}

Considering first the general case in which $\theta$ may take an
arbitrary value, the matter Lagrangian must have the form
(\ref{ewgtspecialcase}), which is equivalent to a
special case of the eWGT matter Lagrangian
(\ref{eweylcovardirac3}). In this case, ${\cal L}_{\rm M}$ must in
fact be the total Lagrangian, since one requires the free
gravitational Lagrangian to vanish for arbitrary $\theta$, as
discussed above. Hence, our conclusions in eWGT regarding the
relationship between the Einstein gauge and scale-invariant variables
presented in Section~\ref{sec:ewgteinstein}, and the motion of test
particles given in Section~\ref{sec:ewgttestp}, must also hold in this
case.

For the special case $\theta=0$, the matter Lagrangian
(\ref{wgtspecialcase}) is an example of the general WGT matter
Lagrangian (\ref{weylcovardirac3}). Similarly, the free gravitational
Lagrangian in the $\theta=0$ case may be {\em any} Lagrangian of the
form $L_{{\cal R}^2}$ in (\ref{lr2}), which is just a special case of
the general WGT free gravitational Lagrangian
(\ref{wgtgravlagdef}). It therefore follows that our conclusions in
WGT regarding the relationship between the Einstein gauge and
scale-invariant variables presented in Section~\ref{sec:wgteinstein},
and the motion of test particles given in
Section~\ref{sec:WGTmassivemotion}, also apply in this
locally-scale-invariant PGT.


\subsection{Local scale-invariance in reduced PGT}
\label{sec:elsipgtnot}
\label{appendix:lsipgtzt}

We now consider the construction of locally scale-invariant reduced
PGTs.  As discussed in Appendix~\ref{app:pgt} (see also
Section~\ref{sec:wgtcovdalt}), in the `reduced' PGT covariant
derivative ${^0}{\cal D}_a$, to which the full PGT covariant
derivative reduces for vanishing PGT torsion, the rotational gauge
field ${A^{ab}}_\mu$ is replaced by ${^0}{A^{ab}}_\mu$, which depends
only on the $h$-field and its first derivatives. This covariant
derivative may be used to construct the `reduced' rotational field
strength (or `curvature') tensor $\zero{{{\cal R}^{ab}}_{cd}}$, which
depends only on the $h$-field and its first and second
derivatives. Moreover, the corresponding reduced translational field
strength (or `torsion') tensor vanishes, since $\czero{{{\cal
      T}^a}_{bc}} \equiv {h_b}^{\mu}{h_c}^{\nu}(\zero{D_\mu} {b^a}_\nu
-\zero{D_\nu}{b^a}_\mu) = 0$. Thus, reduced PGTs can be expressed
entirely in terms of the $h$-field and its derivatives and correspond
to imposing the condition of vanishing PGT torsion directly at the
level of the action.

One may show that, under an extended local dilation,
${^0}{{A'}^{ab}}_\mu = {^0}{A^{ab}}_\mu + {b^a}_\mu{\cal
  P}^b-{b^b}_\mu{\cal P}^a$, which is independent of $\theta$, as is
expected since ${^0}{A^{ab}}_\mu$ depends on only the $h$-field (and
its first derivatives), for which ${{h'}_a}^\mu = e^{-\rho}
{h_a}^\mu$. Thus, the `normal' and `extended' transformations laws for
${^0}{A^{ab}}_\mu$ coincide. The corresponding `reduced' PGT covariant
derivative $\czero{{\cal D}}_a$ transforms inhomogeneously under a
local dilation as
\begin{equation}
\czero{{\cal D}}_c^\prime \vpsi^\prime
= e^{(w-1)\rho}[\czero{{\cal D}}_c \vpsi
+ w {\cal P}_c\vpsi
+{\cal P}^{[b}\delta_c^{a]}\Sigma_{ab}\vpsi],
\label{dzeroprime}
\end{equation}
where $w$ is the Weyl weight of the field $\vpsi$ and $\Sigma_{ab}$
are the generator matrices of the Lorentz group representation to
which $\vpsi$ belongs. Similarly, one finds that the `reduced'
rotational field strength transforms as
\begin{widetext}
\begin{subequations}
\label{rie0primeall}
\begin{eqnarray}
{^0}{{\cal R}^{\prime ab}}_{cd}  & = &  e^{-2\rho}\{{^0}{{\cal R}^{ab}}_{cd} +
2\delta^{[a}_d({^0}{\cal D}_c-{\cal P}_c){\cal P}^{b]} -
2\delta^{[a}_c({^0}{\cal D}_d-{\cal P}_d){\cal P}^{b]} -
2\delta^{[a}_c\delta^{b]}_d {\cal P}^e {\cal P}_e \}, \label{rie0prime}\\
{{^0}{\cal R}^{\prime a}}_{c} & = & e^{-2\rho}\{{^0}{{\cal R}^{a}}_{c} -2\,{^0}{\cal D}_c{\cal P}^a - \delta_c^a\,{^0}{\cal D}_b{\cal P}^b +
2{\cal P}^a{\cal P}_c -2\delta_c^a{\cal P}^2\},\label{ricci0prime}\\
{^0}{\cal R}^\prime & = & e^{-2\rho}\{{^0}{\cal R} -6\,{^0}{\cal D}_a{\cal P}^a-6{\cal P}^2\}, \label{riccis0prime}
\end{eqnarray}
\end{subequations}
\end{widetext}
and $\czero{{{\cal T}^a}_{bc}}$ remains self-consistently zero. It is
worth noting that, under a local dilation, the `reduced' quantities
${^0}{A^{ab}}_\mu$, $\czero{{\cal D}}_c \vpsi$ and ${^0}{{\cal
    R}^{ab}}_{cd}$ all transform independently of $\theta$, and in the
same way as their `full' PGT counterparts ${A^{ab}}_\mu$, ${\cal D}_c
\vpsi$ and ${{\cal R}^{ab}}_{cd}$ do assuming the `special'
transformation law (\ref{specialt}) for the $A$-field $(\theta=1)$,
but with the replacements ${\cal D}_c \to \czero{{\cal D}}_c$ and
${{\cal T}^a}_{bc} \to \czero{{{\cal T}^a}_{bc}} \equiv 0$.

\subsubsection{Free gravitational action}

Following the discussion in Appendix~\ref{app:pgt}, if $L_G$
contains only terms that are at most quadratic in $\zero{{{\cal
      R}^{ab}}_{cd}}$ and its contractions, then it has the general
form (\ref{pgtnotlg}). For $L_G$ to transform covariantly under local
dilations, however, it cannot contain the dimensionful constants
$\kappa$ and $\Lambda$, and so the most general form one need consider
is that given in (\ref{pgtnotlr2}), namely
\begin{equation}
L_{\rm G} = \alpha_1\, \czero{{\cal R}}^2
+ \alpha_2\, \czero{{\cal R}}_{ab}\czero{{\cal R}}^{ab}
+ \alpha_3\, \czero{{\cal R}}_{abcd}\czero{{\cal R}}^{abcd},
\label{pgtnotlg1}
\end{equation}
where the $\alpha_i$ are dimensionless free parameters. As mentioned
in Appendix~\ref{app:pgt}, one can use the Gauss--Bonnet identity
to set to zero any one of the parameters $\alpha_i$, with no loss of
generality (at least classically). For our current purposes, however,
it is more convenient to retain the form (\ref{pgtnotlg1}).

Using the expressions (\ref{rie0prime})--(\ref{riccis0prime}), a short
calculation reveals that $L_G$ transforms covariantly with weight
$w=-4$ under an extended local dilation, provided
$2\alpha_3+\alpha_2=0$ and $3\alpha_1-\alpha_3=0$. Letting
$\alpha=\alpha_3$ for convenience, one thus obtains the
one-parameter free gravitational Lagrangian
\begin{equation}
L_{\rm G} = \alpha(
\czero{{\cal R}}_{abcd}\czero{{\cal R}}^{abcd}
- 2\, \czero{{\cal R}}_{ab}\czero{{\cal R}}^{ab}
+ \tfrac{1}{3}\, \czero{{\cal R}}^2),
\label{weylsq1}
\end{equation}
for which the corresponding action is thus {\em unique} up to surface
terms and an overall scaling.  If one wishes, one may now use the fact
that the Gauss--Bonnet term (\ref{gbid}) contributes a total
derivative to the action (in $D \le 4$ dimensions) to remove terms in
$L_G$. The calculation of the field equations is simplest if one
removes the term proportional to $\czero{{\cal R}}_{abcd}\czero{{\cal
    R}}^{abcd}$ to obtain
\begin{equation}
L_{\rm G} = 2\alpha(\czero{{\cal R}}_{ab}\czero{{\cal R}}^{ab} -
\tfrac{1}{3}\, \czero{{\cal R}}^2),
\label{conformall}
\end{equation}
which changes only by a surface term under local changes of scale, and
hence the equations of motions are covariant under such a
transformation.

When interpreted geometrically, the theory defined by (\ref{weylsq1})
or (\ref{conformall}) coincides with so-called {\em conformal gravity}
theory\cite{mannheim06}. Indeed, as one might suspect, one can
easily show that (\ref{weylsq1}) is proportional to the Weyl-squared
Lagrangian $L_{\rm W} = \czero{{\cal C}}_{abcd} \,\czero{{\cal
    C}}^{abcd}$, where $\czero{{\cal C}}_{abcd}$ is the gauge theory
equivalent of the Weyl tensor, given by
\begin{eqnarray}
\czero{{\cal C}}_{abcd} \!=\! \czero{{\cal R}}_{abcd} &&-\tfrac{1}{2}(
\eta_{ac}\czero{{\cal R}}_{bd}
\!\!-\!\eta_{ad}\czero{{\cal R}}_{bc}
\!\!-\!\eta_{bc}\czero{{\cal R}}_{ad}
\!+\!\eta_{bd}\czero{{\cal R}}_{ac})\nonumber \\
&&+\tfrac{1}{6}(\eta_{ac}\eta_{bd}-\eta_{ad}\eta_{bc})\czero{{\cal R}}.
\label{eqn:weyltdef}
\end{eqnarray}
This tensor represents the traceless part of $\czero{{\cal
    R}}_{abcd}$; it has the same symmetries, but satisfies the extra
condition of being trace-free (contraction on any pair of indices
yields zero). It transforms covariantly as $\czero{{\cal
    C}}^\prime_{abcd} = e^{-2\rho} \,\czero{{\cal C}}_{abcd}$ under a
local change of scale, so that $L_W$ also transforms covariantly with
a Weyl weight $w(L_W)=-4$, and hence the corresponding action is
invariant under a local change of scale, as found above.

It should be noted, however, that the Lagrangian (\ref{weylsq1}) or
(\ref{conformall}) is quadratic in second-order derivatives of the
$h$-field, and so the corresponding field equations will typically be
linear in fourth-order derivatives of $h$. This suggests that
conformal gravity may suffer from Ostrogradsky's instability, although
this needs to be checked in detail. Ostrogradsky's instability is
often reinterpreted as non-unitary behaviour in the associated quantum
theory, but this instability occurs in the classical Hamiltonian and
survives canonical quantisation\cite{woodard15}; it thus seems
unlikely that it can be circumvented by any attempt to redefine the
Fock space of the quantum theory\cite{bender08}.  Such issues may be
problematic for conformal gravity as a fundamental description of the
gravitational field, although further investigation is clearly
necessary.

\subsubsection{Matter action}

To construct an appropriate matter action, we again follow the
rationale presented in Section~\ref{sec:wgtdirac}, but appropriately
modified to use `reduced' PGT quantities relevant to the special case
of vanishing PGT torsion. Thus, in the most general case, one might
consider the Lagrangian (\ref{sipgtlmldefs}), but with the
replacements ${\cal D}_a \to {^0}{\cal D}_a$, ${\cal R} \to {^0}{\cal
  R}$ and without the term proportional to $L_{{\cal T}^2}$,
namely\cite{footnote24}
\begin{eqnarray}
{\cal L}_{\rm M}  =  h^{-1}[\tfrac{1}{2}i\bar{\psi}\gamma^a
\,{\stackrel{\leftrightarrow}{{{^0}{\cal D}_a}}}\psi\!-\! \mu\phi\bar{\psi}\psi
&&+ \tfrac{1}{2}\nu\, ({^0}{\cal D}_a\phi) \,({^0}{\cal D}^a \phi)
\!-\! \lambda\phi^4 \nonumber \\ &&\hspace{12mm}-
  a\phi^2\,{^0}{\cal R}].
\label{sipgt0lm}
\end{eqnarray}

Using (\ref{dzeroprime}) and (\ref{riccis0prime}), one may show that,
under a local change of scale, ${\cal L}_{\rm M}$ transforms as
\begin{eqnarray}
\hspace*{-5mm}{\cal L}_{\rm M}^\prime \!=\! {\cal L}_{\rm M} \!-\! h^{-1}[\tfrac{1}{2}\nu {\cal P}^a
 \,\czero{{\cal D}}_a\phi^2 &&- 6a \phi^2  \,\czero{{\cal D}}_a {\cal
   P}^a\nonumber \\&& +
(\tfrac{1}{2}\nu+6a)\phi^2 {\cal P}_a {\cal P}^a].
\end{eqnarray}
By setting $\tfrac{1}{2}\nu + 6a =0$, however, one obtains
\begin{equation}
{\cal L}_{\rm M}^\prime = {\cal L}_{\rm M} - h^{-1}[\tfrac{1}{2}\nu
\,\czero{{\cal D}}_a(\phi^2 {\cal P}^a)],
\end{equation}
where we recognise the second term on the RHS as a total derivative.
Thus, in this special case, the corresponding field equations are
covariant under local dilations. The resulting matter Lagrangian,
obtained by setting $\tfrac{1}{2}\nu + 6a =0$ in (\ref{sipgt0lm}),
coincides with that typically assumed in conformal
gravity\cite{mannheim06}. It is worth noting that, as found in the
previous section, the dimensionless parameters $\nu$ and $a$ must have
opposite signs, which is different to the usual case in WGT and eWGT,
where both parameters are positive. Thus, the resulting matter
Lagrangian is again unusual in having the same sign for the kinetic
term for $\phi$ and the $\phi^2\,{^0}{\cal R}$ term.

\subsubsection{Einstein gauge, scale-invariant variables and
  motion of particles}

One may show that the definition (\ref{eqn:weyltdef}) of the Weyl
tensor ${^0}{\cal C}_{abcd}$ also holds if one makes either of the
replacements ${^0}{\cal R}_{abcd} \to {^0}{\cal R}^\ast_{abcd}$ or
${^0}{\cal R}_{abcd} \to {^0}{\cal R}^\dagger_{abcd}$ (and similarly
for the associated contractions) on the RHS, thereby replacing the
`reduced' curvature tensor in PGT with its counterpart either in WGT
or eWGT. Thus, the free gravitational Lagrangian (\ref{weylsq1})
retains its form if one makes either set of replacements, and is hence
a special case of the free gravitational
Lagrangians in both reduced WGT and reduced eWGT, namely (\ref{wgtnotlg}) and
(\ref{ewgtnotlg}) respectively.

Similarly, the matter Lagrangian (\ref{sipgt0lm}) with
$\tfrac{1}{2}\nu + 6a=0$ may be shown to equal (\ref{ewgtspecialcase})
with ${\cal D}_a \to {^0}{\cal D}_a$ and ${\cal T}_a\equiv 0$ and
either set of replacements ${\cal D}^\dagger_a \to {^0}{\cal
  D}^\ast_a$, ${\cal R}^\dagger \to {^0}{\cal R}^\ast$ ${\cal
  T}^\dagger_{abc} \to {^0}{\cal T}^\ast_{abc} \equiv 0$ or ${\cal
  D}^\dagger_a \to {^0}{\cal D}^\dagger_a$, ${\cal R}^\dagger \to
{^0}{\cal R}^\dagger$ ${\cal T}^\dagger_{abc} \to {^0}{\cal
  T}^\dagger_{abc} \equiv 0$.  Thus, (\ref{sipgt0lm}) with
$\tfrac{1}{2}\nu + 6a =0$ is equivalent to a special case of the
general matter Lagrangian in both reduced WGT and reduced eWGT, namely
(\ref{weylcovardirac3}) and (\ref{eweylcovardirac3}) respectively,
with vanishing WGT/eWGT torsion.

As a result, our conclusions regarding the equivalence of the Einstein
gauge and scale-invariant variables, and the motion of test particles,
presented in Sections~\ref{sec:wgteinstein} and
\ref{sec:WGTmassivemotion} for WGT and Sections~\ref{sec:ewgteinstein}
and \ref{sec:ewgttestp} for eWGT, also hold in this case.


\section{Discussion and conclusions}
\label{sec:conc}

We have presented a general approach to constructing gauge theories of
gravity in a manner that differs from that usually presented in the
literature. In particular, in keeping with the field theories that
describe the other fundamental interactions, we maintain throughout
the notion of gauge fields in Minkowski spacetime, rather than
adopting the more usual geometric interpretation, which has some
non-trivial consequences. We also work exclusively in terms of finite
local transformations, rather than their usual infinitesimal forms;
this allows, in our view, for a more transparent interpretation of the
resulting theories.

We focus our attention in particular on constructing locally
scale-invariant gauge theories of gravity. This is usually achieved by
extending local Poincar\'e invariance to include invariance under
local changes of scale by gauging Weyl transformations.  The resulting
Weyl gauge theory (WGT) is well known and generalises Poincar\'e gauge
theory (PGT) by providing a systematic gauge-theoretic means of
constructing both matter and gravitational actions that are invariant
under local scale transformations, in addition to GCT and local
Lorentz rotations, by the introduction of a new (vector) gravitational
gauge field related to local dilations and the construction of an
associated and more general covariant derivative. The WGT action also
typically depends on a scalar compensator field $\phi$, which opens up
the possibility of including terms in which $\phi$ is non-minimally
coupled to the gauge field strength tensors, in addition (usually) to
kinetic and self-interaction terms for $\phi$. Although WGT has been
widely studied, we make some observations that differ from typical
accounts of the subject, in particular regarding the interpretation of
the Einstein gauge and also the equations of motion of matter fields
and test
particles. In the Einstein gauge, one simplifies the equations of
motion by using local scale-invariance to set the compensator scalar
field to a constant, $\phi(x)=\phi_0$. This is usually considered as
representing the choice of some definite scale in the theory, and is
often given the physical interpretation of corresponding to some
spontaneous breaking of the scale symmetry. We show, however, that the
equations of motion in the Einstein gauge are identical in form to
those obtained when working in terms of scale-invariant variables,
where the latter involves no breaking of the scale symmetry. This
suggests that one should introduce further scalar fields, in addition
to the compensator field $\phi$, to enable a true breaking of the
scale symmetry. Regarding the trajectories of test particles in WGT
(and PGT as a special case), we construct an action for a classical
spin-$\frac{1}{2}$ point particle, which we use as our model for
`ordinary matter', and show that such particles satisfy an equation of
motion that corresponds to geodesic paths, rather than autoparallels;
we also show that the same is true for photons.

Our main objective, however, is to present a novel alternative to WGT
by considering a more general `extended' transformation law for the
rotational gauge field under local dilations, which includes its
`normal' transformation law in WGT as a special case.  This is
motivated by the observation that the PGT (and WGT) matter actions
both for the Dirac field and the electromagnetic field are already
invariant under local dilations if one assumes this `extended'
transformation law, in the same way as they are for the `normal'
transformation law assumed in WGT. Moreover, under a global scale
transformation, the two transformation laws coincide, and so both may
be considered as equally valid gauging of global Weyl scale
invariance. The key difference between the two sets of transformations
is that, whereas under the `normal' Weyl transformations the PGT gauge
fields ${h_a}^\mu$ and ${A^{ab}}_\mu$ transform covariantly with
weights of $w$, $-1$ and $0$, respectively, under the `extended'
transformations the rotational gauge field ${A^{ab}}_\mu$ transforms
inhomogeneously. This has the consequence that the transformation
properties of the PGT rotational gauge field strength (or `curvature')
${\cal R}_{abcd}$ and translational gauge field strength (or
`torsion') ${\cal T}_{abc}$ are treated in a more balanced manner.
The resulting `extended' Weyl gauge theory (eWGT) again relies on the
introduction of a new (vector) gravitational gauge field $V_\mu$
related to the `extended' local dilations and the construction of an
associated covariant derivative ${\cal D}^\dagger_a$.

Our extended WGT has a number of interesting features, which include
the following.
\begin{itemize}
\item The translational, rotational and dilational field strength
  tensors ${\cal R}^\dagger_{abcd}(h,A,\partial A,V,\partial V)$,
  ${\cal T}^\dagger_{abc}(h,\partial h, A)$ and ${\cal
    H}^\dagger_{ab}(h,\partial h, \partial^2 h,A,\partial A, \partial
  V)$, respectively, depend on the gauge fields ${h_a}^\mu$,
  ${A^{ab}}_\mu$ and $V_\mu$ and their derivatives in a profoundly
  different way to WGT, as indicated, leading to a very different
  dependence of the free gravitational and matter actions
  on the gauge fields.
\item An action that is at most quadratic in the field strength
  tensors ${\cal R}^\dagger_{abcd}$ and ${\cal T}^\dagger_{abc}$,
  respectively, yield equations of motion that are, in general, linear
  in the second derivatives of the gauge fields, and hence
  the corresponding Hamiltonian does not suffer from Ostrogradsky's
  instability. If one includes a term quadratic in ${\cal H}^\dagger_{ab}$,
  however, it contributes terms to the resulting field equations that
  are linear in fourth-order derivatives of the ${h_a}^\mu$
  field. Nonetheless, as we discussed in
  Section~\ref{sec:ewgtsecondorder}, there are grounds for omitting
  the term quadratic in ${\cal H}^\dagger_{ab}$ from the action and, moreover,
  even if this term is included, the theory may still not suffer from
  Ostrogradsky's instability.

\item The trace ${\cal T}^\dagger_a$ of the eWGT translational gauge
  field strength vanishes identically, which has a number of
  consequences. These include: the automatic equivalence of the
  minimal coupling procedure when applied at the level of the action
  or directly in the field equations; a simplified torsion-squared
  part of the free gravitational action; and simplified Bianchi
  identities. In particular, the covariant divergence of the
  translation gauge field strength depends only the rotational and
  dilational gauge field strengths.

\item The energy-momentum tensors $t_{ab}$ and $\tau_{ab}$ derived
  from the free gravitational and matter actions, respectively, are
  not separately covariant under extended local dilations, although
  their non-covariant parts cancel to yield a covariant $h$-field
  equation, as required. Nonetheless, one can construct covariant
  versions $t^\dagger_{ab}$ and $\tau^\dagger_{ab}$ of
  free-gravitational and matter energy momentum tensors; this is most
  naturally achieved by adopting an alternative variational principle
  in which one makes a change of field variables such that
  ${A^{ab}}_\mu$ is replaced by ${A^{\dagger ab}}_\mu \equiv
  {A^{ab}}_\mu + ({\cal V}^a{b^b}_\mu - {\cal V}^b{b^a}_\mu)$.

\item In terms of this alternative set of field variables, the total
  Lagrangian density does not depend explicitly on the dilation gauge
  field $V_\mu$ or its derivatives. This leads to the identification
  of an alternative dilation current $\zeta^\dagger_a$, which vanishes
  identically. As a consequence, the contribution of the free
  gravitational and matter sectors, respectively, to the $V$-field
  equation is merely twice the relevant contraction of their
  contributions to the $A$-field equation.

\item The conservation equations have a very different form to those
  in PGT or WGT. In particular, invariance of the free gravitational
  action under extended local dilations leads to a differential
  conservation law similar to that obtained in WGT, but also an
  additional algebraic condition that the trace ${t^{\dagger a}}_a$ of
  the free-gravitational energy momentum tensor (i.e. the trace of the
  gravitational sector's contribution to the $h$-field equation)
  should vanish. Similarly, in the matter sector, one obtains an
  additional algebraic condition on the trace ${\tau^{\dagger a}}_a$
  of the total matter energy-momentum tensor in terms of the equations
  of motion of the matter field $\vpsi$ and (compensator) scalar field
  $\phi$.

\item The combination of the previous two results leads to eWGT having
  only three independent field equations, namely the $h$-equation, the
  $A$-equation and the $\vpsi$-equation. In this respect, eWGT is more
  similar to PGT than standard WGT.

\item The conclusions given above regarding the interpretation of the
  Einstein gauge and the equations of motion of test particles in WGT
  are also found to hold in eWGT.

\item In the special case of identically vanishing eWGT translational
  gauge field strength, ${\cal T}^\dagger_{abc}\equiv 0$, the eWGT
  covariant derivative ${\cal D}^\dagger_a$ simplifies to a `reduced'
  form ${^0}{\cal D}^\dagger_a$ that has a very different structure to
  that in WGT, since it still depends on the rotational gauge field
  ${A^{ab}}_\mu$.

\item One may use the `reduced' covariant derivative ${^0}{\cal
  D}^\dagger_a$ to construct `reduced' eWGT, which corresponds to
  imposing the condition of vanishing eWGT torsion directly at the
  level of the action. Unlike reduced PGT and WGT, the resulting
  theory, and in particular the reduced rotational field strength
  tensor ${^0}{\cal R}^\dagger_{abcd}(h,\partial h,\partial^2
  h,A,\partial A,V,\partial V)$, depends on all the gauge fields, most
  notably the rotational $A$-gauge field. As in reduced PGT and WGT,
  however, the rotational field strength tensor contains second-order
  derivatives of the $h$-field, so that terms in the action that are
  quadratic in ${^0}{\cal R}^\dagger_{abcd}$ will contribute terms to
  the resulting equations of motion that are linear in fourth-order
  derivatives of $h$, and such theories typically suffer from
  Ostrogradsky's instability. It is important, however, to distinguish
  reduced eWGT from simply imposing the condition ${\cal
    T}^\dagger_{abc}= 0$ in the field equations of eWGT, which yields
  different equations of motion in general, and a theory that does not
  suffer from Ostrogradsky's instability, if the term proportional to
  ${\cal H}^\dagger_{ab}{\cal H}^{\dagger ab}$ is excluded from the
  free gravitational Lagrangian.

\item The geometric interpretation of eWGT is in terms of a new
  spacetime geometry that represents an extension of Weyl--Cartan
  $Y_4$ spacetime.
  
\item The extended transformation law introduced for the rotational
  gauge field ${A^{ab}}_\mu$, and the associated introduction of its
  modified counterpart ${A^{\dagger ab}}_\mu$, implement Weyl scaling
  in a novel way that may be related to gauging of the full conformal
  group, although this remains a topic for future research.

\end{itemize}

We have also explicitly presented a new scale-invariant gauge theory
of gravity, defined by the most general parity-invariant eWGT
Lagrangian that is at most quadratic in the eWGT field strengths and
can accommodate `ordinary' matter. We derive the field equations for
this theory and comment briefly on their structure. The phenomenology
of eWGT and its application to astrophysics and cosmology will be
described in forthcoming papers.

Finally, we consider the construction of PGT actions that are
invariant under local dilations, focussing on the extended
transformation law for the rotational gauge field (which includes as
as special case the normal transformation law usually assumed in
WGT). After considering the general case, we discuss the construction
of locally-scale-invariant reduced PGT, and identify the unique
resulting theory as equivalent to conformal gravity, when interpreted
geometrically, but note that this theory may suffer from
Ostrogradsky's instability. We show that, in general,
locally-scale-invariant (reduced) PGTs are merely special cases of
(reduced) WGT or eWGT, depending on the transformation law assumed for
the rotational gauge field.


\begin{acknowledgments}
The authors thank Steve Gull, Chris Doran and Paul Alexander for
numerous interesting discussions of gravity over many years, and the
anonymous referee for several insightful suggestions.
\end{acknowledgments}

\appendix

\section{Semi-classical model for a spinning point particle}
\label{appendix:dirac}

The dynamics of a fermion is described by the Dirac equation, together
with the quantum-mechanical rules for constructing observables. For
many applications, however, such as determining the motion of a
massive matter particle, it is useful to work with semi-classical and
classical approximations to the full quantum theory, which we
summarise here.

\subsection{Re-expression of the Dirac Lagrangian}

The standard special-relativistic Lagrangian for a classical Dirac
field $\psi$ of mass $m$ is given by
\begin{equation}
L_{\rm D} = \tfrac{1}{2}i
[\bar{\psi}\gamma^\mu\partial_\mu\psi-(\partial_\mu\bar{\psi})\gamma^\mu\psi]
-m\bar{\psi}\psi
\equiv
\Re(i\bar{\psi}\slashed{\partial}\psi)- m\bar{\psi}\psi,
\label{diraclmagain}
\end{equation}
where the kinetic energy term is written so that it is manifestly real
(and the sign ensures the terms contribute positive energy on
quantisation) and $\slashed{\partial} \equiv \gamma^\mu
\partial_\mu$. Our aim here is to rewrite the kinetic term in
(\ref{diraclmagain}) in an alternative form by obtaining an identity
that will allow us to replace $\bar{\psi}$.

We begin by considering the quantity $\bar{\psi}\slashed{J}$, where
$J^\mu =\bar{\psi}\gamma^\mu\psi$ is the Dirac current, which we may
write as
\begin{equation}
\bar{\psi}\slashed{J} = J^\mu\bar{\psi}\gamma_\mu =
\bar{\psi}\gamma^\mu\psi\bar{\psi}\gamma_\mu.
\label{eqn:psijslash}
\end{equation}
The spinor outer product $\psi\bar{\psi}$ may be rewritten using the
Fierz rearrangement formula as
\begin{eqnarray}
\hspace*{-5mm}\psi\bar{\psi} = \tfrac{1}{4}(\bar{\psi}\psi+\bar{\psi}\gamma^\nu\psi\gamma_\nu
-\bar{\psi}\gamma^\nu\gamma^5\psi\gamma_\nu\gamma^5
&&+\tfrac{1}{2}\bar{\psi}\sigma^{\nu\rho}\psi\sigma_{\nu\rho} \nonumber \\&&
+\bar{\psi}\gamma^5\psi\gamma^5),
\label{fierzid}
\end{eqnarray}
where $\gamma^5=\gamma_5=i\gamma^0\gamma^1\gamma^2\gamma^3$ and
$\sigma^{\nu\rho}=\frac{i}{2}[\gamma^\nu,\gamma^\rho]$ (and we have
assumed the spinor components to be normal $c$-numbers, rather than
Grassmann variables). Inserting (\ref{fierzid}) into
(\ref{eqn:psijslash}), one finds after a short calculation that
\begin{equation}
\bar{\psi}\slashed{J}
= \bar{\psi}(\bar{\psi}\psi+\bar{\psi}i\gamma^5\psi i\gamma^5).
\label{psijslash2}
\end{equation}
It is convenient to define the real scalar $\rho \equiv\bar{\psi}\psi$
and real pseudoscalar $\beta \equiv \bar{\psi}i\gamma^5\psi$, so that
(\ref{psijslash2}) reads $\bar{\psi}\slashed{J}=\bar{\psi}(\rho +
\beta i\gamma^5)$. Postmultiplying this expression by $(\rho - \beta
i\gamma^5)$ and rearranging, one finds that $\bar{\psi}$ can be
written as
\begin{equation}
\bar{\psi} = \frac{\bar{\psi}(\rho+\beta i\gamma^5)\slashed{J}}{\rho^2+\beta^2}.
\end{equation}
Inserting this form into the kinetic part of the Dirac Lagrangian
(\ref{diraclmagain}) and writing the result explicitly in terms of
$\psi$ and $\bar{\psi}$, one finally obtains the re-expressed form
\begin{equation}
L_{\rm D} = \frac{\Re[i\bar{\psi}
(\bar{\psi}\psi+\bar{\psi}i\gamma^5\psi i\gamma^5)
\slashed{J}\slashed{\partial}\psi]}{(\bar{\psi}\psi)^2+
(\bar{\psi}i\gamma^5\psi)^2}- m\bar{\psi}\psi.
\label{eqn:ldalt}
\end{equation}

In particular, we note that the quantity
$\slashed{J}\slashed{\partial}\psi$ appearing in (\ref{eqn:ldalt}) may
be written as
\begin{equation}
\slashed{J}\slashed{\partial}\psi
= J^\mu\partial_\mu\psi-i\sigma^{\mu\nu}J_\mu\partial_\nu\psi.
\label{jderivdecomp}
\end{equation}
The quantity $J^\mu\partial_\mu\psi$ is clearly interpreted as a
derivative along a streamline of the Dirac fluid. Similarly, the
remaining term on the RHS of (\ref{jderivdecomp}) is a derivative
along the direction $n^\nu \equiv -i\sigma^{\mu\nu}J_\mu$, which is
orthogonal to $J^\mu$, since $J_\nu n^\nu = -i\sigma^{\mu\nu}J_\mu
J_\nu = 0$.  Thus, $-i\sigma^{\mu\nu}J_\mu\partial_\nu\psi$ is
interpreted as a derivative perpendicular to a streamline of the
Dirac fluid.

\subsection{Action for a spin-$\frac{1}{2}$ point particle}

Our reason for rewriting the Dirac Lagrangian in the alternative form
(\ref{eqn:ldalt}) is that it lends itself to the construction of an
action for a classical spin-$\frac{1}{2}$ point particle, which we use
as our model for `ordinary matter'.  The essential
idea\cite{lasenby98} is to specialise to motion along a single
streamline defined by the Dirac current $J^\mu$.  Thus the particle is
described by a worldline $x^\mu(\lambda)$, together with a `rotor'
$R(\lambda)$, which is a Dirac spinor defined along the worldline that
satisfies the normalisation conditions $\bar{R}R = 1$ and
$\bar{R}i\gamma^5R=0$, and contains information about the velocity and
spin of the particle.

Thus, to obtain the Lagrangian for a point particle, one replaces
$\psi$ by $R$ in (\ref{eqn:ldalt}), restricts derivatives to lie
purely along the particle worldline, and identifies the velocity
$\dot{x}^{\mu} \equiv dx^\mu/d\lambda$ with $\bar{R}\gamma^\mu R$.
The last identification is enforced by including in the action
integral a Lagrange multiplier $p_\mu(\lambda)$, which is identified
with the momentum of the particle; note that, in general, $p_\mu$ and
$\dot{x}^{\mu}$ are not collinear. Finally, an einbein $e(\lambda)$ is
introduced (written as $me(\lambda)$ for later convenience) to ensure
reparameterisation invariance along the particle worldline. The
resulting action may be written as
\begin{equation}
S = \int d\lambda\,
[\Re(i\bar{R}\dot{R})-p_\mu(\dot{x}^{\mu}-me\bar{R}\gamma^\mu R)
  -m^2e].
\label{spinningppaction}
\end{equation}
An equivalent expression has been derived previously\cite{lasenby98},
but there the action is instead expressed in terms of geometric
algebra. Variation of the action with respect to the dynamical
variables $x^\mu(\lambda)$, $R(\lambda)$, $p^\mu(\lambda)$ and
$e(\lambda)$ leads to the semi-classical equations of motion for a
spin-$\frac{1}{2}$ point particle\cite{doran98}.

One proceeds to the full classical approximation by replacing
$\bar{R}\gamma^\mu R$ by $p^\mu/m$ and then neglecting the particle
spin by dropping all terms that contain $R$. Redefining the einbein
$e(\lambda) \to \tfrac{1}{2}e(\lambda)$ for later convenience, this
process leads to the action
\begin{equation}
S = - \int d\lambda\, [p_\mu \dot{x}^{\mu}-\tfrac{1}{2}e(p_\mu p^\mu - m^2)],
\label{pointparticleaction}
\end{equation}
and variation with respect to the remaining dynamical variables
$x^\mu(\lambda)$, $p^\mu(\lambda)$ and $e(\lambda)$ leads to the
classical equations of motion for the point particle.

\section{Poincar\'e gauge theory}
\label{app:pgt}

The discussion of WGT in Section~\ref{sec:wgt} also serves, with
appropriate notational modifications, as an account of PGT, which may
be considered as a special case. In particular, by setting $\rho=0$ in
the global Weyl transformation (\ref{weyl}), one recovers a global
Poincar\'e transformation. Thus, all the subsequent results in
Section~\ref{sec:globalweyl} relating to global Weyl transformations
are also valid for global Poincar\'e transformations provided one sets
$\rho=0$.

In an analogous manner, local Weyl transformations reduce to local
Poincar\'e transformations if $\rho(x)=0$. In this case, it is no
longer necessary to introduce the dilation gauge field $B_\mu$ to
construct the PGT $\Lambda$-covariant derivative $D_\mu\vpsi$. Thus,
all the results given in Sections~\ref{sec:lmwgt} and
\ref{sec:minkintwgt} also hold in PGT if one sets $\rho(x)=0=B_\mu$
and removes all asterisks from derivative operators.  In particular,
it is worth noting that the rotational gauge field ${A^{ab}}_\mu$
transforms identically in PGT and WGT, according to (\ref{atrans}).

Similar considerations apply to field strength tensors. Indeed, all
the results presented in Section~\ref{sec:wgtgfs} are also valid in
PGT if one sets $B_\mu=0$ and removes all asterisks. In particular, it
is worth noting that the rotational field strength tensor
${R^{ab}}_{\mu\nu}$ in (\ref{rfsdef}) has the same form in both PGT
and WGT (and so no distinction is made between them), whereas the
dilation field strength $H_{\mu\nu}$ in (\ref{dfsdef}) vanishes
identically in PGT. Also, the translation field strengths in PGT and
WGT are related by (\ref{tstarfromt}).

Equally, the results presented in Section~\ref{sec:wgtcovdalt}
regarding the construction of a `reduced' covariant derivative also
hold in PGT if one again sets $B_\mu=0$ and removes all asterisks.
Moreover, the decomposition (\ref{afromht}) or (\ref{astarzerodef})
holds with all asterisks removed from the RHS. Consequently, in the
special case where the equations of motion result in ${\cal T}_{abc}$
being independent of ${\cal A}_{abc}$, one obtains an explicit
expression for the $A$-field in terms of just the $h$-gauge field.
Turning to the Bianchi identities, the same prescription of setting
$B_\mu=0$ (so that $H_{\mu\nu}=0$) and removing all asterisks from the
results given in Section~\ref{sec:wgtbianchi} leads to the
corresponding identities in PGT.

In constructing a free gravitational Lagrangian for PGT, one has
considerably more freedom than in the WGT case discussed in
Section~\ref{sec:wgtfga}, since one requires only that $L_{\rm G}$ is
a scalar function of the two PGT field strengths ${\cal R}_{abcd}$ and
${\cal T}_{abc}$ and there is no restriction on its Weyl weight $w$.
Thus, in principle, $L_{\rm G}$ may contain terms of any order in the
field strengths. As discussed in the Introduction, however, if one
demands adequacy of kinematics and dynamics, and consistency with the
standard Einstein general relativity in the macroscopic limit, one is
led naturally to a Lagrangian $L_{\rm G}$ that is, at most, quadratic
the field strength tensors, such that
\begin{equation}
L_{\rm G} = -\kappa^{-1}(\Lambda + a{\cal R}) + L_{{\cal R}^2} + \kappa^{-1}L_{{\cal T}^2},
\label{pgtgravlagdef}
\end{equation}
where ${\cal R}\equiv {{\cal R}^{ab}}_{ab}$, $\Lambda$ is a
cosmological constant, $a$ is a dimensionless free parameter (usually
positive), and $L_{{\cal R}^2}$ and $L_{{\cal T}^2}$ are given by
(\ref{lr2}) and (\ref{lt2}) respectively. As in WGT, the field
strength ${\cal R}_{abcd}$ satisfies a form of the Gauss--Bonnet
identity, such that the combination in (\ref{gbid}) contributes a
total derivative to the action (in $D \le 4$ dimensions). Hence one
may set any one of $\alpha_1$, $\alpha_3$ or $\alpha_6$ in (\ref{lr2})
to zero, without loss of generality (at least classically). The
precise version of PGT under consideration depends on the choice of
the parameters $a$, $\Lambda$, $\{\alpha_i\}$ and $\{\beta_i\}$ in
$L_{\rm G}$. As an illustration, in Appendix~\ref{ECtheory} we give a
brief account of the form of the theory first considered by Kibble, in
which $L_{\rm G} \propto {\cal R}$.

The total action $S_{\rm T}$ is simply the sum of the free
gravitational and matter actions. In the free gravitational sector,
adopting the usual form of the Lagrangian (\ref{pgtgravlagdef})
induces dependencies on both gauge fields $h$ and $A$ (suppressing
indices for brevity), and their first derivatives. In particular, the
dependence on $\partial h$ results from the presence of $L_{{\cal
    T}^2}$ in (\ref{pgtgravlagdef}), which is absent in WGT. In the
matter sector, covariant derivatives of the matter field $\vpsi$
induce a dependence on $\vpsi$, $\partial\vpsi$, $h$ and $A$, but one
typically does not include a compensator scalar field $\phi$ (although
$\phi$ is usually introduced in locally-scale-invariant PGTs, see
Section~\ref{sec:lsipgt}, we will not consider this possibility here). Thus, the
total Lagrangian density is
\begin{equation}
{\cal L}_{\rm T} = {\cal L}_{\rm G}(h,\partial h, A,\partial A)
+ {\cal L}_{\rm M}(\vpsi,\partial\vpsi,h,A),
\end{equation}
where we have indicated the functional dependencies in the most
general case. The general structure of the resulting equations of
motion for the gauge fields is the same as that described for WGT in
Section~\ref{sec:wgtfieldeqns}, except that the $B$-field equation
(\ref{eqn:weylgenfe3}) in WGT is obviously absent in PGT. Similarly,
the equation of motion for the matter field $\vpsi$ has the same form
as (\ref{weylpeqn}) but with all asterisks removed, but the
$\phi$-field equation (\ref{weylpeqnphi}) in WGT is typically absent
in PGT.  The conservation laws in PGT may also be obtained from those
given for WGT in Section~\ref{sec:wgtconslaws}, by removing all
asterisks and setting ${\cal H}_{ab}$, $j^a$, $\zeta^a$ to zero and
$\phi$ to unity in (\ref{eqn:wgtcons1})--(\ref{eqn:wgtcons2}) and
(\ref{wgtlmcons1})--(\ref{wgtlmcons2}). Clearly, the WGT conservation
laws (\ref{eqn:wgtcons3}) and (\ref{wgtlmcons3}) relating to local
dilation invariance are usually absent in PGT.

The coupling of the PGT gravitational gauge fields to the Dirac matter
field and electromagnetic field is analogous to that in WGT, discussed
in Sections~\ref{sec:wgtdirac} and \ref{sec:wgtem}, except that one
does not typically introduce the scalar (compensator) field
$\phi$. Indeed, as mentioned in these discussions, in both cases the
Lagrangian density in PGT is identical to that in WGT, with the slight
modification $\mu\phi\bar{\psi}\psi \to m\bar{\psi}\psi$ for the Dirac
field. Thus, after this change and then following our prescription of
removing all asterisks and setting $B_\mu=0$ and $\phi=1$, all the
results in these sections are also valid in PGT.  Similar
considerations apply to results regarding the motion of test particles
given in Section~\ref{sec:WGTmassivemotion}, with the key conclusion
that they follow the gauge theory equivalent of geodesic motion,
rather than autoparallels.

For `reduced' PGT, an analogous situation to that described in
Section~\ref{sec:reducedwgt} holds. In particular, the results
regarding the construction of `reduced' field strengths also hold in
PGT if one again sets $B_\mu=0$ (and so $H_{\mu\nu}=0$ also) and
removes all asterisks. Since the reduced `translational' field
strength of the $h$-field again vanishes identically, ${^0}{{\cal
    T}^a}_{bc} \equiv 0$, only the reduced `rotational' field strength
${^0}{{\cal R}^{ab}}_{cd}(h,\partial h, \partial^2h)$ remains, which
depends on only the $h$-field and its first and second derivatives, as
indicated. Thus, the entire resulting theory can be written purely in
terms of just the $h$-gauge field.  For the Bianchi identities
involving `reduced' field strengths, it is worth noting that the PGT
versions of (\ref{wgtbi1}) and (\ref{wgtbi2}) reduce to the familiar
differential Bianchi identity and cyclic identity, respectively, of
the Riemann tensor in $V_4$. Also, (\ref{wgtcbi2}) and (\ref{wgtcbi3})
reduce, respectively, to the familiar results in $V_4$ that the
covariant divergence of the Einstein tensor vanishes and the Ricci
tensor is symmetric. The expressions for how the `full' PGT rotational
field strength tensor ${{\cal R}^{ab}}_{cd}$ and its contractions are
related to their `reduced' counterparts are obtained by removing all
asterisks from (\ref{riemann0})--(\ref{randr0}).  It is worth noting
that the PGT version of the last term on the RHS of (\ref{randr0}) may
be rewritten in terms of a total derivative as $\czero{{\cal
    D}_a}{\cal T}^{a} = h\partial_\mu (h^{-1}{h_a}^\mu {\cal T}^a)$,
whereas this is not possible in WGT\cite{footnote25}.

For reduced PGT, the free gravitational Lagrangian analogous to
(\ref{pgtgravlagdef}) is
\begin{equation}
L_{\rm G} = -\kappa^{-1}(\Lambda + a\,\czero{{\cal R}}) + L_{\czero{{\cal R}}^2},
\label{pgtnotlg}
\end{equation}
where $L_{\czero{{\cal R}}^2}$ is based on (\ref{lr2}) with ${{\cal
    R}^{ab}}_{cd}$ replaced by $\czero{{{\cal R}^{ab}}_{cd}}$. Unlike
its WGT counterpart, however, $\czero{{{\cal R}^{ab}}_{cd}}$ obeys all
the usual symmetries of the curvature tensor in a Riemannian $V_4$
spacetime, and so the number of distinct terms in $L_{\czero{{\cal
      R}}^2}$ is reduced to
\begin{equation}
L_{\czero{{\cal R}}^2} = \alpha_1\, \czero{{\cal R}}^2
+ \alpha_2\, \czero{{\cal R}}_{ab}\czero{{\cal R}}^{ab}
+ \alpha_3\, \czero{{\cal R}}_{abcd}\czero{{\cal R}}^{abcd},
\label{pgtnotlr2}
\end{equation}
where the $\alpha_i$ are dimensionless free parameters.  In fact, one
can simplify (\ref{pgtnotlr2}) still further (in $D \le 4$
dimensions), since $\czero{{\cal R}}_{abcd}$ and its contractions also
satisfy a Gauss--Bonnet identity of the form (\ref{gbid}), but with
${\cal R}_{abcd} \to \czero{{\cal R}}_{abcd}$. Thus, one can set to
zero any one of the parameters $\alpha_i$ in (\ref{pgtnotlr2}) with no
loss of generality (at least classically). The calculation of the
corresponding field equations is simplest if one sets $\alpha_3=0$,
which is the preferred choice, and thus yields the modest
four-parameter family of free gravitational Lagrangians
\begin{equation}
L_{\rm G} = -\kappa^{-1}(\Lambda + a\,\czero{{\cal R}}) + \alpha_1\, \czero{{\cal R}}^2
+ \alpha_2\, \czero{{\cal R}}_{ab}\czero{{\cal R}}^{ab}.
\label{pgtnotlgfinal}
\end{equation}
This Lagrangian depends on only the $h$ gauge field and its
derivatives.  In general, however, it is quadratic in second-order
derivatives of the $h$-field, and so the corresponding field equations
will typically be linear in fourth-order derivatives of $h$; such
theories typically suffer from Ostrogradsky's instability, although
this must be investigated on a case-by-case basis. If one demands
the field equations to be at most second-order in derivatives of $h$,
then one must set $\alpha_1=\alpha_2=0$ (at least in $D \le 4$
spacetime dimensions), and on choosing $a=\tfrac{1}{2}$ the resulting
theory corresponds precisely to general relativity with a cosmological
constant, when interpreted geometrically. Another interesting choice
of coefficients in (\ref{pgtnotlgfinal}) is $a=0=\Lambda$,
$\alpha_1=-\tfrac{1}{3}\alpha_2$, which corresponds to the free
gravitational Lagrangian of so-called conformal gravity theory, when
interpreted geometrically\cite{mannheim06}; this theory is
discussed briefly in Section~\ref{appendix:lsipgtzt} as the unique
example of a locally scale-invariant reduced PGT.

Finally, the geometric interpretation of PGT follows from the
corresponding discussion of WGT in Section~\ref{sec:geowgt} by
adopting the usual prescription of removing all asterisks and setting
$B_\mu$ (and therefore $H_{\mu\nu}$) to zero. In particular, it is
worth noting that RHS of the semi-metricity condition
(\ref{wgtsemimet}) vanishes in PGT, which shows that the spacetime
has, in general, a Riemann--Cartan $U_4$ geometry, i.e. the connection
is metric compatible but may exhibit non-zero torsion. Also, the
quantities (\ref{wgtcurvature}) and (\ref{wgttorsion}) become simply
the curvature and (minus) torsion tensor of the $U_4$ spacetime, and
(\ref{wgthtensor}) vanishes identically. For the geometric
interpretation of the `reduced' quantities, one adopts the same
approach as above, which leads one to recognise
$\zero{{R^\rho}_{\sigma\mu\nu}}$ as the standard curvature tensor in a
Riemannian $V_4$ spacetime, which obeys all the familiar symmetries,
plus the cyclic and Bianchi identities. It is worth noting that the
comments made at the end of Section~\ref{sec:geowgt} regarding the
choice to interpret some quantities as geometric properties of the
underlying spacetime and others as fields residing in that spacetime
are equally true in PGT.  As a concrete example, consider the PGT
version of (\ref{randr0}) in the geometric interpretation, namely
\begin{eqnarray}
R = {^0}R+\tfrac{1}{4}T^{\mu\nu\lambda}T_{\mu\nu\lambda} &&+
\tfrac{1}{2}T^{\mu\nu\lambda}T_{\nu\mu\lambda}-T^\mu T_\mu \nonumber \\&&
- \frac{2}{\sqrt{-g}}\partial_\mu (\sqrt{-g}T^\mu),\label{rvsr0geo}
\end{eqnarray}
which relates the Ricci--Cartan scalar $R \equiv
{R^{\mu\nu}}_{\mu\nu}$ in $U_4$ spacetime to the Ricci scalar ${^0}R
\equiv {^0}{R^{\mu\nu}}_{\mu\nu}$ in $V_4$ spacetime. In standard
Einstein--Cartan theory (see Appendix~\ref{appendix:ectheory} for its
interpretation as a gauge theory in Minkowski spacetime), the
gravitational Lagrangian is taken to be $L_{\rm G} = -R/(2\kappa)$ and
the theory is interpreted as a geometric theory in $U_4$
spacetime. One could, however, equally well rewrite the Lagrangian
using (\ref{rvsr0geo}) and interpret the theory as a geometric theory
in $V_4$ spacetime that includes a tensor field $T_{\mu\nu\sigma}$
(obeying the appropriate symmetries) defined on that spacetime, for
which the Lagrangian is given by the terms containing
$T_{\mu\nu\lambda}$ in (\ref{rvsr0geo}) (apart from the final one,
which can be omitted since it contributes only a total derivative to
the integrand of the action).

\section{Einstein--Cartan theory}
\label{ECtheory}
\label{appendix:ectheory}

The precise version of PGT under consideration is determined by the
form of the gravitational Lagrangian density ${\cal L}_{\rm G}$ and
the matter Lagrangian density ${\cal L}_{\rm M}$. As an example, we
here give a brief account of the original such theory, first
considered by Kibble, in which ${\cal L}_{\rm G}$ is chosen to be
proportional to the allowable covariant expression of lowest degree,
namely the linear invariant ${\cal R} = {{\cal R}^{ab}}_{ab}$. Thus,
in Kibble's original theory, one has
\begin{equation}
{\cal L}_{\rm G} = -\frac{1}{2\kappa}h^{-1}{\cal R},
\label{gravlag}
\end{equation}
which corresponds to setting $a=\tfrac{1}{2}$ and $\Lambda=\alpha_i=\beta_i=0$
in (\ref{pgtgravlagdef}); the factor $\tfrac{1}{2}$ is introduced for later
convenience. The resulting theory is known as Einstein--Cartan (EC)
theory\cite{footnote26}, which, when re-interpreted geometrically, is a direct
generalisation of general relativity to include torsion sourced by the
spin-angular-momentum (if any) of the matter field. We will first
consider a general such matter field, but then specialise to the Dirac
field.

\subsection{Field equations}

The above choice for the gravitational Lagrangian density simplifies
the general structure of the PGT gravitational field equations a
little, since ${\cal L}_{\rm G}$ does not depend on derivatives of the
$h$-field.  In particular, one finds that the gravitational field
equations read (after slight rearrangement)
\begin{eqnarray}
{{\cal
    R}^{ac}}_{bc}-{\textstyle\frac{1}{2}}\delta_b^a {\cal R} & = &
\kappa
    h{\tau^a}_b, \label{hfieldeqn}\\
h{b^c}_\mu D_\nu[h^{-1}({h_a}^\mu
      {h_b}^\nu - {h_a}^\nu {h_b}^\mu)] & = & 2\kappa h
    {\sigma_{ab}}^c.
\label{afieldeqn}
\end{eqnarray}
where the energy-momentum and spin-angular-momentum, respectively,
associated with the matter Lagrangian ${\cal L}_{\rm M}$ provide the source terms
on the right-hand sides of the equations.

The first field equation (\ref{hfieldeqn}) is written primarily in
terms of the rotational gauge field strength ${{\cal R}^{ab}}_{cd}$
and clearly resembles the Einstein equations in general relativity,
although with a very different interpretation here. We can, in fact,
rewrite the second field equation (\ref{afieldeqn}) in terms of the
translational field strength ${{\cal T}^a}_{bc}$ to obtain
\begin{equation}
 {{\cal T}^c}_{ab} +
\delta_a^c {\cal T}_b-\delta_b^c{\cal T}_a
=2\kappa h {\sigma_{ab}}^c.
\label{ecfe2a}
\end{equation}
where ${\cal T}_a \equiv {{\cal T}^c}_{ac}$, as defined in
(\ref{htrelation}).  In particular, by contracting (\ref{ecfe2a}) on
the indices $b$ and $c$, we find
\begin{equation}
{\cal T}_a = -\kappa h {\sigma_{ab}}^b.
\label{sigcontract}
\end{equation}
This result may be used to re-write the second field equation
(\ref{ecfe2a}) to give an explicit expression for the translational
gauge field strength, which reads
\begin{equation}
{{\cal T}^c}_{ab} = \kappa h
(2{\sigma_{ab}}^c + \delta_a^c {\sigma_{bd}}^d
- \delta_b^c {\sigma_{ad}}^d).
\label{tfromsigma}
\end{equation}
It is worth noting that this relationship is algebraic, rather than
differential. Thus, as mentioned in the Introduction, the ${\cal
  T}$-field is non-propagating, i.e. it must vanish outside of spin sources.

It should be noted that the second field equation (\ref{ecfe2a}) can,
in principle, be solved for ${A^{ab}}_\mu$ or, equivalently, for its
fully Lorentz-index counterpart ${{\cal A}^{ab}}_c = {h_c}^\mu
{A^{ab}}_\mu$. In particular, combining the PGT form of the expression
(\ref{afromht}) (with all asterisks removed) and (\ref{tfromsigma}),
one quickly finds that
\begin{eqnarray}
\hspace*{-7mm}{\cal A}_{abc} =  &&\tfrac{1}{2}(c_{abc}+c_{bca}-c_{cab}) \nonumber \\&&+
\kappa h(\sigma_{abc}\!-\!\sigma_{bca}\!-\!\sigma_{cab}
+\eta_{ac}{\sigma_{bd}}^d
-\eta_{bc}{\sigma_{ad}}^d),
\label{ECasolution}
\end{eqnarray}
where ${c^c}_{ab}$ was defined in (\ref{riccicoeffsdef}) (again
removing all asterisks).  In the
general case, in which ${\sigma_{ab}}^c \neq 0$, this is not, however,
an explicit solution since ${\cal A}_{abc}$ also occurs on the
right-hand side. Nonetheless, if ${\cal L}_{\rm M}$ is linear in the
derivatives of the matter field(s), then $\sigma_{abc}$ is independent
of ${\cal A}_{abc}$ and so (\ref{ECasolution}) is an explicit solution
for ${\cal A}_{abc}$. In this case, ${\cal A}_{abc}$ is no longer an
independent dynamical variable, but is derivable from the $h$-field,
and so one may express the entire theory in `second-order'
form.

The second field equation (\ref{ecfe2a}) also allows for
straightforward consideration of special cases in which
${\sigma_{ab}}^c$ satisfies given constraints. From
(\ref{sigcontract}), we see that in the special case where
${\sigma_{ab}}^b=0$, then ${\cal T}_a$ also vanishes, thereby greatly
simplifying (\ref{ecfe2a}). Moreover, if the spin-angular-momentum of
the matter field vanishes altogether, such that ${\sigma_{ab}}^c=0$,
then so too must the translational gauge field strength ${{\cal
    T}^c}_{ab}$, and so the $h$-field itself must
satisfy
\begin{equation}
{\cal  D}_b{h_a}^\mu-{\cal D}_a{h_b}^\mu=0.
\label{wedgeeqn}
\end{equation}

\subsection{Dirac matter field}

It is worth noting that, if the matter source in EC theory is a
massive Dirac field $\psi$, such that the matter Lagrangian density is
given by (\ref{diraclm3}) with $\mu\phi \equiv m$, several
simplifications occur. Firstly, the Dirac matter Lagrangian is linear
in the derivatives of $\psi$, so that the spin-angular-momentum tensor
$\sigma_{abc}$ is independent of the rotational gauge field ${\cal
  A}_{abc}$. Thus, as discussed above, the latter may be expressed in
terms of the $h$-field, and is no longer an independent
field. Moreover, from (\ref{sigcontract}), we find that
\begin{equation}
{\cal T}_a  = -\kappa h {\sigma_{ab}}^b = 0,
\label{uvanish}
\end{equation}
since $\sigma_{abc}=\sigma_{[abc]}$ for Dirac matter and so all
contractions of the spin-angular-momentum tensor vanish.  An immediate
consequence of (\ref{uvanish}) is that the second gravitational field
equation (\ref{ecfe2a}) takes the much simpler form ${{\cal T}^c}_{ab}
= 2\kappa h {\sigma_{ab}}^c$. More interestingly, the condition
(\ref{uvanish}) implies that the covariant Dirac equation reduces to
\begin{equation}
i\gamma^a{\cal D}_a\psi-m\psi = 0,
\label{covardiracec}
\end{equation}
which is precisely the form that would be obtained by applying the
minimal-coupling procedure directly at the level of the Dirac
equation, rather than at the level of the action. Thus, we see that in
EC theory, at least, the two approaches are consistent.

\subsection{Geometrical interpretation of EC theory}

Finally, we comment briefly on the geometric interpretation of the
special case of Einstein--Cartan theory, discussed in
Section~\ref{ECtheory}.  In this case, the gravitational field
equations (\ref{hfieldeqn}) and (\ref{tfromsigma}) can be rewritten in
the form
\begin{eqnarray}
\hspace*{-8mm}R_{\mu\nu}-\tfrac{1}{2}g_{\mu\nu}R & = & \kappa
\frac{\tau_{\mu\nu}}{\sqrt{-g}},\label{einsteineqn}\\
{T^\lambda}_{\mu\nu} & = & \frac{\kappa}{2\sqrt{-g}}
(2{\sigma_{\mu\nu}}^\lambda + \delta_\mu^\lambda {\sigma_{\nu\rho}}^\rho
-\delta^\lambda_\nu{\sigma_{\mu\rho}}^\rho),
\end{eqnarray}
where $R_{\mu\nu}\equiv {R^\lambda}_{\mu\lambda\nu}$ and $R\equiv
{R^\mu}_\mu$ have their usual forms, but one must remember that the
affine connection ${\Gamma^\lambda}_{\mu\nu}$ is not, in general,
symmetric in $\mu$ and $\nu$. Also, $\tau_{\mu\nu} = {e^a}_\mu
{e^b}_\nu \tau_{ab}$ and ${\sigma_{\mu\nu}}^\lambda = {e^a}_\mu
{e^b}_\nu {e_c}^\lambda {\sigma_{ab}}^c$, where the former is, in
general, not symmetric.

It is worth noting that, in the case where the spin-angular-momentum
${\sigma_{\mu\nu}}^\lambda$ of the matter field vanishes, then so too
does the torsion ${T^\lambda}_{\mu\nu}$ and hence the affine
connection reduces to the metric connection. In this case,
$R_{\mu\nu}$ reduces to the standard symmetric Ricci tensor
${^0}{R_{\mu\nu}}\equiv R_{\mu\nu}(V_4)$ in a Riemann spacetime,
expressed entirely in terms of the metric $g_{\mu\nu}$. Hence
(\ref{einsteineqn}) reduces to the usual form for the Einstein
equation and so, in this limit, the geometrical interpretation of EC
theory reduces to general relativity.


\section{Dirac gravitational theory}
\label{sec:diracgrav}
\label{appendix:dgt}

The precise form of WGT under consideration depends on the form of the
free gravitational Lagrangian density ${\cal L}_{\rm G}$ and matter
Lagrangian density ${\cal L}_{\rm M}$. As an illustration, we consider
here (an extension of) the first scale-invariant theory of gravity to
be proposed that can accommodate `ordinary' matter, which was explored
by Dirac\cite{dirac73} in the context of attempting to establish a
deeper physical understanding of his `large numbers hypothesis'
relating microscopic (quantum) and macroscopic (gravitational) scales.
Dirac originally cast his theory in wholly geometrical terms, but it
can be naturally considered in the framework of WGT in Minkowski
spacetime, as we show below\cite{canuto77}.  Indeed, the WGT approach
allows for a straightforward generalisation of Dirac's theory to
non-zero torsion, which we will include\cite{kasuya75} (thus Dirac's
original theory was, in fact, an example of a reduced WGT, as
discussed in Section~\ref{sec:reducedwgt}). Our discussion parallels
that carried out by Lewis et al.\cite{footnote27} in the language of
geometric algebra, but is presented here in the context of our more
general formulation of WGT.

\subsection{Free gravitational and matter actions}

In constructing the free gravitational Lagrangrian density ${\cal L}_{\rm G} =
h^{-1}L_{\rm G}$, the requirements of local Weyl symmetry restrict our
choice of $L_{\rm G}$ to the general `${\cal R}^2+{\cal H}^2$'-form
in (\ref{lg2}). The WGT equivalent of the free gravitational Lagrangian
suggested by Dirac has the simple form\cite{footnote28}
\begin{equation}
L_{\rm G} = -\tfrac{1}{4}\xi{\cal H}_{ab}{\cal H}^{ab},
\label{dgtlgfree}
\end{equation}
in which no `${\cal R}^2$'-terms appear, and where $\xi$ is a
dimensionless constant (usually positive); the factor of
$-\tfrac{1}{4}$ is included for later
convenience.

Turning to the matter action, in addition to some original matter
field $\vpsi$, Dirac also added a `compensator' massless scalar field
$\phi$. In particular, he included a kinetic term for $\phi$, a
self-interaction term proportional to $\phi^4$, and a term
proportional to $\phi^2{\cal R}$ that non-minimally (conformally)
couples $\phi$ to the rotational gravitational field strength. Thus,
one has
\begin{eqnarray}
L_{\rm M} &\equiv& L_\vpsi + L_\phi + \phi^2 L_{\cal R} \nonumber \\
&=& L_\vpsi + \tfrac{1}{2}\nu {\cal D}^\ast_a\phi\,{\cal D}^{\ast
  a}\phi -\lambda\phi^4 - \tfrac{1}{2}a\phi^2{\cal R},
\label{diracgravlagm}
\end{eqnarray}
where $\nu$, $\lambda$ and $a$ are dimensionless constants (usually
positive), and factors of $\tfrac{1}{2}$ are
included for later convenience.  The functional dependencies of the
Lagrangian for the matter field $\vpsi$ are
$L_\vpsi(\vpsi,\partial\vpsi,h,A,B,\phi)$; note that $L_\vpsi$ depends
on $\phi$ if it includes some interaction term between $\phi$ and
$\vpsi$. Indeed, Dirac provided the original motivation for such
matter Lagrangians, varying forms of which we have adopted in
Sections~\ref{sec:wgtdirac}, \ref{sec:elsipgt} and
\ref{sec:elsipgtnot}\cite{footnote29}.

The total Lagrangian
density is then ${\cal L}_T=h^{-1}L_T$, where $L_T$ has the form
\begin{eqnarray}
\hspace*{-12mm}L_T &=& L_{\rm M} + L_{\rm G} \nonumber \\
&=& L_\vpsi
\!+\! \tfrac{1}{2}\nu{\cal D}^\ast_a\phi\,{\cal D}^{\ast
  a}\phi \!-\!\lambda\phi^4
\!\!-\! \tfrac{1}{2}a\phi^2{\cal R} \!-\! \tfrac{1}{4}\xi {\cal
  H}_{ab}{\cal H}^{ab}.\phantom{A}
\label{diracgravlag}
\end{eqnarray}
We note again that the Lagrangian $L_\vpsi$ is allowed, in general, to
contain the scalar field $\phi$, and that the total Lagrangian does
not contain any derivatives of the $h$-field.  It is also worth noting
that all the coupling constants $(a,\xi,\nu,\lambda)$ are
dimensionless (and usually positive). In the case of $a$, Dirac (and
others) believed that this implied $\phi^{-2}$ should replace Newton's
gravitational constant $G$. Hence, if $\phi$ changed with time, tied
to cosmological expansion in some way, then the effective $G$ would
also change with time, thereby allowing his `large numbers hypothesis'
(see the Introduction) to remain true at all cosmic epochs.  In
addition, Dirac chose to fix some of the constants in the Lagrangian
so that $L_T$ did not contain the dilation gauge field $B_\mu$
explicitly (except perhaps through $L_\vpsi$). Here, we will instead
retain the more general form (\ref{diracgravlag}).

A more profound difference between our approach and that originally
taken by Dirac is that we will consider the theory in the context of
WGT in Minkowski spacetime, rather than as a metric-based geometrical
theory, although the former is easily interpreted in geometric terms,
as discussed in Section~\ref{sec:geowgt}. Perhaps more importantly, we
will obtain the corresponding field equations by varying the action
(\ref{diracgravlag}) with respect to the four independent fields
${h_a}^\mu$, ${A^{ab}}_\mu$, $B_\mu$ and $\phi$ respectively\cite{kasuya75}, whereas Dirac varied his geometrically-interpreted
action with respect to $B_\mu$, $\phi$ and the spacetime metric
$g_{\mu\nu}$. Our approach is thus closer in spirit to the Palatini
formalism in geometric theories, in which the metric and the
connection are varied independently, as opposed to variation with
respect to the metric alone, assuming a predetermined form of the
connection.  Moreover, as already mentioned, Dirac chose the latter
approach with the connection assumed to have the form given in
(\ref{diracconnection}), so that torsion is zero by construction,
whereas our approach naturally allows for non-zero
torsion.

Unsurprisingly, our approach does lead to some differences in the
final field equations as compared with those obtained by Dirac but, as
we will show, these differences are, in fact, rather minor and the
overall structure of the resulting theory is very similar using either
approach.

\subsection{Field equations}

On varying the action (\ref{diracgravlag}), the $h$-field equation is
found to be
\begin{eqnarray}
a\phi^2 ({{\cal R}^{a}}_{b} -\tfrac{1}{2}\delta_b^a{\cal R}) =
&& h {(\tau_\vpsi)^a}_b + [\nu({\cal D}^{\ast a}\phi) ({\cal
    D}^\ast_b\phi)-\delta^a_b L_\phi] \nonumber \\
&&-\xi ({\cal H}^{ac}{\cal H}_{bc}-\tfrac{1}{4}\delta^a_b {\cal
  H}_{cd}{\cal H}^{cd}),
\label{diracgrav-heqn}
\end{eqnarray}
where ${(\tau_\vpsi)^a}_b$ is the energy-momentum tensor of the matter
field $\vpsi$, and we recognise the second and third terms on the RHS
as the energy-momentum tensors of the scalar field $\phi$ and of the
dilation gauge field $B_\mu$ respectively.

The $A$-field equation is
given by
\begin{equation}
a[\phi^2{{\cal T}^{\ast c}}_{ab}
+\delta_a^c ({\cal D}^\ast_b + {\cal T}^\ast_b)\phi^2
-\delta_b^c ({\cal D}^\ast_a + {\cal T}^\ast_a)\phi^2]
= 2h
{(\sigma_\vpsi)_{ab}}^c,
\label{diracgrav-aeqn}
\end{equation}
where ${(\sigma_\vpsi)_{ab}}^c$ is the spin-angular-momentum tensor of the
$\vpsi$ matter field. It is worth noting that (\ref{diracgrav-aeqn})
still holds if one makes the replacements ${\cal D}^\ast_a \to {\cal
  D}_a$, ${{\cal T}^{\ast c}}_{ab} \to {{\cal T}^{c}}_{ab}$ and ${\cal
  T}^\ast_a \to {\cal T}_a$ to PGT quantities, although the resulting
equation is clearly no longer manifestly WGT covariant.

Varying the action (\ref{diracgravlag}) with respect to the $B$-field,
and rewriting the resulting equation of motion in manifestly
WGT covariant form, one finds
\begin{equation}
\xi[\tfrac{1}{2} {{\cal T}^{\ast c}}_{ab} {\cal H}^{ab} - ({\cal
    D}^\ast_a+{\cal T}^\ast_a){\cal H}^{ac}] - \nu \phi {\cal
  D}^{\ast c} \phi = h (\zeta_\vpsi)^c
\label{diracgrav-beqn}
\end{equation}
where $(\zeta_\vpsi)^c$ is the dilation current of the $\vpsi$ matter
field. It is worth noting that the term proportional to $\xi$ can be
rewritten in terms of PGT quantities by making the replacements ${\cal
  D}^\ast_a \to {\cal D}_a$, ${{\cal T}^{\ast c}}_{ab} \to {{\cal
    T}^{c}}_{ab}$ and ${\cal T}^\ast_a \to {\cal T}_a$.

Finally, the $\phi$-field equation has the form
\begin{equation}
\pd{L_\vpsi}{\phi}-\nu
({\cal D}^\ast_a+{\cal T}^\ast_a){\cal D}^{\ast a}\phi -4\lambda\phi^3
- a\phi{\cal R} = 0.
\label{diracgrav-phieqn}
\end{equation}
%

\subsection{Einstein gauge}

As discussed in Section~\ref{sec:wgteinstein} in the context of WGT
in general, provided $\phi \neq 0$, the set of field equations can be
simplified considerably by recasting the theory entirely in terms of
scale-invariant field variables defined in (\ref{wgtsivardef}). As we
showed, however, the resulting field equations in the new variables
are {\em identical} in form to those in the original variables after
setting $\phi(x)=\phi_0$, which is the standard Einstein gauge
condition.  Since the latter approach is more familiar and
straightforward, we will adopt it here, but bear in mind the above
equivalence throughout.  It is worth noting that this conclusion
precludes Dirac's original hope of substantiating a time-varying
gravitational parameter $G$ and his large numbers hypothesis.  To
obtain the forms of the field equations
(\ref{diracgrav-heqn})--(\ref{diracgrav-phieqn}) in the Einstein
gauge, one merely sets $\phi=\phi_0$ and makes the replacement
$\left.{\cal D}^\ast_a\phi\right|_{\phi=\phi_0} = -\phi_0{\cal B}_a$
throughout.  It is also convenient to set $a = 1/(\kappa\phi_0^2)$,
where $\kappa$ is Einstein's gravitational constant, which is a
dimensionally consistent identification.

In this case, it is
straightforward to show that the $h$-field equation
(\ref{diracgrav-heqn}) becomes
\begin{widetext}
\begin{equation}
{{\cal R}^{a}}_{b} -\tfrac{1}{2}\delta_b^a{\cal R} - \delta^a_b \Lambda =
\kappa \left[\left.h{(\tau_\vpsi)^a}_b\right|_{\phi=\phi_0}
+\nu ({\cal B}^a{\cal B}_b-\tfrac{1}{2}\delta^a_b {\cal
  B}^c{\cal B}_c)
- \xi ({\cal H}^{ac}{\cal H}_{bc}-\tfrac{1}{4}\delta^a_b {\cal
  H}_{cd}{\cal H}^{cd})\right].
\label{dghafe}
\end{equation}
\end{widetext}
where $\Lambda=\lambda\phi_0^2/a=\kappa\lambda$ has the correct
dimensions of $(\mbox{length})^{-2}$ for a physical cosmological
constant.  Thus, one recovers the $h$-field equation (\ref{hfieldeqn})
from standard EC theory, with the addition of an extra term on the LHS
representing a cosmological constant, and extra terms on the RHS
denoting the energy-momentum of the dilation gauge field $B_\mu$,
which we recognise as that of a massive vector field (or Proca field)
of mass-squared $m^2=\nu\phi_0^2/\xi$.

In the Einstein gauge, the $A$-field equation
(\ref{diracgrav-aeqn}) becomes
\begin{equation}
 {{\cal T}^c}_{ab} +
\delta_a^c {\cal T}_b-\delta_b^c{\cal T}_a
=\left.2\kappa h {(\sigma_\vpsi)_{ab}}^c\right|_{\phi=\phi_0},
\label{dgafe}
\end{equation}
where we have written the result in terms of PGT (unstarred)
quantities for later convenience. This equation is identical to the
form (\ref{afieldeqn}) in EC theory. Thus, on imposing the Einstein
gauge, we find that Dirac gravitational theory reduces simply to EC
theory, but with the addition of a cosmological constant and the
massive vector field $B_\mu$. Since the field equation (\ref{dgafe})
is algebraic, rather than differential, we must again conclude that,
as in EC theory, the torsion is non-propagating; this arises from the
free gravitational Lagrangian (\ref{dgtlgfree}) not containing any
`${\cal R}^2$' terms.

Adopting the Einstein gauge, the equation of motion (\ref{diracgrav-beqn})
for the $B$-field reads
\begin{equation}
\xi\left[\tfrac{1}{2} {{\cal T}^{c}}_{ab} {\cal H}^{ab} - ({\cal
    D}_a+{\cal T}_a){\cal H}^{ac}\right] + \nu \phi_0^2 {\cal B}^c =
\left.h (\zeta_\vpsi)^c\right|_{\phi=\phi_0},
\label{einb}
\end{equation}
where we have again rewritten the result in terms of unstarred PGT quantities.

Finally, the $\phi$-field equation (\ref{diracgrav-phieqn}) becomes
\begin{equation}
{\cal R} + 4 \Lambda = \kappa \left[\phi_0\left.\partial_\phi L_\vpsi\right|_{\phi=\phi_0}
+ \nu\phi_0^2
  ({\cal D}_a+{\cal T}_a + {\cal B}_a){\cal B}^a\right]
\label{einphi}
\end{equation}
where we have once more written the result in terms of unstarred PGT
quantities.

It is worth re-iterating at this point that, when written in terms of
the scale-invariant variables defined in (\ref{wgtsivardef}), the full
field equations (\ref{diracgrav-heqn})--(\ref{diracgrav-phieqn}) have
precisely the same form as those in (\ref{dghafe})--(\ref{einphi})
obtained in the Einstein gauge, after replacing (in the latter set)
the matter field $\vpsi$ and the gravitational gauge fields
${h_a}^\mu$, ${A^{ab}}_\mu$ and $B_\mu$ (and hence the PGT covariant
derivative and the field strength tensors ${{\cal R}^{ab}}_{cd}$,
${{\cal T}^a}_{bc} = {{\cal T}^{\ast a}}_{bc} -\delta^a_c{\cal
  B}_b+\delta^a_b{\cal B}_c$ and ${\cal H}_{ab}$ and their
contractions) with their scale-invariant counterparts (notated with
carets) discussed in Section~\ref{sec:wgteinstein}.

\subsection{Dirac matter field}

Let us now consider the particular case where the matter is described
by a Dirac field.  Following the discussion in
Section~\ref{sec:wgtdirac}, an appropriate locally Weyl covariant
Lagrangian for the Dirac field is given by
\begin{equation}
L_\psi = \tfrac{1}{2}i\bar{\psi}\gamma^a
{\stackrel{\leftrightarrow}{{\cal D}_a}}\psi - \mu\phi\bar{\psi}\psi,
\label{dgtdlag}
\end{equation}
where $\mu$ is a dimensionless parameter. It is worth noting that
$\mu\phi$ has the dimensions of mass in natural units.

The energy-momentum and spin-angular-momentum tensors of the Dirac
field $\psi$ are again given by (\ref{diracemt}) and (\ref{diracsam})
respectively, but with $m$ replaced by $\mu\phi$. Since $B_\mu$ does not
appear in (\ref{dgtdlag}), the dilation current $(\zeta_\psi)^\mu=0$ for the
Dirac field. Moreover, since the $A$-field equation (\ref{dgafe}) in
the Einstein gauge is identical to (\ref{ecfe2a}) in EC theory, from
(\ref{covardiracec}), we again have ${\cal T}_a=0$. Thus, the equation
of motion for the Dirac field is simply
\begin{equation}
i\gamma^a{\cal D}_a\psi-\mu\phi\psi = 0,
\end{equation}
which, in turn, ensures that $L_{\rm M}$ evaluates to zero on
shell. As a result, the trace of the energy-momentum tensor of the
Dirac field is easily found to be $\tau_\psi =
h^{-1}\mu\phi\bar{\psi}{\psi}$. Applying the Einstein gauge condition
$\phi=\phi_0$ to the above results is trivial.

\subsection{Dilation gauge field as a dark matter candidate}

In the Einstein gauge, both the $h$-field and $A$-field equations are
identical to those in standard EC theory, with the addition of a
cosmological constant and the dilation gauge field.  It is therefore
worth exploring the $B$-field equation in more detail to determine
what additional physics beyond EC theory (if any) is contained in
Dirac gravitational theory sourced by Dirac matter. Expanding out the
field strengths on the LHS of (\ref{einb}) in terms of the $B$-field,
swapping the order of some covariant derivatives, and employing the
results (\ref{sigcontract}), (\ref{tfromsigma}) and (\ref{uvanish})
from EC theory with Dirac matter, after a long calculation one obtains
\begin{eqnarray}
{\cal D}_a{\cal D}^{a} {\cal B}^{c} &&-{\cal R}^{ac}{\cal B}_a + m^2{\cal B}^c
\nonumber \\&&+ 2\kappa[(\delta_b^c {\cal D}_a-\kappa h
  {\sigma_{ab}}^c)(h \sigma^{abd} {\cal B}_d)]=0,\phantom{AB}
\label{dgbfemid2}
\end{eqnarray}
where we have defined $m^2 \equiv \nu\phi_0^2/\xi$ and
${\sigma_{ab}}^c$ is the spin-angular-momentum tensor of the Dirac
field (evaluated in the Einstein gauge). Thus we see that the massive
vector field ${\cal B}_a$ has a non-minimal coupling to the
gravitational field and only couples to the Dirac matter field through
its spin-angular-momentum. Moreover, in modelling a realistic matter
fluid composed of spin-1/2 particles, one would expect spin-averaging
effects to result in the spin-angular-momentum that the Dirac field
would otherwise represent being massively diluted.  Thus, we expect
the vector field to be only weakly interacting with ordinary matter
and it hence makes an interesting dark matter candidate. Clearly, in the
case where the (effective) spin-angular-momentum of the matter field
vanishes (and thus so does the torsion), then (\ref{dgbfemid2})
reduces to the equation of motion for a free massive vector field of
mass $m$ non-minimally coupled to the gravitational field.

In closing, it is worth pointing out that, in addition to
automatically containing a candidate for dark matter, Dirac
gravitational theory also provides a natural means for introducing a
cosmological constant term in the gravitational field equation
(\ref{dghafe}). Thus, it would seem worthwhile to explore its
phenomenology further, in particular its cosmological consequences.


\section{Alternative derivation of eWGT covariant derivative}
\label{sec:altddagderiv}
\label{appendix:altddagderiv}

We present here an alternative approach to deriving the form of the
covariant derivative (\ref{ewgtcovdiv}) in eWGT, which is
complementary to that presented in Section~\ref{sec:elwi} and was, in
fact, the original route taken by the authors. In particular, in this
approach the Weyl weight $w$ of the field on which the covariant
derivative acts does not need to be inserted `by hand', but instead
arises naturally.

One begins by noting that in the standard WGT covariant derivative
(\ref{weyldmudef}), namely $D^\ast_\mu = D_\mu + w B_\mu$, the value
of $w$ has to change depending on the Weyl weight of the field on
which it acts and this leads to the commutator of two such derivatives
being similarly non-unique. This might all be deemed somewhat
displeasing.  Thus, as an alternative, one might instead consider
introducing a vector dilation gauge field (called $V_\mu$ to
distinguish it from the field $B_\mu$ in WGT) in a different manner,
by including it in a modified rotational gauge field of the form
(\ref{adaggerdef}), namely
\begin{equation}
{A^{\dagger ab}}_\mu  =  {A^{ab}}_\mu + ({\cal V}^a{b^b}_\mu
- {\cal V}^b{b^a}_\mu),
\label{adaggerdefrepeat}
\end{equation}
where ${\cal V}_a = {h_a}^\mu V_\mu$. In turn, this leads one to
introduce the new derivative operators
\begin{equation}
{\cal D}^\natural_a\vpsi \equiv {h_a}^\mu D^\natural_\mu \vpsi \equiv
 {h_a}^\mu (\partial_\mu +
\tfrac{1}{2}{A^{\dagger bc}}_\mu\Sigma_{bc})\vpsi.
\label{dnatdef}
\end{equation}

As mentioned at the start of Section~\ref{sec:ewgt}, the PGT matter
actions for the massless Dirac field and the electromagnetic field are
already invariant under local dilations with the extended
transformation law (\ref{eweylatrans}) for the $A$-field, in the same
way as the are for the normal transformation law
(\ref{weylatrans2}). Indeed, this provides a strong motivation for
considering the extended form. This does mean, however, that one
cannot determine the transformation law under local dilations of the
gauge field $V_\mu$, and hence either derivative operator in
(\ref{dnatdef}), by the usual technique of replacing the PGT covariant
derivative ${\cal D}_a$ with ${\cal D}^\natural_a$ in either the Dirac
or electromagnetic field action, and demanding invariance under local
dilations.

As an alternative, however, one can fix the transformation properties of
$V_\mu$ by demanding instead that the field strength tensor derived
from the commutator of two derivatives transforms covariantly under
local dilations. One finds that the commutator is given by
\begin{equation}
[D^\natural_\mu, D^\natural_\nu]\vpsi = \tfrac{1}{2}{R^{\dagger
    ab}}_{\mu\nu}\Sigma_{ab}\vpsi,
\end{equation}
where ${R^{\dagger ab}}_{\mu\nu}$ is the eWGT rotational gauge field
strength found previously and given in (\ref{eqn:rdaggerdef}). Then,
demanding that ${R^{\dagger ab}}_{\mu\nu}$ transforms covariantly
under local dilations with the extended transformation law for the
$A$-field, one requires that, as found previously, the dilation gauge
field transforms as
\begin{equation}
V_\mu^\prime = V_\mu + \theta P_\mu,
\label{eqn:vtransagain}
\end{equation}
where $P_\mu \equiv \partial_\mu\rho$. This transformation for $V_\mu$
implies that the modified rotational gauge field ${A^{\dagger
    ab}}_\mu$ in (\ref{adaggerdefrepeat}), and hence the field
strength tensor ${R^{\dagger ab}}_{\mu\nu}$, are both invariant
$(w=0)$ under extended local dilations. One can then define ${{\cal
    R}^{\dagger ab}}_{cd} \equiv {h_c}^\mu {h_d}^\nu {R^{\dagger
    ab}}_{\mu\nu}$ as before, which transforms covariantly with weight
$w=-2$.

By adopting the transformation law (\ref{eqn:vtransagain}) for
$V_\mu$, however, one quickly finds that $D^\natural_\mu\vpsi$ does
{\em not} transform covariantly under extended local dilations, hence
explaining our reluctance thus far to call $D^\natural_\mu$ (or ${\cal
  D}^\natural_a$) a covariant derivative. Nonetheless, using the
covariant field strength tensor ${\cal R}^\dagger_{abcd}$ that we have
constructed, one can still define a new scale-invariant theory of
gravity identical to that presented in Section~\ref{sec:sigtg}, but
with $\xi=0$ in the free gravitational Lagrangian $L_{\rm G}$ and
$\nu=\beta_1=\beta_2=0$ in the matter Lagrangian $L_{\rm M}$, since we
have not (yet) defined the quantities appearing in these terms. By
varying the resulting total Lagrangian with respect to the gauge
fields ${h_a}^\mu$ and ${A^{ab}}_\mu$ (and also $V_\mu$ and $\phi$, if
desired, although this does not yield independent field equations),
one obtains precisely the same equations of motion as those presented
in Section~\ref{sec:sigtgfeqns}, but with
$\xi=\nu=\beta_1=\beta_2=0$. These equations of motion are covariant
under extended local dilations, despite ${\cal D}^\natural_a$ not
being a covariant derivative. Indeed, for any covariant field $\chi$
(which may be either a matter field $\vpsi$ or $\phi$, or the rotational
gauge field strength tensor ${\cal R}^\dagger_{abcd}$), inspection of
the equations of motion allows one to {\em identify} the terms that
need to be added to ${\cal D}^\natural_a\chi$ to assemble the
appropriate {\em covariant} derivative ${\cal D}^\dagger_a\chi$. In
each case, one finds that ${\cal D}^\dagger_a\chi = {h_a}^\mu
D^\dagger_\mu\chi$, where $D^\dagger_\mu$ has the form
(\ref{ewgtcovdiv}), but with the appropriate value of the Weyl weight
$w$ of the field $\chi$ picked out automatically.

Similarly, inspection of the equations of motion allows one to
identify the covariant object ${\cal T}^\dagger_{abc}$. Indeed, once
one has picked out the appropriate form of the eWGT covariant
derivative operator ${\cal D}^\dagger_a$, one can then consider its
commutator (\ref{ewgtdacomm}) and identify ${\cal T}^\dagger_{abc}$
and also ${\cal H}^\dagger_{ab}$ as further gauge field strength
tensors, as introduced in Section~\ref{sec:ewgtgfs}. These can then be
added to the total Lagrangian, as before, to arrive back at the full
scale-invariant gauge theory of gravity discussed in Section~\ref{sec:sigtg},
without requiring $\xi=\nu=\beta_1=\beta_2=0$.


\section{Symbols and notation}
\label{appendix:notation}

Since this paper contains a considerable amount of special
notation, for readers' convenience we summarise in
Table~\ref{tab:symbols} those symbols that occur most frequently.

\begingroup
\squeezetable
\begin{table*}
\caption{\label{tab:symbols} Frequently occurring symbols and notation
  used in this paper. We list only those quantities relating to the
  Minkowski spacetime interpretation of the gauge field theories; for
  a discussion of their geometric interpretation in terms of more
  general spacetimes and the associated notation, see
  Sections~\ref{sec:geowgt} and \ref{sec:ewgtgeo} and
  Appendix~\ref{app:pgt}. See also the discussion of the index
  conversion properties of ${h_a}^{\mu}$ and ${b^a}_{\mu}$ in
  Section~\ref{sec:minkintwgt}, and the associated notation; aside
  from the gauge fields themselves, in this table we list primarily
  the covariant versions of quantities, which have exclusively Latin
  indices. Note that scale-invariant (Weyl weight $w=0$) versions of
  the quantities listed below are denoted with a caret, for example
  ${\widehat{h}_a}^{\,\,\,\mu}$ or $\widehat{{\cal D}}^\ast_a$ or
  ${\widehat{{\cal R}}^{ab}}_{\,\,\,\,\,\,cd}$; see Sections~\ref{sec:wgteinstein}
  and \ref{sec:ewgteinstein}.}
\begin{ruledtabular}
\begin{tabular}{ll}
{\em Gravitational gauge field variables} &   \\
${h_a}^{\mu}$
& translational gauge field ($h$-field) \\
$h$
& determinant of ${h_a}^{\mu}$ \\
${b^a}_{\mu}$
& inverse $h$-field \\
${A^{ab}}_\mu$
& rotational gauge field ($A$-field) \\
$B_\mu$
& WGT dilational gauge field ($B$-field)\\
$V_\mu$
& eWGT dilational gauge field ($V$-field) \\
${A^{\dagger ab}}_\mu \equiv {A^{ab}}_\mu + 2\eta^{c[a}{b^{b]}}_\mu
{h_c}^\nu V_\nu$
& eWGT extended rotational gauge field ($A^\dagger$-field)\\
& \\
{\em Derivative operators} & \\
${\cal D}_a \equiv {h_a}^\mu D_\mu \equiv {h_a}^\mu(\partial_\mu +
\tfrac{1}{2}{A^{ab}}_\mu\Sigma_{ab})$ & PGT (generalised) covariant
derivative\footnotemark[1] \\
$\partial^\ast_\mu \equiv \partial_\mu + wB_\mu$ &
WGT `augmented' partial derivative  \\
${\cal D}^\ast_a \equiv {h_a}^\mu D^\ast_\mu \equiv {h_a}^\mu(\partial^\ast_\mu +
\tfrac{1}{2}{A^{ab}}_\mu\Sigma_{ab})$ & WGT (generalised) covariant
derivative\footnotemark[1] \\
$\partial^\dagger_\mu \equiv \partial_\mu - w(V_\mu +\tfrac{1}{3}T_\mu)$ &
eWGT `augmented' partial derivative  \\
${\cal D}^\dagger_a \equiv {h_a}^\mu D^\dagger_\mu \equiv {h_a}^\mu(\partial^\dagger_\mu +
\tfrac{1}{2}{A^{\dagger\,ab}}_\mu\Sigma_{ab})$ & eWGT (generalised) covariant
derivative\footnotemark[1] \\
${\cal D}^\natural_a \equiv {h_a}^\mu D^\natural_\mu \equiv {h_a}^\mu(\partial_\mu +
\tfrac{1}{2}{A^{\dagger\,ab}}_\mu\Sigma_{ab})$ & eWGT (generalised) semi-covariant
derivative\footnotemark[1] \\
& \\
{\em Gauge field strengths} & \\
${{\cal R}^{ab}}_{cd} = {{\cal R}^{\ast ab}}_{cd} = 2{h_a}^\mu {h_b}^\nu(\partial_{[\mu}
  {A^{ab}}_{\nu]}+{A^a}_{c[\mu}{A^{cb}}_{\nu]})$
& PGT and WGT rotational gauge field strength\footnotemark[1],\footnotemark[2] \\
${{\cal R}^{\dagger\,ab}}_{cd} = 2{h_a}^\mu {h_b}^\nu(\partial_{[\mu}
  {A^{\dagger\,ab}}_{\nu]}+{A^{\dagger\,a}}_{c[\mu}{A^{\dagger\,cb}}_{\nu]})$
& eWGT rotational gauge field strength\footnotemark[1],\footnotemark[2] \\
${{\cal T}^a}_{bc} = 2{h_b}^\mu {h_c}^\nu D_{[\mu} {b^a}_{\nu]}$
& PGT translational gauge field
strength\footnotemark[2]\\
${{\cal T}^{\ast a}}_{bc} = 2{h_b}^\mu {h_c}^\nu D^\ast_{[\mu} {b^a}_{\nu]}$
& WGT translational gauge field strength\footnotemark[2] \\
${{\cal T}^{\dagger a}}_{bc} = 2{h_b}^\mu {h_c}^\nu D^\dagger_{[\mu} {b^a}_{\nu]}$
& eWGT translational gauge field strength\footnotemark[2] \\
${{\cal T}^{\natural a}}_{bc} = 2{h_b}^\mu {h_c}^\nu D^\natural_{[\mu} {b^a}_{\nu]}$
& eWGT translational gauge `semi' field strength\footnotemark[1],\footnotemark[2] \\
${\cal H}_{ab} = 2{h_a}^\mu {h_b}^\nu\partial_{[\mu} B_{\nu]}$
& WGT dilational gauge field strength \\
${\cal H}^\dagger_{ab} = 2{h_a}^\mu {h_b}^\nu\partial_{[\mu}
(V_\nu+\tfrac{1}{3}T_{\nu]})$
& eWGT dilational gauge field strength \\
& \\
{\em Reduced $A$-fields and related quantities} & \\
${c^a}_{bc}\equiv 2{h_b}^\mu {h_c}^\nu\partial_{[\mu} {b^a}_{\nu]}$
& PGT Ricci rotation coefficients \\
${c^{\ast a}}_{bc}\equiv 2{h_b}^\mu {h_c}^\nu\partial^\ast_{[\mu} {b^a}_{\nu]}$
& WGT Ricci rotation coefficients \\
${c^{\dagger a}}_{bc}\equiv 2{h_b}^\mu {h_c}^\nu\partial^\dagger_{[\mu} {b^a}_{\nu]}$
& eWGT Ricci rotation coefficients \\
${^{0}A^{ab}}_\mu \equiv \tfrac{1}{2}{b^c}_\mu({c^{ab}}_c + {{c^b}_c}^a
- {c_c}^{ab})$
& PGT reduced $A$-field (when ${{\cal T}^a}_{bc} \equiv 0$)\\
${^{0}A^{\ast ab}}_\mu \equiv \tfrac{1}{2}{b^c}_\mu({c^{\ast ab}}_c +
{{c^{\ast b}}_c}^a
- {c^\ast_c}^{ab})$
& WGT reduced $A$-field (when ${{\cal T}^{\ast a}}_{bc} \equiv 0$)\\
${^{0}A^{\dagger ab}}_\mu \equiv \tfrac{1}{2}{b^c}_\mu({c^{\dagger ab}}_c +
{{c^{\dagger b}}_c}^a
- {c^\dagger_c}^{ab})$
& eWGT reduced $A^\dagger$-field (when ${{\cal T}^{\dagger a}}_{bc} \equiv 0$)\\
& \\
{\em Currents and related quantities} & \\
$L$
& Lagrangian \\
${\cal L} \equiv h^{-1}L$
& Lagrangian density \\
${t^a}_b \equiv {t^a}_\mu {h_b}^\mu \equiv {h_b}^\mu \delta{\cal
  L}_{\rm G}/\delta {h_a}^\mu$
& gravitational sector energy-momentum tensor\footnotemark[3] \\
${s_{ab}}^c \equiv {s_{ab}}^\mu {b^c}_\mu \equiv {b^c}_\mu \delta{\cal L}_{\rm
  G}/\delta {A^{ab}}_\mu$
& gravitational sector spin-angular-momentum tensor\footnotemark[3] \\
$j^a \equiv j^\mu {b^a}_\mu \equiv {b^a}_\mu\delta {\cal
  L}_{\rm G}/\delta B_\mu$
& gravitational sector dilation current\footnotemark[3] \\
${t^{\dagger a}}_b \equiv {t^{a}}_b + 2({s_{cb}}^a{\cal
  V}^c-{s^{ac}}_c{\cal V}_b)$ & gravitational sector eWGT covariant
energy-momentum tensor\footnotemark[3] \\
$j^{\dagger a} \equiv j^a-2{s^{ab}}_b$ & gravitational sector eWGT covariant
dilation current\footnotemark[3] \\
${\tau^a}_b \equiv {\tau^a}_\mu {h_b}^\mu \equiv {h_b}^\mu \delta{\cal
  L}_{\rm M}/\delta {h_a}^\mu$
& matter sector energy-momentum tensor\footnotemark[3] \\
${\sigma_{ab}}^c \equiv {\sigma_{ab}}^\mu {b^c}_\mu \equiv {b^c}_\mu
\delta{\cal L}_{\rm M}/\delta {A^{ab}}_\mu$
& matter sector spin-angular-momentum tensor\footnotemark[3] \\
$\zeta^a \equiv \zeta^\mu {b^a}_\mu \equiv {b^a}_\mu\delta {\cal
  L}_{\rm M}/\delta B_\mu$
& matter sector dilation current\footnotemark[3] \\
${\tau^{\dagger a}}_b \equiv {\tau^{a}}_b + 2({\sigma_{cb}}^a{\cal
  V}^c-{\sigma^{ac}}_c{\cal V}_b)$
& gravitational sector eWGT covariant energy-momentum tensor\footnotemark[3] \\
$\zeta^{\dagger a} \equiv \zeta^a-2{s^{ab}}_b$ & gravitational sector eWGT covariant
dilation current\footnotemark[3] \\
\end{tabular}
\end{ruledtabular}
\footnotetext[1]{Reduced versions of quantities (which
  result when the corresponding translational gauge field strength
  vanishes) are denoted by a preceding zero superscript, for example
  $^{0}{\cal D}_a$ or $^{0}{{\cal R}^{ab}}_{cd}$, and are obtained by
  replacing ${A^{ab}}_\mu$ or ${A^{\dagger ab}}_\mu$ by the
  corresponding reduced $A$-field or $A^\dagger$-field, respectively; see Sections~\ref{sec:reducedwgt} and \ref{sec:reducedewgt} and Appendix~\ref{app:pgt}.}
\footnotetext[2]{Contractions are defined by
${{\cal R}^a}_b \equiv {{\cal R}^{ac}}_{bc}$,
${\cal R} \equiv {{\cal R}^a}_a$ and ${\cal T}_a = {{\cal
      T}^b}_{ab}$, and similarly for other versions.}
\footnotetext[3]{Analogous quantities can be defined separately for
the individual (covariant) parts of the
gravitational and matter Lagrangian densities.}
\end{table*}
\endgroup





\begin{thebibliography}{99}
\bibitem{footnote1}
Note that, while {${h_a}^\mu$} and {${A^{ab}}_\mu$} are relatively
standard notations for the translational and rotational gravitational
gauge fields, respectively, the choice of notation for their
corresponding field strength tensors varies considerably between
authors; we have adopted a notation that anticipates the geometrical
interpretation of these tensors as the curvature and (minus) torsion,
respectively, of a Riemann--Cartan {($U_4$)} spacetime.
\bibitem{footnote1a} If one varies only the fields and not the
  coordinates, as in the derivation of the Euler--Lagrange equations
  resulting from a given action, for example, then the `total' variation
  $\delta\vpsi$ and the `form' variation $\delta_0\vpsi$ clearly
  coincide, and is usually denoted by the former.
\bibitem{footnote1b}
We adopt the following sign conventions: $(+,-,-,-)$ metric signature,
$[{\cal D}_c,{\cal D}_d]\vpsi \equiv \tfrac{1}{2}{{\cal R}^{ab}}_{cd}
\Sigma_{cd} \vpsi - {{\cal T}^a}_{cd}{\cal D}_a\vpsi$, ${{\cal R}^a}_b
\equiv {{\cal R}^{ac}}_{bc}$ and ${\cal T}_a\equiv {{\cal T}^b}_{ab}$.
\bibitem{footnote3}
To keep the notation simple, however, we will still indicate Lorentz
transformations by {$\Lambda$}, rather than {$SL(2,C)$} elements.
\bibitem{footnote4}
A re-expression of the Dirac Lagrangian is presented in
Appendix~\ref{appendix:dirac}, which lends itself to the modelling of
classical point matter particles, which we will discuss in
Section~\ref{sec:WGTmassivemotion}.
\bibitem{footnote5}
It is worth noting that, under a local dilation alone, if the field
{$\vpsi$} has weight {$w$}, then {${\cal D}^\ast_\mu\vpsi$} also transforms
covariantly with weight {$w$}, whereas {${\cal D}^\ast_a\vpsi$} has weight
{$w-1$}.
\bibitem{footnote8}
It is also worth pointing out that, in terms of their counterparts
$\zero{{\cal A}_{abc}}$ and ${\cal K}_{abc}$ in PGT (see
Appendix~\ref{app:pgt}), one has $\zero{{{\cal A}^{\ast ab}}_{c}} =
\zero{{{\cal A}^{ab}}_{c}} + \delta^a_c{\cal B}^b - \delta^b_c{\cal
  B}^a$ and $ {{\cal K}^{\ast ab}}_{c} = {{\cal K}^{ab}}_c -
\delta^a_c{\cal B}^b + \delta^b_c{\cal B}^a$, so that the
decomposition (\ref{afromht}) or (\ref{astarzerodef}) also holds with
all asterisks removed from the RHS.
\bibitem{footnote9}
It is worth noting that the PGT (and WGT) rotational gauge field
strength tensor (or `curvature') ${\cal R}_{abcd}$ is already
covariant under local dilations, whereas the PGT translational gauge
field strength (or `torsion') ${\cal T}_{abc}$ is not; one must
instead introduce the WGT `torsion' (\ref{tstarfromt}).
\bibitem{footnote10}
One could alternatively define ${t^a}_\mu \equiv h\,\delta {\cal
  L}_{\rm G}/\delta {h_a}^\mu $, ${s_{ab}}^\mu \equiv h\,\delta {\cal
  L}_{\rm G}/\delta {A^{ab}}_\mu$, and $j^\mu \equiv h\,\delta {\cal
  L}_{\rm G}/\delta B_\mu$, and similarly for the matter sector. This
choice corresponds more closely to the standard `metrical' versions of
such tensors, but we will not adopt this convention here.
\bibitem{footnote11}
We note that, slightly unusually, in the last two definitions it is
not the $h$-field, but its inverse, that is used to convert a Greek
index into a Roman one; this should be borne in mind when dealing with
these quantities.
\bibitem{footnote12}
To include gravitational effects in the action
(\ref{spinningppaction}) for a spinning point particle one must, in
addition, make the replacement $\dot{R} \equiv \dot{x}^\mu
\partial_\mu R \to \dot{v}^a {\cal D}^\ast_aR$.
\bibitem{footnote12aa} It is worth noting that this form for the
  equation of motion of a test particle appears naturally in
  scalar-tensor gravity theory\cite{brans61,puetzfeld15}. In the
  standard interpretation, the scalar field in scalar-tensor gravity
  in fact plays the same role of introducing a physical scale, in a
  similar manner to the scalar field in the WGT, although see
  Section~\ref{sec:wgteinstein} for a further discussion of this
  issue.
\bibitem{footnote12a} It is easily shown that $\bar{\psi} \gamma^a
  \,{\stackrel{\leftrightarrow}{{^{0}{\cal D}}^\ast_a}} \psi =
  \bar{\psi} \gamma^a\,{\stackrel{\leftrightarrow}{{^{0}{\cal
          D}}_a}}\psi$ (which does not depend on the dilation gauge
  field $B_\mu$), so the kinetic term in the Dirac action in reduced
  PGT is already invariant under local dilations, in the same way as
  its counterpart in `full' PGT, as discussed in
  Section~\ref{sec:wgtdirac}.
\bibitem{footnote13}
Since the $\vect{e}_\mu(x)$ are derived from a coordinate system, they
form a holonomic set of basis vectors, whereas the set
$\hvect{e}_a(x)$ are, in general, non-holonomic.
\bibitem{footnote14}
We need not consider the local Lorentz and GCT parts of the
transformation, since the behaviour of the gravitational gauge fields
${h_a}^\mu$ and ${A^{ab}}_\mu$ under these transformations will remain
the same as in WGT.
\bibitem{footnote14a} The arbitrary parameter $\theta$ could, in
  principle, be promoted to a field $\theta(x)$. Provided it has Weyl
  weight $w=0$, one would obtain the same $theta$-independent form for
  the eWGT covariant derivative $D^\dagger_\mu$ and the same gauge theory.
\bibitem{footnote15}
The dagger on the derivative operator is simply to distinguish the
eWGT covariant derivative from the PGT and WGT covariant derivatives
$D_\mu$ and $D^\ast_\mu$, respectively, and should not be confused
with the operation of Hermitian conjugation. In particular, we note
that the eWGT covariant derivative $D^\dagger_\mu \vpsi = [D_\mu +
  \tfrac{1}{2} ({\cal V}^a{b^b}_\mu- {\cal
    V}^b{b^a}_\mu)\Sigma_{ab}-wV_\mu -\tfrac{1}{3}wT_\mu]\vpsi$
represents a far more substantial modification to $D_\mu\vpsi$ than
the WGT covariant derivative $D_\mu^\ast\vpsi = (D_\mu + w
B_\mu)\vpsi$.
\bibitem{footnote16}
In terms of their counterparts $\zero{{\cal A}_{abc}}$ and ${\cal
  K}_{abc}$ in PGT (see Appendix~\ref{app:pgt}), one has $\zero{{{\cal
      A}^{\dagger ab}}_{c}} = \zero{{{\cal A}^{ab}}_{c}} -
\delta^a_c({\cal V}^b + \tfrac{1}{3}{\cal T}^b) + \delta^b_c ({\cal
  V}^a + \tfrac{1}{3}{\cal T}^a)$ and ${{\cal K}^{\dagger ab}}_{c} =
      {{\cal K}^{ab}}_c + \tfrac{1}{3}\delta^a_c{\cal T}^b -
      \tfrac{1}{3}\delta^b_c{\cal T}^a$, so confirming that the
      decomposition (\ref{adaggerzerodef}) also holds with all daggers
      removed, as discussed in Section~\ref{sec:wgtcovdalt}.
\bibitem{footnote17}
An alternative and more `automated' route for constructing covariant
forms of $t_{ab}$ and $\tau_{ab}$ is given in
Section~\ref{sec:ewgtaltvarp}.
\bibitem{footnote17a}
It is easily shown that $\bar{\psi} \gamma^a
  \,{\stackrel{\leftrightarrow}{{^{0}{\cal D}}^\dagger_a}} \psi =
  \bar{\psi} \gamma^a\,{\stackrel{\leftrightarrow}{{^{0}{\cal
          D}}_a}}\psi$ (which does not depend on the dilation gauge
  field $V_\mu$), so the kinetic term in the Dirac action in reduced
  PGT is already invariant under extended local dilations, in the same way as
  its counterpart in `full' PGT, as discussed in
  Section~\ref{sec:ewgtdirac}.
\bibitem{footnote18}
It is worth noting that the semi-metricity condition
(\ref{ewgtsemimet}) and that given in (\ref{wgtsemimet}) for WGT can
both be written as $\nabla_\sigma g_{\mu\nu} = -\tfrac{1}{2}(
{\Gamma^\lambda}_{\lambda\sigma}
-\christoffel{\lambda}{\lambda}{\sigma})g_{\mu\nu}$, where
$\christoffel{\lambda}{\mu}{\nu}$ is the metric connection; this also
holds in PGT.
\bibitem{footnote19}
One can straightforwardly construct parity-odd terms for inclusion in
the free gravitational Lagrangian in an analogous manner to WGT, but
we do not consider such terms here.
\bibitem{footnote20}
The expression (\ref{ewgtlr2}) can also be written in a form for which
its symmetries are more manifest, namely
%
$
L_{{\cal R}^2}  =  \alpha_1 {\cal R}^{\dagger 2}
+ \alpha'_2 {\cal R}^\dagger_{(ab)}{\cal R}^{\dagger (ab)}
+ \alpha'_3 {\cal R}^\dagger_{[ab]}{\cal  R}^{\dagger [ab]}
+ \alpha'_4 {\cal R}^\dagger_{a(bc)d}{\cal R}^{\dagger a(bc)d}
+ \alpha'_5 {\cal R}^\dagger_{a[bc]d}{\cal R}^{\dagger a[bc]d}
+ \alpha_6 {\cal R}^\dagger_{abcd}{\cal R}^{\dagger cdab},
$
%
where the new coefficients are $\alpha'_2 = \alpha_2+\alpha_3$,
$\alpha'_3 = \alpha_2-\alpha_3$, $\alpha'_4 = \alpha_4+\alpha_5$ and
$\alpha'_5 = \alpha_4-\alpha_5$.  In this case, the eWGT version of
the Gauss--Bonnet identity allows one to set $\alpha_1$ or $\alpha_6$
to zero as before, or set $\alpha'_2=\alpha'_3$.
\bibitem{footnote21}
The expression (\ref{ewgtlt2}) can also be written in the more
manifestly symmetric form $L_{{\cal T}^2} = \beta'_1{\cal
  T}^\dagger_{(ab)c}{\cal T}^{\dagger (ab)c} + \beta'_2 {\cal
  T}^\dagger_{[ab]c}{\cal T}^{\dagger [ab]c}$, where
$\beta'_1=\beta_1+\beta_2$ and $\beta'_2=\beta_1-\beta_2$.
\bibitem{footnote23}
One may also straightforwardly include the electromagnetic field in
${\cal L}_{\rm M}$, coupled both to the Dirac field and to gravity, in
an analogous manner to that discussed at the end of
Section~\ref{sec:wgtem}, but will we not consider this extension here.
\bibitem{footnote24}
Once again, one may also straightforwardly include the electromagnetic
field in ${\cal L}_{\rm M}$, coupled both to the Dirac field and to
gravity, but will we not consider this extension here.
\bibitem{footnote25}
Moreover, after a long but straightforward calculation, one may show
that $\czero{{{\cal R}^{ab}}_{cd}}$ is itself related to the
corresponding WGT quantity $\czero{{{\cal R}^{\ast ab}}_{cd}}$ by
$\czero{{{\cal R}^{\ast ab}}_{cd}} = \czero{{{\cal R}^{ab}}_{cd}} -
2\delta^{[b}_d({^0}{\cal D}_c - {\cal B}_c){\cal B}^{a]} +
2\delta^{[b}_c({^0}{\cal D}_d - {\cal B}_d){\cal B}^{a]} -2 {\cal B}^e
{\cal B}_e \delta^{[a}_c \delta^{b]}_d$, from which one finds that
their contractions are related by ${^0}{{\cal R}^{\ast a}}_c =
{^0}{{\cal R}^{a}}_c - 2 ({^0}{\cal D}_c - {\cal B}_c){\cal B}^a -
({^0}{\cal D}_e + 2{\cal B}_e){\cal B}^e\delta^a_c$ and ${^0}{\cal
  R}^\ast = {^0}{\cal R}-6 ({^0}{\cal D}_e+{\cal B}_e){\cal B}^e$.
Similarly, $\czero{{{\cal R}^{ab}}_{cd}}$ and its contractions are
related to the corresponding eWGT quantities by the same
relationships, but with the replacements $\czero{{{\cal R}^{\ast
      ab}}_{cd}} \to \czero{{{\cal R}^{\dagger ab}}_{cd}}$ and ${\cal
  B}_a \to -({\cal V}_a +\tfrac{1}{3}{\cal T}_a)$.
\bibitem{footnote26}
More appropriately, the theory is also known as
Einstein--Cartan--Kibble--Sciama (ECKS) theory, but we shall adopt the
more usual, shorter naming convention (with apologies to Kibble and
Sciama).
\bibitem{footnote28}
It is probably fair to say that Dirac would not have considered
$L_{\rm G}$ in (\ref{dgtlgfree}) to be his entire gravitational
Lagrangian, and would have also included the term proportional to
$\phi^2{\cal R}$ that appears in (\ref{diracgravlagm}). Since this
term contains the scalar field $\phi$, however, we have instead
included it in the matter Lagrangian.
\bibitem{footnote29}
In fact, Dirac's original theory has vanishing torsion and so he
proposed a matter Lagrangian of the form (\ref{diracgravlagm}), but
with the WGT covariant derivative and Ricci scalar replaced by their
`reduced' versions ${^0}{\cal D}^\ast_a$ and ${^0}{\cal R}^\ast$,
respectively, as described in Section~\ref{sec:wgtcovdalt}. We extend
his approach here to include torsion.
\end{thebibliography}
\end{document}